\documentclass[a4paper,11pt,english]{book}
\usepackage{babel}
\usepackage{fancyhdr}
\usepackage{graphics}
\usepackage{float}
\usepackage{graphicx}
\usepackage{subfigure}
\usepackage{amssymb}
\usepackage{amsmath}
\usepackage{delarray}
\usepackage{wrapfig}

\usepackage{lscape}

\renewcommand{\chaptermark}[1]%
{\markboth{#1}{}}
\renewcommand{\sectionmark}[1]%
{\markright{ \thesection\  #1}}
\usepackage{pstcol}
\makeatletter
\def\LigneVerticale{\vrule height 5cm depth 2cm\hspace{0.1cm}\relax}
\def\LignesVerticales{%
  \let\LV\LigneVerticale\LV\LV\LV\LV\LV\LV\LV\LV\LV\LV}
\def\GrosCarreAvecUnChiffre#1{%
  \rlap{\vrule height 0.8cm width 1cm depth 0.2cm}%
  \rlap{\hbox to 1cm{\hss\mbox{\white #1}\hss}}%
  \vrule height 0pt width 1cm depth 0pt}

\def\@makechapterhead#1{\hbox{%
    \huge 
    \LignesVerticales
    \hspace{-0.5cm}%
    \GrosCarreAvecUnChiffre{\thechapter}
    \hspace{0.2cm}\hbox{#1}%
}\par\vskip 2cm}
\def\@makeschapterhead#1{\hbox{%
    \huge 
    \LignesVerticales
    \hbox{#1}%
}\par\vskip 2cm}

\newcommand{\etal}{{\it et al.}}
\newcommand{\ce}[1]{Eq.~(\ref{#1})}

\newcommand{\cf}[1]{Fig.~\ref{#1}}

\newcommand{\alltrace}{\ensuremath{\mathbb{T}\mathrm{r}}}
\def\thru#1{\mathrel{\mathop{#1\!\!\!/}}}
\newcommand{\beq}{\begin{eqnarray}}
\newcommand{\eeq}{\end{eqnarray}}

\newcommand{\captionfonts}{\footnotesize}
\makeatletter  
\long\def\@makecaption#1#2{%
  \vskip\abovecaptionskip
  \sbox\@tempboxa{{\captionfonts #1: #2}}%
  \ifdim \wd\@tempboxa >\hsize
    {\captionfonts #1: #2\par}
  \else
    \hbox to\hsize{\hfil\box\@tempboxa\hfil}%
  \fi
  \vskip\belowcaptionskip}

\usepackage[pdftex]{hyperref}
\begin{document}
{\newpage
\pagestyle{empty}


\begin{figure}[H]
\centering \includegraphics [height=4cm]{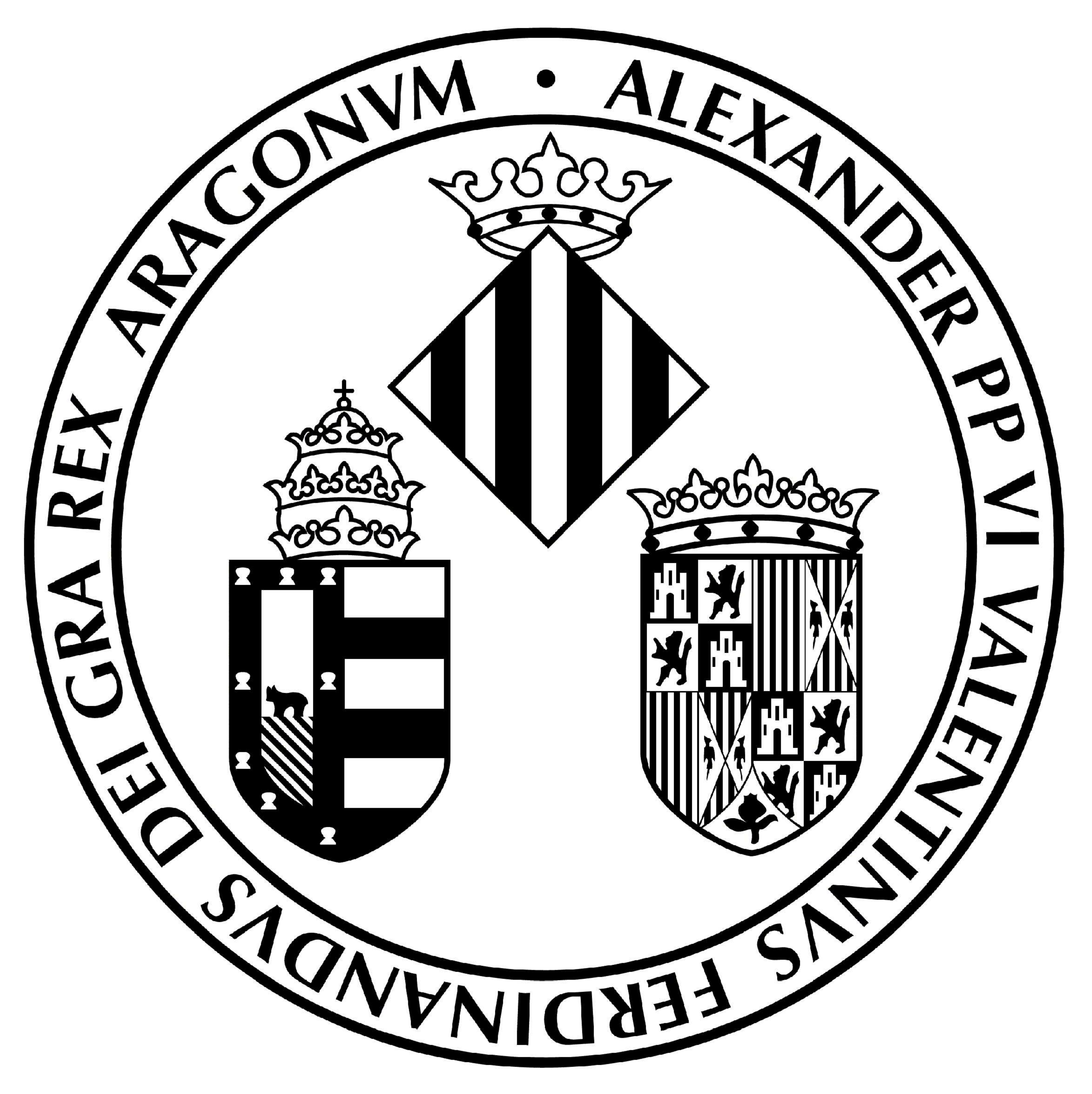}
\end{figure}
 
\vspace*{5cm}

{

\begin{center}
{\bf \Large Generalized Parton Distributions of Pions. }\\
{\bf \Large Spin Structure of Hadrons.}
\end{center}

\vspace*{5cm}

\begin{center}

PhD Thesis\\
\vspace{1cm}
 Author: Aurore Courtoy \\
Advisor: Santiago Noguera\\
\vspace{1cm}
Departamento de F\'isica Te\'orica\\
 Universidad de Valencia
 \\
 \vspace{1cm}
13th of October 2009
\end{center}
}
\newpage

}

{\newpage
\pagestyle{empty}

\vspace*{15cm}

\begin{flushright}
\begin{minipage}{10cm}
\begin{flushright}
\it Tout homme est fou, mais qu'est-ce qu'une destin\'ee humaine
\it  sinon une vie d'effort pour unir ce fou \`a l'Univers.

\vspace{1cm}

\it  Malraux
\end{flushright}
\end{minipage}
\end{flushright}

\newpage
}

{\newpage
\pagestyle{empty}

\vspace*{15cm}

\begin{flushright}
\begin{minipage}{13cm}

I would like to thank my advisor Santiago Noguera as well as my (senior) collaborators Vicente Vento and Sergio Scopetta, for those 4 years working together in Valencia.
I would also like to thank the other PhD students who have shared my office, my lunches, my coffee breaks, my moody days (and whose office, lunches, coffee breaks and moody days I have shared too). Many thanks to the secretaries for their kindness and their helpfulness.
\\

This work has been supported by  the Ministerio de Educaci\'on y Ciencias  (Spain) under the  grant AP2005-5331.

\end{minipage}
\end{flushright}

\newpage
}

\setcounter{page}{0}
\tableofcontents

\setcounter{page}{0}

\chapter*{Introduction}
\addcontentsline{toc}{chapter}{Introduction}

One of the main open question in physics is the understanding of the hadrons as constituted of quarks and gluons.
In effect, the idea that quarks are the building blocks of hadrons goes back to 1964, when Gell-Mann and  Zweig published their famous papers ~\cite{GellMann:1964nj, Zweig:1981pd}.
In 1973, the formulation  of  Quantum ChromoDynamics (QCD), as the theory of the strong interaction between quarks, was proposed by Fritzsch, Gell-Mann and Leutwyler~\cite{Fritzsch:1973pi}. But despite the fact that QCD has been adopted 35 years ago and even though this theory has been verified at high energy,  
we have not been able to  reveal   the detailed structure of hadrons. 
One might wonder why we have not  achieved this aim yet. The answer is that  the  peculiar features of  the strong interactions renders such a task very intricate.

At high energy, the perturbative treatment of QCD, which is justified by  asymptotic freedom~\cite{Gross:1973id, Politzer:1973fx}, 
 allows for
the explanation of the hadronic phenomena,  whose basic ingredients   are the {\it Parton Distributions}. These distributions can be experimentally determined for a relatively high value of the momentum transfer $Q^2$. 
Therefore, we are able to evolve those quantities and give predictions for different values of $Q^2$.

At low energy (typically below $\sim 1$ GeV), there is no justification for such a  perturbative treatment of  QCD and, therefore, we do not longer  have a description of  the relevant low energy observables from QCD. In this regime, hadronic models and effective theories come into play. In these schemes, the parameters are fixed  by phenomenology.
\\

There exists no clear and definite frontier between these two energy regimes.  
Neither do exist definite connections between the relevant degrees of freedom: even though they are the elementary degrees of freedom of QCD, free quarks and gluons have never been observed. This fact has revealed that  quarks and gluons were confined in hadrons. On the other hand, another important feature of the strong interactions is related to the chiral symmetry. In the limit of massless quarks,  chiral symmetry, which is  a property of the QCD lagrangian, is realized in the Goldstone mode, i.e.  is spontaneously broken. 
The fundamental r\^ole played by this symmetry  and its realization has been confirmed by the small mass of the pion as well as properties of the strong processes at intermediate energies like the Partial Conservation of the Axial Current or the Soft Pion Theorems. 
Therefore, the structure of a hadron arises from  an interplay between these two important properties of QCD, namely,  confinement and the realization of the chiral symmetry. The energy regime from which we can gather information on the hadron structure resides in the frontier between the two regimes described above, where both confinement and chiral symmetry  are supposed to be into play.
\\

A way of connecting the perturbative and non-perturbative  {\it worlds} is the study of Parton Distributions using models. The 
scheme that has been proposed runs as follows: 
we  build  models, whose characteristics somehow mimic the QCD's properties that are relevant at the energy range we are considering ($\sim 1$ GeV); we then evaluate the Parton Distributions in these models; finally, we evolve these  distributions
to the  scale of the experiment in order to compare with the data.
\\

The reactions allowing to access the different types of Parton Distributions have been  receiving great attention from the hadronic physics community. 
These reactions are such that they enable us to
 look with a good resolution inside the hadron and allow us to resolve the very short distances, i.e. small configurations of quarks and gluons. Since at short distances the interactions between quarks and gluons become weak, this part of the process is described through {\it perturbative} QCD.
A resolution of such short distances is  obtained  with the help of non-strongly interacting probes.  Such a probe, typically  a photon, is provided by  hard reactions, like Deep Inelastic Scattering, Semi-Inclusive Deep Inelastic Scattering, Deeply Virtual Compton Scattering,~\ldots. 
In that scheme, the Parton Distributions  reflect  how the target reacts to the probe, or how  the quarks and gluons are distributed inside the target. The insight into the structure of hadrons is reached at that stage: 
 the large virtuality of the photon, $Q^{2}$, involved in such processes  allows for the factorization of the hard ({\it perturbative}) and soft ({\it non-perturbative}) contributions in their amplitudes.  
 It is this {\it non-perturbative} part of the process that will interest us all along this thesis.
\\

The first steps towards the understanding of the structure of hadrons were taken  by studying the total cross sections of fully inclusive processes or longitudinal asymmetries,  which receive
simple interpretations in  parton models~\cite{Jaffe:1974nj}. 

 From a theoretical point of view, Parton Distributions are related to diagonal matrix elements of  a bilocal operator on the light-cone from the initial hadron state to the same final hadron state, i.e. the same particle with the same momentum. 
They have  simple probabilistic interpretations. For collinear quarks, the Parton Distributions  $q(x)$ receive the name of number densities because they reflect the probability density of finding a quark with a fraction $x$ of the longitudinal momentum of the parent hadron, regardless of its spin orientation. Also, the definition of other appealing quantities arise, such as the helicity distribution (i.e. longitudinal polarization), as well as the more mysterious transversity distribution. Now,  the distribution of transverse polarization is not observable in fully inclusive processes, thus, it requires the analyses from other hard processes, e.g. semi-inclusive processes. 

A first proliferation of distributions was the consequence of considering  non-collinear quarks. It basically means that  the intrinsic motion of quark could hardly be  understood through the three previous densities. Nowadays, Transverse Momentum, $\vec{k}_T$, Dependent parton distributions  are being thoroughly examined in the context of Spin Physics.  Since they embody an additional internal degree of freedom, they can be involved in non-trivial correlations with spin  that might lead to a deeper understanding of the spin structure of  hadrons. 
\\
\\

In these days, the experimental facilities allow to access not only inclusive processes like Deep Inelastic Scattering but also exclusive deep  processes like the Deeply Virtual Compton Scattering (DVCS) $\gamma^{\ast}p\to \gamma p$ or Hard Exclusive Meson ($M$) Production (HEMP) $\gamma^{\ast}p\to M p$. The first project of a facility devoted to the study of exclusive processes in the scaling regime was proposed as the ELFE project~\cite{Anselmino:2001qu, Laget:2001ds}, which unfortunately did not succeed. 
In the last few years, experimental data related to the structure of hadrons, like asymmetries and DVCS or HEMP cross sections,  have been obtained from different collaborations. We mention  the HERMES, e.g. \cite{Airapetian:2001yk}, as well as  the H1, e.g. \cite{Aktas:2005ty}, and the ZEUS, e.g. \cite{Chekanov:2003ya}, collaborations at DESY;  the CLAS collaboration, e.g. \cite{Stepanyan:2001sm}, at the Hall B of JLab.

The theoretical description of these processes requires the generalization of the Parton Distributions to  {\it Generalized Parton Distributions}~\cite{Ji:1996ek, Mueller:1998fv, Radyushkin:1996nd}. These distributions  are matrix elements of non-diagonal distributions, namely, those which have a final state with different momentum than the initial one (for reviews, see, e.g., Refs.~\cite{Belitsky:2005qn, Boffi:2007yc, Diehl:2003ny, Goeke:2001tz, Ji:1998pc}). The Generalized Parton Distributions describe non-forward matrix elements of
bilocal operators on the light-cone. They measure the response of the internal structure of the hadrons to the probes in the above-mentioned processes.
These {\it new} distributions have been shown to be interesting theoretical tools for the study of hadrons;  they connect, through sum rules, to the hadronic Form Factors. Also their forward limit connects to the usual Parton Distributions.  The properties of such matrix elements that ensue from their analysis from {\it first principles} are presented in Chapter~\ref{sec:df}. Those properties are used as constraints on the modeling procedure that has been described above for the Parton Distributions.

Already the prediction of some phenomenological models have been verified to be in agreement with the data mentioned above. This was taken as an encouraging sign that the framework of exclusive deep processes was adapted to our purposes. 
However, a detailed mapping as well as the extraction of GPDs from the experiments will require more data in order to constrain the models. In a near future, we expect new data on DVCS from the CLAS and  HERMES  collaborations (see references of proposals in~\cite{Guidal:2008zza}). The improvement up to $12$ GeV of the CEBAF accelerator at JLab foreseen for 2012~\cite{prop-jlab} is already involving much of the hadronic physics community in proposals for the access of the involved distributions as a way to explore the  structure of the hadrons. So is the COMPASS experiment at CERN.
\\
\\

Also, collisions of a real photon and a highly virtual photon can be an useful tool for studying fundamental aspects of QCD. Inside this class of processes, the exclusive meson pair production in $\gamma^{\ast}\gamma$ scattering has been
analyzed in Ref.~\cite{Pire:2004ie} introducing of a new kind of distribution amplitudes, called {\it Transition Distribution Amplitude} (TDA).
They represent a  generalization of parton distributions to the case where the initial and
final states correspond to different particles. For obvious practical reasons, the first transitions that have been studied are the  (mesonic) pion-photon TDA, alternatively governing  processes like $\pi^{+}\pi^{-}\rightarrow
\gamma^{\ast}\gamma$ or $\gamma^{\ast}\pi^{+}\rightarrow\gamma\pi^{+}$ in the
kinematical regime where the virtual photon is highly virtual but with small
momentum transfer. Nevertheless, it has also been proposed in Ref.~\cite{Pire:2004ie} baryonic TDAs
that have been analyzed for the $\pi\to N$ transition in Ref.~\cite{Lansberg:2007ec}. 
\\
\\

This thesis aims to contribute to the understanding of the hadron structure by gathering  the present knowledge on  the pion-photon {\it Transition Distribution Amplitudes}.

For that purpose, we present the results for  the pion-photon TDAs in a field theoretic scheme treating the pion as a bound state in a fully covariant manner using the Bethe-Salpeter equation, with the pion structure described by the Nambu~-~Jona
Lasinio (NJL) model~\cite{Courtoy:2007vy, Courtoy:2008af, Courtoy:2008ij}. Actually, the NJL model is the most realistic model for the pion, based on a local quantum field theory built with quarks. It respects the realizations of chiral symmetry and gives a good description of the low energy physics of the pion \cite{Klevansky:1992qe}.
The NJL model is a non-renormalizable field theory and therefore a cut-off
procedure has to be defined. The Pauli-Villars regularization procedure has been chosen because of the symmetries it respects. Also, the NJL model together with its regularization procedure, is regarded as an effective theory of QCD. 
\\
In terms of the physical processes, the NJL model is used to describe the soft (non perturbative) part of
the deep processes, while for the hard part conventional perturbative QCD must be used.
\\

For the sake of clarity, we will develop this topic step by step, following the  theoretical improvements described above. Namely, 
in order to illustrate the framework of Parton Distributions as well as the formalism developed to calculate such quantities, the detailed calculation of the pion {\it Distribution Amplitudes} in the NJL model is presented in Chapter~\ref{sec:pionda}. In the same Chapter, the link between pion Distribution Amplitudes and Parton Distributions are displayed through the Soft Pion Theorem.
We also show the consistency of the model we use by emphasizing that, once QCD evolution is
taken into account, a good agreement is reached between the calculated  Parton Distributions and
the experimental one. 
\\

 According to the same scheme  we extend our considerations to the pion {\it Generalized Parton Distributions} in Chapter~\ref{sec:gpd}. The results for the calculation of the latter in the formalism of Chapter~\ref{sec:pionda} are shown~\cite{Theussl:2002xp}. One could wonder whether the results, even if constrained by the {\it first principles} properties, are consistent with other formalisms. With that question in mind, we display the parameterization of the GPDs through the so-called {\it Double Distributions}.
The last step of the analysis is the application of the  QCD evolution equations. The effect of  evolution on the
pion  GPD calculated in the NJL model has been  studied in
\cite{Broniowski:2007si}, whose results we will be discussed in detail.
\\

These previous developments enable a better prehension of the Transition Distribution Amplitudes as non-perturbative objects matching with some well-known constraints. A peculiarity of these transitions consists in the connection of their structure with the one given by the pion radiative decay $\pi^{+}\rightarrow\gamma e^{+}\nu$\ process~\cite{ Bryman:1982et, Moreno:1977kx}, which stresses the importance of the r\^ole played by the Partial Conservation of the Axial Current.
A detailed analysis of the matrix element of bilocal current defining the pion-photon TDAs is presented in Chapter~\ref{sec:tda}.  Their properties are displayed and studied. The results of the calculation in the NJL model are subsequently given.

Different studies of the axial and vector pion-photon TDA have been performed using
different quark models ~\cite{ Broniowski:2007fs,Courtoy:2007vy, Kotko:2008gy, Tiburzi:2005nj}. A qualitative comparison  as well as the parameterization through Double Distributions are also presented, which allow us to conclude that there is a qualitative agreement between all the different approaches.
\\

The direct application of the results obtained for the TDAs to phenomenology is obviously the estimation of the cross section for the processes they govern.   Chapter~\ref{sec:cross} is devoted to the study of the cross sections for the exclusive meson  production in $\gamma^{\ast}\gamma$ scattering~\cite{Courtoy:2008nf, Lansberg:2006fv}, where, in  Ref.~\cite{Courtoy:2008nf},  the evolution of the  TDA is taken into account.
\\
\\

Up to this point, we have proposed to consider, in this thesis, the  structure of the pion.  
Within the study of the hadron structure that we have carried out, we wanted to extend our analyses to the spin structure of the proton. 
This is why we briefly report some recent improvements in Spin Physics. 
From the phenomenology, it is known that
Semi-Inclusive DIS off a transversely polarized target shows azimuthal asymmetries, e.g. \cite{hermes, compass}, that could be understood through non-trivial correlations between  intrinsic motion of the quarks and transverse spin. A hint on the spin structure of the proton could be given through the understanding of the modulation of the number density of unpolarized quarks in a polarized proton. This modulation  is due to the correlation between the intrinsic transverse momentum of the quarks and the transverse component of the proton spin. The results for this so-called Sivers function, calculated in two different models~\cite{Courtoy:2008vi, Courtoy:2008dn}, are exposed in Chapter~\ref{sec:sivers}. 
\\
\\

In the final Chapter, we conclude by gathering all the questions and all the ideas that emerged in this work, which represent good starting points for future research.

\chapter{Hadron Structure from Distribution Functions}
\label{sec:df}
\pagestyle{fancy}

An intuitive approach of parton distribution can be provided in  the context of the parton model. Feynman \cite{Feynman:1969ej} introduced the parton distributions as phenomenological quantities describing the properties of the nucleon in high-energy scattering.
Actually a nucleon moving at the speed of light can be seen as made of noninteracting massless and collinear partons. In such a scheme the parton distributions are
the density of partons as a function of the fraction of the nucleon longitudinal momentum they carry.

Nevertheless the nucleon does not exactly consits of free partons. Knowledge of the internal structure of the nucleon, or the pion, is required
in order to properly study the parton distributions as a nonperturbative object.

From the phenomenological point of view, the study of Deep Inelastic Scattering processes, where individual partons are resolved, revealed itself to
 be an essential tool in this context.
The unpolarized parton distributions have been extracted with good precision from various high-energy scattering data \cite{Gluck:1994uf,Lai:1996mg}.
Those distributions have provided important clues about the hadron structure but yet important pieces of information are missed out in these quantities.
\\

\begin{wrapfigure}[10]{l}{70mm}
\centering
 \includegraphics [height=3.5cm]{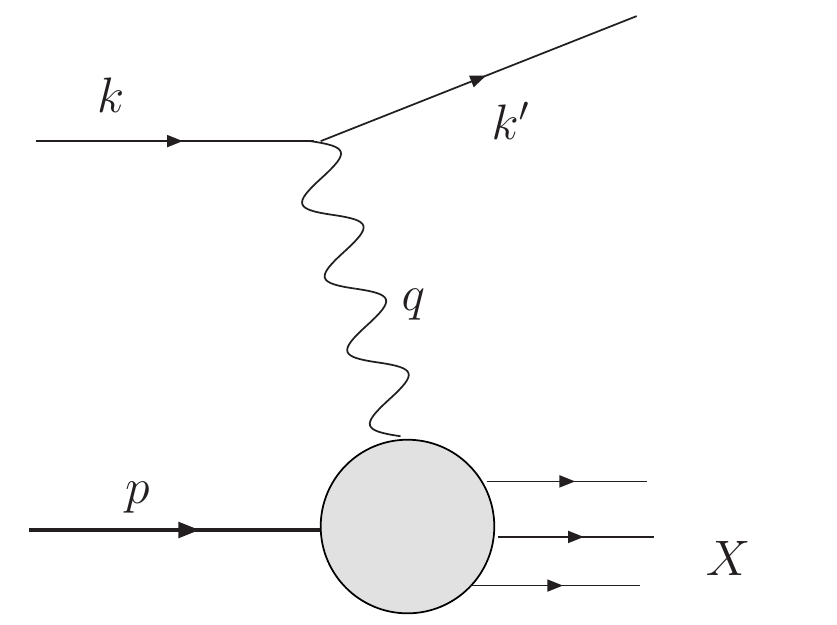}
\caption{\small{DIS: Kinematics of lepton-hadron scattering.}}
\label{DIS}
\end{wrapfigure}
In order to study the properties of hadrons we will examine the properties of the hadronic matrix element defining them. The adapted  framework for these purposes  is any Deep Inelastic Process, characterized by the Bjorken limit. The cross section for those processes~\ref{DIS} can be  separated into a leptonic $ l_{\mu\nu}$ and a hadronic tensors $W^{\mu\nu}$.
While the leptonic tensor is known completely, $W^{\mu\nu}$, which describes the internal structure of the nucleon, depends on non perturbative strong interaction dynamics. It is expressed as the commutator of two currents $J^{\mu}$ defined at 2 different space-time points,
\beq
W^{\mu\nu}&=&\frac{1}{4\pi}\, \int d^4 z\,e^{iq\cdot z} \langle p, S\vert [ J^{\mu}(z),J^{\nu}(0)]\vert p,S\rangle\quad,
\label{wmunu}
\eeq
and can be related to the imaginary part of the forward Virtual Compton Scattering through the Optical Theorem. 

\section{On the Light Cone }

Without loss of generality, any 4-vector could be expressed as, see Appendix.~\ref{sec:convention},
\begin{eqnarray}
a^{\mu}&=&(a^+,\vec{a}^{\perp},a^-)
\quad,\label{4-vec-lc}
\end{eqnarray}
with $\vec{a}^{\perp}=(a^1,a^2)$ and with  the components  $a^{\pm}= \frac{1}{\sqrt{2}}(a^0\pm a^3)$.
\\
\\
We define two vectors whose norm is zero
\begin{eqnarray}
\bar{p}^{\mu}&=&\frac{p^+}{\sqrt{2}}(1,0,0,1)=p^+(1,\vec{0}^{\perp},0)\quad,\nonumber\\
n^{\mu}&=&\frac{1}{p^+\,\sqrt{2}}(1,0,0,-1)=\frac{1}{p^+}(0,\vec{0}^{\perp},1) \quad.
\label{ll-vec}
\end{eqnarray}
Those vectors follow the properties~(\ref{light-cone-vector}).
\\
The Sudakov decomposition for a vector
\begin{eqnarray}
a^{\mu}&=& \alpha \bar{p}^{\mu}+a^{\perp \mu}+\beta n^{\mu}\nonumber\quad,\\
&=& (a.n)\bar{p}^{\mu}+a^{\perp \mu}+(a.\bar{p})n^{\mu}\quad,
\nonumber
\end{eqnarray}
with $a^{\perp \mu}=(0,\vec{a}^{\perp \mu},0)$, can be adopted for the vectors $q, z$, with $q$ constrained by the DIS kinematics.
\\
In terms of the Sudakov vectors, the nucleon momentum $p$ and the momentum of the virtual photon $q$ can be written as
\beq
p^{\mu}&=& \bar p^{\mu}+\frac{M^2}{2}\, n^{\mu}\quad,\nonumber\\
q^{\mu}&\stackrel{Bj}\longrightarrow& (p\cdot q)\, n^{\mu}-x\, \bar p^{\mu}\quad,
\nonumber
\eeq
with $x\equiv Q^2/2 p\cdot q$ the Bjorken variable.
\\

One could inquire about the restriction imposed by the kinematical regime of DIS on the domain of validity of the expression of the hadronic tensor~(\ref{wmunu}). The argument of the exponential is crucial.
For a generic 4-vector $z^{\mu}=\alpha_z \bar p^{\mu}+ z^{\perp\mu}+\beta_z n^{\mu} $, the product $q\cdot z$ reads, in the Bjorken limit, 
\beq
q\cdot z &\stackrel{Bj}\longrightarrow&  (p\cdot q) \, \alpha_z-x\, \beta_z\quad,
\nonumber
\eeq
  When $(p\cdot q)$ goes to infinity, the oscillations produced by the exponential term in~(\ref{wmunu}) imply that 
 $ \alpha_z\lesssim 1/(p\cdot q)$.
 Also, causality implies that the commutator $[ J^{\mu}(z),J^{\nu}(0)]$ vanishes unless $z^2> 0$, therefore $\vec{z}^{\perp 2}\lesssim 2\,\alpha_z \beta_z$. These two restrictions together imply that the hadronic tensor~(\ref{wmunu}) in the Bjorken regime is dominated by light-like distances, i.e. $z^2\sim 0$.
\\

Since the dominant contribution to DIS comes from the light-cone, it is natural to consider a dynamical formulation in which the light-cone plays a special r\^ole, namely: the light-cone dynamical form
~\cite{Bjorken:1970ah, Kogut:1969xa}.
~\footnote{Following Dirac~\cite{Dirac:1949cp} there are three independent parameterizations of space and time, namely, three different forms of dynamics which are completely equivalent; namely the instant, front and point forms. 
The most familiar form is the {\it instant form}, where the quantization is specified on a space-time hypersurface at a given {\it instant time} $x^0=0$. 
 In the light-cone approach, called the {\it front form}, the theory is quantized at a particular value of the light-cone time $x^+$. The quantization conditions are specified on a hypersurface that is a tangent plane to the light-cone defined at a light-cone time $x^+=0$.}
 The vector decomposition~(\ref{4-vec-lc}) actually corresponds to the light-cone decomposition, with the light-like vectors~(\ref{ll-vec}).

\section{Ordering}

The domain of validity of the approach has been defined. So has been the treatment of the dynamics on the light-cone. We can now start to investigate on the consequences/advantages of these definitions by analyzing the operator describing the hadronic tensor, $W^{\mu\nu}$~(\ref{wmunu}).

Since  DIS processes are light-cone dominated,~\footnote{Light-cone dominated does not mean short distance, but the light-cone domination can be led to short distances.} we can apply
the Wilson's Operator Product Expansion (OPE). A product of local operators $\hat{A}(x)$ and  $\hat{B}(y)$ at short distance, i.e. $x-y\sim 0$, can be expanded in a serie of well-defined local operators
$\hat{O}_i(x)$ with singular $c$-number coefficients $C_i(x)$ for ($i=0,1,2,3,\ldots$) \cite{Muta:1998vi}
\begin{eqnarray}
\hat{A}(x)\hat{B}(y)&=& \sum_{i=0}^{\infty}\,C_i(x-y)\hat{O}_i(\frac{x+y}{2})\quad.
\label{ope1}
\end{eqnarray}

The OPE basically represents a separation between short-distance, i.e. the singular coefficients, and large-distance, i.e. the local operators, objects. It  is a formal step towards the factorization property.
For instance, the Operator Product Expansion \ce{ope1} of simple  bilocal currents  is, see for example \cite{Muta:1998vi},
\begin{eqnarray}
j(z)j(0)&=& \sum_{i,n} C^{(i)}_n (z^2)\,z^{\mu_1}\ldots z^{\mu_n}\, {\cal O}_{\mu_1\ldots \mu_n} (0)\quad,
\label{bilocur}
\end{eqnarray}
with $C^{(i)}_n (z^2)$ singular $c$-numbers called Wilson coefficients. The terms in the expression~(\ref{bilocur}) are arranged in the order of decreasing singularity, the leading contribution being the first term, providing for an expansion parameter.

The strength of light-cone singularity can be found by simple dimensional analysis. If we call $d_j$ the dimension of the currents, $d_{\cal O}^{(i)}(n)$ the dimension of the operator ${\cal O}_{\mu_1\ldots \mu_n} (0)$ and $n$ the spin, the dimension of the
singular coefficients is
\begin{eqnarray}
C^{(i)}_n (z^2)&\sim& (z^2)^{-d_j+{d_{\cal O}^{(i)}(n)\over 2} -\frac{n}{2}}\quad.
\end{eqnarray}
The strength of the singularity is then governed by $\tau^i_n\equiv d_{\cal O}^{(i)}(n)-n $ and we call $\tau^i_n$ the twist.
The smaller the value of the twist, the higher the light-cone singularity: we can use the twist as an expansion parameter.
In QCD the lowest twist operator has $\tau^i_n=2$.
A more informal definition of the twist is often used \cite{Jaffe:1996zw}. The twist is the order in $M/Q$, $M$ being the target mass, at which an operator
contributes to DIS processes. In other words we have $\left(M/Q\right)^{\tau^i_n-2}$.
\\
\begin{wrapfigure}[14]{r}{65mm}
\centering
 \includegraphics [height=4.2cm]{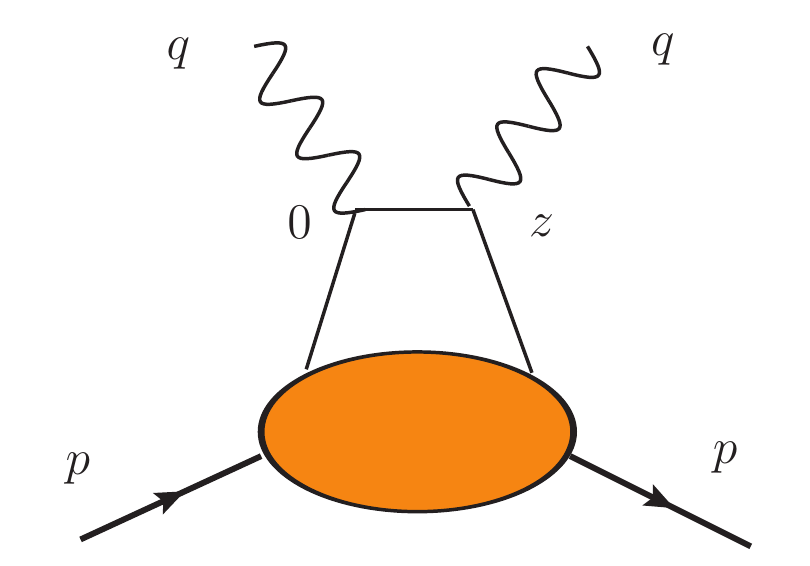}
 \caption{Handbag diagram. Most of the Feynman diagrams in this thesis have been drawn using JaxoDraw~\cite{Binosi:2008ig}.}
\label{hand}
\end{wrapfigure}
An alternative way of isolating the leading-twist contribution of the hadronic tensor $W^{\mu\nu}$~(\ref{wmunu}) is obtained 
by decomposing the hard and soft parts like in Ref.~\cite{Jaffe:1996zw}. The vector current associated with a fermion field is
\beq
J^{\mu}(z)&=& : \bar \psi (z)\gamma^{\mu}\psi(z):\quad.
\nonumber
\eeq
with $:\hat O:$ the normal ordering. We will need  the identity $\left[ \bar  \psi_1 \psi_1, \bar \psi_2 \psi_2\right]= \bar \psi_1 \{ \psi_1, \bar \psi_2\} \psi_2 -\bar \psi_2 \{\psi_2, \bar \psi_1\} \psi_1 $, as well as  the following one, for massless fields,
\beq
\{ \psi(z),\bar \psi(0) \}&=& \frac{1}{2\pi}\, \thru\partial \epsilon (z_0)\, \delta(z^2)\quad.
\nonumber
\eeq
The commutator of two currents appearing in~(\ref{wmunu}) can be expressed in terms of bilocal operators
\beq
[ J^{\mu}(z),J^{\nu}(0)]&=&- \frac{1}{4\pi}\,\left (\partial_{\rho} \epsilon (z_0)\, \delta(z^2)\right)
\Big[
\frac{1}{2} \mbox{tr}\left(\gamma^{\mu} \gamma^{\rho} \gamma^{\nu} \gamma^{\alpha}\right)\, \left( 
\bar \psi (z)\,\gamma_{\alpha}\, \psi (0)- \bar \psi (0)\,\gamma_{\alpha}\, \psi (z)\right)\nonumber\\
 &&-i\epsilon^{\mu\rho\nu\alpha} \left( 
\bar \psi (z)\,\gamma_{\alpha}\,\gamma_5\, \psi (0)+\bar \psi (0)\,\gamma_{\alpha}\,\gamma_5\, \psi (z)\right)
\Big]\quad.
\label{facto-hadrotens-curr}
\eeq
\\
The hadronic matrix element, decomposed in the following way,
\beq
W^{\mu\nu}
&=& -\left( \frac{1}{4\pi}\ \right)^2\, \int d^4 z\,e^{iq\cdot z}\, \partial_{\rho} \left( \delta (z^2)\varepsilon (z^0) \right)\, \nonumber\\
&&\hspace{1cm}
\Big[
 \mbox{tr}\left(\gamma^{\mu} \gamma^{\rho} \gamma^{\nu} \gamma^{\alpha}\right)\, \langle p, S\vert 
\bar \psi (z)\,\gamma_{\alpha}\, \psi (0)
\vert p,S\rangle +\ldots  
\Big]
\quad,
\label{facto-hadrotens}
\eeq
with the ellipses representing the other terms present in~(\ref{facto-hadrotens-curr}), 
gives rise to the {\it handbag} diagram depicted on Fig.~\ref{hand}. The hard part comes from the quark propagator between $0$ and $z$, whereas the soft part is parameterized through a bilocal matrix element represented by the lower blob. The latter object is related to the {\it Parton Distributions}.

The expression~(\ref{facto-hadrotens}) has been shown without writing explicitly the color indices. However, the latter are implicit and, in order to preserve gauge invariance, one has to take into account the propagation of the quarks in the gluon background by linking the two quark fields through the {\it Wilson link}, i.e.
\beq
\bar \psi (z)\,  \Gamma \psi (0) \longrightarrow \bar \psi (z)\, {\cal P} \, \left(  \exp{i\int_0^z\, d\zeta^{\mu}\, A_{\mu} (\zeta) } \right) \Gamma \psi (0)\quad.
\nonumber
\eeq
Since the $\delta$-function in~(\ref{facto-hadrotens}) selects the light-cone, we can expand the gauge link and keep only the leading-twist contribution. In the usual parton distribution, the leading-twist contribution involves the plus-component of the gauge field. An appropriate choice of the gauge reduces the Wilson link to unity: the light-cone gauge $A^+=0$ is  chosen that allows the interpretation of the parton distributions as probability densities.~\footnote{
The light-cone gauge is not always the most adapted  choice. For instance, leading-twist Final State Interactions, i.e. the exchange of a gluon between the active quark and one of the "spectator", after scattering by the virtual photon would rather be studied in a non-singular gauge~\cite{bjy}. This comment will make sense in Chapter~\ref{sec:sivers}.}
\\

We assume that the handbag diagram contribution to the hadronic tensor is dominated by small values of the transverse momentum of the quark. We therefore consider collinear quarks.
Hence, to leading-twist, the (unpolarized) Parton Distributions read
\beq
\int \frac{dz^-}{2\pi}\, e^{iz^-p^+ x }\, \langle p\vert  \bar \psi (0, z^-, \vec{0}_{\perp}) \,
\gamma^{+} \psi (0) \vert p\rangle &=& \frac{1}{p^+} q(x)\quad.
\label{PDF-intro}
\eeq

\section{Deeply Virtual Compton Scattering}
\label{sec:dvcs}

\begin{wrapfigure}[12]{l}{65mm}
\centering
 \includegraphics [height=3.5cm]{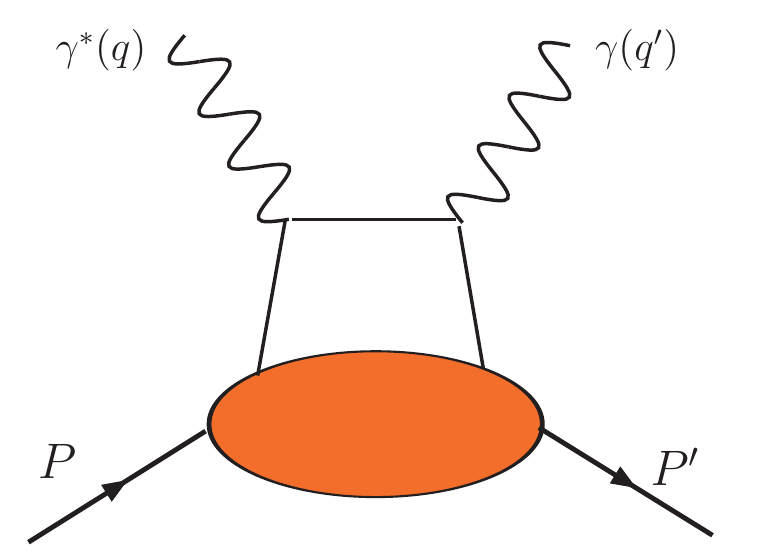}
 \caption{\small {Virtual Compton Scattering}}
\label{VCS}
\end{wrapfigure}

The Forward Virtual Compton Scattering (FVCS) 
 is related to the hadronic part of DIS through the optical theorem.
What generally refers to as Virtual Compton Scattering (VCS) is any process where 2 photons are involved and where at least one of them is virtual.
Virtual Compton Scattering  in the Bjorken region was proposed by Ji \cite{Ji:1996ek} and Radyushkin \cite{Radyushkin:1996nd} as a tool to extract new structure 
functions of the nucleon.
Since in this limit the momentum transfer is large, we will refer to VCS in the Bjorken region as Deeply-Virtual Compton Scattering (DVCS).
The DVCS provides for a new ground to explore the quark and the gluon structure of the nucleon. 
\\

A new kind of distributions, namely non-diagonal distributions,
 called off-forward, skewed or generalized
parton distributions (GPDs),  were  introduced by Ji \cite{Ji:1996ek} and Radyushkin \cite{Radyushkin:1996nd} from the DVCS amplitude
\begin{eqnarray}
T^{\mu\nu}(P+P',q,P-P')&=& i\int d^4z\,e^{-iq.z}\langle P'|Tj^{\mu}(z)j^{\nu}(0)|P\rangle\qquad,
\label{DVCSampl}
\end{eqnarray}
whose sum rule  are related to the Form Factors of the target. 
\\

Following the example of the analysis of DIS in the previous Section, we would like to determine the dominant contributions to DVCS.

\subsection{Definition of the Generalized Parton Distributions}

The kinematics are similar to the DIS case. Nevertheless the off-forward nature of the new distributions requires an additional variable for the description of the initial and final parton states, called
the skewness variable, $\xi$, which reflects the longitudinal momentum asymmetry. Hence the Generalized Parton Distributions  are functions of two more variables
in comparison to the usual Parton Distribution Functions, namely they depend on  $x$,
 in this context the variable $x$ is just the Fourier variable of the light-cone distance,
$\xi$ and the four-momentum transfer (squared) $t$.
\\
In this work, we choose to follow Ji's notations instead of Radyushkin's\footnote{For a comparison of the variables see \cite{Golec-Biernat:1998ja}.}.
Hence we introduce the average 4-momentum  $p^{\mu}=(P+P')^{\mu}/2$ which has to be collinear to $q^{\mu}$ in the $z$ direction. The decomposition
on the light-cone is  \cite{Guichon:1998xv}
\begin{eqnarray}
p^{\mu}&=& \bar{p}^{\mu}+\frac{1}{2}\left(M^2-\frac{t}{4}\right)n^{\mu}\qquad,\nonumber\\
q^{\mu}&=& -2\xi\,\bar{p}^{\mu}+\frac{Q^2}{4\xi}n^{\mu}\qquad,\\
\Delta^{\mu}&=& -2\xi\,\bar{p}^{\mu}+\xi\left(M^2-\frac{t}{4}\right)n^{\mu} +\Delta^{\mu}_{\perp}\qquad.\nonumber
\label{lightconedecomp}
\end{eqnarray}
The skewness variable is defined as $\xi\simeq Q^2/(4\,p\cdot q)$ and is related to the Bjorken variable $x$ in the Bjorken limit. Here the momentum transfer $t=\Delta^2$  is negative.

Similarly to the expression~(\ref{facto-hadrotens}) obtained for the DIS process, 
factorization for DVCS can be studied~\cite{Ji:1998xh}. 
 Therefore the DVCS amplitude \ce{DVCSampl}  fulfills its r\^ole in  being expressed as a convolution of  hard and soft parts \cite{Ji:1996ek}
\begin{eqnarray}
 T^{\mu\nu}(p,q,\Delta) 
           & = & -{1\over 2} (p^\mu n^\nu+p^\nu n^\mu-g^{\mu\nu})
             \int_{-1}^1 dx \left({1\over x-\xi + i\epsilon} 
            + {1\over x+\xi + i\epsilon}\right) \nonumber 
              \\ &&  \left[ H(x,\xi,t)
                     \bar u(P') {\thru n} u(P)
             + E(x, \xi,t) 
                   \bar u(P') {i\sigma^{\alpha\beta}n_\alpha
              \Delta_\beta \over 2M}u(P) \right] \nonumber \\ &&
              - {i\over 2}\epsilon^{\mu\nu\alpha\beta}
                p_\alpha n_\beta
                 \int_{-1}^1 dx \left({1\over x-\xi + i\epsilon} 
            - {1\over x+\xi + i\epsilon}\right)  \nonumber \\ 
            && 
           \left[ \tilde H(x,\xi,t)\bar u(P')
           \thru n \gamma_5 u(P) + \tilde E(x,\xi,t)
                  {\Delta\cdot n\over 2M}\bar u(P')\gamma_5 u(P) \right] \qquad,
\end{eqnarray}
\begin{wrapfigure}[11]{r}{65mm}
\centering
 \includegraphics [height=2.6cm]{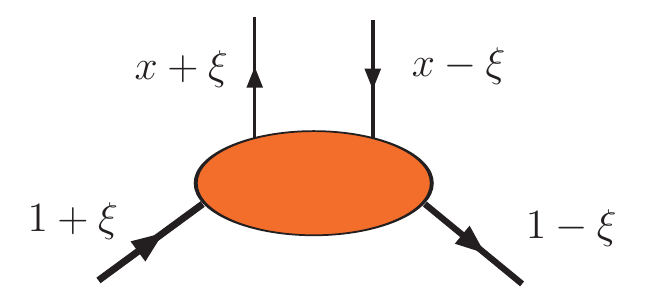}
 \caption{\small {Kinematics of the Deeply-Virtual Compton Scattering \cite{Ji:1996ek}.}}
\label{DVCS}
\end{wrapfigure}
with $u(P)$ the nucleon spinor. 
\\
This decomposition corresponds to the {\it handbag} diagram depicted on Fig.~\ref{VCS}.
The dominant  subprocess \cf{DVCS} consists on  a virtual photon with momentum $q^{\mu}$  absorbed by a single quark with momentum $k^{\mu}$. Subsequently the
 quark radiates a real photon, whose momentum is now $q^{\mu}-\Delta^{\mu}$, and falls back with a resulting momentum $k^{\mu}+\Delta^{\mu}$ in the nucleon. The momentum of the recoil nucleon is  $P^{' \mu}~=~P^{\mu}~+~\Delta^{\mu}$.
\\

The four distributions $H(x,\xi,t)$, $E(x,\xi, t) $, $\tilde H(x,\xi,t)$ and $\tilde E(x,\xi,t)$ represent the soft part, namely the {\it Generalized Parton Distributions}.
They are defined through the matrix elements of quark operators at a light-like separation. To leading twist, we have
\begin{eqnarray}
 && \frac{1}{2}\int  {dz^- \over 2\pi} e^{i xz^-p^+}
 \langle P'|\bar\psi(-\frac{z}{2})\gamma^+
 \psi(\frac{z}{2})|P \rangle \Big | _{z^+=\vec{z}^{\perp}=0} \nonumber\\
 &=& \frac{1}{2p^+}\Big[H(x,\xi,t) \bar u(P')\gamma^+ u(P)  + E(x,\xi,t) \bar u(P'){i\sigma^{+\nu}
     \Delta_{\nu}\over 2M}u(P)\Big]\qquad,  \nonumber \\
 \nonumber \\
&& \frac{1}{2}  \int  {dz^- \over 2\pi} e^{i xz^-p^+}
 \langle P'|\bar\psi(-\frac{z}{2})\gamma^+\gamma_5
 \psi(\frac{z}{2})|P \rangle \Big | _{z^+=\vec{z}^{\perp}=0} \nonumber\\
 & =&\frac{1}{2p^+}\Big[ \tilde H(x,\xi,t) 
 \bar u(P')\gamma^+ \gamma_5 u(P)  + \tilde E(x,\xi,t) \bar u(P')
      {\gamma_5\Delta^+\over 2M}u(P)\Big]\qquad.
\label{GPD}
\end{eqnarray}
The two distributions $H(x,\xi,t)$ and $\tilde H(x,\xi,t)$ conserve the nucleon helicity while the two others are said to be (nucleon) helicity-flipping.

\subsection{Properties}
\label{sec:gpdprop}

Let us overview the properties of these new objects that follow from the analysis from {\it first principles}.

\subsubsection{Support}

The support in $x$ of the GPDs is $[-1,1]$. A negative momentum fraction would correspond to an antiquark. In \cf{DVCS}, the active quark  has a  ($+$)-component momentum 
fraction $x+\xi$.
 From the definition of the skewness variable,
\begin{eqnarray}
\xi&=&\frac{P^+-P^{'+}}{P^++P^{'+}}=-\frac{\Delta^+}{2p^+}\qquad,
\end{eqnarray}
and from the fact that ($+$)-momenta of physical states  cannot be negative, it follows that
the physical region for $\xi$ is the interval $[-1,1]$. However, from the light-cone decomposition, we find that $\xi$ is bounded by
\begin{eqnarray}
 0\leq\xi\leq\frac{\sqrt{-t}}{2\sqrt{M^2-\frac{t}{4}}}<1\qquad.
\end{eqnarray}
\\
Given the intervals in $x$ and in $\xi$, we can distinguish 3 regions in $x$, 

\begin{itemize}
\item In the region $x \in [\xi,1]$ both momentum fractions  $x+\xi$ and $x-\xi$ are positive. In this region the GPD describes the emission and reabsorption of a quark.

\item The behavior is similar in the region $x \in [-1,-\xi]$. Both momentum fractions $x+\xi$ and $x-\xi$ are negative and therefore the GPD describes the emission
and reabsorption of an antiquark, respectively with momentum fractions $\xi-x$ and $-\xi-x$. This is shown on the extreme sides of \cf{regions}.

\vspace*{0.5cm}

\begin{figure}[ptb]
\centering
 \includegraphics [height=3cm]{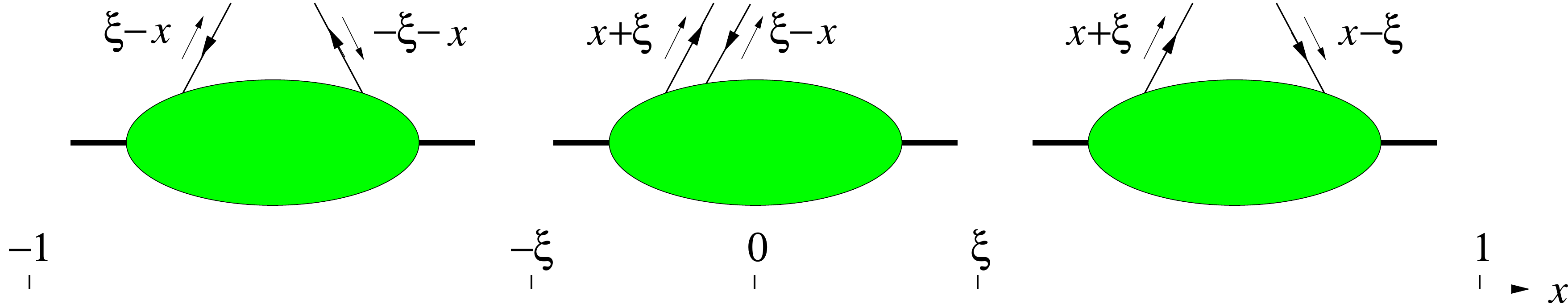}
 \caption[GPD's support.]{\small {The support of the GPDs. The parton interpretation of GPDs in the $x$-intervals $[-1,-\xi]$, $[-\xi,\xi]$ and $[\xi,1]$. Figure taken from Ref.~\cite{Diehl:2003ny}.}}
\label{regions}
\end{figure}

\vspace*{0.5cm}

\item In the third region, i.e. $x \in [-\xi,\xi]$, one has that $x+\xi$ is positive but $x-\xi$ negative. This can be interpreted as the emission of a quark with momentum
fraction $x+\xi$ and the emission of an antiquark with momentum fraction $\xi-x$ both emitted from the initial nucleon.
\end{itemize}

\subsubsection{Forward Limit}

When $P'=P$, the skewness variable $\xi$ is equal to zero.
As a consequence GPDs, in the regions corresponding to emission/reabsorption of a quark or an antiquark, i.e. $[-1,-\xi]$ and $[\xi,1]$, have an equivalence
with usual PDFs. This is not true for the region of emission of a quark and an antiquark since  the interval $[-\xi,\xi]$ reduces 
to zero in the same limit.
This means that, in the forward limit,  when 
 the initial and final states are equal, i.e. equal helicities and  $P'=P$,  
 Generalized Parton Distributions as defined in  \ce{GPD} are expected to reduce to ordinary 
Parton Distributions \ce{PDF-intro}, 
\begin{eqnarray}
H^q(x,0,0)&=&q(x)\qquad,\nonumber\\
H^q(x,0,0)&=&-\bar q(-x)\qquad,
\end{eqnarray}
for $x>0$ and $x<0$ respectively. 
\\

When considering the hadronic tensor in DIS~(\ref{facto-hadrotens}) we have omitted the spin dependent distribution function as well
as the gluon distribution functions. However it seems useful to notice that, in the forward limit, the second helicity-conserving distribution 
is related to the spin dependent parton distribution: $\tilde H(x,0,0)=\Delta q (x)$ for $x>0$.

\subsubsection{Symmetries}

Time reversal interchanges $P$ with $P'$ and its invariance ensures the distribution will not change. By so doing we change the sign of $\xi$ and find
that the GPDs are symmetric in $\xi$
\begin{eqnarray}
H(x,\xi,t)&=& H(x,-\xi,t)\qquad.
\label{xi-xi}
\end{eqnarray}
Quark distributions are neither even nor odd in $x$.

\subsubsection{Sum Rules}

On the other hand, forming the first moment of the GPDs by integrating over the momentum fraction $x$, we get the sum rules
\begin{eqnarray}
\int dx\,H(x,\xi,t)= F_1(t)\qquad && \int dx\,E(x,\xi,t)= F_2(t)\qquad,\nonumber\\
\int dx\,\tilde H(x,\xi,t)= g_A(t)\qquad && \int dx\,\tilde E(x,\xi,t)= g_P(t)\qquad,
\end{eqnarray}
where the dependence on $\xi$ drops out. The integration over $x$ of the matrix element \ce{GPD} removes all reference to the particular light-cone direction
with respect to which $\xi$ is defined. As a consequence the Form Factors are $\xi$-independent.

The Form Factors
 $F_1(t)$  and $F_2(t)$ are respectively the Dirac and Pauli Form Factors 
\begin{eqnarray}
\langle P'| \bar \psi(0)\gamma^{\mu}\,\psi(0)|P\rangle&=& \bar u(P')\Big[
F_1(t)\gamma^{\mu}+F_2(t)\frac{i\sigma^{\mu\alpha}\Delta_{\alpha}}{2M}
\Big]u(P)\qquad,
\end{eqnarray}
and $g_A(t)$ and $g_P(t)$ the axial and pseudoscalar ones
\begin{eqnarray}
\langle P'| \bar \psi(0)\gamma^{\mu}\gamma_5\,\psi(0)|P\rangle&=& \bar u(P')\Big[
g_a(t)\gamma^{\mu}\gamma_5+g_p(t)\frac{\gamma_5\Delta^{\mu}}{2M}
\Big]u(P)\qquad,
\end{eqnarray}
defined for each separate flavor.
\\

\subsubsection{Gluon Distributions and the Ji's Sum Rule}

We can define the gluon GPDs in a similar way as the quark GPDs
\begin{eqnarray}
 && \frac{1}{p^+}\int  {dz^- \over 2\pi} e^{i xz^-p^+}
 \langle P'|G^{+}_{\mu}(-\frac{z}{2})\
G^{+}_{\mu}(\frac{z}{2})|P \rangle \Big | _{z^+=\vec{z}^{\perp}=0} \nonumber\\
 &=& \frac{1}{2p^+}\Big[H^g(x,\xi,t) \bar u(P')\gamma^+ u(P)  + E^g(x,\xi,t) \bar u(P'){i\sigma^{+\nu}
     \Delta_{\nu}\over 2M}u(P)\Big]\qquad,  \nonumber \\
 \nonumber \\
&& \frac{-i}{p^+}  \int  {dz^- \over 2\pi} e^{i xz^-p^+}
 \langle P'|G^{+}_{\mu}(-\frac{z}{2})\
\tilde G^{+}_{\mu}(\frac{z}{2})|P \rangle \Big | _{z^+=\vec{z}^{\perp}=0} \nonumber\\
 & =&\frac{1}{2p^+}\Big[ \tilde H^g(x,\xi,t) 
 \bar u(P')\gamma^+ \gamma_5 u(P)  + \tilde E^g(x,\xi,t) \bar u(P')
      {\gamma_5\Delta^+\over 2M}u(P)\Big]\qquad.
\label{GPD-gluon}
\end{eqnarray}
\\
		
An interesting sum rule concerns the quark and gluon spin contribution to the nucleon spin as it  has  been one of the main motivation for GPDs.
The second moment of both $H(x,\xi,t)$ and $E(x,\xi,t)$, the helicity conserving distributions as well as the first moments of the same gluonic distributions
are directly related to the total  angular momentum contribution to the nucleon spin $J$ \cite{Ji:1996ek}
\begin{eqnarray}
\langle J^3\rangle&=& \langle J^3_q\rangle + \langle J^3_g\rangle\quad,\nonumber\\
\frac{1}{2}&=&\frac{1}{2}\,\sum_q\, \int_{-1}^1 dx\,x\,[H^q(x,\xi,0)+E^q(x,\xi,0)]\, + \frac{1}{2}\,\int_0^1 dx\,[H^g(x,\xi,0)+E^g(x,\xi,0)]
\quad.
\label{sumrulespin}
\end{eqnarray}
As a consequence,  
the total angular momentum carried by the quarks or the gluons  inside a proton could be known if one takes the particular value of the desired Form Factors at the  forward limit $t=0$. Since the spin part of the angular momentum is found through the helicity distributions $\Delta q(x)$, the Ji's sum rule enables to access the orbital angular momentum of the quarks and gluons.

\subsection{Polynomiality}
\label{subsec:poly}

 By writing down the most general expression in terms 
of relevant vectors,  the first moments are immediately related to their Form Factors.
The higher-moments of the GPDs are accessed through the tower of twist-2 operators, generalization of the current, of the type
\begin{eqnarray}
O_V^{\mu\mu_1\ldots \mu_{n-1}}&=& \bar \psi \gamma^{{\large\{} \mu}\,i\stackrel{\leftrightarrow}{\cal D}^{\mu_1}\ldots 
i\stackrel{\leftrightarrow}{\cal D}^{\mu_n{\large\}}}\psi\quad,
\label{vec-twist-2}
\end{eqnarray}
where the action of $\{\ldots\}$ on Lorentz indices produces the symmetric, traceless part of the tensor.
In a general way, the matrix element of  towers of twist-two operators appear in the operator product expansion. %
\\

 Using symmetries, we can write down the general form for the matrix element of the latter tower between states of unequal 
momenta.
  All possible Form Factors therefore arise, and 
  for the vector operator, we obtain~\cite{Ji:1998pc}
\begin{eqnarray}
\langle P'|O_V^{\mu\mu_1\ldots \mu_{n-1}}|P\rangle&=& \bar{u}(P')\gamma^{{\large\{} \mu}u(P)\,\sum_{i=0}^{[\frac{n-1}{2}]}\,A_{n,2i}(t)\,\Delta^{\mu_1}\ldots \Delta^{\mu_{2i}}
p^{\mu_{2i+1}}\ldots p^{\mu_{n-1} {\large\}}}\nonumber\\
&+& \bar{u}(P')\frac{i\sigma^{ {\large\{}\mu\alpha}\Delta_{\alpha}}{2M}u(P)\,\sum_{i=0}^{[\frac{n-1}{2}]}\,B_{n,2i}(t)\,\Delta^{\mu_1}\ldots \Delta^{\mu_{2i}}
p^{\mu_{2i+1}}\ldots p^{\mu_{n-1} {\large\}}}\nonumber\\
&+&C_n(t)\,\frac{1}{2}(1+(-1)^{n})\, \frac{1}{M}\, \bar{u}(P')u(P)\Delta^{ {\large\{}\mu}  \Delta^{\mu_1}\ldots \Delta^{\mu_{n-1}{\large\}}}\quad.
\label{towert2}
\end{eqnarray}
Time-reversal invariance imposes  that only even powers in $\Delta$ appears.

We project the result (\ref{towert2}) on the light-front by multiplying by $n_{\mu}n_{\mu_1}\ldots n_{\mu_{n-1}} $ and define the off-forward parton distributions through 
the Form Factors 
\begin{eqnarray}
n_{\mu} n_{\mu_1}\ldots n_{\mu_{n-1}}\,\langle P'|O_V^{\mu_1\ldots \mu_n}|P\rangle&=& \bar{u}(P')\gamma^+u(P)\,H_n(\xi,t)\,+\,
\bar{u}(P')\frac{i\sigma^{+\alpha}\Delta_{\alpha}}{2M}u(P)\,E_n(\xi,t)\qquad,
\end{eqnarray}
with $H_n(\xi,t)$ and $E_n(\xi,t)$ the moments of the helicity-conserving quark GPDs which are defined as
\begin{eqnarray}
H_n(\xi,t)&=& \int_{-1}^1 dx\,x^{n-1} H(x,\xi,t)\nonumber\\
&=&\sum_{i=0}^{[\frac{n-1}{2}]}\,A_{n,2i}(t)\,(-2\xi)^{2i}+C_n(t)\,\frac{1}{2}(1+(-1)^{n})\, (-2\xi)^n\qquad,\nonumber\\
E_n(\xi,t)&=& \int_{-1}^1 dx\,x^{n-1} E(x,\xi,t)\nonumber\\
&=&\sum_{i=0}^{[\frac{n-1}{2}]}\,B_{n,2i}(t)\,(-2\xi)^{2i}-C_n(t)\,\frac{1}{2}(1+(-1)^{n})\, (-2\xi)^n\qquad.
\label{poly-gpd-theo}
\end{eqnarray}
Basically it means that the moments of the GPDs are polynomials of the skewness variable $\xi$ of order at most $n$.
\\

We have here presented the properties of the bilocal matrix elements defining the GPDs of the proton. The latter can be easily adapted to, e.g., the pion.

\section{Transverse Momentum Dependent Distribution Functions}

All along this Chapter the quantities have been  defined  considering  only collinear quarks. %

Let us now account for the transverse motion of quarks~\cite{Barone:2001sp}. The number of Parton Distributions at leading-twist increases from the three usual (i.e. number density $q(x)$, helicity density $\Delta q(x)$ and transversity $\Delta_T q(x)$) to eight distributions;
\beq
q, \Delta q, \Delta_T q \,, \, \overbrace{ g_{1T}, h_{1L}^{\perp},  h_{1T}^{\perp}\underbrace{f_{1T}^{\perp}, h_1^{\perp}}_{T\mbox{\small -odd}}}^{k_T\mbox{\small -dependent}}
\nonumber
\eeq

 For the description of
the quark content of the proton, the following quantity~\footnote{It is given in
the light-cone gauge $A^+ = 0$.} is relevant,
\beq
\Phi_{ij}(x,\vec{k}_T) =\int \frac{dz^-d^2\vec{z}_T}{(2\pi)^3} 
\ e^{i(xp^+ z^- -\vec{k}_T\cdot \vec{z}_T)} \,\langle p,S \vert \overline \psi_j (0)\, 
\psi_i(0,z^-, \vec{z}_T) \vert p,S \rangle \quad,
\eeq
depending on the light-cone fraction of the quark momentum, $x = k^+/p^+$ 
and the transverse momentum component $\vec{k}_T$.
Using Lorentz invariance, hermiticity, and parity invariance one finds that
 up to 
leading order in $1/Q$~\cite{Boer:1997nt}
\beq
\Phi(x,\vec{k}_T)&=& 
\frac{1}{2}\,\Biggl\{
q(x, k_T)\,\thru n_+
+ f_{1T}^\perp (x, k_T)\, \frac{\epsilon_{\mu \nu \rho \sigma}
\gamma^\mu n_+^\nu k_T^\rho S_T^\sigma}{M}
+g_{1s} (x, k_T)\, \gamma_5\thru n_+
\nonumber \\ 
& & \qquad
+\Delta_T q(x, k_T)\,i\sigma_{\mu\nu}\gamma_5 n_+^\mu S_T^\nu 
+h_{1s}^\perp (x, k_T)\,\frac{i\sigma_{\mu\nu}\gamma_5 
n_+^\mu k_T^\nu}{M} 
+ h_1^\perp (x, k_T)\,\frac{\sigma_{\mu \nu} k_T^\mu n_+^\nu}{M}
\Biggr\}\, ,
\nonumber\\
\label{decom-bm}
\eeq
where $S$ is the spin of the proton target.
 The quantities $g_{1s}$ and  $h_{1s}^\perp$ are shorthand for 
\beq
g_{1s}(x, k_T) &= &
\lambda\,\Delta q(x,k_T) + \frac{\vec{k}_T \cdot \vec{S}_T }{M}
\,g_{1T}(x, k_T)\, ,\nonumber\\
\nonumber\\
h_{1s}^\perp(x, k_T) &= &
\lambda\,h_{1L}^\perp(x,k_T) + \frac{\vec{k}_T \cdot \vec{S}_T }{M}
\,h_{1T}^\perp(x, k_T),
\eeq
with $M$ the mass of the proton, $\lambda = M\,S^+/p^+$ the light-cone helicity, 
and $\vec{S}_T$ the transverse spin of the target hadron.

The three distribution functions $q(x, k_T), \Delta q(x,k_T) , \Delta_T q(x,k_T) $ reduce to their analogs  under integration over $k_T$, whereas the 5 others vanish if integrated over the transverse momentum.  
\\

The two {\it na\"ively $T$-odd} distribution functions $f_{1T}^{\perp}, h_1^{\perp}$ are respectively the Sivers and the Boer-Mulders functions. Their r\^ole in the left-right asymmetries as well as their contribution to the understanding of the spin structure of hadrons will be highlighted in Chapters~\ref{sec:sivers} $\&$  \ref{sec:out-gpd}.
\chapter{Pion Distribution Amplitude}
\label{sec:pionda}
\pagestyle{fancy}

This Chapter is devoted to the pion Distribution Amplitude (DA).  The DA $\phi$ is the probability amplitude for finding,  in the pion,  the valence quarks sufficiently near the light cone. 
Its behavior as well as its evolution in $Q^2$ has been extensively studied in Refs.~\cite{Broniowski:2007si, Lepage:1980fj}. We also study the pion Parton Distribution Function (PDF). Nowadays accurate experimental data allow for a comparison from the model calculations.

Since there are a well-known quantities, the pion DA and PDF represent a perfect framework for illustrating our approach. Also, the properties of the pion DA and PDF are placed under scrutiny. We consider sum rules, support and soft pion theorem; the latter linking the DA to the PDF or to the Generalized Distribution Amplitudes.  The discrete symmetries as well as isospin relation are used to relate the different pion DAs. We  draw the way to QCD evolution of the DA and the PDF. We eventually give the expression of the electromagnetic Form Factor and its well-known asymptotic limit; namely the Brodsky-Lepage Form Factor.
\\

\section{The Pion DA in the NJL Model}
 \label{sec:da-cal}

In this Section, we illustrate our approach by performing the complete calculation of a simple quantity: the pion DA $\phi(x)$. 
This will fix the conventions and allow us to give the parameters of the model. 
Also, the result of the calculation will tell us whether the model is appropriate for the calculation of such quantity.  In particular, in the next Sections, we will see  that  the normalization condition  as well as the support in $x$ are respected.

\begin{wrapfigure}[9]{r}{6cm}
\centering
\includegraphics[height=3cm]{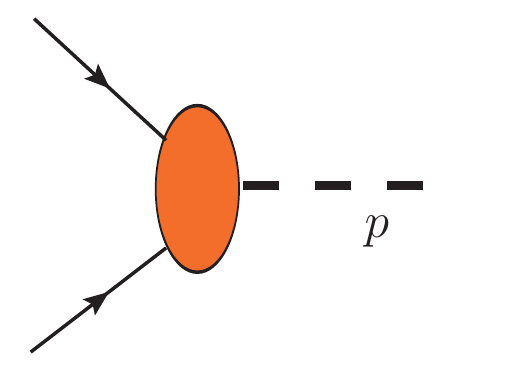}
\caption{The pion DA.}%
\label{bs-da}%
\end{wrapfigure}

By definition, the pion DA is, as depicted in Fig.~\ref{bs-da},
\begin{eqnarray}
\int\frac{dz^{-}}{2\pi}e^{i(x-\frac{1}{2})p^{+}z^{-}}\left.
\left\langle 0\right\vert
\bar{q}\left(  -\frac{z}{2}\right)  \rlap{$/$}n\hspace*{-0.05cm}\gamma_{5}%
\tau^{-}q\left(  \frac{z}{2}\right)  \left\vert \pi\left( p\right)
\right\rangle \right\vert
_{z^{+}=\vec{z}^{\bot}=0}=\frac{1}{p^{+}}i\sqrt{2}f_{\pi }\phi\left(
x\right)  \quad.
\label{pda}
\end{eqnarray}
The DA has a support in $x\in [0,1]$ and obeys the normalization condition
\beq
\int_0^1\, dx \,\phi(x)&=& 1\quad.
\label{normda}
\eeq

The field theoretical approach we propose to use here has been first developed in Ref.~\cite{Theussl:2002xp}. The pion is described as a bound state in the sense of Bethe-Salpeter. In order to have an exact solution for the bound state, the kernel has to be chosen carefully. The Nambu~-~Jona-Lasinio interaction being point-like, it leads to such a solution. Furthermore, as it is extensively explained in Appendix~\ref{sec:njl}, the NJL model respects all the required symmetries of the problem. The pions appear to be the three Goldstone modes coming from the chiral symmetry (dynamical) breaking. 
\\
The shortcomings of the model are the absence of confinement and the non-renormalizability due to the point-like interaction. A regularization scheme should be chosen indeed. Also, the choice of this scheme defines the model per se. The NJL model completed with its regularization scheme is considered as an effective theory of QCD; even if built with quarks.  
Lorentz covariance and gauge invariance will assure the recovering of the properties of the parton distributions; so that the Pauli-Villars scheme is chosen that fulfills these conditions (see Section~\ref{sec:regunjl}).
The regularization parameters, namely the  cutoff $\Lambda$ as well as the coupling strength  $G$, are determined by calculating the pion decay constant $f_{\pi}$ and the quark condensate.  For $f_{\pi}=93$ MeV and $\langle \bar u u \rangle=\langle \bar d d \rangle=-(250\,\mbox{MeV})^3$, we obtain~\cite{Klevansky:1992qe}
\begin{eqnarray}
 \Lambda= 859\quad\mbox{MeV}\quad&,&\quad G\Lambda^2=2.84\quad.
\label{param}
\end{eqnarray}
From the gap equation Eq.~(\ref{gap}), we also fix the value of the constituent quark mass to 
\beq
m= 241 \quad\mbox{MeV}\quad.
\eeq
This model for the pion is used in the description of the bound state.
The Bethe-Salpeter (BS) amplitude for the bound state  $\pi$ with four-momentum $p$ is given by
\begin{eqnarray}
\vec{\chi}_{\beta\alpha}(x_1,x_2;p)&=& \langle 0|T q_{\beta}(x_1)\bar q_{\alpha}(x_2)|\vec{\pi}(p)\rangle  \qquad.
\label{bsampl}
\end{eqnarray}
The NJL model contains  a four-fermion interaction. Direct and exchange diagrams are related to each other by a Fierz transformation. Since they go like $1/N_c$, the exchange diagrams are subleading and we may only consider direct diagrams.  We also work in  the ladder approximation, which consists in considering the iteration of the simplest closed loop with the kernel given by the model. The NJL interaction, giving rise to the kernel
\beq
V_{\alpha\beta, \delta\gamma}(k,k';p)&=&2i G ( i\,\gamma_5 \vec{\tau}^{\pi})_{ \delta\gamma}  ( i\,\gamma_5 \vec{\tau}^{\pi})_{\alpha\beta}\quad,
\label{kernel-v}
\eeq
renders the BS equation for the bound state easy to handle.%
\\
Precisely, the integral equation generated by the chain of diagrams  is the Bethe-Salpeter equation \cite{Itzykson:1980rh} for the bound state.
 Solving the Bethe-Salpeter equation in the ladder approximation with the kernel $V(k,k';p)$,\footnote{ See Eqs.~(\ref{bs}, \ref{bs-njl}) in Appendix~\ref{sec:njl}.}
we find the expression for the Bethe-Salpeter (BS) amplitude 
\begin{equation}
\vec{\chi}\left(  k;p\right)  =iS\left(  k+\frac{p}{2}\right)
\,\vec{\Phi}\left( k,p\right)  \,iS\left(  k-\frac{p}{2}\right) \quad,
\end{equation}
with $S^{-1}(k)=\thru k-m$.
The quark-pion vertex function  is 
\beq
\vec{\Phi}\left( k,p\right) &=&i\,g_{\pi qq} \,i\gamma_5 \,\vec{\tau}^{\,\pi}\quad,
\label{vertex}
\eeq
with the quark-pion coupling constant $g_{\pi qq} $ determined by the standard normalization of the BS equation.
\\

 In the theoretical scheme defined above, the pion DA is evaluated by developing the matrix element on the r.h.s. of Eq.~(\ref{pda}) and rearrange it in such a way that we can identify the BS amplitudes $\chi (k; p)$.
Therefore the  Eq.~(\ref{pda}) becomes
\begin{eqnarray}
i\frac{\sqrt{2}f_{\pi}}{p^+}\phi(x)&=& -\int\frac{dz^-}{2\pi}\,
e^{ip^+(x-\frac{1}{2}) z^-} \langle 0 \vert q_{\beta}\left(
\frac{z}{2}\right)\bar q_{\alpha}\left( -\frac{z}{2}\right)\vert
\pi(p)\rangle \left(\thru n \gamma_5
\tau^-\right)_{\alpha \beta}\quad;\nonumber\\
&=&-\int\frac{dz^-}{2\pi}\, e^{ip^+(x-\frac{1}{2}) z^-}
\chi_{\beta\alpha}\left( \frac{z}{2},-
\frac{z}{2}\right)\left(\thru n \gamma_5
\tau^-\right)_{\alpha \beta}\quad.
\label{da-bs}
\end{eqnarray}
We define the Fourier transform, Ref.~\cite{Theussl:2002xp},
\beq
\vec{\chi}_P(x_1,x_2)&=& e^{-i P X}\int\, \frac{d^4k}{(2\pi)^4}\, e^{-ik\cdot r}\vec{\chi}(k, P)\quad,
\label{ft}
\eeq
with the center of mass and relative coordinates $X=\mu_1x_1+\mu_2 x_2$ and $r=x_1-x_2$ and where $\mu_{1,2}=m_{1,2}/(m_1+m_2)$.
So that
\beq
i\frac{\sqrt{2}f_{\pi}}{p^+}\phi(x)&=&
-\int\frac{dz^-}{2\pi}\, e^{ip^+ (x-\frac{1}{2})z^-}\,\int
\frac{d^4 k}{(2\pi)^4} e^{-ik\cdot z}\alltrace
[\chi(k,p)\thru n\gamma_5 \tau^-]\quad,\nonumber\\
&=& -(-ig_{\pi qq})N_c
\int\frac{dz^-}{2\pi}\, e^{ip^+ (x-\frac{1}{2})z^-}\nonumber\\
&&\int
\frac{d^4 k}{(2\pi)^4} e^{-ik\cdot z} \mbox{tr}
[S(k+\frac{p}{2}) i \gamma_5\tau^{\pi^+}
S(k-\frac{p}{2})\thru n\gamma_5 \tau^-]\quad.
\end{eqnarray}
The symbol $\alltrace$  at the first line is the trace over the Dirac, isospin, flavor and color spaces. Going one step further in the calculation, we  perform the trace over isospin for a $\pi^+$.
By making the change of
variable $k\rightarrow -k+{p/2}$, we find 
\begin{eqnarray}
&&i\,\sqrt{2}f_{\pi}\,\phi^{\pi^+}(x)\nonumber\\
&&= -i (-4m)p^+\,(-ig_{\pi
qq})N_c \sqrt{2}\,\int \frac{d^4 k}{(2\pi)^4} \,
\frac{\delta(p^+(x-1)+k^+)}{\left((-k)^2-m^2+i\epsilon\right)\left((-k+p)^2-m^2+i\epsilon\right)}\,.\nonumber\\
\label{2-prop}
\end{eqnarray}
The $\delta$-function imposes a decomposition on light-cone components in order to perform the $d^4 k$ integral
\begin{eqnarray}
\int \frac{d^4 k}{(2\pi)^4}=\frac{1}{(2\pi)^2}\int\frac{d^2 k^{\perp}}{(2\pi)^2}\,\int dk^-\,\int dk^+\quad.\nonumber
\end{eqnarray}
The integral over $dk^+$ is trivial: the $\delta$-function imposes the $+$-component of the momentum carried by the outgoing quark.
The remaining integral over $d^2k^{\perp}$ in Eq.~(\ref{2-prop}) is divergent. Since it has been obtained with the NJL model interaction, the integral over  $d^2k^{\perp}$ has to be consistently regularized using the Pauli-Villars regularization, which is part of the model.

The result  Eq.~(\ref{2-prop}) is proportional to the two-propagator integral $\tilde{I}_{2,P}(x, \xi)$ defined by Eq.~(\ref{i2p})
\beq
i\sqrt{2}f_{\pi}\,\phi^{\pi^+}(x)&=& -4m\, i g_{\pi qq}\, N_c\sqrt{2} \, \tilde{I}_{2,P}(x, 0)\quad.
\label{2-prop-da}
\eeq
\begin{figure}[tb]
\centering
\includegraphics [height=5cm]{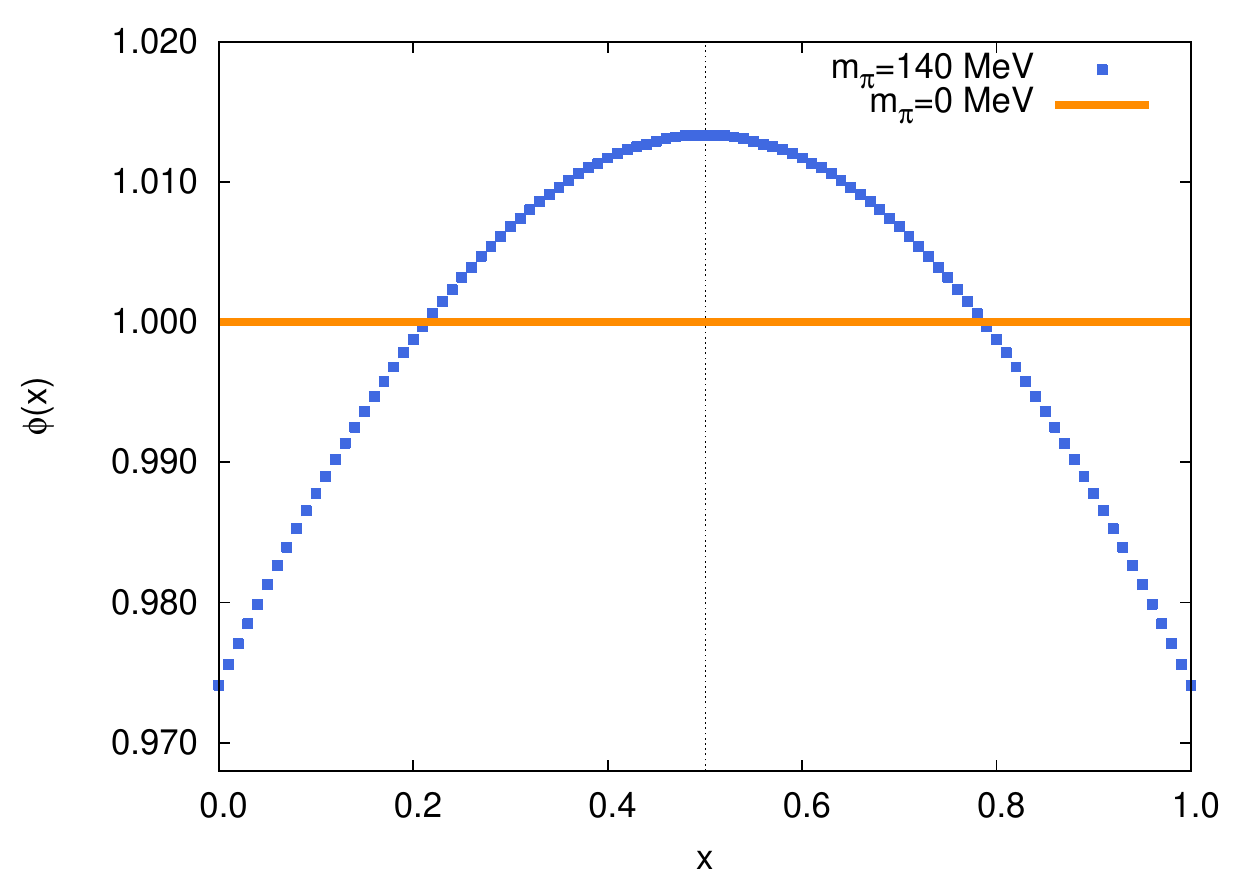}
\caption{The pion DA in NJL. The dotted blue line represents the DA for the  physical value of the pion mass while the plain orange line represents the chiral limit. }
\label{da-plot}%
\end{figure}
In doing so, the normalization condition of the DA Eq.~(\ref{normda}) is recovered for any value of the pion mass.

The result in the  NJL  model is shown on Fig.~\ref{da-plot} for both the physical pion mass and the chiral limit.  In the latter case, the pion DA is a constant which is equal to $1$. This result is obtained by plugging $f_{\pi}$ as given in Eq.~(\ref{piondecaycst}) into Eq.~(\ref{2-prop-da}). In the physical case, the DA deviates by  less than $3\%$ from 1.~\footnote{Mind the scale in Fig.~\ref{da-plot}!}
The distribution amplitude does manifestly not vanish at the end points for none of the pion masses.
This result is obviously characteristic of the model within which the calculation has been performed. It is not a QCD result, and in fact, QCD evolution, as will be shown later on, will push the DA to zero at the end points.

\section{Symmetries}

In the previous paragraph, we have considered the case of a $\pi^+$. Using discrete symmetries we can relate, in a model-independent way,  the $\pi^+$ DA to the, e.g.,  $\pi^-$ DA.

For a generic particle ${\cal P}$ with momentum $\vec{p}$, spin projection $s_z$ and charge $Q_i$, we write the particle state $|{\cal P}(\vec{p}, s_z, Q_i) \rangle$. Parity, time reversal invariance and charge conjugation act on this generic state in the following way
\beq
P|{\cal P}(\vec{p}, s_z, Q_i) \rangle &=& \eta_P |{\cal P}(-\vec{p}, s_z, Q_i) \rangle\quad,\nonumber\\
\nonumber\\
T|{\cal P}(\vec{p}, s_z, Q_i) \rangle &=& \eta_T \,(-1)^{s-s_z} |{\cal P}(-\vec{p}, -s_z, Q_i) \rangle\quad,\nonumber\\
\nonumber\\
C|{\cal P}(\vec{p}, s_z, Q_i) \rangle &=& \eta_C  |{\cal P}(\vec{p}, s_z, -Q_i) \rangle\quad,
\label{CPT}
\eeq
with $\eta_P, \eta_T, \eta_C$ the respective phases. For the $\pi^+$, which is  $J^{PC}=0^{-+}$, are
$\eta_P=\eta_T=-1,\eta_C=1 $.
\\
In the light-front, the $T$ symmetry as defined by the second expression of Eq.~(\ref{CPT}) changes $x^+$ to $-x^-$ and vice versa. It is therefore useful to define the $V$ symmetry has the combination $P_z T$, what gives
\beq
V|{\cal P}(\vec{p}, s_z, Q_i) \rangle &=& \eta_V  \,(-1)^{s-s_z}|{\cal P}(\vec{p}, -s_z, Q_i) \rangle\quad,
\eeq
with $\eta_V=\eta_P\, \eta_T$.

In Appendix~\ref{sec:sym} are developed in details the symmetry relations for any type of distributions. 
We here relate such relations for the pion DA as given by Eq.~(\ref{pda}).
 Using the relations Eq.~(\ref{CPT}), Time reversal invariance leads to
 \beq
\left ( \phi^{\pi^+}(x) \right)^{\ast} &=& \phi^{\pi^+}(x)\quad;
 \eeq
 charge conjugation to
  \beq
 \phi^{\pi^-}(x) &=& \phi^{\pi^+}(1-x)\quad;
 \label{CC-phi}
 \eeq
and $CPT$ to
 \beq
\left ( \phi^{\pi^-}(1-x) \right)^{\ast} &=& \phi^{\pi^+}(x)\quad.
 \eeq
 
As perfectly illustrated on Fig.~\ref{da-plot}, the pion DA is symmetric around $x=1/2$ so that $\phi (x)=\phi(1-x)$ as required by isospin: it is easily seen from Eq.~(\ref{k+int}) that it is implied by the fact that the quarks carry  the  same constituent mass. By Eq.~(\ref{CC-phi}), it means that there is no difference between the DA for a $\pi^+$ or a $\pi^-$. This also corresponds to the isospin relation: the distribution amplitude for a $u$-quark is related by the change $x\to 1-x$ to the amplitude for a $d$-quark. 

Those relations remain unchanged under evolution.


\section{The Pion Parton Distribution Function}

\begin{wrapfigure}[9]{l}{7.5cm}
\centering
\includegraphics [height=2.3cm]{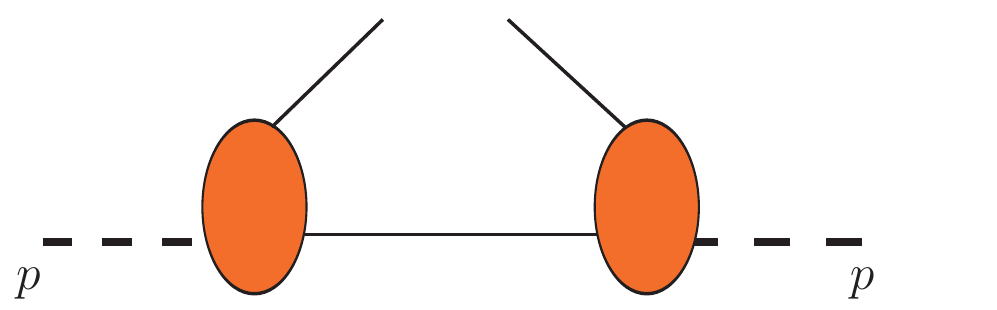}
\caption{The pion Parton Distribution Function. The orange blobs represent the pions.}
\label{bs-pdf}%
\end{wrapfigure}

The nomenclature of distribution functions includes a huge variety of transitions. As overviewed in the Introduction, the names are mostly related to the nature of the final and initial states. For instance, the  DA is a transition from a physical state to the vacuum.
Also, by definition, the pion Parton Distribution Function is the probability density to find a quark carrying a fraction $x$ of the parent pion's momentum, i.e. 
\beq
 \int\, \frac{dz^-}{2\pi} \, e^{i\, (x-\frac{1}{2})\,z^-p^+}
 \langle \pi^+(p) \vert \bar{q}\left (-\frac{z}{2}\right)\,\thru n\,\frac{1}{2}(1+\tau^{3})\,\,q\left (\frac{z}{2}\right) \vert \pi^+(p) \rangle
&=& \frac{1}{p^+}\,q(x)\quad,
\eeq
as illustrated on Fig.~\ref{bs-pdf}.
In the NJL model we obtain
\beq
q(x)&=& 4\, N_c g_{\pi qq}^2\, \left[
-\frac{1}{2}\tilde I_{2,P'}-\frac{1}{2}\tilde I_{2,P}(x, 0)+ m_{\pi}^2\, x\, \tilde I_3^{GPD} (x, 0, 0)
\right]\quad,
\eeq
with the $2$- and $3$-propagator integrals defined Eqs.~(\ref{i2ppr}, \ref{i2p}, \ref{i3pgpd-a}).
In the chiral limit, $q(x)$ is equal to~$1$.
 The NJL model has been applied to the study of pion Parton Distribution in different occasions~\cite{Davidson:1994uv,
Davidson:2001cc, Ruiz Arriola:2002bp}. 
More elaborated studies of pion
Parton Distribution have been performed in the Instanton Liquid Model \cite{Anikin:2000rq}, and in
lattice calculation inspired  nonlocal
Lagrangian models~\cite{ Noguera:2005ej, Noguera:2005cc}, which confirm that the result obtained in
the NJL model for the Parton Distribution is a good approximation. 
\\

A low-energy  theorem  based on  PCAC\footnote{Partial Conservation of the Axial Current; see Section~\ref{sec:pcac} in Appendix~\ref{sec:njl}.} links the pion DA and the  distributions of  two physical pion states.
\\
Following the same steps as in Ref.~\cite{Donoghue:1992dd}, we  hereafter derive this relation between the pion DA $\phi(x)$ and the pion PDF $ q(x)\vert_{p^{\mu}_{\pi_2}\to 0}$, which leads to the particularly interesting result
\beq
\phi_{\chi}(x)=q_{\chi}(x)\quad,
\label{softpion-chi}
\eeq
where $\chi$ means $m_{\pi}=0$  MeV for the pion under scrutiny.
\\
Let us take the matrix element of a pion-pion transition with momentum transfer $p_1-p_2$; we call it $f^{ab}$.
Such a matrix element obeys the following isospin decomposition ~\cite{Polyakov:1998ze} 
 \beq
 f^{ab}&=& \delta^{ab}\,  f^{I=0}+\frac{1}{2}\, \mbox{tr}([\tau^a,\tau^b]\,  \tau^c)\, f^{I=1}\quad.
 \nonumber
 \eeq
 It is related to the $u-$ and $d-$quarks distributions by $f^u+f^{\bar d}$.  We consider the isovector  contribution of such a
  matrix element
 \beq
\left \langle \pi^b(p_2) \left\vert \bar{q}(x) \,\hat n\,\,\frac{\tau^{3}}{2}\,\,q(0)  \right\vert \pi^a(p_1) \right\rangle\quad.
\label{pi-pi-matrix}
 \eeq
\\
Using the Lehmann-Symanzik-Zimmermann reduction formula we can write
 \beq
&& \langle \pi^b(p_2)\vert \bar{q}(x) \,\hat n\,\,\frac{\tau^{3}}{2}\,\,q(0)  \vert \pi^a(p_1)\rangle\nonumber\\
&&= i\,\int d^4x\, e^{-i p_2\cdot x} \, \left (  \Box_y^2+m_{\pi}^2 \right) \, \langle 0 \vert \, T\, \left \{\pi^b(y)\,\bar{q}(x) \,\hat n\,\,\frac{\tau^{3}}{2}\,\,q(0) \right\} \vert \pi^a(p_1) \rangle\quad,\nonumber
 \eeq
where $\pi^j(x)$ is the pion interpolating field and is given by PCAC,  Eq.~(\ref{pcac}). We now integrate twice by parts and use the relation for the derivative of the Time-ordered product~$\partial_{t'} T\{ \varphi(x')\varphi^{\dagger}(x)\}=T\{ (\partial_{t'}\varphi(x'))\varphi^{\dagger}(x)\} +\delta(t'-t)\,[ \varphi(x'),\varphi^{\dagger}(x)]$. %
\\
Taking the soft pion limit, i.e. $p_2^{\mu}\to 0$, the matrix element (\ref{pi-pi-matrix}) becomes
\beq
 &&\lim_{p_2^{\mu}\to 0}\langle \pi^b(p_2)\vert \bar{q}(x) \,\hat n\,\,\frac{\tau^{3}}{2}\,\,q(0)  \vert \pi^a(p_1)\rangle\nonumber\\
&&\hspace{2cm} = \frac {-i }{f_{\pi}} \int d^4x
\left[
 \delta(y_0-x_0)\langle 0 \vert [A^{0 b} (y), \bar{q}(x)] \,\hat n\,\,\frac{\tau^{3}}{2}\,\,q(0)\vert \pi^a(p_1) \rangle \right. \nonumber\\
&&\hspace{4.3cm} 
\left . +\delta(y_0)\langle 0 \vert \bar{q}(x) \,\hat n\,\,\frac{\tau^{3}}{2}\, [A^{0 b} (y),q(0)]\vert \pi^a(p_1) \rangle\right]\quad.\nonumber\\
\label{softpion}
\eeq
The axial current gives rise to the axial charge 
$Q^a_5=\int d^4 x\,  \delta(x_0 )A^a_0(x)$ with the equal-time commutators obeying 
%
$[Q_5^a,q(x)]= i/2\,\tau^{\pi^a}\, \gamma_5\, q(x)$. %
\\
The expression (\ref{softpion}) becomes
\beq
\lim_{p_2^{\mu}\to 0}\langle \pi^b(p_2)\vert \bar{q}(x) \,\hat n\,\,\frac{\tau^{3}}{2}\,\,q(0)  \vert \pi^a(p_1)\rangle
&=& \frac {1 }{f_{\pi}} \left[ \langle 0 \vert \bar q(x)\, \sqrt{2}\frac{\tau^b}{2}\gamma_5\,\hat n\,\,\frac{\tau^{3}}{2}\,\,q(0)\vert \pi^a(p_1) \rangle \right. 
\nonumber\\
&&\hspace{1.5cm} 
\left . +\langle 0 \vert \bar{q}(x) \,\hat n\,\,\frac{\tau^{3}}{2}\,\,\sqrt{2} \frac{\tau^b}{2}\gamma_5\,q(0)\vert \pi^a(p_1) \rangle\right]\quad,\nonumber\\
&=& \frac {i\,\epsilon^{3bc} }{f_{\pi}} \,\langle 0 \vert \bar{q}(x) \,\hat n\, \sqrt{2}\frac{\tau^c}{2}\gamma_5\,q(0)\vert \pi^a(p_1) \rangle\quad,
\eeq
where we have used the algebra $[{\tau^{3}/2}\, , \,{\tau^b/2}]=i\,\epsilon^{3bc}\,{\tau^c/2}$.
\\

In order to go to the pion Distribution Functions, we Fourier transform Eq.~(\ref{pi-pi-matrix}) for quarks at a light-light separation and replace $p_1$ by $p$ and make the transition diagonal in isospin and in momentum.
In the soft pion limit, the isovector pion PDF is exactly the pion DA
 \beq
&&\lim_{p_{\pi_2}^{\mu}\to 0} \int\, \frac{dz^-}{2\pi} \, e^{i\, (x-\frac{1}{2})\,z^-p^+}
 \langle \pi_2^+(p)\vert \bar{q}\left (-\frac{z}{2}\right)\,\gamma^+\,\frac{\tau^{3}}{2}\,\,q\left (\frac{z}{2}\right) \vert \pi_1^+(p) \rangle\nonumber\\
\nonumber\\
&&\hspace{1cm}= \frac {-i}{\sqrt{2}\,f_{\pi}}\int\, \frac{dz^-}{2\pi} \, e^{i\, (x-\frac{1}{2})\,z^-p^+} 
 \langle 0 \vert \bar{q} \left(-\frac{z}{2}\right) \,\gamma^+\gamma_5\, \tau^-\,q \left(\frac{z}{2}\right) \vert \pi_1^+(p) \rangle\quad,
\label{pdf}
 \eeq
 what corresponds to the definition of the pion DA, Eq.~(\ref{pda}).
 This results takes all its physical meaning in the chiral limit $m_{\pi_1}=m_{\pi_2}=0$, where both pions are chosen to have zero mass by consistency. The PDF actually has to be diagonal in state. If one of the pion is massless ($p_{\pi_2}^{\mu}\to 0$ implies that $p_{\pi_2}^2=0$), the other pion has to be massless as well. We find the expected result of Eq.~(\ref{softpion-chi}).\\

\section{Towards the  Support Problem}
\label{sec:support}

We now focus on the support property that is easy to illustrate in the context of  2-propagator integrals. Nevertheless the discussion that follows is extended to 3-propagator integrals. \\

The support in $x$ is determined by  the pole structure in $k^-$ of the function given in Eq.~(\ref{2-prop-da}).
Developping this integral, Eq.~(\ref{i2p}), leads to
\begin{eqnarray}
&&i\,\sqrt{2}f_{\pi}\phi(x)\nonumber\\
&&=-i\, 4m\,p^+\,ig_{\pi
qq} N_c \sqrt{2}\,\frac{1}{(2\pi)^2}\int \frac{d^2
\vec{k}_{\perp}}{(2\pi)^2} \,dk^-\,\nonumber\\
&&\frac{1}{2p^+ (1-x)\left(k^--\frac{\vec{k}_{\perp}^2+m^2}{2p^+
(1-x)}-\frac{i\epsilon}{2p^+ (1-x)}\right) \, 2p^+ x \left(p^--k^-
-\frac{(\vec{p}_{\perp}-\vec{k}_{\perp})^2+m^2}{2p^+
x}+\frac{i\epsilon}{2p^+ x}\right)}\quad .\nonumber\\
\label{k+int}
\end{eqnarray}
The integral over $dk^-$ is performed in the complex plane making use of the Cauchy's theorem.
Would $x \thru \in [0,1]$, both the poles would be on the same side of the real axis and the integral would be identically zero.
The definition Eq. (\ref{pda}) therefore leads to the
required support in $x$, i.e.  $x \in [0,1]$. 

It is true as long as the regularization scheme that we use is covariant. This condition is necessary but not sufficient: the regularization procedure should preserve the symmetries of the light-cone instead.  Regularizing Eq.~(\ref{k+int}) through a 3-momentum cutoff would break covariance, spoiling the required support. However, it is obvious that a cutoff of the type $\vec{\Lambda}=(\Lambda^-, \vec{\Lambda}_{\perp})$, with $\Lambda^-< \infty$ prevent us from closing the contour in the complex plane of $k^-$. This leads to support corrections of the order of $1/\Lambda^-$.
Therefore the loss of light-cone invariance  here produces  the loss of the correct support. 
\\

Many calculations of PDFs show problems with the support properties.
In particular, the support problem occurs in calculations performed in non relativistic (NR) Constituent Quark Models (CQM), e.g. Ref.~\cite{Traini:1997jz}. In such NR calculations, the light-cone coordinates are  defined from the 3-momentum and the energy, $k^2_0=m^2+\vec{k}^2$, so that the dependence on $k^-$ is not the genuine one. A slight support violation  is observed. In the same Reference~\cite{Traini:1997jz}, the inclusion of Final State Interactions is proposed to restore translational invariance and therefore the support, which in turn will restore the number of particle  conservation.  A trivial solution to the support problem would be to change the normalization in order to rescale the support to compensate for the problem.
In the same line, we  also mention that a slight violation of the support is  found in  the MIT bag model  e.g. Ref.~\cite{Signal:1989yc}.  
\\

Let us now comment on the role of the interaction on the support problem, staying in the description of the hadron as a bound state in the sense of Bethe-Salpeter.

The vertex function $\Phi (k, p)$  defines the BS amplitudes, which are afterwards plugged into the  distribution in the BS approach~(\ref{da-bs}). The former is defined within a model with its given symmetries; whereas the latter requires covariance on the light-cone. It is important to realize that, in order to preserve the support property,  one has to go to the light-front without breaking its covariance~\cite{Noguera:2005cc}, a constraint that is respected by NJL model.
 
 The BS equation for the pion has been studied using different vertices, e.g.,~\cite{Choi:2001fc, Choi:2002ic, Tiburzi:2001je, Tiburzi:2002sw, Tiburzi:2002sx,VanDyck:2008ap}.  In the five first References, use has been made of a reduction of the BS amplitudes to light-cone wave functions by projecting on the light-cone. In Refs.~\cite{Choi:2001fc, Choi:2002ic}, the light-front BS vertex functions were replaced by wave functions obtained in a light-front CQM. In Refs.~\cite{Tiburzi:2001je, Tiburzi:2002sw, Tiburzi:2002sx}, the scalar Wick-Cutkosky model was adopted and successfully applied to the calculation of scalar meson GPDs. The correct support has been found. It is however not the case in Ref.~\cite{annelies, VanDyck:2008ap}  where the calculation is performed in a covariant CQM, i.e., the Bonn model.  Once more, the problem  is initially non covariant but  the authors  attempted to restore Lorentz invariance by  applying a boost;  attempt which was nonetheless either too schematic or incomplete.
 In Ref.~\cite{VanDyck:2007jt} a detailed study of the problem suggests that the dependence upon the energy, $k^-$, of the vertex is responsible for the introduction of additional dynamical poles.  
 As a counter example we mention the non-local Lagrangian~\cite{Noguera:2005ej} used to calculate PDFs in Ref.~\cite{Noguera:2005ej, Noguera:2005cc}. The non-locality of the interaction obviously leads to a $k^-$-dependent vertex, whose definition is chosen consistently with the definition of the mass $m(p)$, given there by lattice QCD. The problem is totally covariant and the support is recovered in the physical cases. Since the calculation in non-local models is performed in the Euclidean space,  it is the Wick rotation  that, in this case,  might spoil the support property.
 
 It is therefore tempting to explain the fulfillment  of the support property by the conservation of  covariance on the light-cone.

\section{QCD Evolution}
\label{sec:evol-da}

In a previous Section, we have proven that the pion DA and the PDF coincide in the chiral limit. Nevertheless this result based on PCAC must be strongly broken by QCD evolution when we move from low to high $Q^2$. 
We will now  discuss how evolution changes the PDFs and the DAs. Our first objective will be to determine the scale of the model, which can be fixed from our knowledge of PDFs.

The Parton Distributions here obtained are valid at  a low scale where scaling still holds and our model is still defined. However, one would like to link the results of a model calculation  with QCD. In other words, one would like to know the corrections when going to higher energy.
The start of the scaling violations is given by the so-called evolution equations, which schematically describe the dependence of the distribution functions on $Q^2$
\beq
\phi(x)\stackrel{?} \longrightarrow \phi(x,Q)\quad.\nonumber
\eeq
The pioneer works on the behavior of  Parton Distribution Functions (PDFs) with $Q^2$ were done by  Dokshitzer~\cite{Dokshitzer:1977sg}, Gribov and Lipatov~\cite{Gribov:1972ri, Lipatov:1974qm} as well as by Altarelli and Parisi~\cite{Altarelli:1977zs}. The QCD evolution of the Distribution Amplitudes has been studied by Efremov and Radyushkin~\cite{Efremov:1979qk} as well as by Brodsky and Lepage~\cite{Lepage:1980fj}. This gave rise to the so-called ERBL evolution equations.

The r\^ole of the model calculations is  to provide for the initial conditions  to the evolution equations.

\subsubsection{The initial scale from the NJL calculation}

The scale of validity of the model calculation is still to be found. One needs to fix the value of $Q_0$ for which  the quark distributions obtained in the NJL model are considered to be a good approximation of the QCD quark distributions. For the purpose of evolving the parton distributions, we use the code of Freund
 and McDermott \cite{FreundMcDermott}.
 
 The only information we have at hand is the momentum sum rule. Obviously momentum conservation holds and we know that the sum of the momentum fractions carried by each constituent, namely the valence quarks ($v$), the gluons ($g$) and the sea quarks ($s$),   i.e. the second moment of the PDF $f(x,Q)$
 \beq
 \langle x \rangle_y(Q)\equiv \int_0^1 dx\, x \, f(x,Q)\quad
 \eeq
 has to be conserved. In the models we are interested in, only valence quarks are taken into account, so that, 
\beq
\langle x \rangle_v(Q_0)=1\quad,\quad&&\quad \langle x \rangle_s(Q_0)+\langle x \rangle_g(Q_0)=0\quad.
\eeq

To leading order (LO),  scaling violations are manifested in a logarithmical dependence, of the PDFs, on the scale through the running coupling constant
\beq
\alpha(Q^2)&=&\frac{4\pi}{\beta\, \mbox{ln} (Q^2/\Lambda^2)}\quad,
\eeq
with $\beta=11/3\, N_c-2/3 \,N_f$ and $\Lambda$ is the scale of QCD. In order to find the value of $Q_0$, we need to fix the value of $\Lambda$ consistently with the evolution code; i.e. we choose
$
\Lambda=0.174\,\mbox{GeV}$.  Following  the NLO evolution, one is to use $\Lambda=0.246\,  \mbox{GeV}$.

Knowing that the
momentum fraction of each valence quark at $Q=2$ GeV is $0.235$ \cite{Sutton:1991ay}, we  fix the initial point of the evolution
in such a way that the evolution of the second moment of
the pion Parton Distribution  reproduces this result. This condition is fulfilled
at a rather low value, i.e. 
\beq
Q_{0}=0.29 \mbox{GeV}\,\,&,& \quad\mbox{for the LO evolution}\, ;\nonumber\\
Q_{0}=0.43  \mbox{GeV}\,\,&,& \quad\mbox{for the NLO evolution}\, .\nonumber\\
\label{modelscale}
\eeq
\begin{wrapfigure}[18]{l}{7.5cm}
\begin{flushleft}
\includegraphics[height=6.cm]{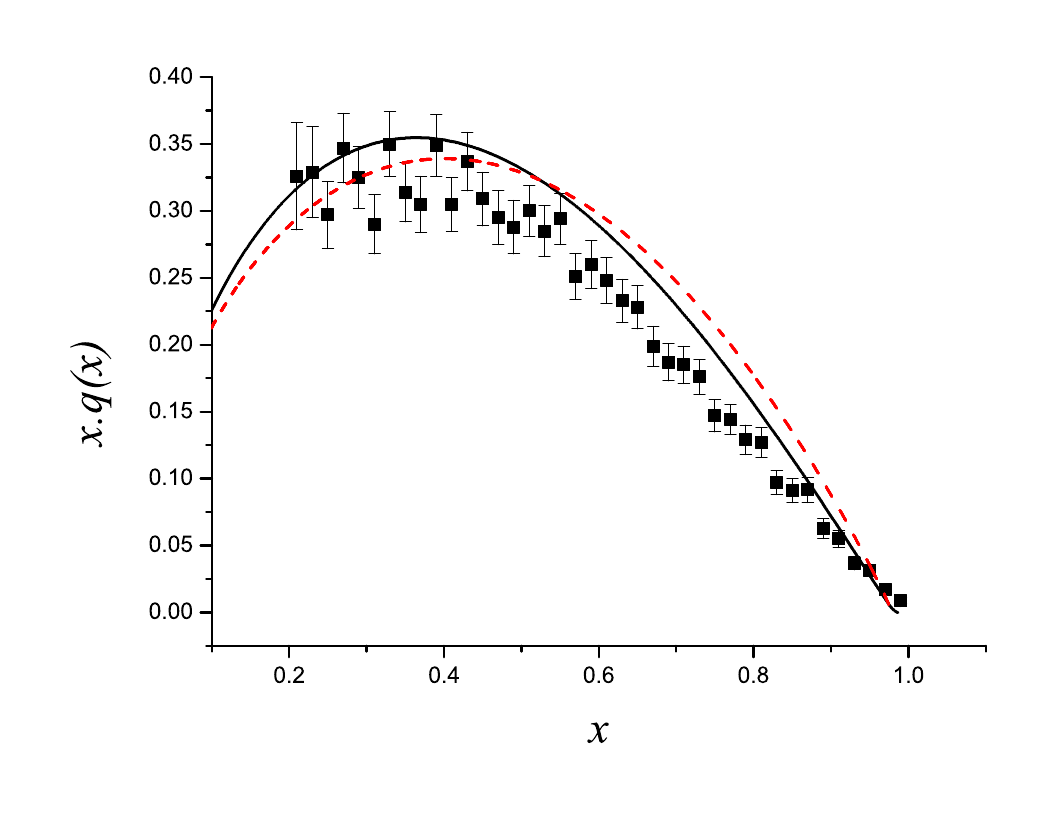}%
\caption{Pion parton distribution. See text. }
\label{PionPDF}%
\end{flushleft}
\end{wrapfigure}
The first result that we obtained is that the  PDF resulting from the NLO evolution is basically unimproved with respect to the one evolved to LO.
 In order to illustrate the latter statement,
 we have depicted the  pion PD evolved at both LO and
 NLO in Fig.~\ref{PionPDF}. The solid (black) line of the Figure corresponds to the LO
evolution of the NJL model prediction while the dashed (red) line  represents the NLO. Both evolved results are compared to the experimental
data~\cite{Conway:1989fs}. The agreement with the data is not better in one or the other case.
 The effect of the NLO evolution is compensated in the LO evolution
going to a lower value of $Q_{0}$, a result that has already been noticed
in  proton Parton Distributions \cite{Traini:1997jz}.
It is therefore obvious that, for numerical reasons, we will prefer to evolve the distributions to LO.

 The method used here is also applied by the authors of Ref.~\cite{Golec-Biernat:1998ja}.  There exists other ways of fixing the scale, using different data or comparing with lattice data. They are reviewed in the latter Reference.

\subsubsection{The QCD evolution of the Pion DA}
\label{sec:evol-da}

The Distribution Amplitudes  have a logarithmic dependence in $Q$ which is completely determined by the QCD evolution equations derived in Ref.~\cite{Lepage:1979zb, Lepage:1979zb-2, Lepage:1980fj}.  
The QCD evolution equations for the distribution amplitudes can be expressed, to leading order,  in terms of Gegenbauer polynomials
\beq
\phi(x,Q)&=& x(1-x) \sum_{\substack{n=0\\ even}}^{\infty}\, a_n \, C^{3/2}_n(2x-1)\,\left( \mbox{ln}\frac{Q^2}{\Lambda^2}\right)^{-\gamma_n}\quad,
\label{evol-da}
\eeq	
where only even $n$ contribute since $\phi(x,Q)=\phi(1-x,Q)$ is required by isospin.
The anomalous dimensions are 
\beq
\gamma_n&=& \frac{C_F}{\beta}\, \left(  1+4\sum_2^{n+1}\frac{1}{k}\, -\frac{2}{(n+1)(n+2)}\right)\quad,
\eeq
where $C_F=(N_c^2-1)/(2N_c)$ and $\beta=11/3\, N_c-2/3 \,N_f$.
\\
The evolution equations would be completely known if the coefficients $a_n$ were determined. These coefficients  can be determined from  the initial conditions which are provided by the model calculation.  The evaluation of the  DA in the model provides us for the $\phi(x_i,Q_0)$. By using this initial distribution together with 
the orthogonality relations for the Gegenbauer polynomials,
\beq
a_n\, \left( \mbox{ln}\frac{Q_0^2}{\Lambda^2}\right)^{-\gamma_n}= 4\,\frac {2n+3}{(n+1)(n+2)}\, \int^1_0 dx\, C^{3/2}_n(2x-1)\,\phi(x, Q_0)\quad, \quad n=\mbox{even}\quad,
\label{ortho-gegen}
\eeq
one might be able to find the $a_n$ coefficients, namely, to determined the evolution equations.

\begin{figure}[H]
\centering
\includegraphics [height=6.5cm]{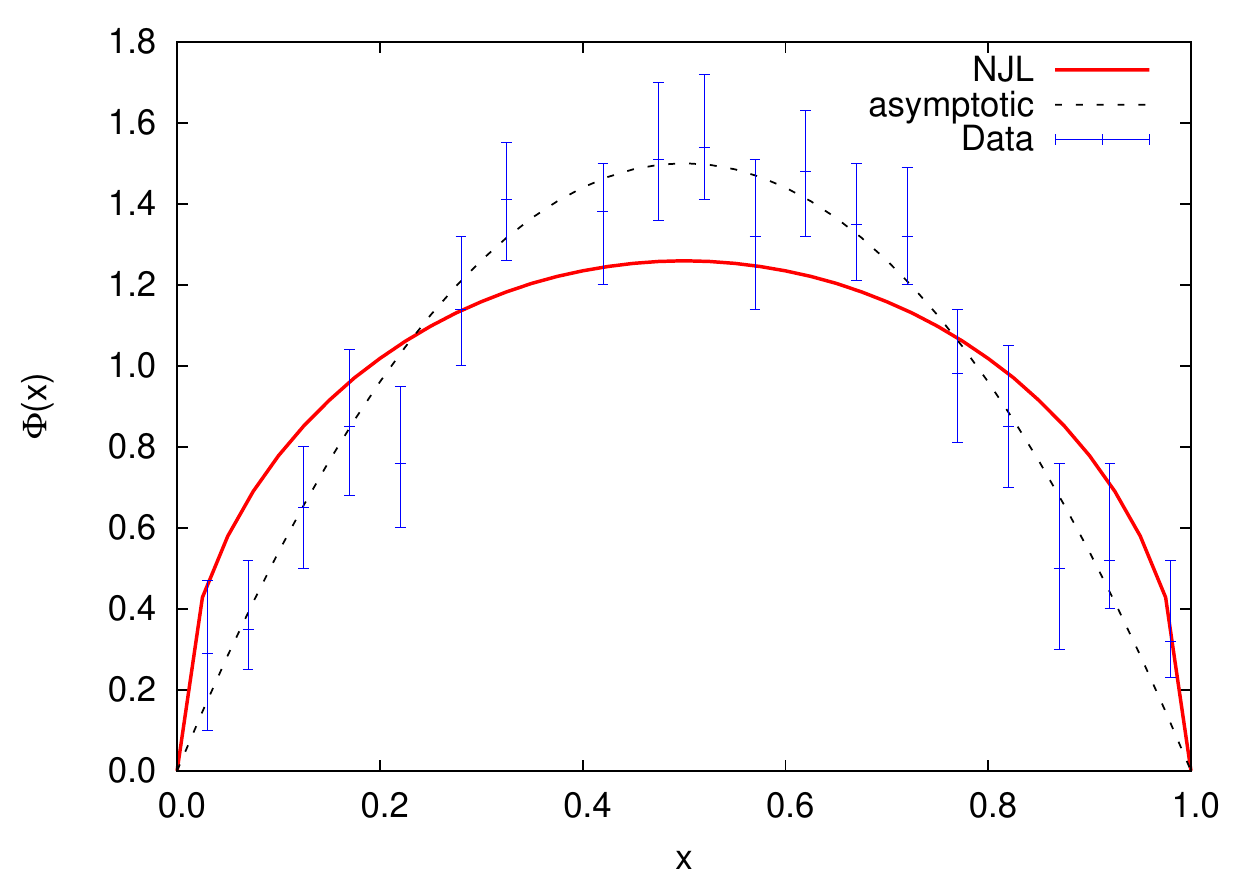}
\caption{ The pion DA in the NJL model  evolved to the scale $Q=2$ GeV (red curve) and compared to E971 di-jet measurement~\cite{Aitala:2000hb} after proper normalization of the data Ref.~\cite{Broniowski:2007si}. The dashed line represent the asymptotic DA.}
\label{da-evol-rab}%
\end{figure}
\begin{wrapfigure}[31]{l}{90mm}
\centering
\includegraphics [height=5.5cm]{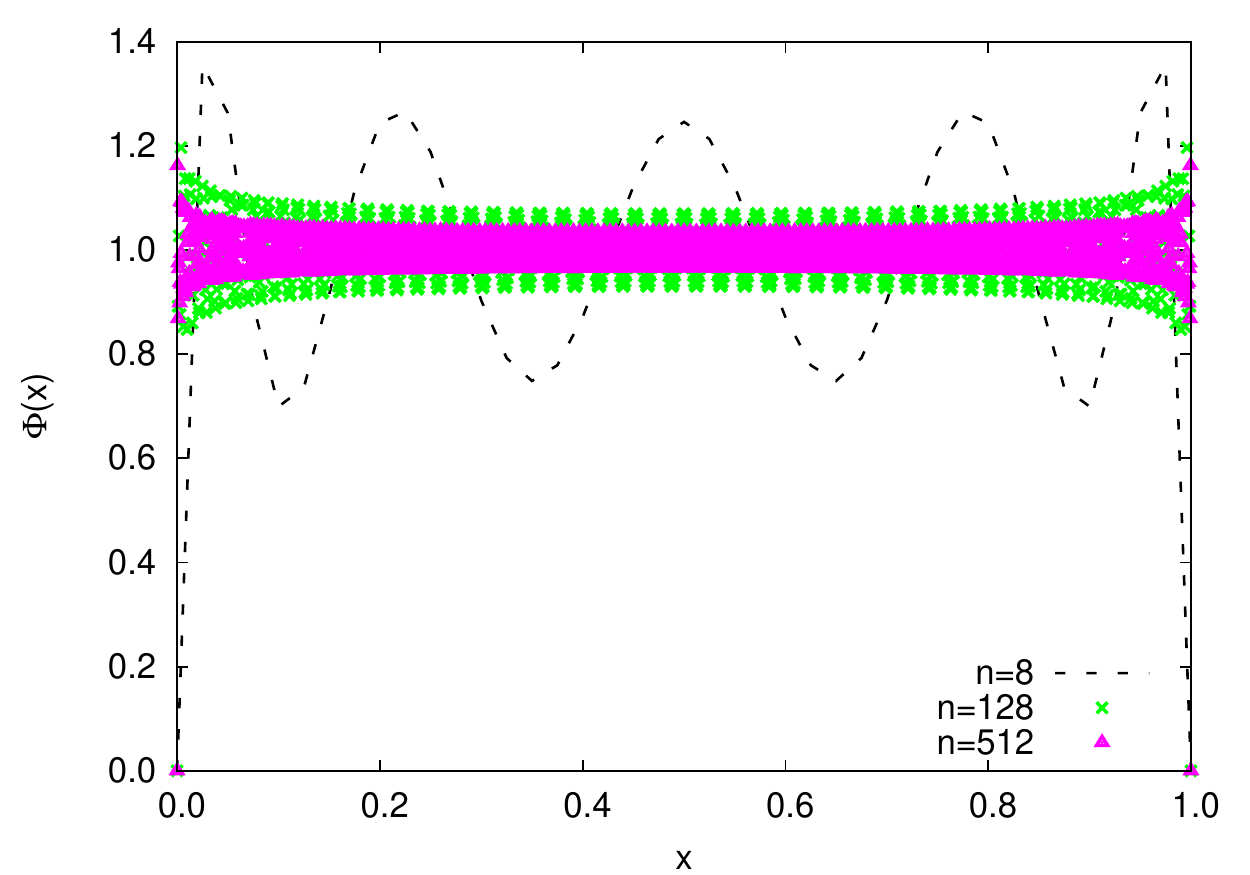}
\caption{ The truncated series~(\ref{evol-da-njl}) for  the pion DA in  the NJL model  at  $Q_0$.}
\label{da-evol-q0}%
%
%
%
%
\vspace{.1cm}
\centering
\includegraphics [height=5.5cm]{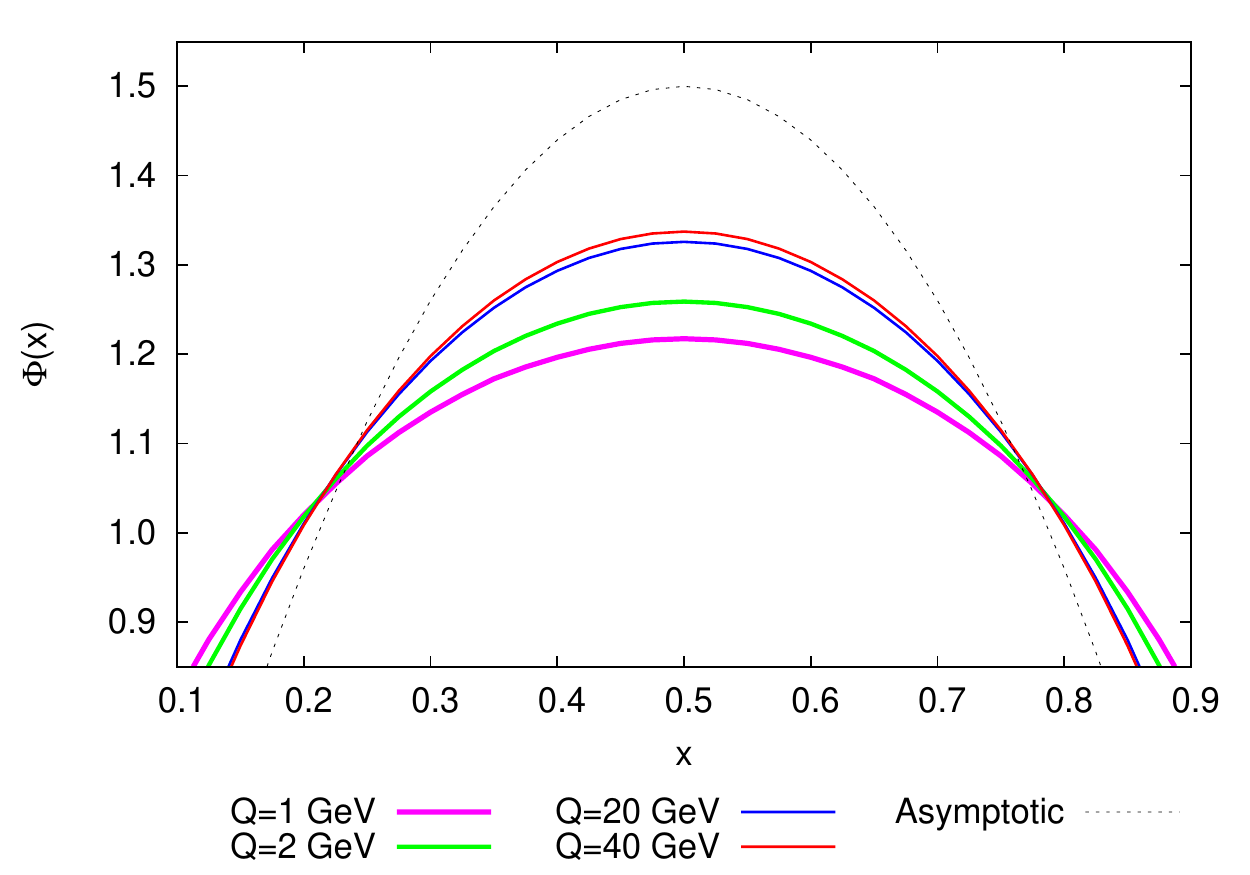}
\caption{ The pion DA in  the NJL model  evolved to the scales $Q=1,2 , 20\, \& \, 40$ GeV as well as the asymptotic form. }
\label{da-evol-us}%
\end{wrapfigure}

All the model calculations, after evolution, should agree with the theoretical asymptotic behavior of the DA, that is given by Eq.~(\ref{evol-da}) with $Q^2\to \infty$. Since $\gamma_0=0$ and  $\gamma_n > 0$ for positive $n$, only the first term, i.e. $n=0$, survives
\beq
\phi (x, Q)&\stackrel{Q^2\to \infty}=& a_0\, x(1-x)\quad.
\label{phi-asympt}
\eeq
Using the orthogonality relations for the Gegenbauer polynomials Eq.~(\ref{ortho-gegen}) and the renormalization of the charged weak current, it is found in Ref.~\cite{Lepage:1980fj} that $a_0=6$.
We can find the other coefficients through the initial conditions given by our result in the NJL model.
As mentioned in Section~\ref{sec:da-cal}, in the chiral limit, the pion DA  is a constant normalized to $1$
\beq
\phi (x, Q_0)&=&1\quad,
\label{phi0}
\eeq
what holds at the scale of the model, $Q_0$.
It is then easy to apply the above-described formalism and  determine  the $a_n$ for our model calculation.
Plugging Eq.~(\ref{phi0}) into the relation (\ref{ortho-gegen}), we find\footnote{$\int^1_0 dx\, C^{3/2}_n(2x-1)=1$ for $n=$ even.}
\beq
\phi(x,Q)&=& 4\,x(1-x) \sum_{\substack{n=0\\ even}}^{\infty}\,\frac {2n+3}{(n+1)(n+2)}\, C^{3/2}_n(2x-1)\,\left(\frac{ \mbox{ln}(Q^2/\Lambda^2)}{ \mbox{ln}(Q_0^2/\Lambda^2)}\right)^{-\gamma_n}\quad.
\label{evol-da-njl}
\eeq
\\
At  a scale $Q_0$, the series of the previous result should give $1$. However, it must be realized that this result is reached for $n\to\infty$ which is numerically not feasible.  At a low scale $Q_0< Q\lesssim 1$ GeV, where changes due to evolution are more important, one should include a high number of terms of the series. As proven by Fig.~\ref{da-evol-q0}, the desired result is reached by increasing the $n$ value. It nevertheless takes an infinite number of terms to compensate for $0$ going to $1$ at the end-points.
\\
On the other hand, once QCD is switched on, the end-point problems for the pion DA calculated in the NJL model in the chiral limit is solved, showing the need for QCD evolution.

The pion DA evolved to a scale of $Q=2$ GeV  is shown in  Fig.~\ref{da-evol-rab} for the NJL model with  Pauli-Villars regularization. It is compared to the data of the E791 measurement~\cite{Aitala:2000hb} once properly normalized~\cite{Broniowski:2007si}. The asymptotic DA is also plotted that seems to be in a better agreement with the data. 

In our version of the NJL model, the scale of the model is found to be as low as $Q_0=290$ MeV at LO, see Eq.~(\ref{modelscale}). At high $Q$, one could take into account only the 4th first terms of the series. This approximation is justified by the fact that the coefficients $a_n$ decrease with increasing $n$ and by the asymptotic behaviour in $Q$.
However, for a better resolution and smoothness of the plots, we choose to truncate the series at $n=512$. 
\\
In order to illustrate the result obtained in Eq.~(\ref{evol-da-njl}), we plot  the series in Fig.~\ref{da-evol-us} for different, high  values of $Q$. 
For $Q$ higher than $1$ GeV (pink curve), the effect of evolution starts being less important. The green, blue and red lines represent the LO order evolved at, respectively, $Q=2$ GeV, $Q=20$ GeV and $Q=40$ GeV. The result changes slowly showing that no big improvement to reach the asymptotic value can be done by this model calculation at least at LO evolution. 
One can however  easily check that the analytical asymptotic behavior for the series.~(\ref{evol-da-njl}) is $\phi (x, Q)=6\, x(1-x)$, as required.

\section{End-points}
\label{sec:endpoints}

Although its r\^ole seems to be only that it provides for the initial conditions,  a model calculation is a calculation that has a physical meaning.  As a consequence, the results in the model are supposed to obey all the properties of the initial object.
Turning the problem over:  the results of model calculations are not complete until evolution is performed and most of the properties of Parton Distributions that were respected at the quark-model scale, e.g. isospin relations, will still be after evolution. On the other hand, we will show that other intrinsic properties can be restored. 
\\

In particular, one could evoke the problem of  the distribution functions at the end-points. 
Effectively, in the free parton model, PDFs are expected to vanish at the boundaries, i.e. $x=0, 1$.
 For on-shell particles, at $x=1$ one of the quarks carries the whole $(+)$-component of the momentum and the other quark has zero $(+)$-component of the momentum, i.e. on-shellness implies an infinite $(-)$-component of the momentum. Since the $(-)$-component of the momentum appears in the propagator, the PDF vanishes. Also, this end-point behavior has been  analytically predicted in perturbative QCD, e.g. \cite{Brodsky:1994kg}.

 It is however not what happens for the results in lots of models such as  the NJL model   illustrated in, e.g.,  Fig.~\ref{da-plot}.
This {\it unwelcome} behavior comes from the misunderstanding of the Nature of the quarks beyond the na\"\i ve  parton model. 
 It is  effectively unclear how the constituent quarks of the quark models can be linked to the current quarks of QCD. We are willing phenomenological evidences to shed some light on the  constituent density distributions. The scenario suggested by   Altarelli, Cabibbo, Maiani and Petronzio (ACMP)~\cite{Altarelli:1973ff} pictures the constituent quark as a complex system of point-like partons. This ansatz is so far one of the only for those distributions. A variation of this scenario has been applied in,   e.g., Refs.~\cite{Noguera:2004jq, Noguera:2005cc, Scopetta:1997wk, Scopetta:1998sg}, showing a huge improvement.

 As for the calculation in the  NJL model, the end-point problem shows the limits of validity of our model calculation.
 The simple considerations given above do not hold for the bound-state quarks in the NJL model; and as a matter of fact we cannot expect the Parton Distributions to vanish at the boundary.
This behavior has been shown to arise in a regularization-independent way as they appear in both the results of Refs.~\cite{Broniowski:2007si, Theussl:2002xp}.

Turning our attention to the chiral limit, the question could also be formulated as follows:  what is, in  the NJL model,  the probability density of finding  -massless(current)- quarks with momentum $x$ inside a massless pion?  The quarks  can, a priori, not carry a particular fraction of the massless pion momentum, so that the probability density does not depend on $x$, explaining at the same time the end-point behavior and the result $q(x)=1$. 
This problem is  the opposite of the  zero bounding energy problem, where the mass of the pion is chosen to be twice that of the constituent quarks. In that case, the distribution is rather peaked in $x=1/2$ as shown by a quick analysis of the kinematics.

As for the treatment of the quarks in the calculation given here, our choice has been to maintain consistency with the model in itself. The relation between the current quarks of the NJL model, with mass $m_0$, and the constituent quarks, carrying a mass $m$, is the one given by the Bogoliubov-Valatin transformation which is proper to the model.~\footnote{See Appendix~\ref{sec:njl}.} Therefore an additional convolution of the type ACMP is not necessary.

We conclude by highlighting the fact that the obtained values about  $q(x)=1$ in the vicinity of $x=0, 1$ enable a good reproduction of the evolved PDF at the end-points, see Fig.~\ref{PionPDF}. On the other hand, similar considerations for the pion DA, e.g.  Fig.~\ref{da-evol-rab}, indicate that the obtained DA is too high at the end-points.

\section{The Pion Electromagnetic Form Factor}
\label{sec:pion-ff}

\begin{figure}[tb]
\centering
\includegraphics[height=4cm]{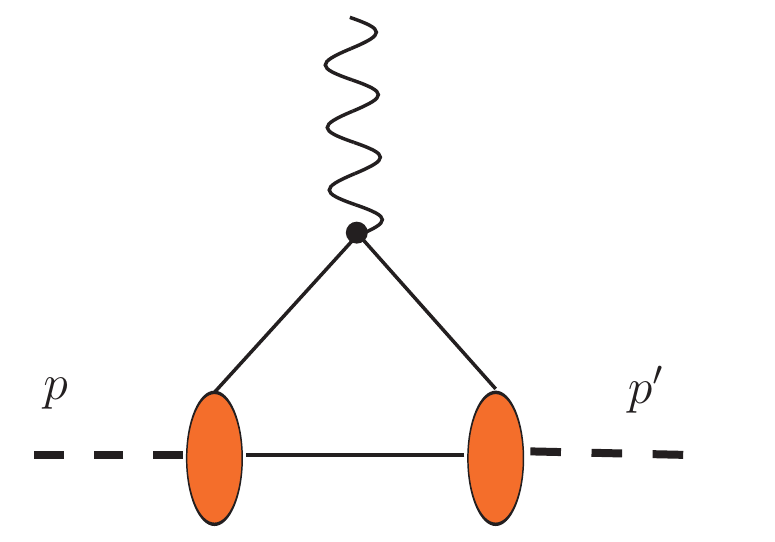}
\caption{The Feynman diagram representing the elastic scattering.}%
\label{elmff}%
\end{figure}

In this Chapter we have examined the meson Distribution Amplitudes as an illustration of the method we will use throughout this thesis. 
The DAs contribute to both elastic and deep inelastic processes, what made the example interesting. In particular, they can contribute to the elastic scattering process Fig.~\ref{elmff}. It is, however,  not the simplest situation for the study of elastic processes:
The simplest objects that allow for the  study of hadron properties are not the meson DAs but  the elastic Form Factors. 
In effect, the Form Factor is  the amplitude for the, e.g., pion to absorb large transverse momentum while remaining intact.
In this Section, we aim to link both quantities and to discuss the asymptotic - and model-independent -  limits of the pion electromagnetic Form Factor. 
\\

For a charged pion $\pi^+$ the electromagnetic Form Factor  $F_{\pi} $ is defined as
\beq
\langle \pi^+(p')| J^{elm}_{\mu}(0)| \pi^+(p)\rangle &=& (p+p')_{\mu} \, e\, F_{\pi} (q^2)\quad,\nonumber
\eeq
with $q^2=(p'-p)^2$; and its charge radius by 
\beq
\langle r^2\rangle _{\pi}&=& -6 \frac{d^2F_{\pi}(q^2)}{d q^2} \Big|_{q^2=0}\quad.\nonumber
\eeq
We evaluate the diagram on Fig.~\ref{elmff} in the NJL model, with both incoming and outgoing particles being pions as well as its crossed version, applying the technique of the Section~\ref{sec:da-cal}. We find
\beq
 F_{\pi} (q^2)&=&\frac{4\,N_c\, g_{\pi qq}^2}{m_{\pi}^2+p\cdot p'}\,\left( -m_{\pi}^2 \,I_2(p)-  p\cdot p'\, I_2(p-p')+ m^4_{\pi}\, I_3(p,p')\right)\quad,
 \label{elm-ff-njl}
\eeq
with the 2- and 3-propagator integrals defined in Appendix~\ref{sec:integral}.
The charge radius evaluated in the NJL model is  found to be $\langle r^2\rangle _{\pi}= 0.31\,\mbox{fm}^2$, which has to be compared to the experimental value of $0.44$ fm$^2$. It must be realized that our model does not include the vector mesons, which  also contribute to the charge radius of the pion, see e.g.~\cite{dorokhov}.
\\

Here, we have used an effective theory of QCD in order to calculate the electromagnetic quantity. The expression (\ref{elm-ff-njl}) is a useful check for the sum rules for GPDs  and will be used in the next Chapter.

\subsubsection{The Asymptotic Limit: the Brodsky-Lepage Form Factor}

The electromagnetic Form Factor  can also be seen as the convolution of three probability amplitudes. Basically, those probability amplitudes are the probability of finding a 2-quark valence state in the incoming pion, i.e. the pion DA $\phi$; a hard amplitude $T_H$ for this 2-quark-state to scatter with the photon and, therefore, to produce 2 quarks which can  reform in similar 2-quark state~\cite{Lepage:1980fj}. The latter is described by a pion DA $\phi^{\ast}$;\footnote{$-q^2=Q^2$.}
\beq
F_{\pi}(Q^2)&=& \int ^1_0 dx_i \, dy_i \, \delta(1-\Sigma_i x_i)\, \delta(1-\Sigma_i y_i) \,\phi^{\ast}(y_i,Q)\,T_H(x_i, y_i, Q)\, \phi(x_i,Q)\quad,
\label{bl-ff}
\eeq
where $T_H$ contains all the 2-particle irreducible amplitude for $\gamma^{\ast}+q\bar q \to q\bar q$.
This approach is similar to the previous one in that the pion DA is well-described by the Bethe-Salpeter amplitude on the light-front. However, the pion DA calculated within this formalism does not explicitly   depend on the scale unless we switch on QCD. Now, the effects of QCD evolution need to be taken into account into the convolution~(\ref{bl-ff}), so that we need the evolved pion DA.

Using the light-cone perturbation theory together with the light-cone gauge, Brodsky and Lepage~ \cite{Lepage:1979zb, Lepage:1979zb-2} obtained the evolution equations of Section~\ref{sec:evol-da}. Those equation can in turn be applied to the calculation of the Form Factors. 
The hard amplitude gives the behavior in powers of $Q^2$. 
For instance, plugging Eq.~(\ref{evol-da}) with $T_H$ evaluated in the same light-cone perturbation theory, they found, to leading logarithmic corrections
\beq
F_{\pi}(Q^2)&=&\frac{4\pi\, C_F\, \alpha_s(Q^2)}{Q^2}\,\left | \sum_{n=0}^{\infty} \, \frac{a_n}{2}\,\frac{f_{\pi}}{\sqrt{N_c}}\, \left(  \mbox{log} \frac{Q^2}{\Lambda^2}\right)^{-\gamma_n}  \right |^2 \,\left [ 1+{\cal O} \left(\alpha_s(Q^2), m/Q\right) \right]\quad,\nonumber\\
\eeq
which shows the most important dynamical feature of the Form Factors, namely, their power-law fall-off. 
The asymptotic limit is given by Eq.~(\ref{phi-asympt}) with known $a_0$
\beq
F_{\pi}(Q^2)&\stackrel{Q^2\to \infty}=& \frac{4\pi\, C_F\, \alpha_s(Q^2)}{Q^2}\,\frac{9\,f_{\pi}^2}{N_c}\quad,\nonumber\\
&\stackrel{Q^2\to \infty}=&16\,\pi \alpha_s(Q^2) \frac{f_{\pi}^2}{Q^2}\quad,
\label{bl-ff}
\eeq
with $N_c=3$.

\chapter{Pion Generalized Parton Distributions}
\label{sec:gpd}
\pagestyle{fancy}

The adjective {\it generalized} in the name Generalized Parton Distributions (GPDs) appeared in the early nomenclature of distribution functions~\cite{Ji:1996ek, Ji:1996nm}. 
Their introduction through the factorization theorems for deep exclusive processes demonstrated the wide applicability of  the diagonal transitions obtained with inclusive processes. 
Deep exclusive processes have also been discussed in terms of  {\it off-forward} parton distributions~\cite{Radyushkin:1996nd} for they are non-diagonal in momentum, i.e. there is a momentum transfer between the initial and final states. The off-forward distributions are univocally related to the GPDs.
The initial and final states we will nevertheless  consider as corresponding to the same quantum numbers when referring to {\it Generalized} Parton Distributions. Such distributions,  with states not corresponding to the same particles,  are called Transition GPDs or Transition Distribution Amplitudes.
 The study of the latter quantities is actually the main topic of this thesis and will be fully developed in Chapter~\ref{sec:tda}.
 \\

In Chapter \ref{sec:df} we have introduced the Generalized Parton Distributions for the nucleons. Nevertheless such bilocal matrix elements can also be defined for mesonic states, e.g. pions. Let us now focus on the study of the pion GPDs applying the rules developed in the NJL model which have been given in the previous Chapter about the Pion Distribution Amplitude. 
\\

The set of distributions proposed for the proton  in Refs.~\cite{ Diehl:2001pm, Diehl:2005jf} is adapted to the pion case including both the chiral-even and the chiral-odd GPDs.
The tensorial structure for the $\gamma^{\ast}\pi \rightarrow \gamma\pi$ process allows a single twist-2 chiral-even GPD. In the light-front, the  tensorial decomposition of this GPD is found, using the recipe of Ref.~\cite{Anikin:2000em, Diehl:2001pm}, from the vector current
\begin{eqnarray}
F^{\pi}(x, \xi, t)&=& \frac{1}{2}\int \,\frac{dz^-}{2\pi}\,e^{ixz^-p^+}\left\langle \pi(P')\left	\vert\bar q\left(-\frac{z}{2}\right)\,\gamma^+\,q\left(\frac{z}{2}\right)
\right\vert \pi(P)\right\rangle |_{\vec{z}^{\perp}=z^+=0}\quad,\nonumber\\
\nonumber\\
&=& \frac{1}{2\,p^+}\,H^{\pi} (x, \xi, t)\,\,(P+P')\cdot n= \frac{1}{p^+} H^{\pi} (x, \xi, t) 
\quad,
\label{GPDNJL}
\end{eqnarray}
with $q(x)$ the quark field and where $2p^{\mu}=(P'+P)^{\mu}$.
On the other hand, the axial current gives rise to a twist-3 GPD, involving the $\gamma^{\perp}\gamma_5$ current.~\footnote{ The tensor structure of the twist-3 GPD is given by
\beq
\tilde F^{\pi} (x, \xi, t)&=& \frac{1}{2}\int \,\frac{dz^-}{2\pi}\,e^{ixz^-p^+}\left\langle \pi(P')\left	\vert\bar q\left(-\frac{z}{2}\right)\,\gamma^{\perp}\gamma_5\,q\left(\frac{z}{2}\right)
\right\vert \pi(P)\right\rangle |_{\vec{z}^{\perp}=z^+=0}\quad,\nonumber\\
\nonumber\\
&=&
\frac{1}{2\,p^+}  \frac{i\epsilon^{\perp\nu\rho\sigma} p_{\nu} n_{\rho}\Delta_{\sigma}}{m_{\pi}^2}\tilde H^{\pi} (x, \xi, t) \quad,
\nonumber
\label{twist3-gpd}
\eeq
where by $\perp$ we understand the component $1$ or $2$.
}
This distribution function will not be studied in this thesis. The effect  of twist-3 distributions is small but, for nucleons, their inclusion is useful for the restoration of  electromagnetic gauge invariance in DVCS, e.g. Ref.~\cite{Anikin:2000em}.  In the pion case, the twist-3 GPD can be associated to a Single-Spin Asymmetry~\cite{Anikin:2001zv}. 
\\

Taking into account  chiral-odd, namely  helicity-flip,  GPDs makes  the number of twist-2 pion GPDs increase. This case will be overviewed in the {\it Future Perspectives} in Chapter~\ref{sec:spinpion}.
\\

 In the literature, different approaches of the non-diagonality -in momentum- have been proposed for the hadronic matrix element
 \beq
\left\langle \pi(P')\left	\vert\bar q\left(-\frac{z}{2}\right)\,\Gamma^{\mu}\,\,q\left(\frac{z}{2}\right)
\right\vert \pi(P)\right\rangle\quad.
\label{hadr-ma}
 \eeq
 
 One bases its analysis of the longitudinal asymmetry on the kinematics of the process. Since the dependence upon the skewness variable is hence explicit,  this approach leads to the so-called {\it skewed} parton distributions~\cite{Ji:1996ek,Ji:1996nm, Radyushkin:1996nd, Radyushkin:1997ki}. Another approach is to write the spectral decomposition of the matrix element of the bilocal operator at a light-like distance without kinematical assumptions on the skewness variable, allowing one to make statements about this very dependence. This second approach leads to the so-called Double Distributions (DD), which have been proposed in Refs.~\cite{Mueller:1998fv, Radyushkin:1996nd, Radyushkin:1997ki}. 
\\

The Nambu - Jona-Lasinio model is an appropriate choice in the context of model calculation of both  the pion  {\it skewed}  PDs and DDs. As above-mentioned the pion can be treated as a bound-state in a fully covariant manner following the Bethe-Salpeter 
approach inside the NJL model. Given the tools exposed in Chapter~\ref{sec:pionda} we are now able to extend the application of the model for the pion to the definition of  GPDs. So, in this Chapter we will aim to analyze the twist-2 pion  {\it skewed} PD following the lines of Ref.~\cite{Theussl:2002xp} and link it to the DD approach~\cite{Broniowski:2007si}.  In particular, semi-kinematical approaches deriving from the original DD as well as the resonance exchange contribution will be investigated.  The QCD evolution already described in the context of Distribution Amplitudes will be overviewed. 

\section{Kinematics for Skewed Parton Distributions}

We define the kinematics of the GPDs following the conventions of Ref.~\cite{Guichon:1998xv}. 
As it has been explained in Chapter~\ref{sec:df},   the momenta in deep processes are expressed in light-cone coordinates.  

\begin{figure}[H]
\centering
 \includegraphics [height=2cm]{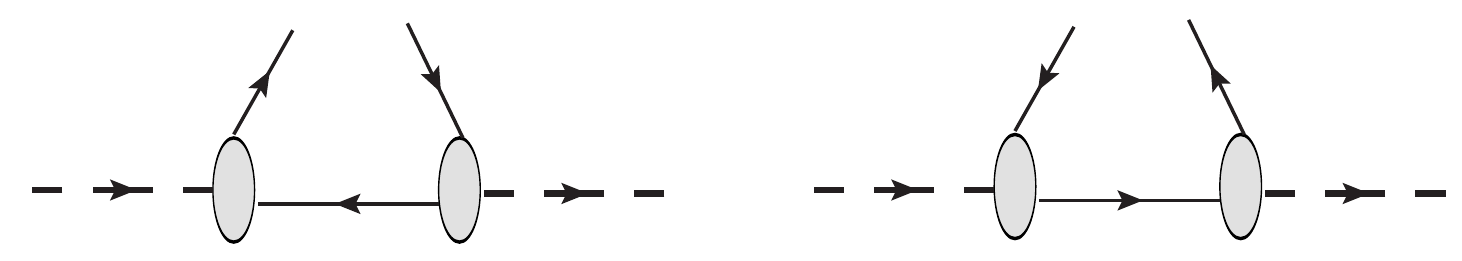}
 \caption{\small{$u$ and $d$-quark contributing diagrams to the  GPD in the NJL model.} }
\label{gpdnjl}
\end{figure}

The GPDs depend on three kinematical variables: the Fourier transform variable $x$, the momentum transfer $t$ and the longitudinal momentum asymmetry, also called skewness variable, $\xi$. The latter is defined, for an incoming momentum $P$ and outgoing one $P'$, by 
\beq
\xi&=& \frac{P^+-P^{'+}}{P^++P^{'+}}\quad.\nonumber
\eeq
The momentum $P$ of the incoming pion  and the momentum $P'$ of the outgoing pion are therefore defined, see Fig.~\ref{gpdnjl}, 
\beq
P^{\mu}&=&(1+\xi) \,\bar p^{\mu}-\,\frac{\Delta^{\perp\mu}}{2}+\,\frac{1-\xi}{2} \bar M^2\, n^{\mu}\quad,\nonumber\\
P^{' \mu}&=&(1-\xi) \,\bar p^{\mu}+\,\frac{\Delta^{\perp\mu}}{2}+\,\frac{1+\xi}{2} \bar M^2\, n^{\mu}\quad,
\eeq
with 
\beq
\bar M^2 &=&\frac{1}{1-\xi^2}\, \left( m_{\pi}^2+\frac{\Delta^{\perp 2}}{4}\right)=m_{\pi}^2-\frac{t}{4}\quad,\nonumber
\eeq
and where we have used the on-shellness conditions $P^2=P^{'2}=m_{\pi}^2$.
$\Delta^{\mu}$  is related to  the momentum transfer $t$ by $\Delta^2=t$, and defined as follows
\beq
\Delta^{\mu}=(P'-P)^{\mu}&=& -2\xi \,\bar p^{\mu}+\Delta^{\perp\mu}+\xi \bar M^2\,n^{\mu}\quad,\nonumber\\
\nonumber\\
t&=& -4\,\xi^2\,\bar M^2-\,\Delta^{\perp 2}\quad;
\label{t-kin}
\eeq
\beq
p^{\mu}=\frac{(P+P')}{2}^{\mu}&=&\bar p^{\mu}+  \,\frac{\bar M^2}{2}\,n^{\mu}\quad. \nonumber
\eeq
From the relation~(\ref{t-kin}) and the fact that $\Delta^{\perp 2}$ is a positive quantity, we can deduce the interval of kinematically allowed values for the skewness variable $\xi$
\beq
0\leq \xi \leq \frac{\sqrt{-t}}{2\,\sqrt{\bar M^2}}\quad.
\eeq

\section{The Chiral-Even  Pion GPD}
\label{sec:gpdpioneven}

The calculation of $H^{\pi}$ in the NJL model  has  been performed in Ref.~\cite{Theussl:2002xp}. In this Section, we comment these results.

For the $\pi^+$, the two contributing diagrams are depicted in \cf{gpdnjl}. The bilocal  $u$-quark current in the BS approach is given by
\begin{eqnarray}
&&\langle \pi^j(P')|\bar u\left(x'\right)\,\gamma^{\mu}\,u\left(x\right)|\pi^j(P)\rangle\nonumber\\ 
&&\hspace*{2cm}=\int d^4 x_2 \, \alltrace\, \Big\{
\bar \chi_{P'}^{j}(x',x_2)\frac{1}{2}(1+\tau_3)\,\gamma^{\mu}\Big[\chi^{j}_P(x, x_2)(i\overset{\leftarrow }{\thru \partial}{}^{(2)}- m_2)\Big] \Big\}\nonumber\\
&&\hspace*{2.3cm}+\int d^4 x_1\alltrace
\Big\{ \bar{\chi}_{P^{\prime }}^{j}(x_{1},x)
(i\overset{\rightarrow }{\thru{\partial}}{}^{(1)}-m_{1})
\chi _{P}^{j}(x_{1},x^{\prime })\frac{1}{2}
( 1+\tau _{3}) \gamma _{\mu }\Big\}\quad,
\label{elmtmatr}
\end{eqnarray}
the $j$ index selects the isospin: $j=\pi^+, \pi^-$ or $\pi^0$. Here the projector  $\frac{1}{2}(1+\tau_3)$ selects  the contributions from the $u$-quark present in \ce{elmtmatr}. 
The two terms on the r.h.s. correspond each one to one of the 2 components of the bound-state, namely the $u$ or $d$-quark, as active quark.~\footnote{We refer to Section~\ref{sec:da-cal} for the details of the technique.}
\\

We now set the quark masses equal to the constituent quark mass, i.e. $m_1=m_2=m$. 
 For the first diagram of \cf{gpdnjl}, namely the active $u$-quark,
the variables become $p_1=P-k$ and $p_2=k$ while for the active $d$-quark the variables are $p_1=-P+k$ and $p_2=-k$. That is, due to isospin symmetry also, we
can relate both diagrams 
\begin{eqnarray}
H_u^{\pi^{\pm}}(x, \xi,t)&=&-H_{d}^{\pi^{\pm}}(-x, \xi,t)\quad.
\label{iso-gpd}
\end{eqnarray}
 We can then concentrate on the calculation of the $u$-quark distribution.
The GPD~(\ref{GPDNJL}) becomes
\begin{eqnarray}
&&H_u^{\pi^+}(x, \xi,t)\nonumber\\
&&= \frac{1}{2}\int\,\frac{d^4 k}{(2\pi)^4}\,\,\,\delta\left(x -1+\frac{k^+}{p^+}\right)\,\alltrace\,\Big\{
\bar \chi^{\pi^+}_{P'}(P'-k)\,\frac{1}{2}(1+\tau_3)\,\gamma^+ \,\chi^{\pi^+}_P(P-k)(-\thru k  -m)
\Big\}\quad.\nonumber\\
\end{eqnarray}
\\
The BS amplitudes are replaced by their expression in terms of Feynman propagators and the pion-quark vertex function~\ce{chi}.

After having performed the traces and integrating over $k^+$,  we can analyze the poles in $k^-$ similarly to the case of 2-propagator integrals in Section~\ref{sec:support}. In the present case it is the interplay between 3 poles that defines the support property: the integral is non-vanishing if there are poles on both sides. Thus, two kinematical regions are found that correspond to a GPD for the $u$-quark of supports $x\in [\xi,1]$ and $x\in [-\xi, \xi]$. From Fig.~\ref{regions}, it is easily seen that the $u$-quark GPD contributes to the region of emission and reabsorption of a quark as well as to the region of emission of a quark-antiquark pair.
 \\
 
We obtain a combination of 3-propagator, as it might be expected, as well as 2-propagator integrals;
\begin{eqnarray}
H_u^{\pi^+}(x, \xi,t)&=& 4\,\,N_c\,\,g^2_{\pi qq}\,\Big[
-\frac{1}{2}(1+\xi)\tilde I_{2,P}(x, \xi)-\frac{1}{2}(1-\xi)\tilde I_{2,P'}- x \tilde I_{2,\Delta}(x, \xi,t)\nonumber\\
&&\qquad\qquad \qquad +(m_{\pi}^2x+\frac{t}{2}(1-x))\tilde I_{3}^{\mbox{\tiny GPD}} (x, \xi,t)
\Big]\quad,
\label{hu-1st}
\end{eqnarray}
with the light-front  integrals given in Appendix \ref{sec:integral}.
\\

For the pion GPD, one should take into account the isoscalar meson-exchange in the $t$-channel: in the ERBL region, the two quarks can couple with quantum numbers corresponding to the $\sigma$-meson. 
Since the NJL model describes the pion as well as the  $\sigma$-meson, the diagrams~\ref{gpdnjl} may already contain this particular exchange. Moreover, in the region $-\xi\leq x\leq \xi$, a new diagram~\cf{sigmau} should be considered as the model's expression of this $\sigma$-exchange.
 This meson-exchange, which  is an explicit manifestation of chiral symmetry, can be calculated in the Random Phase Approximation;
 \begin{eqnarray}
iU(k^2)&=&\,\Big[2iG+2iG(-i\Pi_{s}(k^2))2iG+\ldots\Big]\quad,\nonumber\\
&=&\,\frac{2iG}{1-2G\Pi_{s}(k^2)}\,\quad,
\label{rpa}
\end{eqnarray}
where $\Pi_{s}(k^2)$ is the scalar proper polarization given in \ce{polarscal}. Notice that the expression of the interaction is similar to the one obtained with the Bethe-Salpeter equation in the ladder approximation \ce{ladderint}. The central part of \cf{sigmau} is given exactly by~\ce{rpa}.
\\

\begin{wrapfigure}[16]{l}{6.7cm}
\centering
 \includegraphics [height=4.5cm]{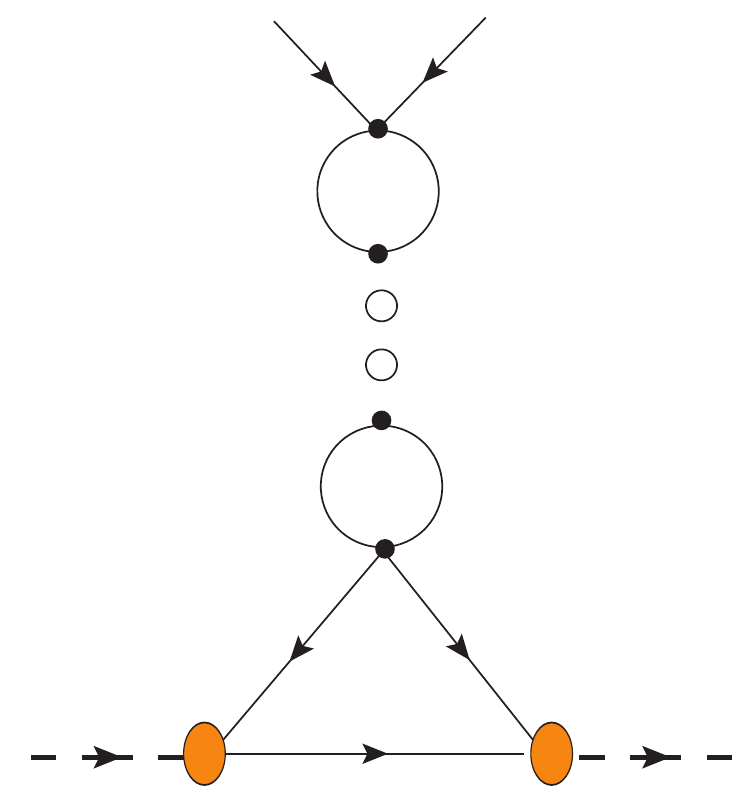}
 \caption[$\sigma$ coupling]{\small{$u$-quark diagram contributing to the GPD in the NJL model coming from the $\sigma$ coupling to the two photons in the $-\xi<x<\xi$
region. The corresponding diagram for the $ d$-quark must be taken into account.}}
\label{sigmau}
\end{wrapfigure}
\begin{figure}[tb]
\centering
 \includegraphics [height=1.5cm]{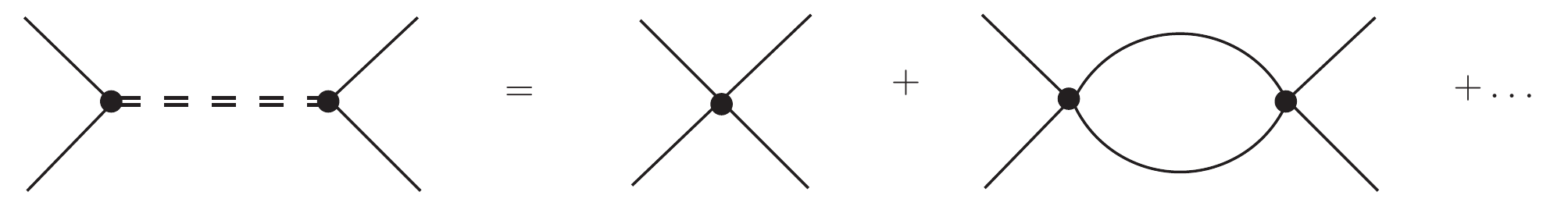}
\caption[RPA]{\small{Effective interaction in the Random Phase Approximation where only the direct terms are considered.}}
\label{rpafig2}
\end{figure}

The second contribution to the GPD for the $u$-quark is then
\begin{eqnarray}
H^{\pi^+}_{u\,\sigma}(x, \xi,t)&=& 4N_cg^2_{\pi qq}\,x\,\tilde I_{2,\Delta}(x,\xi,t)\,C(t)\quad,\nonumber\\
\label{hu-sig}
\end{eqnarray}
with
\begin{eqnarray}
&&C(t)\nonumber\\
&&=(-4m^2)\, \frac{P.P' I_3(m,P,P')-I_2(m, -\Delta)}{m_{\pi}^2 I_2(m,m_{\pi}^2)+\left(4m^2-t\right)I_2(m,\Delta)}\quad.\nonumber
\end{eqnarray}

We automatically find that  the resonance exchange obeys
\beq
H^{\pi^+}_{\sigma}(x, \xi,t)&=& -H^{\pi}_{\sigma}(-x, \xi,t)\quad,
\label{symm-sigma}
\eeq
what corresponds to the symmetry  imposed by Time-reversal and hermiticity. 

As it will be shown, the $\sigma$-exchange term contributes to  both the sum rule and the polynomiality property. Such a resonance exchange should always be taken into account. It reflects the interpretation of GPDs as depending on the interval in $x$.
In the ERBL region, the GPD can no longer be considered as a pure quark or an antiquark distribution but rather as a combination of a $q$ or $\bar q$ distribution as well as a meson Distribution Amplitude; what arises naturally from the diagrammatic description of the {\it skewed} parton distributions, i.e. Fig.~\ref{regions}.
\\

The total distributions for the $u$- and $d$-quarks are then
\begin{eqnarray}
H^{\pi^+\,\mbox{\tiny tot}}_u(x, \xi,t)&=& H_u^{\pi^+}(x, \xi,t)+H^{\pi^+}_{\sigma}(x, \xi,t)\quad,\nonumber\\
H^{\pi^+\,\mbox{\tiny tot}}_{d}(x, \xi,t)&=& H_{d}^{\pi^+}(x, \xi,t)-H^{\pi^+}_{\sigma}(-x, \xi,t)=-H^{\pi^+\,\mbox{\tiny tot}}_u(-x, \xi,t)\quad,
\label{u-d-contr}
\end{eqnarray}
with the $u$-distribution being defined with the support $x\in [-\xi, 1]$ and the $d$-distribution $x\in [-1,\xi]$, as required. The support property has been brought up in Section~\ref{sec:support}. So was the discussion about the support violation; therefore we do not need to develop this argument here.

\begin{figure}
\begin{center}
 \mbox{\subfigure[]{
 \includegraphics [height=5.45cm]{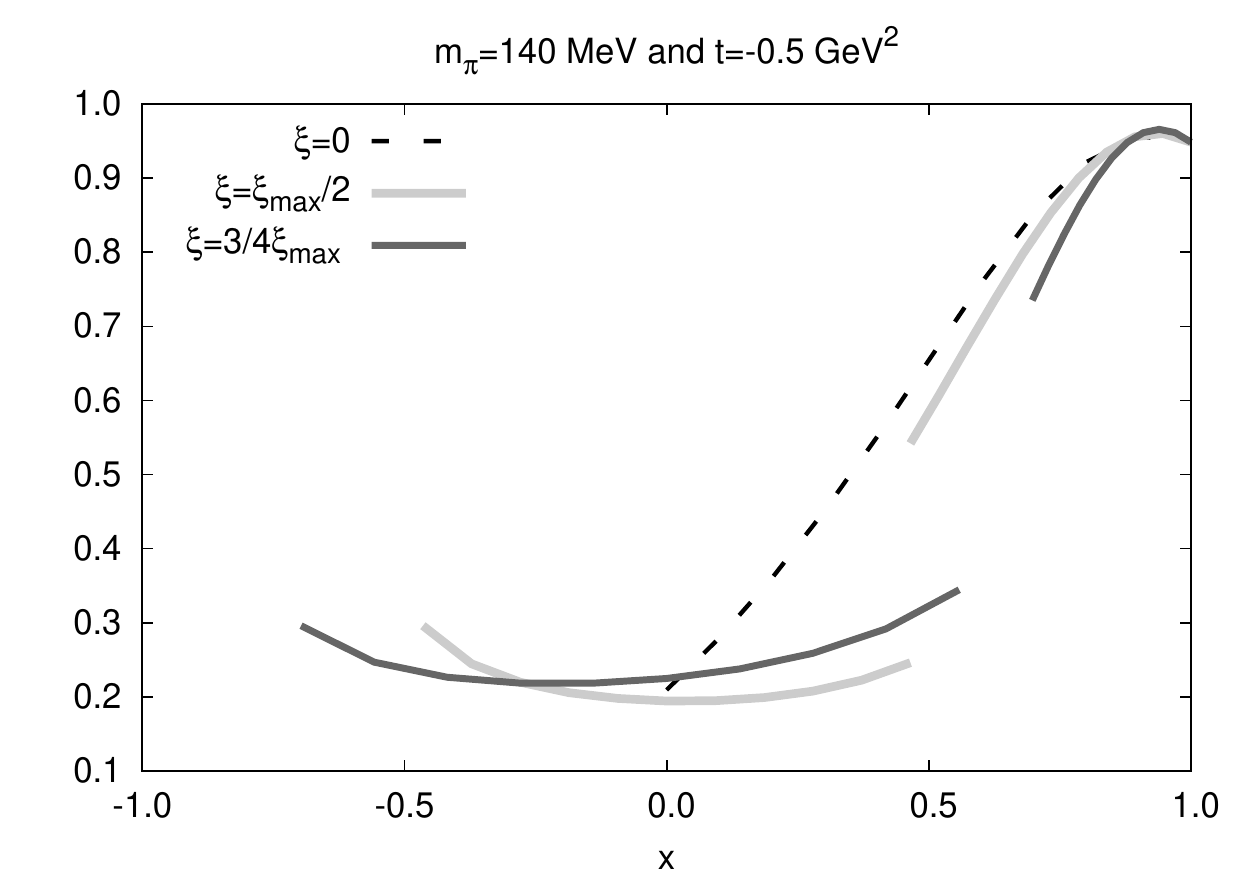}}
 \subfigure[]{
 \includegraphics [height=5.45cm]{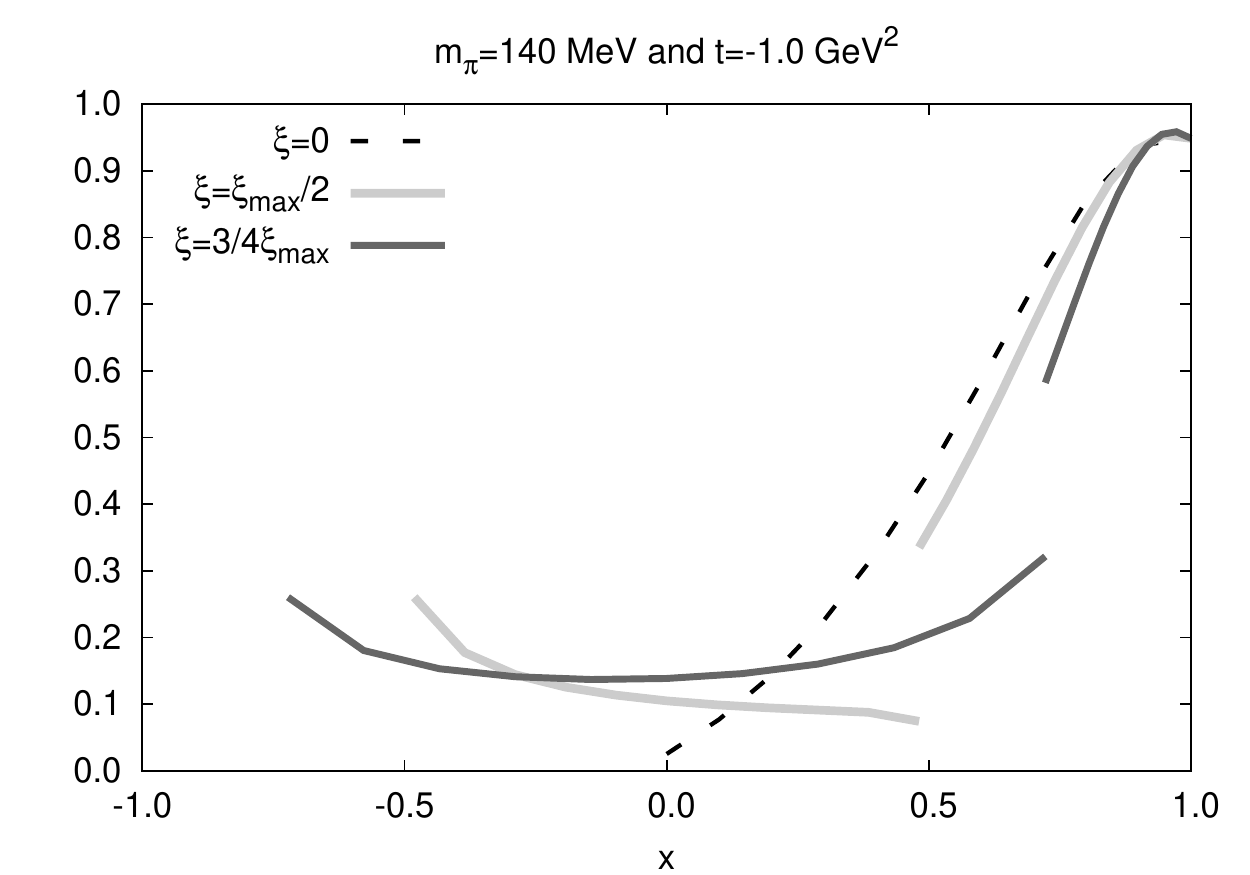}}}
 \caption{{Total GPD for the $u$-quark in the NJL model in the case where,  respectively,  $(a)$ $t=-0.5$ ~GeV$^2$
 and $(b)$  $t=-1$~GeV$^2$ .}}
\label{gpdmpi140t}
 \end{center}
\end{figure}

\subsubsection{Isospin Decomposition}

As it was the case for the PDFs,  the total $\pi^+$ distribution  can be decomposed in an isoscalar and an isovector parts. The two terms are correspondingly defined as, in terms of operators,
\beq
\delta_{ab}\, H^{I=0}(x,\xi, t)&=&\int \frac{dz^-}{4\pi}\, e^{ixp^+ z^-} \langle \pi^a(P')\vert \bar q\left(-\frac{z}{2}\right)\,\gamma^+\,q\left(\frac{z}{2}\right)
\vert \pi^b(P)\rangle |_{z^{\perp}=z^+=0}\quad,\nonumber\\
\nonumber\\
i\epsilon_{3ab} \, H^{I=1}(x,\xi, t)&=&\int \frac{dz^-}{4\pi}\, e^{ixp^+ z^-} \langle \pi^a(P')\vert \bar q\left(-\frac{z}{2}\right)\,\gamma^+\,\tau_3\,q\left(\frac{z}{2}\right) \vert \pi^b(P)\rangle |_{z^{\perp}=z^+=0}\quad,\nonumber\\
\eeq
with the pion field $\vert \pi^{+,-}\rangle=\left( \vert \pi^{1}\rangle\pm i \vert \pi^{2}\rangle\right)/\sqrt{2}$ and $\vert \pi^0\rangle =\vert \pi^3\rangle$.
\\

Using the isospin relation~(\ref{iso-gpd}) and Eq.~(\ref{u-d-contr}) it is easily seen that the $\sigma$-meson  exchange does  only contribute to the isoscalar pion GPD.
Effectively, the isoscalar decomposition  is antisymmetric in $x$ 
\beq
H^{I=0}(x, \xi,t)&=& H^{\pi^+}_u(x, \xi,t)-H^{\pi^+}_u(-x, \xi,t)+2\,H_{u\,\sigma}^{\pi}(x,\xi,t)=-H^{I=0}(-x, \xi,t)\quad.\nonumber\\
\label{isosca}
\eeq
Also, the isovector decomposition is hence symmetric in $x$ and corresponds to the valence  quark distribution
\beq
H^{I=1}(x, \xi,t)&=& H^{\pi^+}_u(x, \xi,t)+H^{\pi^+}_u(-x, \xi,t)=H^{I=1}(-x, \xi,t)\quad.\nonumber\\
\label{isovec}
\eeq
We obtain the corresponding relation for the pions
\beq
H_u^{\pi^0}(x, \xi, t)&=&\frac{1}{2} \,H^{I=0}(x, \xi,t)\quad,\nonumber\\
H_u^{\pi^+}(x, \xi, t)&=&\frac{1}{2}\left(H^{I=0}(x, \xi, t)+ H^{I=1}(x, \xi, t)\right)\quad,\nonumber\\
H_u^{\pi^-}(x, \xi, t)&=&\frac{1}{2}\left(H^{I=0}(x, \xi, t)- H^{I=1}(x, \xi, t)\right)\quad.
\eeq
\\

Let us now comment on the numerical results for the pionic GPDs in the NJL model.
The GPD for the $u$-quark in the NJL model  is discontinuous at $x=\xi$, see \cf{gpdmpi140t}.
So that the isoscalar GPD is discontinuous at $x=-\xi$ and $\xi$.
The $|\xi|$ discontinuity comes from the formulation of the $\sigma$-exchange in the NJL model, as a realization of the chiral symmetry.  This could be understood by considering the $\sigma$-exchange  as a separate DA. In effect, the Distribution Amplitude in the NJL model calculation does not vanish at the end-points unlike its asymptotic form, leading to discontinuities at its end-points. This result has also been obtained in the Chiral Soliton Quark Model when the dependence of the dynamical quark mass is not taken into account~\cite{Polyakov:1999gs}, a situation that is similar to the one presented here. 
However, this peculiarity  of the model calculation will disappear when switching on QCD just like for the Distribution Amplitudes as it has been discussed in  Section~\ref{sec:endpoints}.

\begin{figure}
\centering
 \includegraphics [height=5.5cm]{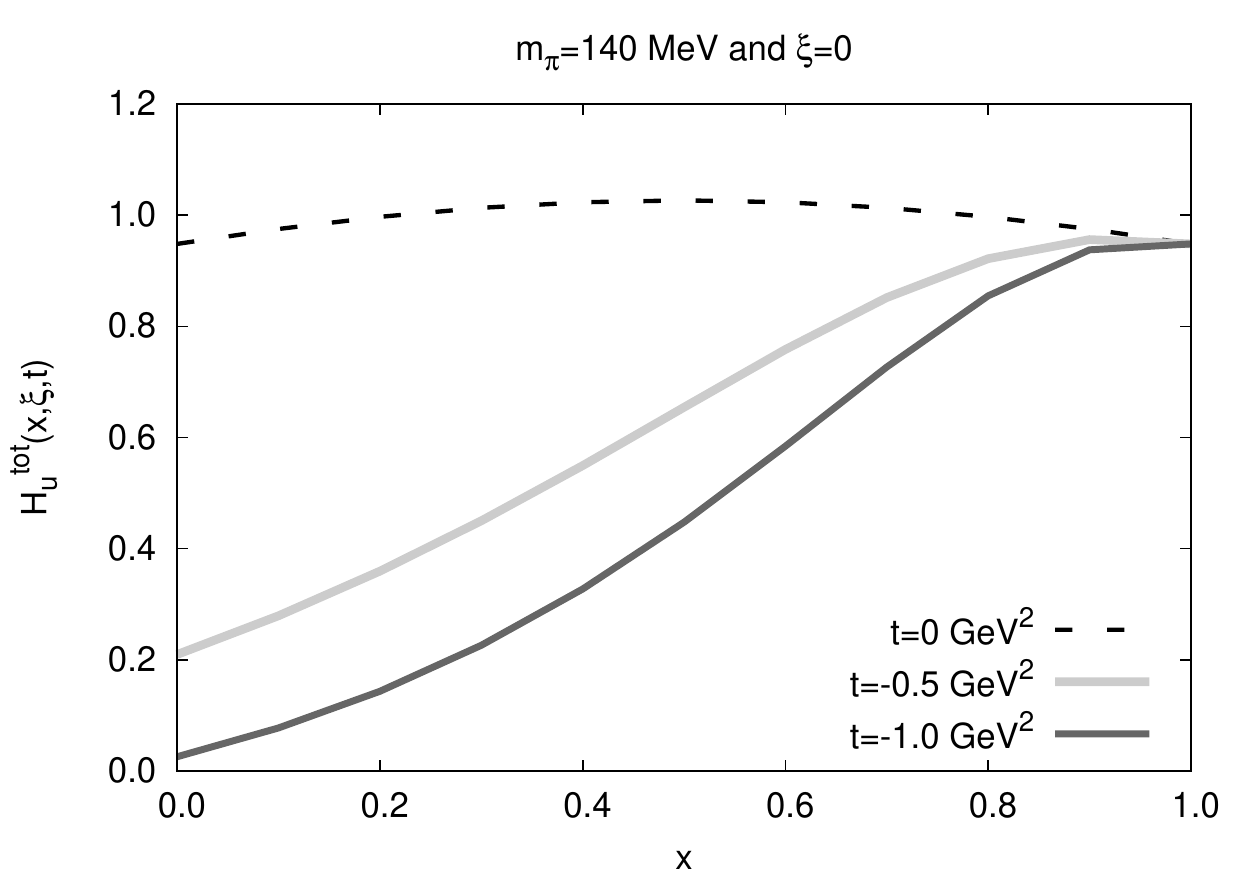}
 \caption{{GPDs in the NJL model in the case $m_{\pi}=140$ MeV and $\xi=0$ .}}
\label{gpdxi0mpi140}
\end{figure}

\subsubsection{Sum Rule and Polynomiality}

We can now check that the properties of GPDs are respected in this model calculation.
The sum rule for the isovector GPD reads~\footnote{The association $\int_{-1}^1 dx\, \sum_q Q_q H_q^{\pi^+} (x, \xi, t)=
\int_{-1}^1 dx\, H^{I=1}(x, \xi,t)$ is valid under the integral over $dx$ only.}
\begin{eqnarray}
\int_{-1}^1 dx\, \sum_q Q_q H_q^{\pi^+} (x, \xi, t)=
\int_{-1}^1 dx\, H^{I=1}(x, \xi,t)&=& 2\, F_{\pi}(t)\quad,
\label{isovec-sr}
\end{eqnarray}
with $F_{\pi}(t)$ the electromagnetic Form Factor given by Eq.~(\ref{elm-ff-njl}). This sum rule is respected and has been analytically verified.
Regarding the isoscalar combination, the sum rule involves the gravitational Form Factors of the pion $\theta_{1,2}(t)$ in the following way,
\beq
\int_{-1}^1 dx\,x\, H^{I=0}(x, \xi,t)&=& \theta_1(t)-\xi^2\, \theta_2(t)\quad.
\label{isosca-sr}
\eeq
Since this sum rule refers to the momentum carried by each quark when taken at $t=0$ GeV$^2$, we expect, in the chiral limit, $\theta_1(0)=\theta_2(0)=1$. This condition has been fulfilled as it can be seen in Table~\ref{tablepoly}.\footnote{Notice that Table~\ref{tablepoly} is given $H^{\pi^+ tot}_u$  which is defined between $[-\xi, 1]$ so that the results have to be mutliplied by 2 when considering $H^{I=0}$ and we extend the integral  to $[-1,1]$.}

The polynomiality property is also respected,
\begin{eqnarray}
\int_{-\xi}^1 dx\,x^{n-1}\,H_u^{\pi\,\mbox{\tiny tot}}(x, \xi,t)&=& F_N(\xi,t)\quad,\nonumber\\
&=& \sum_{i=0}^{[\frac{n}{2}]} \,A_{n,2i}(t)\,\xi^{2i}\quad,
\end{eqnarray}
where at the second line we have explicitly expressed the $F_N(\xi,t)$ as polynomials in $\xi$ of order not higher than $n$.  On virtue of the relation
\beq
H_q(x,\xi,t)&=&H_q(x,-\xi,t)\quad,
\nonumber
\eeq
given by Time-reversal invariance, only even powers in the skewness variable are allowed in the previous expansion in polynomials.
 The numerical results for the coefficients in both cases $m_{\pi}=0$ MeV and $m_{\pi}=140$ MeV are given in Table~\ref{tablepoly}.
\begin{table}[h]
\centering
\small
\begin{tabular}{|c|c|c|c|c|}
\hline
$n$ & 1 & 2 & 3 & 4 \\ \hline 
& $A_{1,0}\left( t\right) $ & $%
\begin{array}{rr}
A_{2,0}\left( t\right)  & A_{2,2}\left( t\right)
\end{array}%
$ & $%
\begin{array}{rr}
A_{3,0}\left( t\right)  & A_{3,2}\left( t\right)
\end{array}%
$ & $%
\begin{array}{rrr}
A_{4,0}\left( t\right)  & A_{4,2}\left( t\right)  & A_{4,4}\left( t\right)
\end{array}%
$ \\ \hline
$%
\begin{array}{c}
m_{\pi }=0\,\,\,\,\,%
\end{array}
$%
\\ \hline
$\begin{array}{l}
t=0 \\
t=-1 \\
t=-10%
\end{array}%
$ & $%
\begin{array}{l}
1. \\
0.487 \\
\multicolumn{1}{r}{0.091}%
\end{array}%
$ & $%
\begin{array}{ll}
0.5 & -0.5 \\
0.336 & -0.240 \\
\multicolumn{1}{r}{0.119} & \multicolumn{1}{r}{-0.093}%
\end{array}%
$ & $%
\begin{array}{ll}
0.333 & 0. \\
0.257 & -0.049 \\
\multicolumn{1}{r}{0.118} & \multicolumn{1}{r}{-0.058}%
\end{array}%
$ & $%
\begin{array}{lll}
0.25 & 0. & -0.25 \\
0.208 & -0.030 & -0.113 \\
\multicolumn{1}{r}{0.111} & \multicolumn{1}{r}{-0.046} & \multicolumn{1}{r}{
-0.044}%
\end{array}%
$ \\ \hline
$%
\begin{array}{c}
m_{\pi }=140%
\end{array}
$%
\\ \hline
$\begin{array}{l}
t=0[-10^{-5}] \\
t=-1 \\
t=-10%
\end{array}%
$ & $%
\begin{array}{l}
1. \\
\multicolumn{1}{r}{0.482} \\
\multicolumn{1}{r}{0.090}%
\end{array}%
$ & $%
\begin{array}{ll}
0.5 & [-0.473] \\
\multicolumn{1}{r}{0.332} & \multicolumn{1}{r}{-0.230} \\
\multicolumn{1}{r}{0.116} & \multicolumn{1}{r}{-0.088}%
\end{array}%
$ & $%
\begin{array}{ll}
0.332 & [0.005] \\
\multicolumn{1}{r}{0.253} & \multicolumn{1}{r}{-0.045} \\
\multicolumn{1}{r}{0.114} & \multicolumn{1}{r}{-0.055}%
\end{array}%
$ & $%
\begin{array}{lll}
0.247 & [0.005] & [-0.236] \\
0.203 & -0.026 & -0.109 \\
\multicolumn{1}{r}{0.108} & \multicolumn{1}{r}{-0.043} & \multicolumn{1}{r}{
-0.042}%
\end{array}%
$ \\ \hline

\end{tabular}
\caption{Coefficients of the polynomial expansion. The pion mass is expressed in MeV
and $t$ is expressed in GeV$^2$.
Values between brackets correspond to $t=-10^{-5}$ GeV$^2$ \cite{Theussl:2002xp}.
\label{tablepoly}}
\end{table}%

The chiral limit, where $m_{\pi}=0$, and with  zero momentum transfer, is perfectly defined in a regularization-independent way.
It makes sense to study separately the $\xi$ dependence of the contributions coming from the {\it direct}  and  {\it crossed} diagrams of Fig.~\ref{gpdnjl} and the $\sigma$-exchange of Fig.~\ref{sigmau};
\beq
H_u^{\pi}(x, \xi,0)&=&\frac{1}{2}\,\left(1-\frac{x}{2\xi}\right)\,\theta(\xi-x)\theta(\xi+x)\,+\,\theta(x-\xi)\theta(1-x)\quad,\nonumber\\
H^{\pi}_{u\,\sigma}(x, \xi,0)&=&\frac{1}{2}\,\frac{x}{2\xi}\,\theta(\xi-x)\theta(\xi+x)\quad.
\label{chiral-u}
\eeq
The total  GPD for the $u$-quark in a $\pi^+$ reads
\begin{eqnarray}
H^{\pi \,\mbox{\tiny tot}}_u(x, \xi,0)&=&\frac{1}{2}\,\theta(\xi-x)\theta(\xi+x)\,+\,\theta(x-\xi)\theta(1-x)\quad.
\label{utot-chi}
\end{eqnarray}
Evaluated for the {\it skewed} PD, the expression of the $\sigma$-resonance is model dependent. However, since the interpretation of the $\sigma$-exchange is unique in the limit $m_{\pi}\to 0$, chiral symmetry implies that the diagram~\ref{sigmau} cancels out the "$\sigma$-exchange like" contribution contained in $H_{\pi}^q$. In other words, we recover the result $H^{I=0}(x,1,0)=0$ predicted by the chiral symmetry~\cite{Polyakov:1999gs}.
\begin{wrapfigure}[9]{r}{75mm}
\centering
\includegraphics [height=5cm]{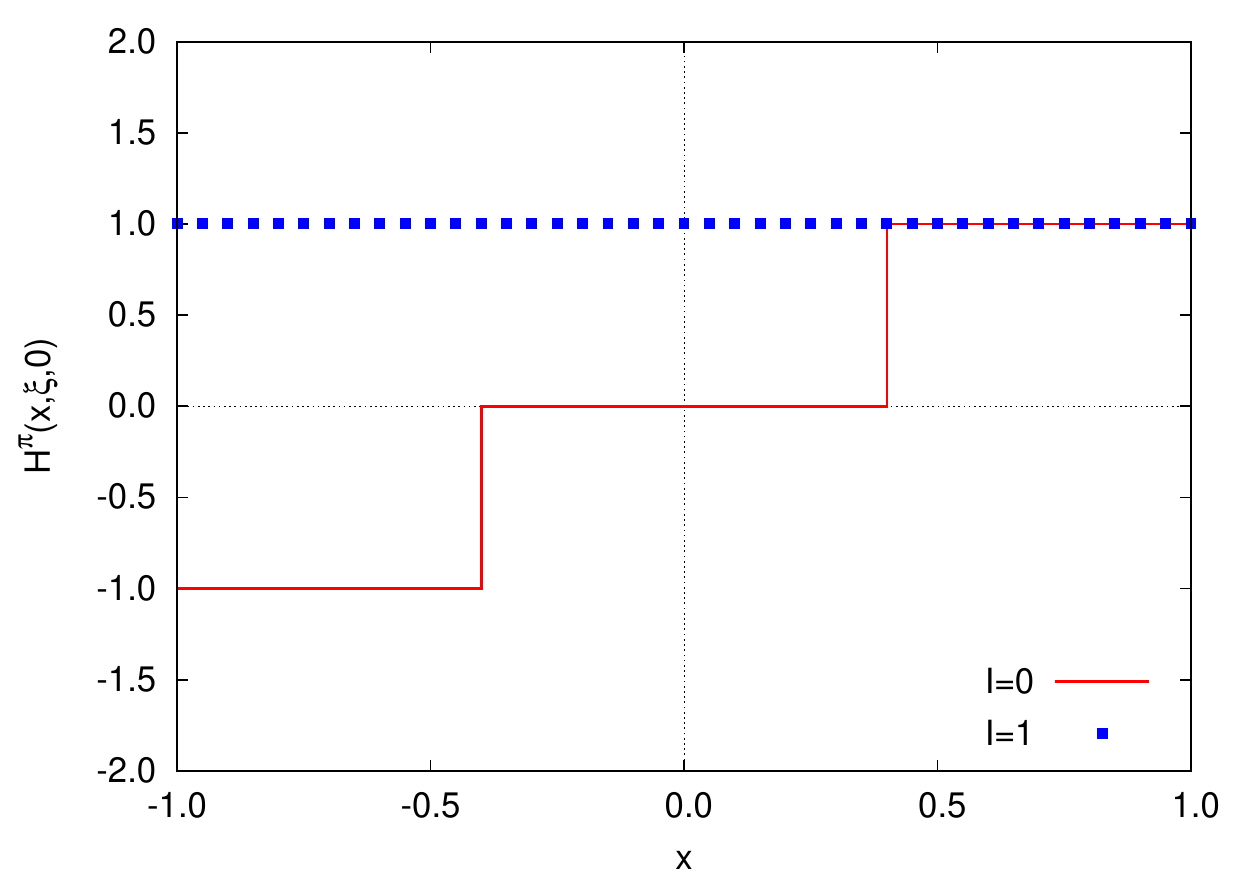}
\caption{Chiral limit for $\xi=0.4$.}
\label{gpdxi4}%
\end{wrapfigure}
 Also, the isoscalar and isovector combinations read
 \beq
&& H^{I=0}(x, \xi,0)\nonumber\\
&&=-\theta(-x-\xi)\theta(1+x)\,+\,\theta(x-\xi)\theta(1-x)\quad,\nonumber\\
\nonumber\\
  &&H^{I=1}(x, \xi,0)\nonumber\\
  &&=\theta(-x-\xi)\theta(1+x)\,\nonumber\\
  &&+\, \theta(\xi-x)\theta(\xi+x)\,+\,\theta(x-\xi)\theta(1-x)\quad,\nonumber
 \eeq
  what is consistent with previous results \cite{Polyakov:1999gs}. This result is illustrated on Fig.~\ref{gpdxi4}.
\\

We  notice also that, in  the forward limit, i.e. for $\xi=t=0$, we find the 
quark distribution function
\begin{eqnarray}
 H^{\pi\,\mbox{\tiny tot}}_u(x, 0,0)&=&q(x)=1\quad,\nonumber
\end{eqnarray}
where, as expected, the $\sigma$-exchange does not contribute.

\vskip 1cm

To this point, we have presented the calculation of the chiral-even twist-2  GPD of the pion in the NJL model using a fully covariant Bethe-Salpeter approach.
The only approximation employed has been the ladder approximation. We have solved the BS equation   exactly thanks to the point-like interaction of the NJL model.
The usual properties of GPDs have been recovered.
Nevertheless discontinuities are encountered,  at the boundaries and are associated to
the off-shellness of the constituent quark. Also the expression of the $\sigma$-exchange in the NJL model  introduces discontinuities at the $x$ values $-\xi$ and $\xi$. 
\\

\subsection{Double Distributions and the $D$-term}
\label{sec:dd-dt}

An alternative way to describe the hadronic matrix elements defining the GPDs is through the so-called Double Distribution (DD) parameterization. By Lorentz invariance 
the matrix element (\ref{hadr-ma}) can only depend on the two independent variables $p\cdot z$ and $\Delta \cdot z$.
Hence the pion DD~\cite{Radyushkin:1996nd,Radyushkin:1997ki}  is defined as a 2-dimensional Fourier transform in these two independent variables. The analysis of diagrams of the type~\ref{gpdnjl} $\&$ \ref{sigmau} yields the behavior of these variables;
\beq
{\cal M}_q^{\pi}\left( p\cdot z, \Delta \cdot z; t \right)
&=&\langle \pi^+(P')\vert \,\bar q\left(-\frac{z}{2}\right)\,\thru z\,q\left(\frac{z}{2}\right)\vert \pi^+(P)\rangle \nonumber\\
&=&2\, (p\cdot z)\,    \int_{-1}^1  d\beta \int_{-(1-|\beta|)}^{1-|\beta|}d\alpha  \, e^{-i\beta (p\cdot  z)+i\alpha (\Delta \cdot z)/2} \,{\cal F}^{\pi}_q ( \beta, \alpha, t)\nonumber\\
&&-(\Delta \cdot  z)\, \int_{-1}^{1} d\alpha  \, e^{i \alpha(\Delta\cdot z)/2}\, {\cal D}^{\pi}_q (\alpha, t)\quad,
\label{ddpion}
\label{dd-sin-dt}
\eeq
for the whole matrix element.
\\

\begin{wrapfigure}[12]{l}{95mm}
\centering
\includegraphics [height=3.2cm]{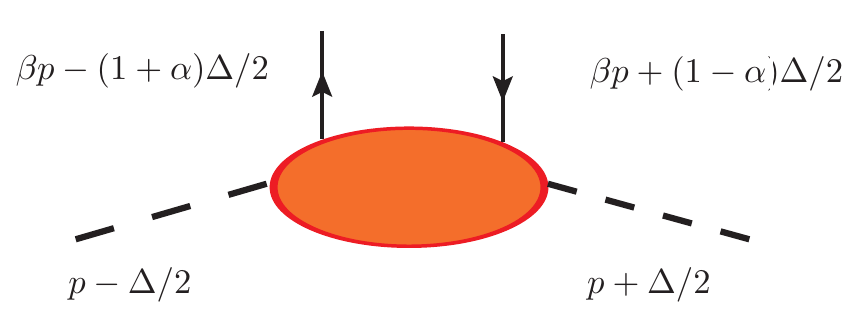}
\caption{Momenta associated with the quarks and hadrons in the Double Distribution parameterization.}
\label{dd-kin}%
\end{wrapfigure}
 Since the  variables $p\cdot z$ and $\Delta\cdot z$ are treated as independent, the DD refers to fixed momentum transfer $t$ but with unfixed individual hadron momenta $P$ and $P'$. In other words, these momenta do not depend on the skewness variable. 
 The interpretation of the latter being  process dependent, one can say that the matrix element on the r.h.s. of Eq.~(\ref{dd-sin-dt}) contains a process-independent information and, hence, has a quite general nature.

The dependence on $\beta$ is of the parton distribution type, while the one on the parameter $\alpha$ resembles the dependence of a distribution amplitude.  This feature  is illustrated in Fig.~\ref{dd-kin}.
\\
\begin{wrapfigure}[11]{r}{90mm}
\centering
\includegraphics [height=3.5cm]{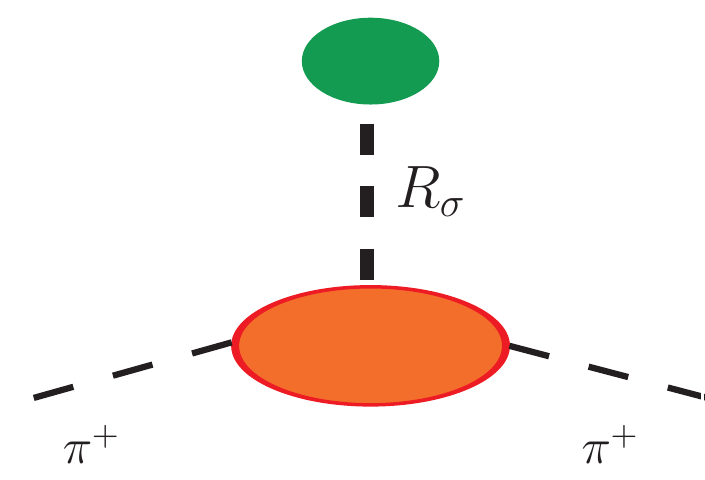}
\caption{The $D$-term as a $\sigma$-resonance in the $t$-channel.}
\label{sigma-pole}%
\end{wrapfigure}
From the na\" ive analysis of GPDs as parameterized through the DD, one would write the whole contribution coming from diagrams such as~\ref{gpdnjl} or~\ref{dd-kin} in the form $f^{\pi}_q(\beta, \alpha, t)$; i.e.~(\ref{dd-sin-dt}). 
From our experience in model calculations, we know that  diagrams of this type may contain a resonance exchange in the $t$-channel whose tensorial structure corresponds exactly to the one of the remaining non-resonant part of the diagram.
The behavior of this meson-exchange contribution with respect to the two independent variables  $p\cdot z$ and $\Delta \cdot z$ is manifestly different. 

As a matter of fact, it has been shown in Ref.~\cite{Polyakov:1999gs} that another twist-2 "DD structure" has to be considered indeed. The origin of this new structure is the subtraction from the matrix element of  those resonances in the $t$-channel (Fig.~\ref{sigma-pole}). Namely they are terms depending solely on the $\Delta \cdot z$ variable, i.e. non-vanishing terms in the limit 
\beq
p\cdot z\to 0\quad.
\label{lim-pz}
\eeq
The function ${\cal D}(\alpha; t)$ is therefore defined  taking the latter limit and subtracting this result from the $p\cdot z$-dependent contribution, i.e.
\beq
{\cal M}_q^{\pi}\left( p\cdot z, \Delta \cdot z; t \right)&=& \underbrace{
{\cal M}_q^{\pi}\left( p\cdot z\neq 0, \Delta \cdot z; t \right)-{\cal M}_q^{\pi}\left( 0, \Delta \cdot z; t \right)}
_{\rightarrow {\cal F}_q^{\pi}(\beta, \alpha; t)}
+\underbrace{
{\cal M}_q^{\pi}\left( 0, \Delta \cdot z; t \right)}
_{\rightarrow {\cal D}_q^{\pi}( \alpha; t)}\quad.
\nonumber
\eeq
\\
The meson-exchanges must have the characteristics of  Distribution Amplitudes so that we expect them to  be parametrized by $\alpha$. In terms of DD, we call the corresponding function the $D$-term~\cite{Polyakov:1999gs}.
The $D$-term is uniquely defined for a given hadronic matrix element.
\\

When applying the limit~(\ref{lim-pz}) to the matrix element as expressed in the NJL model calculation~(\ref{elmtmatr}) taking into account all diagrams, this resonance exchange corresponds in part to the term going like $\tilde I_{2,\Delta}$ of the integral decomposition~(\ref{hu-1st}) as well as to the term coming from the second diagram Fig.~\ref{sigmau}. The whole resonance exchange also includes a $x$-odd part of  the $\tilde I_{2,P}$ as well as of the $\tilde I_3$  integrals in~(\ref{hu-1st}). 
\\

The Double Distributions can be reduced to the usual {\it skewed} parton distribution by taking a one-dimensional cut, i.e. imposing a relation between $p\cdot z$ and $\Delta \cdot z$. This reduction formula  reads
\beq
H^{\pi}_q(x,\xi,t) &=&   \int_{-1}^1  d\beta \int_{-(1-|\beta|)}^{1-|\beta|} d\alpha \, \delta(x-\beta-\xi \alpha) {\cal F}_q^{\pi}(\beta, \alpha, t)+\mbox{sgn}(\xi) D^{\pi}_q \left(\frac{x}{\xi}, t \right)\quad.
\nonumber\\
\label{reduc-form}
\eeq
The $D$-term obeys the symmetries of an isoscalar resonance~(\ref{symm-sigma}). Therefore it contributes solely to the isoscalar GPD as expected.
\\

The parameterization through Double Distribution applies at the level of the Fourier transform of the matrix element. 
This  renders the evaluation of the DD model independent.
 Nonetheless this statement is quite misleading in the sense that the relevant diagrams  depend on the model considered. 
  So, from a technical point of view, one has to evaluate the diagrams that apply, and, with the  Feynman rules resulting from the features of the model. The GPDs appear to be a combination of two- and three-point functions, e.g. in the NJL model Eqs.~(\ref{hu-1st}, \ref{hu-sig}),  which, in turn, can be expressed in terms of Double Distributions through the reduction formula.
 An explicit example  would be the expression of the 3-propagator integral
\beq
\tilde I_3( x, \xi, t)&=&\int_{-1}^1 d\beta\, \int_{-(1-|\beta|)}^{1-|\beta|} d\alpha\,\delta\left(x-\beta -\alpha \xi\right)\, {\cal I}_{3 }(\beta, \alpha,  t)\quad;
\label{reduc-3p}
\eeq
with ${\cal I}_{3}$ a Double Distribution  given in~(\ref{dd-3pf}). The 2-propagator integrals that do not behave like DAs can also be included in the form ${\cal I}(\beta, \alpha,t)$. 
This reduction is totally model independent. 
\\

The model's description of the $\sigma$-exchange, illustrated in Fig.~\ref{sigmau},  also accounts for a formal and general description on the form of the r.h.s. of Eq.~(\ref{dd-sin-dt}). In particular, it is of a pure $D$-term structure. The total description of the matrix element simply results from the sum of the two contributions; namely $H^{\pi}_q$ as given by the reduction formula~(\ref{reduc-form}) plus the corresponding formula for $H^{\pi}_{q \, \sigma}$.

\begin{wrapfigure}[19]{l}{8.8cm}
\centering
 \includegraphics [height=6cm]{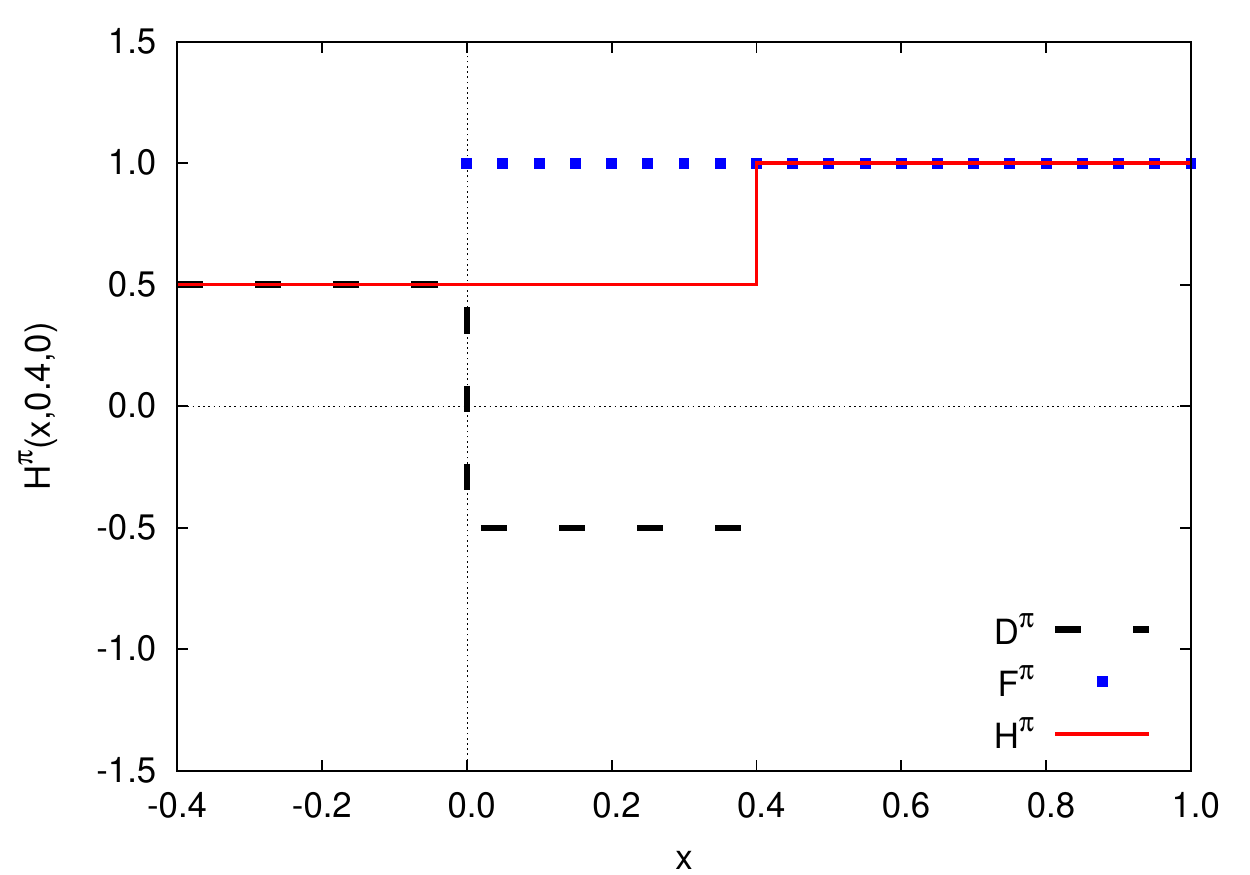}
 \caption{The contributions from the double distribution $F^{\pi}$ (blue dotted line), from the $D$-term (dashed line) and the total distribution $H^{\pi}$ (red curve).}
\label{dd-dt-tot}
\end{wrapfigure}

As already mentioned in Section~\ref{sec:da-cal}, the  difference between the quark models is manifest  in the regularization scheme.  
The chosen regularization scheme is an integrant part of the model since it depends on the model's interaction and conveniences for the adaptation to the problem under scrutiny.
Concretely the Double Distributions are reduced to the physical propagator integrals by imposing a regularization scheme. In the NJL model as presented in the previous Chapter~\ref{sec:pionda}, it would mean that the propagator integrals 
 follow, in the case of 3-propagator integrals, the expression~(\ref{reduc-3p})
to which we impose the Pauli-Villars regularization scheme, i.e. Eq.~(\ref{PV}).

In Ref.~\cite{Broniowski:2007si} the Double Distributions of the pion in the NJL model are given using  a Pauli-Villars regularization in the twice-subtracted version of Ref.~\cite{RuizArriola:2002wr}, as well as in the Spectral Quark Model~\cite{RuizArriola:2003bs}. Both calculations yield results comparable to the calculation of Ref.~\cite{Theussl:2002xp}.
\\

One of the main advantages of the DD parameterization is that the expression (\ref{reduc-form})  
manifestly leads to the correct polynomiality properties of the GPDs.  Effectively, when taking the Mellin moments of the reduction formula, one immediately finds
\beq
&&\int_{-1}^1 dx\,  x^{n-1}\,H^{\pi}_q(x,\xi,t) \nonumber\\
&&= \int_{-1}^1 dx\,  x^{n-1}\Big[ \int_{-1}^{1} d\beta \int_{-(1-|\beta|)}^{1-|\beta|}  d\alpha \,\delta(x-\beta-\xi \alpha) {\cal F}_q^{\pi}(\beta, \alpha, t)+\mbox{sgn}(\xi) D^{\pi}_q \left(\frac{x}{\xi}, t\right)\Big]\quad,
\nonumber\\
\eeq
where the first term on the r.h.s. is a polynomial at a power of order at most $n-1$ of the skewness variable $\xi$. The $D$-term instead gives the highest power in $\xi$, i.e. $n$. \footnote{The latter statement can be visualized by performing the change of variable $x/\xi=\nu$; which carries a $\xi^n$ outside the integral over, then, $\nu$.} \\

Concerning the  chiral limit, the DD approach yields the same results~(\ref{chiral-u}) as the  {\it skewed} PD evaluation~\cite{Polyakov:1999gs}. In this limit, the $D$-term is therefore reduced to the zeroth order term of the resonance exchange coming from the $\tilde I_{2,P}$ term.

Let us illustrate the previous statements.  In the chiral limit,  the $D$-term reads, for the $u$-quark,
\beq
D^{\pi}_u\left(\frac{x}{\xi},0\right)&=& \frac{1}{2}\,\theta\left(x+\xi\right)\theta\left( -x\right)-\frac{1}{2}\,\theta\left(x-\xi\right)\theta\left( x\right)\quad.
\label{dterm-chi}
\eeq
The remaining part is found through the reduction formula, providing that 
\beq
H^{\pi}_u(x,\xi,t)&=&  F_u^{\pi}(x,\xi,t)+D^{\pi}_u\left(\frac{x}{\xi},t\right)\quad.
\label{f+dterm}
\eeq
Using~(\ref{utot-chi}), we obtain
\beq
F^{\pi}_u(x,\xi,0)&=&\theta\left(x\right)\theta\left(1-x\right)\quad,
\nonumber
\eeq
which can obviously not satisfy the polynomiality property. By inserting the $D$-term~(\ref{dterm-chi}), we obtain the expression for the total GPD~(\ref{f+dterm}). In particular, we recover the expression~(\ref{utot-chi}) which, in turn, fulfills the required conditions. These results are illustrated in Fig.~\ref{dd-dt-tot}.
\\

There is considerable freedom in the tensorial decomposition of the matrix element ${\cal M}_q^{\pi}$.
An alternative decomposition to the expression~(\ref{ddpion}) is given in Ref.~\cite{Tiburzi:2002tq},
\beq
{\cal M}_q^{\pi}\left( p\cdot z, \Delta \cdot z; t \right)
&=&\langle \pi^+(P')\vert \,\bar q\left(-\frac{z}{2}\right)\,\thru z\,q\left(\frac{z}{2}\right)\vert \pi^+(P)\rangle \nonumber\\
&=&2\, (p\cdot z)\,    \int_{-1}^1  d\beta \int_{-(1-|\beta|)}^{1-|\beta|}d\alpha  \, e^{-i\beta (p\cdot  z)+i\alpha (\Delta \cdot z)/2} \,{\cal K}^{\pi}_q ( \beta, \alpha, t)\nonumber\\
&&-(\Delta \cdot  z)\, \int_{-1}^1  d\beta \int_{-(1-|\beta|)}^{1-|\beta|}d\alpha  \, e^{-i\beta (p\cdot  z)+i\alpha (\Delta \cdot z)/2} \,{\cal G}^{\pi}_q ( \beta, \alpha, t)\quad,\nonumber\\
\label{dd-milltib}
\eeq
with the function ${\cal K}^{\pi}_q ( \beta, \alpha, t)$ even in $\alpha$ and the function ${\cal G}^{\pi}_q ( \beta, \alpha, t)$ $\alpha$-odd, as a consequence of  Time-reversal invariance.
The corresponding reduction formula reads
\beq
H^{\pi}_q(x,\xi,t) &=&   \int_{-1}^1  d\beta \int_{-(1-|\beta|)}^{1-|\beta|} d\alpha \, \delta(x-\beta-\xi \alpha) \left[{\cal K}_q^{\pi}(\beta, \alpha, t)+\xi \,{\cal G}_q^{\pi}(\beta, \alpha, t)  \right]\quad.
\eeq
The sum rule for ${\cal K}^{\pi}$ gives the electromagnetic pion Form Factor, as expected; whereas the sum rule for ${\cal G}^{\pi}$ gives $0$ since it is an odd function of $\alpha$.

One could relate the above given expression to the {\it usual} DD decomposition by writing all the $\beta$ dependence of ${\cal G}_q^{\pi}(\beta, \alpha, t)$ as a ${\cal K}_q^{\pi}(\beta, \alpha, t)$, leaving a pure $\Delta \cdot z$ in the second term.
The decomposition~(\ref{dd-milltib}) is however calculationally the most useful, since it does not evoke the cumbersome limit~(\ref{lim-pz}). On the other hand the expression of the meson-exchanges is less explicit.
\\

The distribution amplitude character of the $D$-term implies it obeys to the ERBL evolution equations, that has been introduced in the previous Chapter. The total GPD does however not have the same behavior when going the higher $Q^2$.
From Fig.~\ref{regions}, we have defined the three regions as DGLAP-regions for the support in $x$ being either $[\xi,1]$ or $[-1,-\xi]$ and ERBL-region for the $x\in [-\xi,\xi]$ region. Those regions were named after the evolution equations they obey.
 
 \subsection{QCD Evolution}

The form of the Leading-Order (LO) QCD evolution has been given, for the standard PDF evolution, by Dokshitzer~\cite{Dokshitzer:1977sg}, Gribov and Lipatov~\cite{Gribov:1972ri, Lipatov:1974qm} as well as by Altarelli and Parisi~\cite{Altarelli:1977zs}. It gave rise to the DGLAP equations, whose evolution will be followed by the two regions of emission and reabsorption of a quark or antiquark, respectively. 
On the other hand, the equations describing the evolution of the distribution amplitudes were given by Efremov and Radyushkin~\cite{Efremov:1979qk} as well as by Brodsky and Lepage~\cite{Lepage:1980fj}. 
\\

The form of the integral-equation defining the evolution equations depend upon  the nature of the distributions.
In  Eqs.~(\ref{isosca}, \ref{isovec}),  the singlet or sea distribution  have been defined as well as the non-singlet or valence distribution
\beq
H^{S}(x, \xi,t)&=& \sum_q H_q^{I=0}(x, \xi,t)\quad;
\label{singlet}\\
\nonumber\\
H_q^{NS}(x, \xi,t)&=&  H_q^{I=1}(x, \xi,t)\quad.
\label{non-singlet}
\eeq
Still, one main ingredient of QCD evolution has to appear: the gluon distribution, obeying the following symmetry
\beq
xH_g(x,\xi,t)&=&-xH_g(-x,\xi,t)\quad;
\nonumber
\eeq
with the gluon distribution being zero at the scale of the model, i.e. $H_g(x,\xi,t, Q_0^2)=0$.
\\
It is important to notice that the non-singlet distribution does not mix with the singlet under LO evolution. However, the singlet and gluon distribution do mix under evolution.

\begin{figure}
\centering
\includegraphics [height=5.6cm]{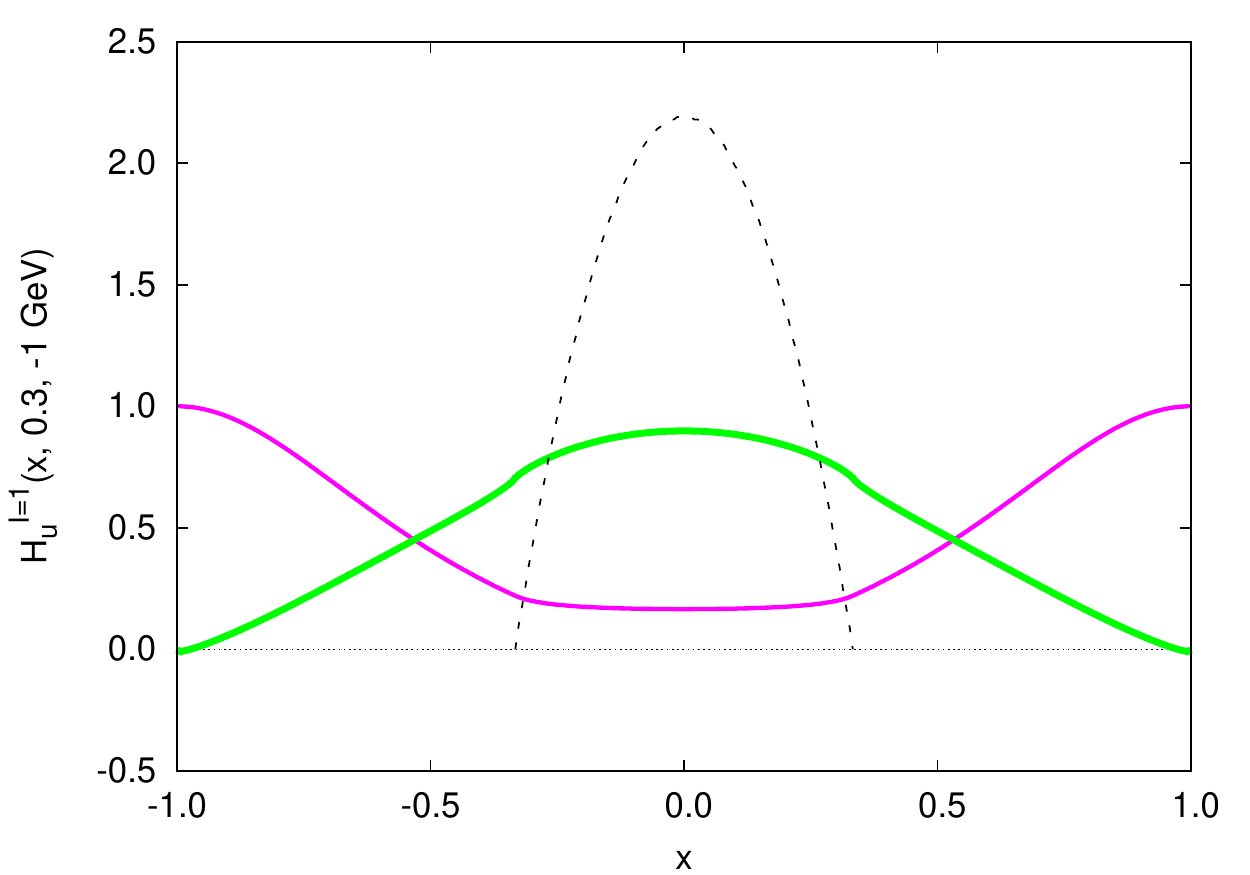}
\includegraphics [height=5.6cm]{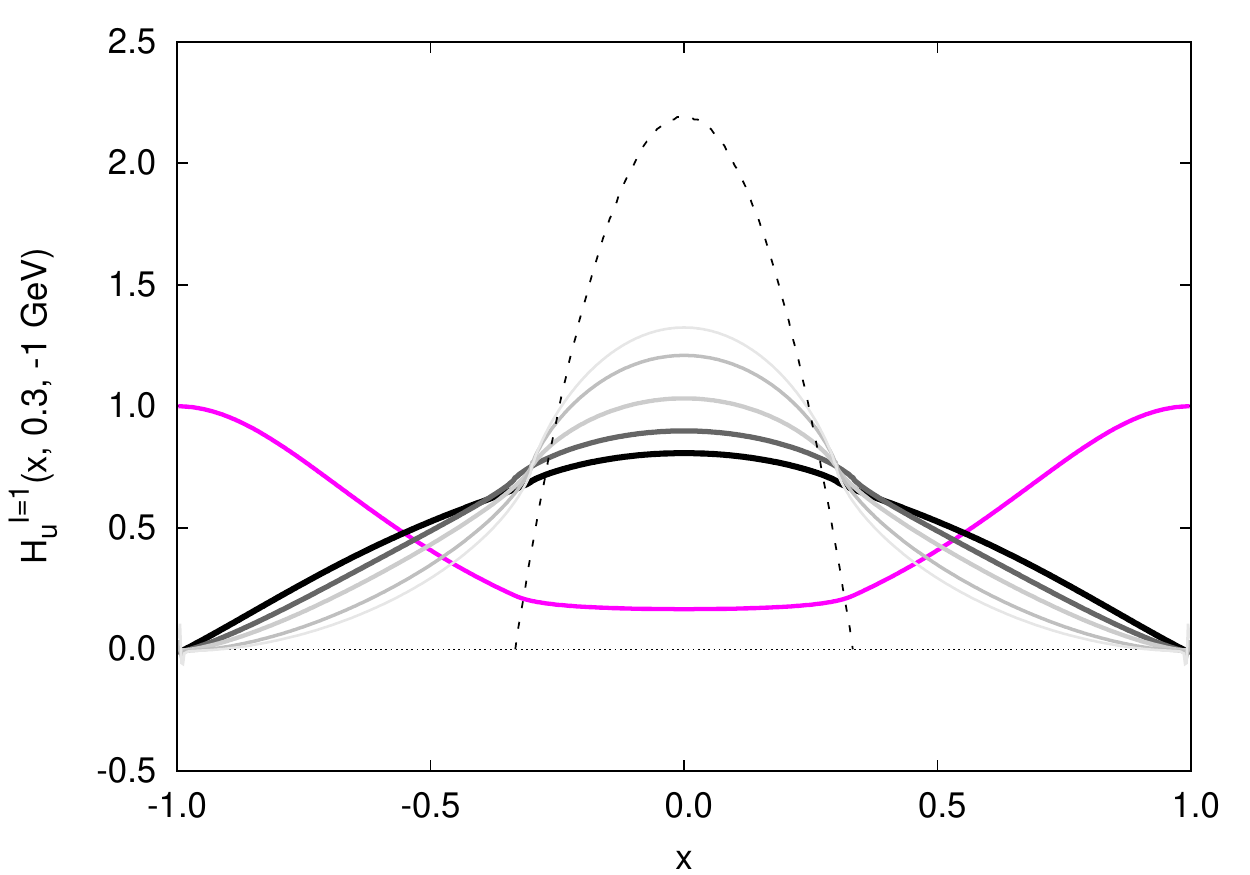}\\
 \includegraphics [height=5.6cm]{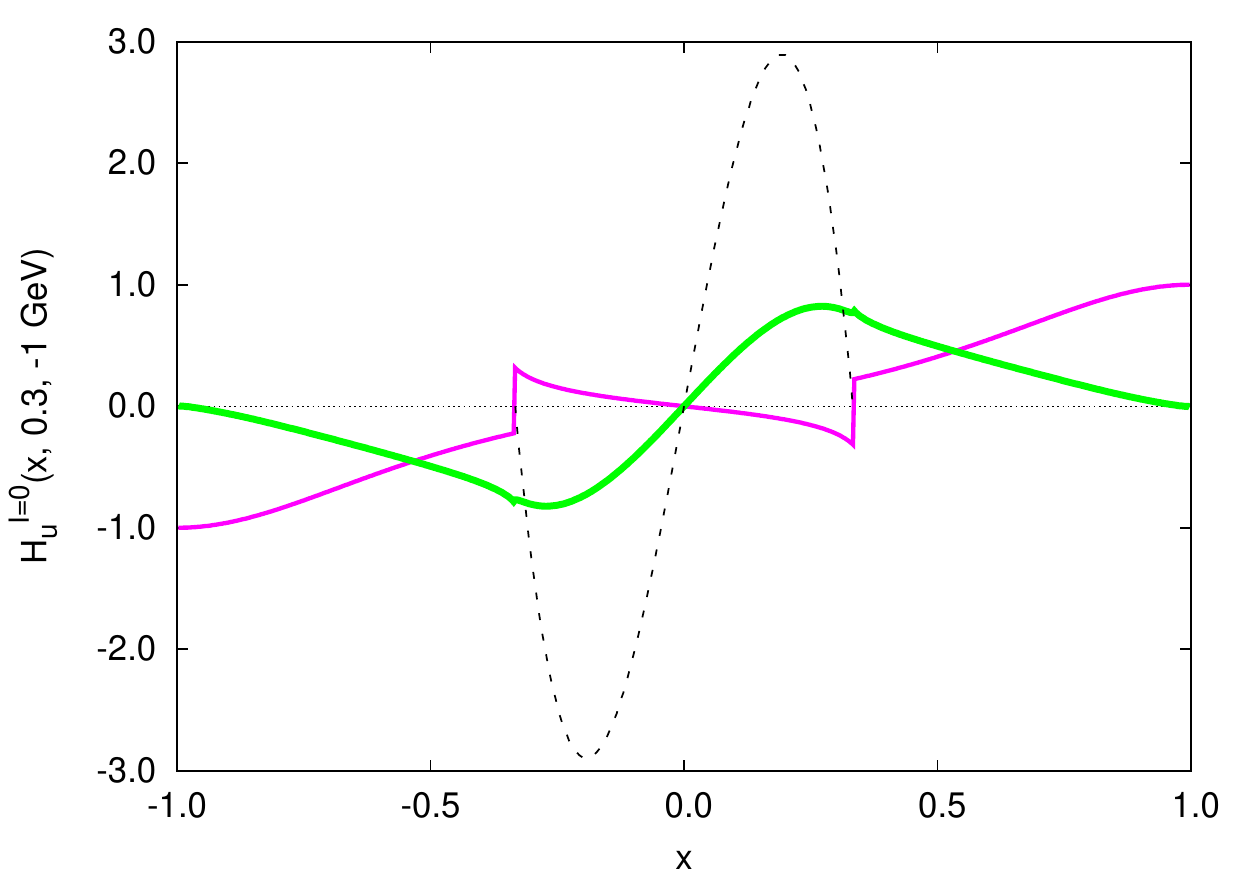}
 \includegraphics [height=5.6cm]{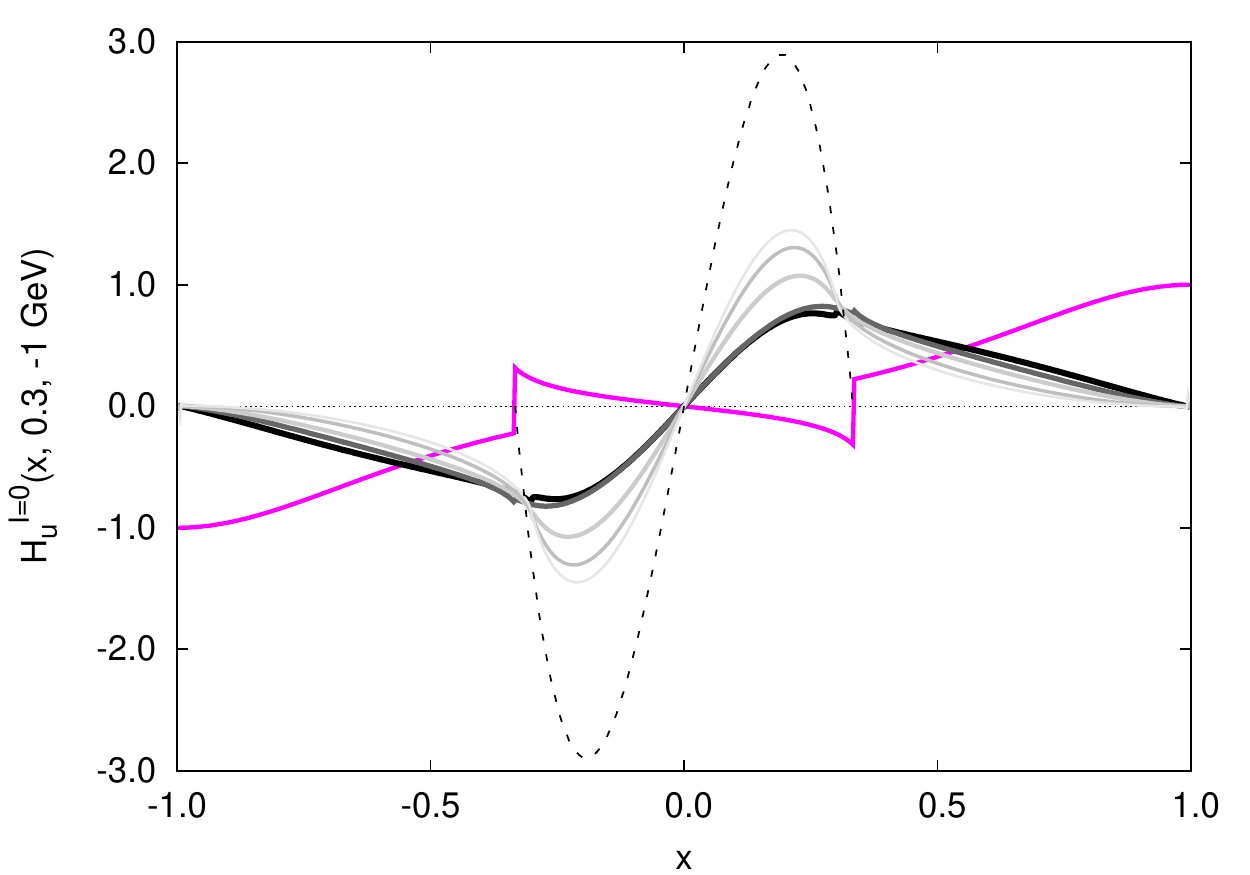}\\
\includegraphics [height=5.6cm]{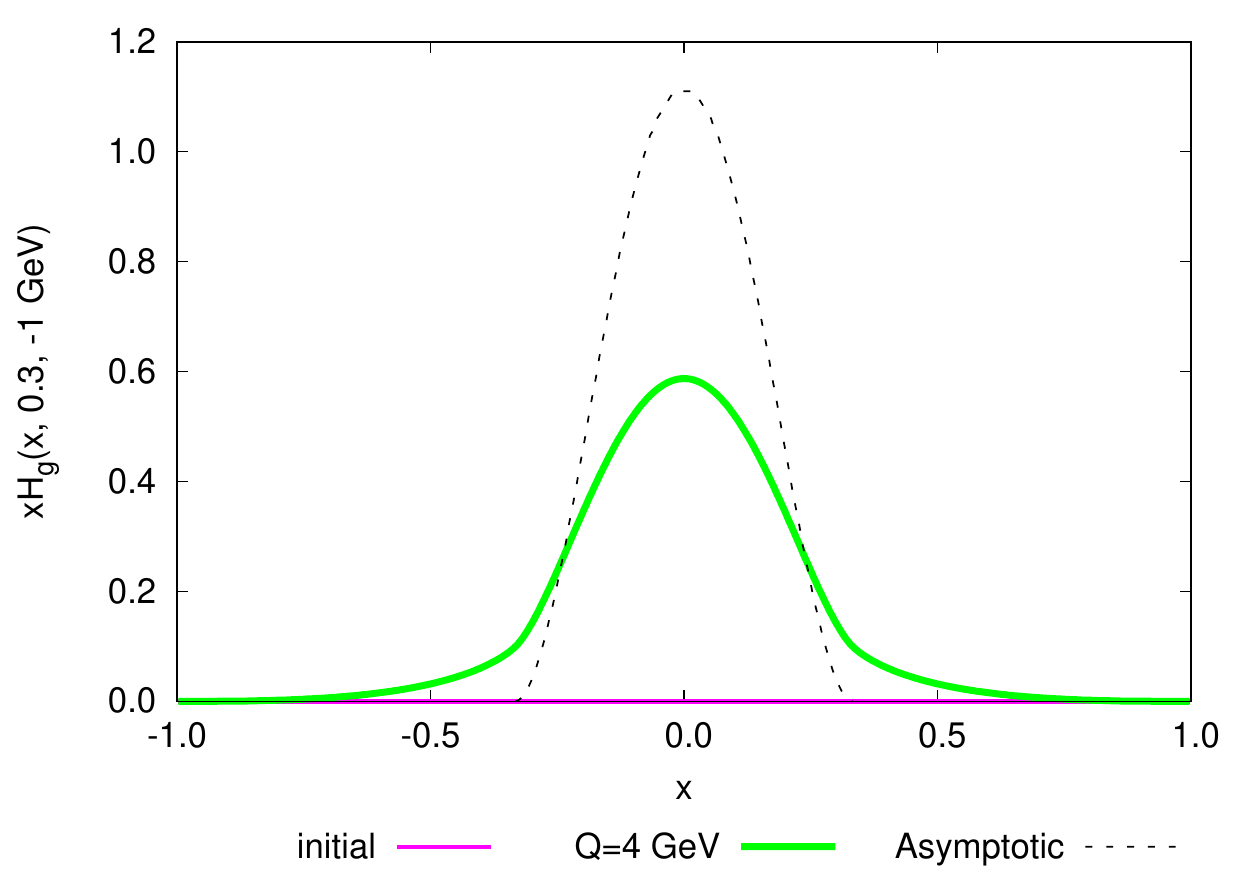}
\includegraphics [height=5.6cm]{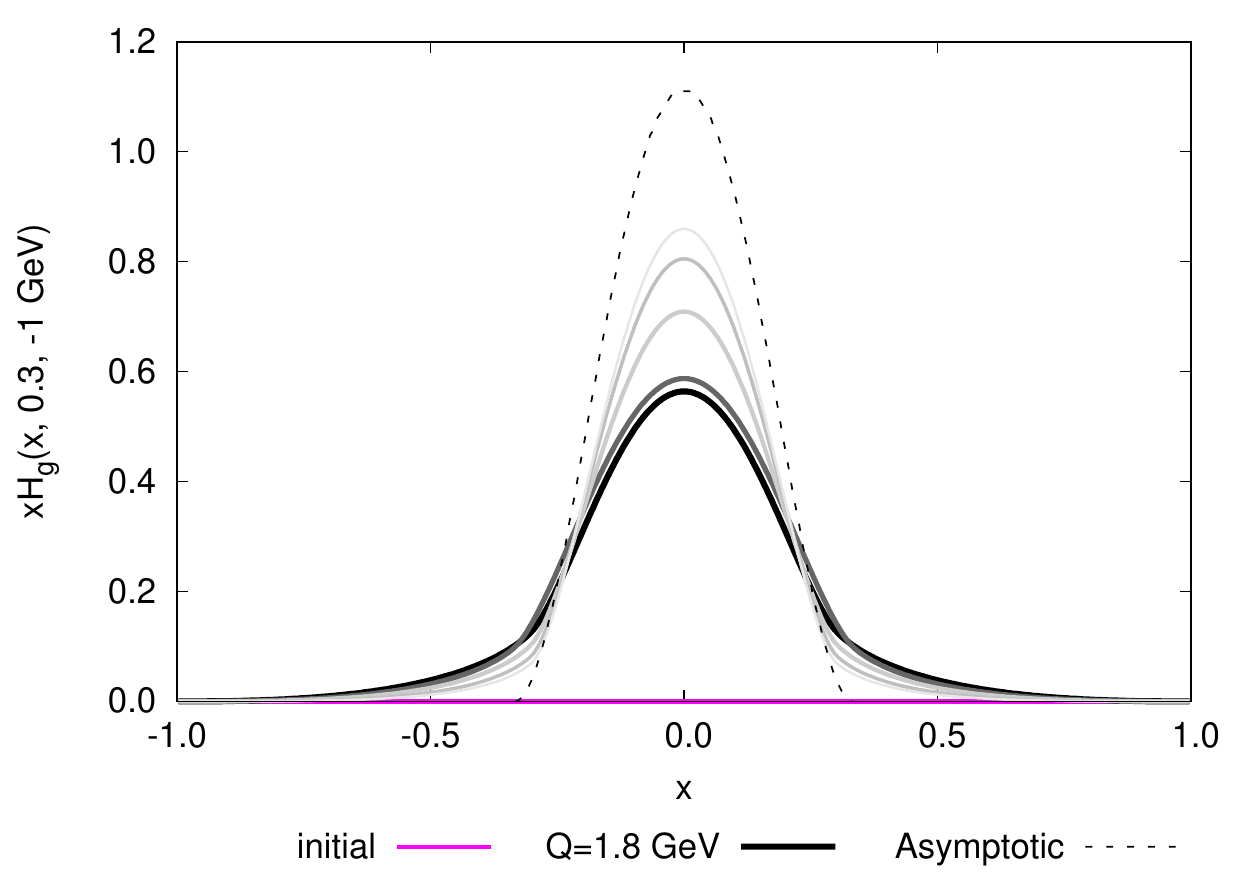}
 \caption{ Results of the LO QCD evolution from the  NJL model  for several values of $t=-1$ GeV$^2$ and for $\xi=1/3$.  From top to bottom, respectively, the non-singlet, the singlet and the gluon distributions. The solid pink lines represent the initial conditions. Left panel: The thick green lines are the evolved result to $Q=4$ GeV and the dashed lines the asymptotic forms.  Right panel: The grey lines are the evolved result to different values of $Q$ ($1,8$; $4$; $10$; $100$ and $1000$ GeV)  and the dotted red line the asymptotic form.  }
 \label{njl-gpd-evol}
\end{figure}

The explicit form of the LO QCD evolution equations for the GPDs can be found in Refs.~\cite{Golec-Biernat:1998ja, Ji:1996nm,Kivel:1999wa, Kivel:1999sk,Mueller:1998fv,Radyushkin:1997ki}.
The effect of LO evolution on the non-singlet, singlet and gluon distributions is depicted on Fig.~\ref{njl-gpd-evol}. The evolution code of Freund and McDermott~\cite{FreundMcDermott} has been used with the value for $Q_0$ and $\Lambda$ as given in (\ref{modelscale}) and in the paragraph preceding the latter expression.
On both panels the LO evolution is performed for the initial scale given by NJL model as given by the expression~(\ref{modelscale}). 
The results presented here are in agreement with the evolution for the NJL model as introduced in Ref.~\cite{Broniowski:2007si}.

The initial distributions are represented by the  thin pink curves. We can see that the isovector GPD (on the top) differs in the ERBL-region from the isoscalar distribution (middle of the panel), where the $D$-term, or $\sigma$-resonance, contributes due to its antisymmetry.

On the left panel, the thick green lines represent the distributions evolved at $Q=4$ GeV. One is immediately struck by the fact that evolution changes completely the end-point behavior as well as smoothes the discontinuities between the two regions, i.e. at $x=\pm \xi$. Even if the evolution seems strong here, it is mentioned in Ref.~\cite{Broniowski:2007si} that the  desired continuity of the functions at the end-points is achieved already at values of  $Q$ infinitesimally larger than $Q_0$. Once more, we refer to the case of the distribution amplitude where a numerical analysis of the truncation of the solution of the evolution equation has been driven, see Fig.~\ref{da-evol-q0}.
\\

An important feature of the evolution in $Q^2$ of the GPDs is that the evolution equations in the ERBL-region depend on the value of the GPDs in both the ERBL-region and the DGLAP-region, while the DGLAP evolution equation do only depend on the value of the GPDs in its very own region.
In other words, the GPDs in the DGLAP-regions feed the ERBL-region. Moreover, as for the DA, evolution pushes the distributions towards low $x$'s. This can be observed in both the left and right panels of Fig.~\ref{njl-gpd-evol}.

As a matter of fact the asymptotic forms for the evolved distributions are solely defined in the ERBL-region. The (anti)symmetry of the  distributions around the point $x=0$ shown in Eqs.~(\ref{singlet}, \ref{non-singlet}) remains unchanged under evolution. The analytic asymptotic solutions were found in Ref.~\cite{Radyushkin:1997ki}. We adopt the forms given in Ref.~\cite{Broniowski:2007si}
\beq
H_q^{NS}(x,\xi)&=&\frac{3}{2\,\xi}  \frac{\xi^2-x^2}{\xi^2}\, F_{\pi}(t)\quad,\nonumber\\
\nonumber\\
H^{S}(x,\xi)&=& \frac{15}{2\,\xi^2} \,\frac{N_f}{4C_F+N_f}\, \frac{x}{\xi}\,\frac{\xi^2-x^2}{\xi^2}\, \frac{1}{2}\left[\theta_1(t)-\xi^2\,\theta_2(t) \right]\quad,\nonumber\\
\nonumber\\
xH^g(x,\xi)&=&\frac{15}{8\,\xi} \,\frac{4\, C_F}{4C_F+N_f}\, \left(\frac{\xi^2-x^2}{\xi^2}\right)^2\, \frac{1}{2}\left[\theta_1(t)-\xi^2\,\theta_2(t) \right] \quad,
\label{asym-gpd}
\eeq
for $-\xi\leq x\leq \xi$ as well as for fixed $t$-values. The $t$-behavior  is given by the sum rules~(\ref{isovec-sr}, \ref{isosca-sr}); a first principles property that remains unchanged after evolution. The presence of the gluon distribution nevertheless affects the isoscalar sum rule~(\ref{isosca-sr})
\beq
\int_{-1}^1 dx\,\left( x\, H^{S}(x, \xi,t, Q^2)+x\, H_g(x, \xi,t, Q^2)\right)&=& \theta_1(t, Q^2)-\xi^2\, \theta_2(t, Q^2)\quad.
\eeq
The values of the gravitational Form Factors for different $t$ values are given in Table~\ref{tablepoly}.

The asymptotic behavior described by Eqs.~(\ref{asym-gpd}) is illustrated on Fig.~\ref{njl-gpd-evol} by the dashed lines. On the right panel, it is clearly shown that the evolution is slow in reaching the solution for $Q\to\infty$ of the GPDs. On the other hand, the evolution is faster at low values of $Q$. This result  we could expect as it happens for the pion DA, which obeys the ERBL evolution equations. In Fig.~\ref{da-evol-us} is depicted the slow evolution of the pion DA at high $Q$-values.
\\

Also, in order to improve in reaching the asymptotic behavior, one should start the evolution from a different value of $Q_0$. However, as it has been explained in Section~\ref{sec:evol-da}, the scale is fixed by the model's results for the second moment of the PDFs. Even going to the NLO could not change such a peculiarity of the model calculations. Effectively, the initial scale for a NLO evolution is found to be higher than for the LO one. However, this lower $Q_0$ for the LO evolution compensates for the effect of NLO evolution: the results are consistent one to the other.
\\

\subsection{From the Experimental Side}

From the experimental point of view the  GPDs of the pion are difficult to determine. Pions decay into muons or photons and therefore the pion Distribution
 Functions cannot be obtained from direct DIS experiments. Rather they have been inferred from Drell-Yan process \cite{Badier:1983mj, Betev:1985pg, Conway:1989fs}
 and direct photon production~\cite{Aurenche:1989sx}  in pion-nucleon 
and pion-nucleus collisions. 
In a near future, the CLAS++ detector with the CEBAF accelerator at 12 GeV  at JLab should allow the observation of pion GPDs through large angle Compton scattering~\cite{prop-jlab}. 

It has been recently proposed, in Ref.~\cite{Amrath:2008vx}, to access the pion GPDs by studying the DVCS on a virtual pion that is emitted by a proton, i.e. the process $e\,p\to e\, \gamma\, \pi^+\,n$.

\chapter{ Transition Distribution Amplitudes}
\label{sec:tda}
\pagestyle{fancy}

Here comes the heart of the thesis. 
The study of Generalized Parton Distribution is extended to distribution amplitudes for  transitions.
\\

Collisions of a real photon and a highly virtual photon are an useful tool for studying fundamental aspects of QCD. Inside this class of processes, the exclusive meson pair production in $\gamma^{\ast}\gamma$ scattering,
\begin{equation}
\gamma_{L}^{\ast}\gamma\rightarrow\pi^{+}\pi^{-}\ \ ,\quad\gamma_{L}^{\ast}
\gamma\rightarrow\rho^{+}\pi^{-} \quad,
\label{processes-tda}%
\end{equation}
 has been
analyzed in Ref.~\cite{Lansberg:2006fv, Pire:2004ie}. 
This process is particularly interesting since it implies, in different kinematical regions, different mechanisms.

\begin{wrapfigure}[18]{r}{75mm}
\begin{center}
\includegraphics[height=5.cm]%
{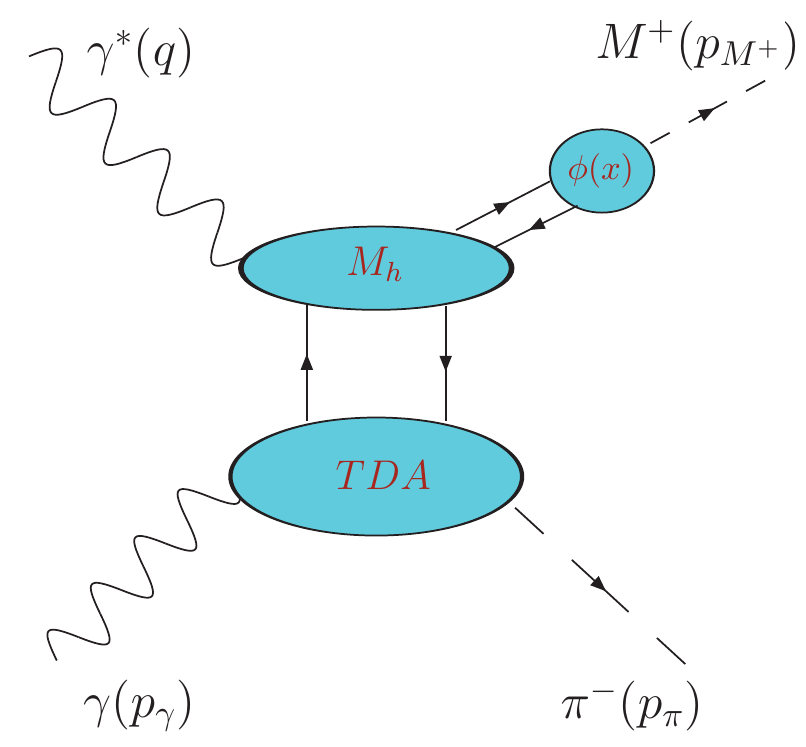}%
\caption{Factorization diagram defining the TDA for the process $\gamma^{\ast
}\gamma\rightarrow\pi M$ at small momentum transfer, $t=\left(  q-p_{M}%
\right)  ^{2},$ and large invariant mass, $s=\left(  p_{\gamma}+q\right)
^{2}$.}%
\label{facto}%
\end{center}
\end{wrapfigure}
 At small momentum transfer $t$ and in the kinematical regime where the photon is highly virtual, a separation between the perturbative and the nonperturbative regimes is assumed to be valid. Through the factorization theorems the amplitude for such reactions can be written as a convolution of a hard part $M_{h}$, a meson
Distribution Amplitude $\phi_{M}$ and another soft part, describing the photon-pion transition by means of the {\it Transition Distribution Amplitudes}.

On the other hand, at small $s$, the factorization corresponds to the GDA's mechanism. The interplay between the two mechanisms in the kinematical region $s, t\ll Q^2$ may lead to an interesting {\it duality} property. This will be considered in the Future Perspectives. Also, the kinematical regime where both $t, s\sim Q^2$ corresponds to large angle scattering {\it \`a la} Brodsky-Lepage and will not be considered here.

The accurate  definition of these objects was  first presented for the mesonic TDAs, by simplicity. In particular, it has been first analyzed  for the pion-photon transitions~\cite{Pire:2004ie} involved in related processes, like hadron annihilation into two photons $\pi^{+}\pi^{-}\rightarrow\gamma^{\ast}\gamma$ and backward virtual Compton scattering $\gamma^{\ast}\pi^{+}\rightarrow\gamma\pi^{+}$.
\\

The concept  of transition distribution  was first introduced by Frankfurt, Polyakov, Strikman and Pobylitsa~\cite{Frankfurt:1998jq, Frankfurt:1999fp} 
 as a study of the transitions $\gamma^{\ast}+p\to Baryon + Meson$ through {\it Skewed} Distribution Amplitudes.

\begin{wrapfigure}[12]{l}{75mm}
\centering
\includegraphics[height=2.6cm]{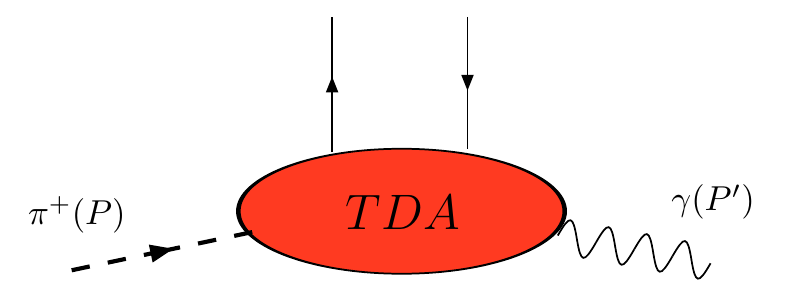}
\caption{The $\pi^+ \to \gamma$ Transition Distribution Amplitude.}%
\label{tdaseul}%
\end{wrapfigure}

In this Chapter, we compile the results of our publications on mesonic Transition Distribution Amplitudes calculated in the NJL model  \cite{Courtoy:2007vy,Courtoy:2008ij}.

We establish the connection between the TDA and the vector and axial pion Form Factors, $F_{V}$
and $F_{A}$. 
The definition of the $\pi\to\gamma$ TDAs are amplified to $\gamma\to\pi$ transitions. 
The calculation and the results in the NJL model are given in details for the {\it skewed} distribution approach. The way to Double Distributions is given that enables us to  compare our results with both phenomenological parameterizations and other model calculations.

In the next Chapter, as an application of the numerical results, we will use the defined quantities to estimate the cross sections for exclusive meson production in  $\gamma^{\ast}\gamma$ scattering~(\ref{processes-tda}) in the same kinematical regime.
\\

\section{The Pion Transition Form Factors}
\label{sec:ffpiondecay}

\begin{figure}[ptb]
\centering
\includegraphics[height=3.3cm]{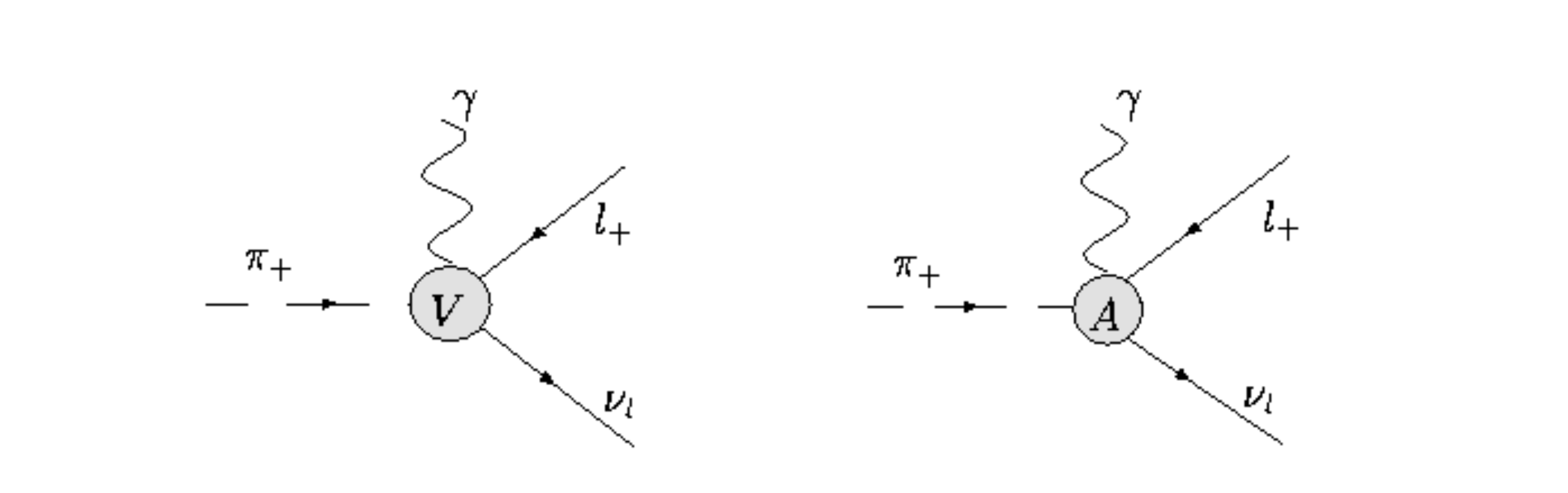}
\caption{Diagrams contributing to the process $\pi^{+}\rightarrow\gamma e^{+}\nu$: the structure-dependent contributions governed, respectively, on the l.h.s. by the vector and, on the r.h.s., the axial Form Factor.}%
\label{ff}%
\end{figure}

\begin{figure}[ptb]
\centering
\includegraphics[height=3.3cm]{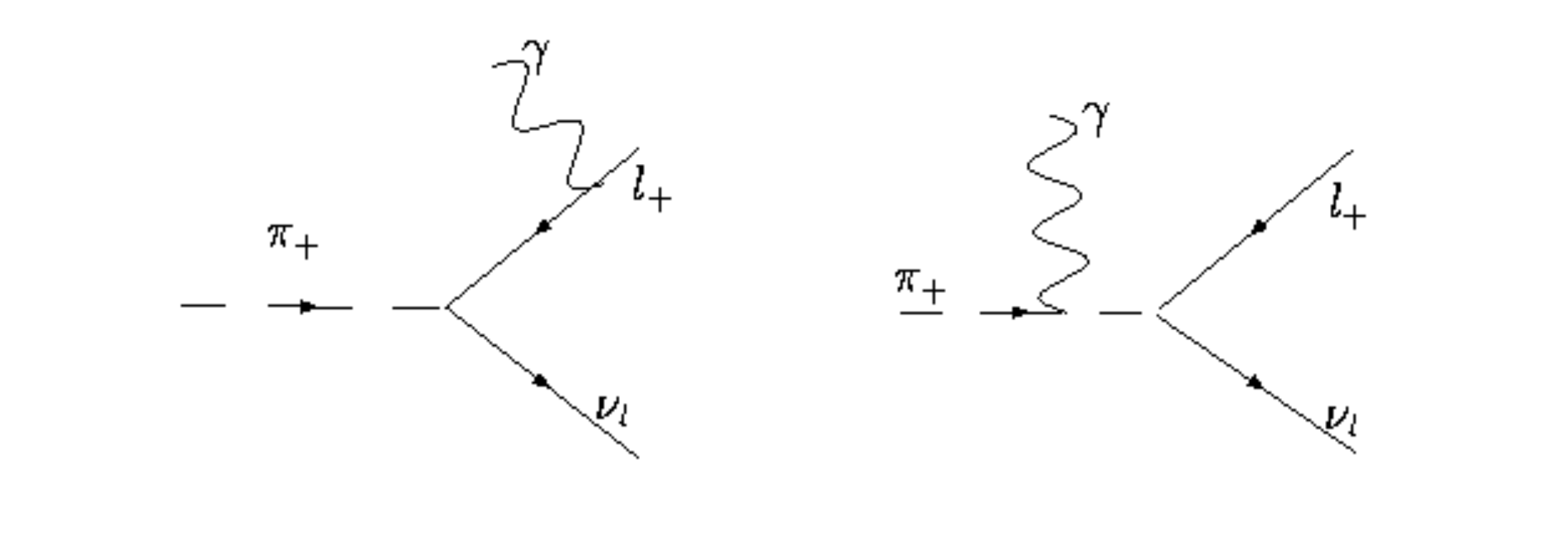}
\caption{Diagrams contributing to the process $\pi^{+}\rightarrow\gamma e^{+}\nu$: the electron (l.h.s.) and the pion (r.h.s.) inner Bremsstrahlung. The electron Bremsstrahlung does not contribute to the $\pi\to\gamma$ hadronic current.}%
\label{brem}%
\end{figure}

\begin{wrapfigure}[14]{r}{7.5cm}
\begin{center}
\includegraphics[
height=2.1741cm,
width=4.9071cm
]%
{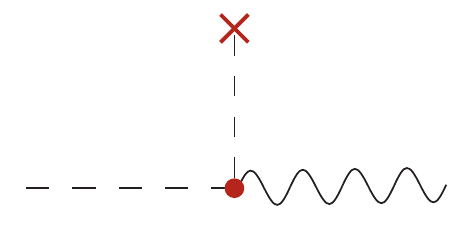}%
\caption{Pion pole contribution between the axial current (represented by a
cross) and the photon-external pion vertex associated to the last contribution
of Eq.~(\ref{2.02}).}%
\label{Fig1}%
\end{center}
\end{wrapfigure}

The pion-photon TDAs are connected, through sum rules, to the vector and
axial-vector pion Form Factors, $F_{V}$ and $F_{A}$. Before giving a proper
definition of the TDAs let us recall the definition of these Form Factors.
They appear in the vector and axial vector hadronic currents contributing to
the decay amplitude of the process $\pi^{+}\rightarrow\gamma e^{+}\nu$ depicted in Figs.~\ref{ff} $ \&$~\ref{brem}. The
precise definition of these currents is \cite{Moreno:1977kx, Bryman:1982et},
\begin{equation}
\left\langle \gamma\left(  p_{\gamma}\right)  \right\vert \bar{q}\left(
0\right)  \gamma_{\mu}\tau^{-}q\left(  0\right)  \left\vert \pi\left(
p_{\pi}\right)  \right\rangle =-i\,e\,\varepsilon^{\nu}\,\epsilon_{\mu\nu\rho\sigma
}\,p^{\rho}_{\gamma}\,p_{\pi}^{\sigma}\,\frac{F_{V}\left(  t\right)  }{m_{\pi}}~,
\label{2.01}%
\end{equation}%
\begin{align}
\left\langle \gamma\left(  p_{\gamma}\right)  \right\vert \bar{q}\left(
0\right)  \gamma_{\mu}\gamma_{5}\tau^{-}q\left(  0\right)  \left\vert
\pi\left(  p_{\pi}\right)  \right\rangle  &  =e\,\varepsilon^{\nu}\left(  p_{\gamma\,\mu}\,p_{\pi\, \nu}-g_{\mu\nu}\,p_{\gamma}\cdot p_{\pi}\right)  \frac{F_{A}\left(
t\right)  }{m_{\pi}}\nonumber\\
&  +e\,\varepsilon^{\nu}\left(  \left(  p_{\gamma}-p_{\pi}\right)  _{\mu}\,p_{\pi\,\nu
}\frac{2\sqrt{2}f_{\pi}}{m_{\pi}^{2}-t}-\sqrt{2}f_{\pi}\,g_{\mu\nu}\right)  ~,
\label{2.02}%
\end{align}
\\
with $f_{\pi}=93
\operatorname{MeV}%
,$ $\varepsilon^{0123}=1$ and $\tau^{-}=\left(  \tau_{1}-i\,\tau_{2}\right)
/2$ as given in Appendix~\ref{sec:convention}. All the structure of the decaying pion is included in the Form Factors $F_{V}$ and $F_{A}$.  We observe that the vector current only contains a
Lorentz structure associated with the $F_{V}$ Form Factor.
  On the other hand, the axial current
is composed of two terms. The first one, defining $F_{A}$, gives the structure
of the pion and is gauge invariant. 
The second one corresponds to the axial current for a point-like
pion and is not gauge invariant. This term coming from the pion inner Bremsstrahlung has, in turn, two different contributions. The first one corresponds to a
point-like coupling between the incoming pion, the outgoing photon and a
virtual pion which is coupled to the axial current. It is depicted in the
diagram of Fig. \ref{Fig1} and can be seen as a result of PCAC, because the
axial current must be coupled to the pion. It isolates the pion pole
contribution from the axial current in a model independent way. The second
contribution of this term, proportional to $f_{\pi}\,g_{\mu\nu}$, corresponds
to a pion-photon-axial current contact term. With these definitions, all the
structure of the pion remains in the Form Factor $F_{A}$. %
The total gauge invariance of the amplitude for the pion radiative decay is ensured by the electron Bremsstrahlung contribution, whose amplitude does not contribute to the hadronic currents~(\ref{2.01},\ref{2.02}).

\vskip 10pt

\section{From Hadronic Currents to Transition Distribution Amplitudes}
\label{sec:kin-tda}

Let us go now to TDAs. \\
For their definition we introduce the light-cone
coordinates as defined in  Appendix~\ref{sec:convention}.

The skewness variable describes the loss of plus momentum of the incident pion, i.e. 
\beq	
\xi&=&{ \left(p_{\pi}-p_{\gamma} \right)  ^{+} \over 2p^{+}}\quad,
\label{skew}
\eeq
with $p^+= {\left(p_{\pi}+p_{\gamma}\right)  ^{+}/2}$.
\\
The explicit expressions for the pion and photon momenta in terms
of the light-front vectors are
\begin{align}
p^{\mu}_{\pi}  &  =\left(  1+\xi\right)  \,\bar{p}^{\mu}-\frac{1}{2}\,\Delta
^{\bot\mu}+\frac{1}{2}\left[  p^{2}\,\left(  1-\xi\right)  +\frac{1}%
{2}\,m_{\pi}^{2}\right]  \,n^{\mu}\quad, \nonumber\\
p^{\mu}_{\gamma} &  =\left(  1-\xi\right)  \,\bar{p}^{\mu}+\frac{1}{2}%
\,\Delta^{\bot\mu}+\frac{1}{2}\left[  p^{2}\,\left(  1+\xi\right)  -\frac
{1}{2}\,m_{\pi}^{2}\right]  \,n^{\mu}\quad. 
\label{pplc}%
\end{align}
Here we have 
\beq
p^{2}=\left[  (p_{\pi}+p_{\gamma})/2\right]  ^{2}=m_{\pi}^{2}/2-t/4\, , &&\quad
\Delta^{\mu}=\left(  p_{\gamma}-p_{\pi}\right)  ^{\mu}\quad, \nonumber\\
\nonumber\\
\Delta^{\bot\mu}=\left(0,\Delta^{1},\Delta^{2},0\right)  =\left(  0,\vec{\Delta}^{\bot},0\right)  \, ,
&&\quad\Delta^{2}=t\quad.
\label{kin-tda}
\eeq
Given the kinematics here above,  it is easy to determine $\vec{\Delta}^{\perp 2}$
\begin{eqnarray}
\vec{\Delta}^{\perp 2}=2\,\xi\, (1-\xi)\,m_{\pi}^2\,-\, (1-\xi^2)\,t\quad,
\end{eqnarray}
which gives the expression for $t$
\begin{eqnarray}
t&=&\frac{2\xi(1-\xi)m_{\pi}^2-\vec{\Delta}^{\perp 2}}{1-\xi^2}\quad.
\end{eqnarray}
The value of the skewness variable is constrained  to range between
 \beq
{t \over\left(  2m_{\pi}^{2}-t\right)  }<\xi<1\quad.
\label{xi-lim-tda}
\eeq
Actually
there is no symmetry relating the distributions for negative and positive
$\xi$ which could have constrained the values of the skewness variable to be
positive, like for GPDs.
\\

The polarization vector of the real photon, $\varepsilon,$
must satisfy the transverse condition, 
\beq
\varepsilon.p_{\gamma}=0\quad,
\nonumber
\eeq
 as well as an
additional gauge fixing condition. 
When going from the hadronic currents to the parton distribution amplitudes, we need
$\varepsilon.n=\varepsilon^{+}/p^{+}$ to kinematically become higher-twist,
i.e. 
\beq
\varepsilon.n\rightarrow0 \quad \mbox{ when} \quad p^{+}\rightarrow\infty\quad .
\nonumber
\eeq
The standard gauge fixing conditions, namely $\varepsilon^{0}=0$ or $\varepsilon
^{+}=0,$ both satisfy the previous requirement. In fact, all what we need is that
the components of the polarization vector remain finite when $p^{+}$ goes to infinity. 

A general expression
for $\varepsilon$ is 
\beq
\varepsilon&=&\vec{\varepsilon}^{\perp}+\alpha \frac{\bar p}{p^+} +\beta \,p^+\, n\quad.
\label{gen-gauge}
\eeq
The transverse condition for the real photon polarization provides us for a first constraint, e.g. on $\beta$,
\beq
\beta\, p^+&=&\frac{1}{1-\xi}\, \left[ \frac{1}{2}\, \vec{\varepsilon}^{\perp}\cdot \vec{\Delta}^{\perp} -\frac{\alpha}{2\, p^+}\, \left(\frac{m^2_{\pi}}{2}\xi-\frac{t}{4}\, (1+\xi) \right)\right]\quad.
\nonumber
\eeq 
Therefore the scalar product  $\varepsilon\cdot \Delta$ is
\beq
\varepsilon\cdot \Delta&=&-\frac{\vec{\varepsilon}^{\perp}.\vec{\Delta}^{\perp}}{1-\xi}+{\cal O}\left(\frac{\alpha}{p^+}\right)\quad,
\eeq
 where the gauge dependence inclosed in $\alpha$ is kinematically higher-twist, i.e. goes like $1/p^+$.  Also,  the decomposition $\varepsilon\cdot \Delta$ depends on the kinematics, what is explicit through the $1-\xi$ in the latter expression. Henceforth the tensor decomposition  would rather be chosen to be $\vec{\varepsilon}^{\perp}.\vec{\Delta}^{\perp}$.
\\
\\

 The vector and axial TDAs are the Fourier transform
of the matrix element of the bilocal currents separated by a
light-like distance
\beq
\bar{q}\left(  -z/2\right) \gamma_{\mu}\left[  \gamma_{5}\right]  q\left(  z/2\right)  \quad.
\nonumber
\eeq
Then, they are directly related to the currents defined
in Eqs. (\ref{2.01}-\ref{2.02}) through the sum rules
\beq
&&\int_{-1}^{1}dx\int\frac{dz^{-}}{2\pi}e^{ixp^{+}z^{-}}\left.  \left\langle
\gamma\left(  p_{\gamma}\right)  \right\vert \bar{q}\left(  -\frac{z}%
{2}\right)  \gamma_{\mu}\left[  \gamma_{5}\right]  \tau^{-}q\left(  \frac
{z}{2}\right)  \left\vert \pi\left(  p_{\pi}\right)  \right\rangle \right\vert
_{z^{+}=z^{\bot}=0}\nonumber\\
&&=\frac{1}{p^{+}}\left\langle \gamma\left(  p_{\gamma
}\right)  \right\vert \bar{q}\left(  0\right)  \gamma_{\mu}\left[  \gamma
_{5}\right]  \tau^{-}q\left(  0\right)  \left\vert \pi\left(  p_{\pi}\right)
\right\rangle \quad. 
\label{2.03}%
\eeq
With this connection we can introduce the leading-twist decomposition of the
bilocal currents. For that we need the light-front vectors $\bar{p}^{\mu
}$ and $n^{\mu}$ defined in Eq.~(\ref{lightvector}), which leads to the expressions~(\ref{pplc}-\ref{kin-tda}). 
\\
To leading-twist, the $\pi^+\to\gamma$ TDAs are defined
\\
\beq
&&\hspace{-1cm}\int\frac{dz^{-}}{2\pi}e^{ixp^{+}z^{-}}\left.  \left\langle \gamma\left(
p_{\gamma}\right)  \right\vert \bar{q}\left(  -\frac{z}{2}\right)
\rlap{$/$}n\hspace*{-0.05cm}\tau^{-}q\left(  \frac{z}{2}\right)  \left\vert
\pi^+\left(  p_{\pi}\right)  \right\rangle \right\vert _{z^{+}=z^{\bot}=0}\nonumber\\
\nonumber\\
&&=\frac{i}{p^{+}}\,e\,\varepsilon^{\nu}\,\epsilon_{\mu\nu\rho\sigma}\,n^{\mu
}\,p^{\rho}\,\Delta^{\sigma}\,\frac{V^{\pi^{+}}\left(  x,\xi,t\right)  }
{\sqrt{2}f_{\pi}}\quad, 
\label{2.04}
\\
\nonumber\\
\nonumber\\
%
%
&&\hspace{-1cm}\int\frac{dz^{-}}{2\pi}e^{ixp^{+}z^{-}}\left.  \left\langle \gamma\left(
p_{\gamma}\right)  \right\vert \bar{q}\left(  -\frac{z}{2}\right)
\rlap{$/$}n\hspace*{-0.05cm}\gamma_{5}\tau^{-}q\left(  \frac{z}{2}\right)
\left\vert \pi^+\left(  p_{\pi}\right)  \right\rangle \right\vert _{z^{+}=z^{\bot}=0}
\nonumber\\
\nonumber\\
&&  =\frac{1}{p^{+}}e\,\left(  \vec{\varepsilon}^{\bot}\cdot\vec{\Delta}^{\bot
}\right)  \frac{A^{\pi^{+}}\left(  x,\xi,t\right)  }{\sqrt{2}f_{\pi}%
}
 +\frac{1}{p^{+}}e\,\left(  \varepsilon\cdot\Delta\right)  \frac{2\sqrt
{2}f_{\pi}}{m_{\pi}^{2}-t}~\epsilon\left(  \xi\right)  ~\phi\left(
\frac{x+\xi}{2\xi}\right)  \quad, \label{2.05}%
\eeq
\\
where $\epsilon\left(  \xi\right)  $ is equal to $1$ for $\xi>0,$ and equal to
$-1$ for $\xi<0$. Here $V\left(  x,\xi,t\right)  $\ and $A\left(
x,\xi,t\right)  $ are respectively the vector and axial TDAs. They are defined
as dimensionless quantities. From the condition (\ref{2.03}) we observe that
they obey the following sum rules,
\begin{equation}
\int_{-1}^{1}dx~V^{\pi^{+}}\left(  x,\xi,t\right)  =\frac{\sqrt{2}f_{\pi}%
}{m_{\pi}}F_{V}\left(  t\right)  ~, \label{2.06}%
\end{equation}%
\begin{equation}
\int_{-1}^{1}dx~A^{\pi^{+}}\left(  x,\xi,t\right)  =\frac{\sqrt{2}f_{\pi}%
}{m_{\pi}}F_{A}\left(  t\right)  \quad.
 \label{2.07}%
\end{equation}
\\

In the second term of Eq. (\ref{2.05}), we have introduced the pion
Distribution Amplitude (DA) $\phi\left(  x\right) $, whose expression in given by Eq.~(\ref{pda}). 
The pion DA $\phi\left(  x\right) $ vanishes outside the region $x\in\left[  0,1\right]  $ and satisfies
the normalization condition~(\ref{normda}).
\\
\begin{wrapfigure}[16]{r}{7cm}
\begin{center}
\includegraphics[height=4.cm]%
{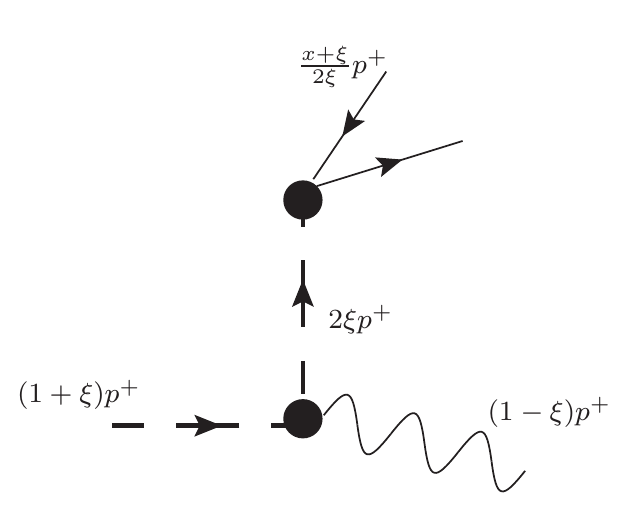}%
\caption{Pion pole contribution to the axial bilocal current corresponding to
the last term of equation (\ref{2.05}).}%
\label{Fig2}%
\end{center}
\end{wrapfigure}
This second term has been introduced in order to isolate the pion pole
contribution of the axial current in a model independent way, as we have done
in Eq. (\ref{2.02}) for the $\pi^{+}\rightarrow\gamma e^{+}\nu$ process.
Therefore, all the structure of the pion remains in the TDA $A\left(
x,\xi,t\right)  $. It can be seen as a result of PCAC, because the axial
current must be coupled to the pion. Therefore, this term is not a peculiarity
of the pion-photon TDAs. A similar pion term will be present in the Lorentz
decomposition in terms of distribution amplitudes of the axial current for any
pair of external particles. We emphasize that it is a model
independent definition, because we have defined the numerator of the pion pole
term as the residue at the pole $t=m_{\pi}^{2}$.  With this definition, all the
structure dependence related to the outgoing $\pi^{\pm}$ is included in
$A\left(  x,\xi,t\right)$. Moreover, the pion pole contribution can be
estimated in a phenomenological way, as we will see later on.
\\

A pion exchange contribution was
analyzed in \cite{Mankiewicz:1998kg, Penttinen:1999th} for the axial
helicity-flip GPD and, in \cite{Tiburzi:2005nj}, a similar structure for the
axial current was obtained using different arguments. %
This term we have represented in Fig.~\ref{Fig2} is only non-vanishing in the ERBL region, i.e. the $x\in\left[
-\xi,\xi\right]  $ region. The kinematics of this region allow the emission or
absorption of a pion from the initial state, which is described through the
pion DA. And it can be seen from Fig. \ref{Fig2} that positive values of $\xi$
corresponds to an outcoming virtual pion, whereas negative values of $\xi$
describe an incoming virtual pion. The latter is related to the matrix element
$\left\langle \pi\left(  p_{\pi}\right)  \right\vert \bar{q}\rlap{$/$}n\gamma
_{5}\tau^{-}q\left\vert 0\right\rangle ,$ instead of the one present in Eq.
(\ref{pda}), what gives rise to the minus sign included in $\epsilon\left(
\xi\right)  .$
\\

In order to make  possible the connection with other works,  it must be realized that  the axial TDA must be defined from an electromagnetic gauge independent tensorial structure. As above-mentioned, the product $\varepsilon\cdot \Delta$ contains gauge dependent terms. Since the TDAs are gauge independent quantities, the tensorial structure defining them has to be itself gauge independent.
The only possibility is hence, see  Eq.~(\ref{2.05}),
\beq
\vec{\varepsilon}^{\bot}.\vec{\Delta}^{\bot} \quad,
\eeq
which is equal to, e.g.,  $\left(  1-\xi\right)\left(  \varepsilon.\Delta\right) $ in the $\varepsilon^0=0$ gauge we have chosen to work in.

In many papers present in the literature, the r.h.s. of Eq.~(\ref{2.05})
 contains only the $A$ term. In a general case the $A$ term
and the pion pole terms have different tensor structures as shown by Eq.~(\ref{2.02}). However we can fix the
gauge convention and the frame conventions in such a way that these two tensorial structures coincide.
In other words, should we have chosen another gauge and a frame in which all the transverse component is in the pion momentum, i.e. $\vec{p}_{\gamma}^{\perp}=0$, the tensorial structure of the axial TDA and the pion pole would have coincided: we have the freedom to choose $\varepsilon_{\mu}$ with only transverse component. In that case,
\beq
\varepsilon\cdot \Delta
&=&-\vec{\varepsilon}^{\perp} \cdot \vec{\Delta}^{\perp}\quad.
\nonumber
\eeq
Only with this particular definition,  we can change our definition of the axial TDA to
\beq
&&\hspace{-1cm}\int\frac{dz^{-}}{2\pi}e^{ixp^{+}z^{-}}\left.  \left\langle  \gamma\left(  p_{\gamma}\varepsilon\right)   \right\vert \bar{q}
\left(  -\frac{z}{2}\right)  \gamma^{+}\hspace*{-0.05cm}\gamma_{5}\ \tau^{-}\ q\left(  \frac{z}{2}\right)
\left\vert\pi^{+}\left(p_{\pi}\right)  \right\rangle\right\vert _{z^{+}=z^{\bot}=0}\nonumber\\
&=&\ \frac{1}{p^{+}}\left[  \ e\,\left(
\vec{\varepsilon}^{\bot}\cdot(\vec{p}_{\gamma}^{\bot}-\vec{p}_{\pi}^{\bot
})\right)  \frac{\bar{A}^{\pi^{+}\rightarrow\gamma}(x,\xi,t)}{\sqrt{2}%
f_{\pi}}\right]\,,
\label{alter-decomp-tens}
\eeq
with%
\begin{equation}
\bar{A}^{\pi^{+}\rightarrow\gamma}(x,\xi,t)=A^{\pi^{+}\rightarrow\gamma}(x,\xi,t)-\frac{4f_{\pi}^{2}}{m_{\pi}^{2}-t}~\epsilon(\xi)~\phi^{\pi}\left(
\frac{x+\xi}{2\xi}\right)  \quad.
\label{Abarra-A}%
\end{equation}
The latter expression shows that the pion pole contribution to the axial TDA
is closely related to the $D$-term of the Generalized Parton Distributions %
\cite{Polyakov:1999gs}. %
However it must be realized that in this case it
gives an explicit contribution to the sum rule
\begin{equation}
\int_{-1}^{1}dx\ \bar{A}^{\pi^{+}\rightarrow\gamma}\left(  x,\xi,t\right)
=\frac{\sqrt{2}\ f_{\pi}}{m_{\pi}}\ F_{A}\left(  t\right)  -\frac{4f_{\pi}
^{2}}{m_{\pi}^{2}-t}\ \left(  p_{\gamma}-p_{\pi}\right)  \cdot n\quad.
\label{alt-sr-tda}
\end{equation}
\\
In what follows,  the definitions derived from the tensorial decomposition~(\ref{2.04}, \ref{2.05}) of the hadronic matrix element for the pion radiative decay will mainly be used.

\section{A Field Theoretical Approach to the Pion-Photon TDA}
\label{sec:tda-njl}

\begin{wrapfigure}{r}{70mm}
\centering
\includegraphics [height=3.5cm]{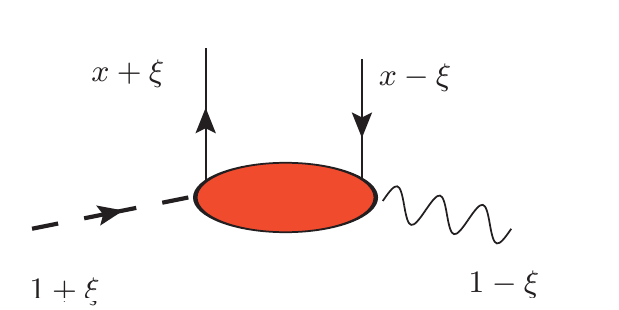}
\caption{Diagram representing the $\pi\to\gamma$ TDAs.}
\label{Fig3bis}%
\end{wrapfigure}
 In Chapter \ref{sec:da-cal}  we have defined a method of calculation for the
pion DA in a field theoretical scheme, treating the pion as a bound state of
quarks and antiquarks in a fully covariant manner using the Bethe-Salpeter
equation. We then applied  the same method for evaluating the pion GPDs in Chapter~\ref{sec:gpd}. The same is done here for the evaluation of
 the pion-photon TDAs~\cite{Courtoy:2007vy}.
This method has enormous advantages because it preserves all the physical
invariances of the problem. Therefore, any property as sum rules or
polynomiality is expected to be preserved.
\\

\begin{figure}[tb]
\centering
 \mbox{
\subfigure[]{\includegraphics [height=3.5cm]{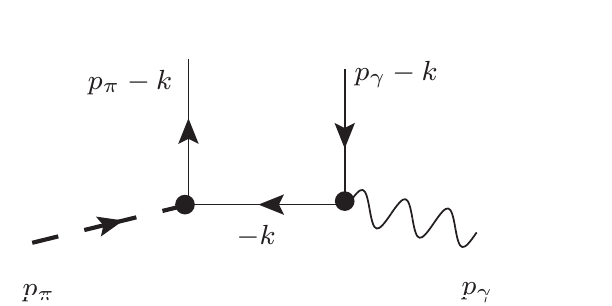}}
\subfigure[]{\includegraphics [height=5.5cm]{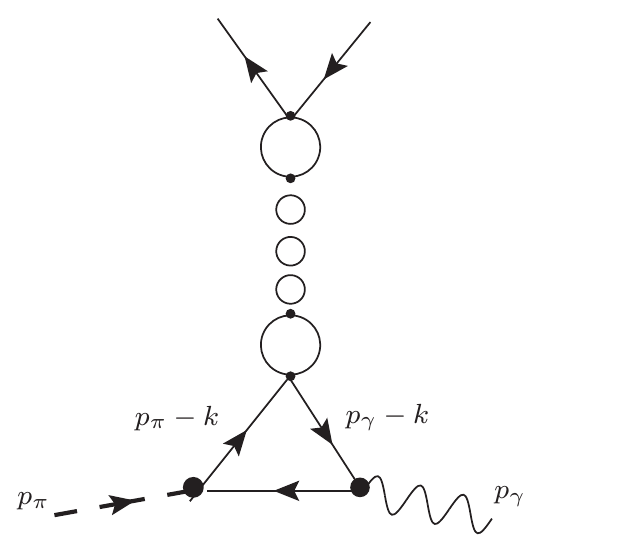}}}
\caption{Diagrams contributing to the TDA. We have depicted diagrams in which
a quark $u$ is changed into a quark $d$ by the bilocal current$.$ There are
similar diagrams in which the antiquark $\bar{d}$ is changed into a $\bar{u}$}%
\label{Fig3}%
\end{figure}

As usual, we consider that the process is dominated by the handbag diagram.
Each TDA has two related contributions, depending on which quark ($u$ or $d$)
of the pion is scattered off by the deeply virtual photon. In Fig. \ref{Fig3} we have
depicted the diagrams in which the photon scatters off the $u$-quark. We
observe that there are two kinds of contributing diagrams. In the first one
the ${d}$-antiquark appears as the intermediate state, while in the second
the bi-local current couples to a quark-antiquark pair coupled in the pion
channel. The latter is present only for the axial current and includes the
pion pole contribution.%

The details of the method of calculation are given in Ref.
\cite{Theussl:2002xp}. In the present case we obtain, from the first kind of
diagram of Fig. \ref{Fig3}, the following contributions~\cite{Courtoy:2007vy} 
\begin{align}
&  \int\frac{dz^{-}}{2\pi}e^{ixp^{+}z^{-}}\left.  \left\langle \gamma\left(
p_{\gamma},\varepsilon\right)  \left\vert \bar{q}\left(  -\frac{z}{2}\right)
\gamma^{\mu}\left[  \gamma_{5}\right]  \tau^{-}q\left(  \frac{z}{2}\right)
\right\vert \pi^{+}\left(  p_{\pi}\right)  \right\rangle \right\vert _{z^{+}%
=z^{\bot}=0}
\nonumber\\
&  =-e\int\frac{d^{4}k}{\left(  2\pi\right)  ^{4}}\varepsilon_{\nu}\left\{
-\delta\left(  xp^{+}-\frac{1}{2}\left(  p_{\gamma}^{+}+p_{\pi}^{+}-2k^{+}\right)
\right)  \right. \nonumber\\
& \hspace{.6cm} \mathbb{T}\mathrm{r}\left[  \frac{1}{2}\left(  \frac{1}{3}+\tau_{3}\right)
\,\gamma^{\nu}\,iS\left(  p_{\gamma}-k\right)  \,\gamma^{\mu}\left[
\gamma_{5}\right]  \tau^{-}\,iS\left(  p_{\pi}-k\right)  \,\Phi^{\pi^{+}}\left(
k,p_{\pi}\right)  \,iS\left(  -k\right)  \right] \nonumber\\
& \hspace{.6cm}  -\delta\left(  xp^{+}-\frac{1}{2}\left(  2k^{+}-p_{\pi}^{+}-p_{\gamma}^{+}\right)
\right) \nonumber\\
& \hspace{.6cm}  \left.  \mathbb{T}\mathrm{r}\left[  \frac{1}{2}\left(  \frac{1}{3}+\tau
_{3}\right)  \,\gamma^{\nu}\,iS\left(  k\right)  \,\Phi^{\pi^{+}}\left(
k,p_{\pi}\right)  \,iS\left(  k-p_{\pi}\right)  \,\gamma^{\mu}\left[  \gamma_{5}\right]
\tau^{-}\,iS\left(  k-p_{\gamma}\right)  \right]  \right\}  \quad,
\label{3.01}%
\end{align}
where $S\left(  k\right)  $ is the Feynman propagator of the quark and
$\Phi^{\pi^{+}}\left(  k,p_{\pi}\right)  $ is the Bethe-Salpeter amplitude for the
pion. Here $\mathbb{T}\mathrm{r}()$ represents the trace over spinor, color,  flavor and Dirac indices. 

The first contribution in Eq. (\ref{3.01}) is the one
depicted in the first diagram of Fig. \ref{Fig3}. The second contribution
corresponds to a similar diagram but changing quarks $u$ and $d$. In the NJL
model, $\Phi^{\pi^{+}}\left(  k,p_{\pi}\right)  $ is as simple as,
\begin{equation}
\Phi^{\pi^{+}}\left(  k,p_{\pi}\right)  =ig_{\pi qq}\,i\gamma_{5}\,\tau^{\pi^+}\quad,
\label{3.02}%
\end{equation}
where $g_{\pi qq}$ is the pion-quark coupling constant defined in Eq.(
\ref{gpiqq}).

\subsection{Vector TDA}

The vector TDA has contribution only from this first kind of diagrams. We can
express $V\left(  x,\xi,t\right)  $ as the sum of the active $u$-quark and the
active ${d}$-quark distributions. The first contribution will be
proportional to the $d$'s charge, and the second contribution to the $u$'s
charge. Therefore, we can write%
\begin{equation}
V^{\pi^{+}}\left(  x,\xi,t\right)  = Q_{d} v_{u\rightarrow d}^{\pi^{+}%
}\left(  x,\xi,t\right)  +Q_{u} v_{\bar{d}\rightarrow\bar{u}}^{\pi^{+}%
}\left(  x,\xi,t\right)  \quad. \label{3.03}%
\end{equation}
Isospin relates these two contributions, 
\beq
v_{u\rightarrow d}^{\pi^{+}}\left(x,\xi,t\right)  &=&v\left(x,\xi,t\right) \quad,\nonumber\\
v_{\bar{d}\rightarrow\bar{u}}^{\pi^{+}}\left(x,\xi,t\right) & =& v\left(-x,\xi,t\right) \quad.
\nonumber
\eeq
A direct calculation gives%
\begin{equation}
v\left(  x,\xi,t\right)  =8\,N_{c}\,f_{\pi}\,g_{\pi
qq}\,m\,\tilde{I}_{3v}\left(  x,\xi,t\right)  \quad, 
\label{3.04}%
\end{equation}
with %
\beq
&&\tilde{I}_{3}^{TDA}\left(  x,\xi,t\right)  \nonumber\\
&&=i\int\frac{d^{4}k}{\left(  2\pi\right)
^{4}}\,\delta\left(  x-1+\frac{k^{+}}{p^{+}}\right)  \frac{1}{\left(
k^{2}-m^{2}+i\epsilon\right)  \,\left[  \left(  p_{\gamma}-k\right)
^{2}-m^{2}+i\epsilon\right]  \,\left[  \left(  p_{\pi}-k\right)  ^{2}-m^{2}%
+i\epsilon\right]  } \quad,
\nonumber\\
 \label{3.05}%
\eeq
given by the expression~(\ref{i3tda}) in Appendix.
\\

In this integral, we first perform the integration over $k^{-}$. The pole
structure of the integrand fixes two non-vanishing contributions to
$v \left(  x,\xi,t\right)  $, the first one in the
region $\xi<x<1$, corresponding to the quark contribution, and the second in
the region $-\xi<x<\xi,$ corresponding to a quark-antiquark contribution.
Given the relation (\ref{3.03}), the support of the entire vector TDA,
$V^{\pi^{+}}\left(  x,\xi,t\right)  $, is therefore $x\in\lbrack-1,1]$. The
analytical expression for (\ref{3.05}) is given by Eqs. (\ref{i3tda}-\ref{i3tda-neg}). 
\\

For the $\pi^{0}\to \gamma^{\ast}\gamma$ process, the contributions of the first type of diagram (Fig.~\ref{Fig3})
can be related to $v\left(  x,\xi,t\right)  $%
\begin{align}
V_{u}^{\pi^{0}}  &  =\frac{Q_{u}}{\sqrt{2}}\left(  v \left(  x,\xi,t\right)  +v\left(-x,\xi,t\right)  \right)  \quad,\nonumber\\
V_{d}^{\pi^{0}}  &  =\frac{Q_{d}}{\sqrt{2}}\left(  v\left(  x,\xi,t\right)  +v\left(-x,\xi,t\right)  \right)  \quad. 
\label{vpi0}%
\end{align}

\subsection{Axial TDA}

Turning our attention to the axial TDA, we find a new contribution arising
from the second diagram of Fig. \ref{Fig3}. This second contribution comes
from the re-scattering of a $q\bar{q}$ pair in the pion channel. Therefore%
\begin{align}
&  \int\frac{dz^{-}}{2\pi}e^{ixp^{+}z^{-}}\left.  \left\langle \gamma\left(
p_{\gamma},\varepsilon\right)  \left\vert \bar{q}\left(  -\frac{z}{2}\right)
\gamma^{\mu}\gamma_{5}\tau^{-}q\left(  \frac{z}{2}\right)  \right\vert \pi
^{+}\left(  p_{\pi}\right)  \right\rangle \right\vert _{z^{+}=z^{\bot}=0}\nonumber\\
&  =(\text{\ref{3.01}})+\sum_{i}M^{i}\frac{2ig}{1-2g\Pi_{ps}\left(  t\right)
}N^{i}\quad, \label{3.06}%
\end{align}
where $i$ is an isospin index,%
\begin{align}
M^{i}  &  =-e\int\frac{d^{4}k^{\prime}}{\left(  2\pi\right)  ^{4}}%
\varepsilon_{\nu}\left\{  -\mathbb{T}\mathrm{r}\left[  \phi^{\pi^{+}}\left(
k^{\prime},p_{\pi}\right)  \,iS\left(  -k^{\prime}\right)  \frac{1}{2}\left(
\frac{1}{3}+\tau_{3}\right)  \,\gamma^{\nu}\,iS\left(  p_{\gamma}-k^{\prime
}\right)  \,i\gamma_{5}\tau^{i}\,iS\left(  p_{\pi}-k^{\prime}\right)  \,\right]
\right. \nonumber\\
&  \left.  -\mathbb{T}\mathrm{r}\left[  \,\phi^{\pi^{+}}\left(  k^{\prime
},p_{\pi}\right)  \,iS\left(  k^{\prime}-p_{\pi}\right)  \,i\gamma_{5}\tau^{i}\,iS\left(
k^{\prime}-p_{\gamma}\right)  \frac{1}{2}\left(  \frac{1}{3}+\tau_{3}\right)
\,\gamma^{\nu}\,iS\left(  k^{\prime}\right)  \right]  \right\}  ~,
\label{3.07}%
\end{align}%
\begin{equation}
N^{i}=-\int\frac{d^{4}k}{\left(  2\pi\right)  ^{4}}\delta\left(  xp^{+}%
-\frac{1}{2}\left(  p_{\gamma}^{+}+p_{\pi}^{+}-2k^{+}\right)  \right)  \mathbb{T}%
\mathrm{r}\left[  iS\left(  p_{\gamma}-k\right)  \,\gamma^{\mu}\gamma_{5}%
\tau^{-}\,iS\left(  p_{\pi}-k\right)  \,i\gamma_{5}\tau^{i}\right]  \quad,
\end{equation}
and $\Pi_{ps}$ is the pseudoscalar polarization, see Fig~\ref{properpol}, described by Eq.~(\ref{polar}),
\begin{equation}
\Pi_{ps}\left(  \Delta^{2}\right)  =-i\int\frac{d^{4}k}{\left(  2\pi\right)
^{4}}\left\{  \mathbb{T}\mathrm{r}\left[  i\gamma_{5}\,iS\left(  k\right)
\,i\gamma_{5}\,iS\left(  \Delta-k\right)  \right]  \right.  \quad;
\label{3.08}%
\end{equation}
\\

We can now evaluate the axial current in a straightforward way.~\footnote{ The pion pole structure of this contribution can be explicitly understood if we observe that
\beq
\sum_{i}M^{i}\frac{2ig}{1-2g\Pi_{ps}\left(  t\right)}N^{i}\vert_{t\simeq m^2_{\pi}}
&=& -8N_c\,f_{\pi} g_{\pi qq}\,m\,\frac{4\xi}{1-\xi}\frac{1}{m_{\pi}^2-t}\,\tilde{I}_{2,\Delta}(x,\xi,t)\quad.
\eeq
}
 Nevertheless,
in order to extract the axial TDA we must subtract the pion pole contribution.
We need for that the pion amplitude studied in Chapter~\ref{sec:da-cal}. In the NJL model, the pion DA is given by Eq.~(\ref{2-prop-da})
 in the $[-\xi,\xi]$ region,
\begin{eqnarray}
\frac{1}{2\xi}\,\phi\left(\frac{x+\xi}{2\xi}\right)&=& -\,\frac{g_{\pi qq}\,4N_c m }{\sqrt{2} f_{\pi}}\,\tilde{I}_{2,\Delta}(x,\xi,m^2_{\pi})\quad.
\end{eqnarray}
\\

The trace over the Dirac space involves a $k$-dependent 3-propagator integral that renders a bit more tricky the calculation.
After a long but direct calculation, we obtain the expression for $A\left(  x,\xi,t\right)  .$
As the vector TDA, $A\left(  x,\xi,t\right)  $ can be expressed as a sum of
the contributions coming from the active $u$-quark and the active $\bar{d}%
$-quark. The first one will be proportional to the $d$'s charge and the second
to the $u$'s charge,
\begin{equation}
A^{\pi^{+}}\left(  x,\xi,t\right)  =Q_{d}\, a_{u\rightarrow d}^{\pi^{+}%
}\left(  x,\xi,t\right)  +Q_{u} \,a_{\bar{d}\rightarrow \bar{u}}^{\pi^{+}%
}\left(  x,\xi,t\right)  \quad. 
\label{3.10}%
\end{equation}
\\
Isospin relates these two contributions, 
\beq
a_{u\rightarrow d}^{\pi^{+}}\left(x,\xi,t\right)&=&a\left(x,\xi,t\right) \quad,\nonumber\\
a_{\bar{d}\rightarrow \bar{u}}^{\pi^{+}}\left( x,\xi,t\right)  &=&-a\left(-x,\xi,t\right) \quad,\nonumber
\eeq
 where the minus sign is originated in the change in
helicity produced by the $\gamma_{5}$ operator. The expression for
$a_{u\rightarrow d}^{\pi^{+}}\left(  x,\xi,t\right)  $ depends on the sign of
$\xi.$ 
\\
In the $\left\vert \xi\right\vert <x<1$ region we have%
\begin{align}
a
\left(  x,\xi,t\right)   &  =-8\,N_{c}\,f_{\pi
}\,g_{\pi qq}\,m\left\{  \left(  1+\frac{m_{\pi}^{2}\left(  x-\xi\right)
+\left(  1-x\right)  t}{2\xi m_{\pi}^{2}-t\left(  1+\xi\right)  }\frac{1+\xi
}{1-\xi}\right)  \tilde{I}_{3v}\left(  x,\xi,t\right)  \right. \nonumber\\
&  +\frac{1}{2\xi m_{\pi}^{2}-t\left(  1+\xi\right)  }\left.  \frac{1+\xi
}{1-\xi}\frac{1}{16\pi^{2}}\sum_{i=1}^{2}c_{i}\left[  \log\frac{m^{2}\left(
m_{i}^{2}-\bar{z}\,m_{\pi}^{2}\right)  }{m_{i}^{2}\left(  m^{2}-\bar
{z}\,m_{\pi}^{2}\right)  }\right]  \right\}  \quad,
\label{a-dglap}
\end{align}
with the abbreviations $\bar{z}=\left(  1-x\right)  \left(  \xi+x\right)
/\left(  1+\xi\right)  ^{2}$. And in the $-\left\vert \xi\right\vert
<x<\left\vert \xi\right\vert $ region we have
\begin{align}
a\left(  x,\xi,t\right)   &  =-8\,N_{c}\,f_{\pi
}\,g_{\pi qq}\,m\left\{  \left(  1+\frac{m_{\pi}^{2}\left(  x-\xi\right)
+\left(  1-x\right)  t}{2\xi m_{\pi}^{2}-t\left(  1+\xi\right)  }\frac{1+\xi
}{1-\xi}\right)  \tilde{I}_{3v}\left(  x,\xi,t\right)  \right. \nonumber\\
&  +\frac{\epsilon\left(  \xi\right)  }{1-\xi}\frac{1}{16\pi^{2}}\sum
_{i=1}^{2}c_{i}\left[  \frac{1+\xi}{2\xi m_{\pi}^{2}-t\left(  1+\xi\right)
}\log\frac{\left(  4\xi^{2}m^{2}-\bar{x}\,t\right)  \left(  m_{i}^{2}-\bar
{y}\,m_{\pi}^{2}\right)  }{\left(  4\xi^{2}m_{i}^{2}-\bar{x}\,t\right)
\left(  m^{2}-\bar{y}\,m_{\pi}^{2}\right)  }\right. \nonumber\\
&  +\left.  \frac{2}{t-m_{\pi}^{2}}\left.  \log\frac{\left(  4\xi^{2}%
m^{2}-\bar{x}\,t\right)  \left(  4\xi^{2}m_{i}^{2}-\bar{x}\,m_{\pi}%
^{2}\right)  }{\left(  4\xi^{2}m_{i}^{2}-\bar{x}\,t\right)  \left(  4\xi
^{2}m^{2}-\bar{x}\,m_{\pi}^{2}\right)  }\right]  \right\}  \quad, \label{3.11}%
\end{align}
where $\bar{x}=\left(  \xi^{2}-x^{2}\right)  $ and $\bar{y}=\bar{z}$ for
$\xi>0$ and $\bar{y}=0$ for $\xi<0$.
\\
\vskip 1cm

\subsection{The Chiral Limit}

The expressions for both the vector and axial TDAs in the chiral limit, i.e.
$m_{\pi}=0$, are well defined. In particular, for $t=0$, we find the following
simple expression for $v(x,\xi,t)$
\begin{equation}
v(x,\xi,0)=\sqrt{2}\,f_{\pi}\,6\,F_{V}^{\pi^{+}%
\chi}(0)\,\left[  \theta(\xi^{2}-x^{2})\,\,\frac{x+\left\vert \xi\right\vert
}{2\left\vert \xi\right\vert (1+\left\vert \xi\right\vert )}\,+\,\theta
(x-\left\vert \xi\right\vert )\theta(1-x)\frac{1-x}{1-\xi^{2}}\right]  \quad,
\label{chiv}%
\end{equation}
where $F_{V}^{\pi^{+}\chi}(0)$ is the chiral limit of the vector pion form
factor at zero momentum transfer
\beq
F_{V}^{\pi^{+}\chi}(0)=\lim_{m_{\pi
}\rightarrow0}\frac{F_{V}^{\pi^{+}}(0)}{m_{\pi}}=0.17%
\operatorname{GeV}%
^{-1}\quad.
\nonumber
\eeq

A similar expression is obtained for $a(x,\xi,t)$
in the chiral limit for $t=0$
\begin{align}
a(x,\xi,0)  &  =-\sqrt{2}\,f_{\pi}\,6\,F_{A}%
^{\pi^{+}\chi}(0)\,\left[  \theta(\xi^{2}-x^{2})\,\epsilon\left(  \xi\right)
\,\frac{\left(  \xi-x\right)  }{4\xi^{2}(1+\left\vert \xi\right\vert )}\left(
x+\xi+\left(  x-\xi\right)  \frac{\left\vert \xi\right\vert -\xi
}{(1+\left\vert \xi\right\vert )}\right)  \right.  \,\,\nonumber\\
& \hspace{.6cm} +\,\left.  \theta(x-\left\vert \xi\right\vert )\theta(1-x)\frac
{(1-x)(x-\xi)}{(1-\xi^{2})(1-\xi)}\right]  \quad, \nonumber\\
\label{chia}%
\end{align}
with $F_{A}^{\pi^{+}\chi}(0)$ the axial Form Factor at $t=0$ in the chiral
limit 
\beq
F_{A}^{\pi^{+}\chi}(0)=\lim_{m_{\pi}\rightarrow0}F_{A}^{\pi^{+}%
}(0)/m_{\pi}=F_{V}^{\pi^{+}\chi}(0)\quad.
\nonumber
\eeq

\subsection{Sum Rules and Polynomiality}
\label{sec:sr-pol}

\label{SecIVSRP}The vector TDA of the $\pi^{+}$ must obey the sum rule given
in Eq.~(\ref{2.06}) with the expression of the vector pion Form Factor in the
NJL model given by Eq.~(\ref{fv}). We have numerically recovered the sum rule
for different $t$ values. In particular we obtain the value $F_{V}^{\pi^{+}%
}(0)=0.0242$ for the vector Form Factor at $t=0$, which is in agreement with
the experimental value 
\beq
F_{V}(0)=0.017\pm0.008\nonumber\quad,
\eeq
 given in \cite{Amsler:2008zz}.
\\
\begin{figure}[tb]
\centering
\includegraphics [height=6cm]{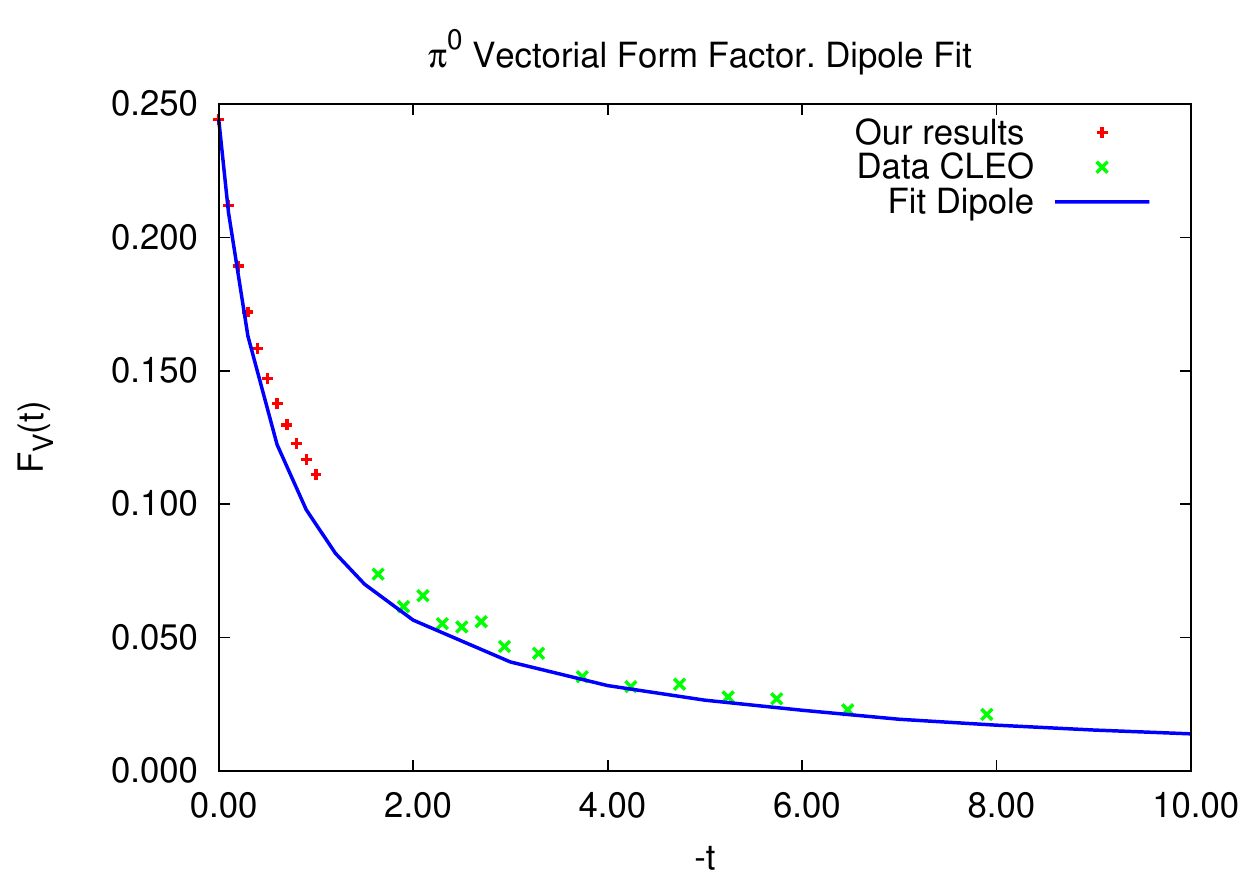}
\caption{The vector Form Factor for the $\pi^0$ compared to the experimental data and to the dipole parameterization based on it~\cite{Gronberg:1997fj}.}%
\label{ffpi}%
\end{figure}

The $\pi^{0}$ distribution must satisfy the following sum rule
\cite{Pire:2004ie}
\begin{equation}
\int_{-1}^{1}dx\,\left(  Q_{u}\,V_{u}^{\pi^{0}}(x,\xi,t)-Q_{d}\,V_{d}^{\pi
^{0}}(x,\xi,t)\right)  =\sqrt{2}\,f_{\pi}\,F_{\pi\gamma^{\ast}\gamma}%
(t)\qquad. \label{4.01}%
\end{equation}
A theoretical prediction of the $\pi^{0}$ Form Factor value is given in Ref.~\cite{Brodsky:1981rp}. In particular, at $t=0$, the value $F_{\pi\gamma^{\ast
}\gamma}(0)=0.272$ GeV$^{-1}$ is found. The neutral pion Form Factor is
directly related to the $\pi^{+}$ vector Form Factor so that the sum rule is
satisfied. We obtain the value $F_{\pi\gamma^{\ast}\gamma}(0)=0.244~$%
GeV$^{-1}$. The difference with the well-known result from the anomaly is due to the regularization parameters, see~(\ref{ff-t0}). In Ref.~\cite{Gronberg:1997fj}, a dipole parameterization based on
experimental data is proposed for the $t$-dependence of $F_{\pi\gamma^{\ast
}\gamma}\left(  t\right)  $,
\beq
F_{\pi\gamma^{\ast}\gamma}\left(  t\right) &=&\frac{F_{\pi\gamma^{\ast}\gamma}\left(  0\right)}{1-t/\Lambda^2}\quad,
\nonumber
\eeq
obtaining for the dipole mass $\Lambda=0.77$ GeV. %
In Fig.~\ref{ffpi} are shown the experimental results of the CLEO experiment~\cite{Gronberg:1997fj} (green diagonal crosses) obtained for higher values of $t$ than our calculation allows.~\footnote{Notice that the recent measurements of the BABAR collaboration for $Q^2> 10$ GeV$^2$ show surprising results~\cite{Aubert:2009mc}: the measured neutral pion Form Factor exceeds the asymptotic limit $f_{\pi}/Q^2$ and contradicts most of the models for the pion DA.} On the same plot, the red crosses represent our results for the neutral pion Form Factor. We have found that, for small values of $t,$ the NJL neutral pion Form Factor
can be parametrized in a dipole form with $\Lambda=0.81$ GeV.  The blue line represents the fit with the latter value for the free parameter.
\\

The axial TDA obeys the sum rules given by Eq.~(\ref{2.07}) with the axial
Form Factor given in the NJL model by Eq.~(\ref{faalltwists}). This sum rule
is satisfied for different $t$ values. The numerical results also coincide. In
particular we obtain $F_{A}^{\pi^{+}}(0)=0.0239$ for the axial Form Factor at
$t=0$, which is about twice the value 
\beq
F_{A}^{\pi^{+}}(0)=0.0115\pm0.0005\quad,
\nonumber
\eeq
given by the Particle Data Group \cite{Amsler:2008zz}.
\\

We expect TDAs to obey the polynomiality condition. However no time reversal
invariance enforces the polynomials to be even in the $\Delta$-momenta like
for GPDs. That means that the polynomials should be \textquotedblleft
complete\textquotedblright, i.e. they should include all powers in $\xi$,
\begin{equation}
\int_{-1}^{1}dx\,x^{n-1}\,V(x,\xi,t)=\sum_{i=0}^{n-1}\,C_{n,i}(t)\,\xi
^{i}\quad. \label{oddeven}%
\end{equation}

We have numerically tested the polynomiality
and obtained it. We observe that, for the vector TDA, the coefficients for the odd powers in $\xi$
are of one order of magnitude smaller than those for the even powers in $\xi$.
In particular we have numerically proved that, in the chiral limit, the
polynomials only contain even powers in $\xi$ for any value of $t,$%
\begin{equation}
\int_{-1}^{1}dx\,x^{n-1}\,[\lim_{m_{\pi}\rightarrow0}V(x,\xi,t)]=\sum
_{i=0}^{[\frac{n-1}{2}]}\,C_{n,2i}^{\chi}(t)\,\xi^{2i}\quad.
 \label{even}%
\end{equation}
We have analytically found that the odd powers in $\xi$ go to zero for the polynomial expansion of
the vector TDA. A study of the polynomiality in the limit given in
Eq.~(\ref{chiv}) leads to the following analytical expression for the
coefficients $C_{n,2i}^{\chi}(t)$
\begin{equation}
C_{n,2i}^{\chi}(0)=\sqrt{2}\,f_{\pi}\,\,2\,F_{V}^{\pi^{+}\chi}(0)\,\left(
-1+2(-1)^{n-1}\right)  \,\left(  \frac{1}{n}-\frac{1}{n+1}\right)  \quad.
\label{4.03}%
\end{equation}
Notice they do not depend on $i$. 
The coefficients in the chiral limit, Eq.~(\ref{even}), as well as the
coefficients for $m_{\pi}=140$ MeV, Eq.~(\ref{oddeven}), we numerically
obtained are given in Table~\ref{tablepolyv}.
\\

The $\xi$-dependence of the moments of the axial TDA  also has a
polynomial form. Those polynomials contain all the powers in $\xi$, i.e. even
and odd,
\begin{equation}
\int_{-1}^{1}dx\,x^{n-1}\,A(x,\xi,t)=\sum_{i=0}^{n-1}\,C_{n,i}^{\prime
}(t)\,\xi^{i}\quad.
\label{oddevenax}
\end{equation}

An analytic study of the polynomiality in the limit given in Eq.~(\ref{chia})
confirms that all the powers in $\xi$ have to be present. The analytic values
for the coefficients in this specific limit are
\begin{align}
C_{n,2i}^{\prime\chi}(0)  &  = \sqrt{2}\,f_{\pi}\,\,2\,F_{A}^{\pi^{+}\chi
}(0)\,\left(  1+2(-1)^{n-1} \right)  \,\frac{n-2i}{n(n+1)(n+2)} \quad
,\nonumber\\
C_{n,2i+1}^{\prime\chi}(0)  &  =\sqrt{2}\,f_{\pi}\,\,2\,F_{A}^{\pi^{+}\chi
}(0)\,\left(  1+2(-1)^{n-1} \right)  \,\frac{-2(i+1)}{n(n+1)(n+2)}\quad,
\label{coefaxial}%
\end{align}
\begin{landscape}
\begin{table}
\centering{ {
\begin{tabular}
[c]{|c|c|c|c|c|}\hline
$n$ & 1 & 2 & 3 & 4\\\hline
& {\scriptsize {$C_{1,0}\left(  t\right)  $}} & {\scriptsize {$%
\begin{array}
[c]{rr}%
C_{2,0}\left(  t\right)  & C_{2,1}\left(  t\right)
\end{array}
$ }} & {\scriptsize { $%
\begin{array}
[c]{rrr}%
C_{3,0}\left(  t\right)  & C_{3,1}\left(  t\right)  & C_{3,2}\left(  t\right)
\end{array}
$}} & {\scriptsize {$%
\begin{array}
[c]{rrrr}%
C_{4,0}\left(  t\right)  & C_{4,1}\left(  t\right)  & C_{4,2}\left(  t\right)
& C_{4,3}\left(  t\right)
\end{array}
$}}\\\hline\hline
{\footnotesize {$%
\begin{array}
[c]{c}%
m_{\pi}=0
\end{array}
$}} &  &  &  & \\\hline
{\footnotesize {$%
\begin{array}
[c]{l}%
t=0\\
t=-0.5\\
t=-1.0
\end{array}
$}} & $%
\begin{array}
[c]{l}%
22.6\\
13.7\\
\multicolumn{1}{r}{10.4}%
\end{array}
$ & $%
\begin{array}
[c]{ll}%
-22.6 & \hspace*{0.1cm}0.0\\
-16.3 & \hspace*{0.1cm}0.0\\
\multicolumn{1}{r}{-13.4} & \multicolumn{1}{r}{0.0}%
\end{array}
$ & $%
\begin{array}
[c]{lll}%
3.77 & \hspace*{0.3cm}0.0 & \hspace*{0.3cm}3.77\\
\multicolumn{1}{r}{3.00} & \hspace*{0.3cm}0.0 & \multicolumn{1}{r}{2.43}\\
\multicolumn{1}{r}{2.57} & \multicolumn{1}{r}{0.0} & \multicolumn{1}{r}{1.90}%
\end{array}
$ & $%
\begin{array}
[c]{llll}%
-6.7 & \hspace*{0.3cm}0.0 & \hspace*{0.3cm}-6.7 & \hspace*{0.3cm}0.0\\
-5.7 & \hspace*{0.3cm}0.0 & \hspace*{0.3cm}-4.9 & \hspace*{0.3cm}0.0\\
-5.0 & \multicolumn{1}{r}{0.0} & \multicolumn{1}{r}{-4.0} &
\multicolumn{1}{r}{0.0}%
\end{array}
$\\\hline\hline
{\footnotesize {$%
\begin{array}
[c]{c}%
m_{\pi}=140
\end{array}
$}} &  &  &  & \\\hline
{\footnotesize {$%
\begin{array}
[c]{l}%
t=0\\
t=-0.5\\
t=-1.0
\end{array}
$}} & $%
\begin{array}
[c]{l}%
22.7\\
\multicolumn{1}{r}{13.7}\\
\multicolumn{1}{r}{10.3}%
\end{array}
$ & $%
\begin{array}
[c]{ll}%
-22.9 & \hspace*{0.3cm}0.44\\
\multicolumn{1}{r}{-16.4} & \multicolumn{1}{r}{0.25}\\
\multicolumn{1}{r}{-13.4} & \multicolumn{1}{r}{0.19}%
\end{array}
$ & $%
\begin{array}
[c]{lll}%
3.80 & \hspace*{0.3cm}-0.10 & \hspace*{0.3cm}3.67\\
\multicolumn{1}{r}{3.00} & \multicolumn{1}{r}{-0.06} &
\multicolumn{1}{r}{2.43}\\
\multicolumn{1}{r}{2.57} & \multicolumn{1}{r}{-0.05} &
\multicolumn{1}{r}{1.87}%
\end{array}
$ & $%
\begin{array}
[c]{llll}%
-6.8 & \hspace*{0.2cm}0.19 & \hspace*{0.3cm}-6.8 & \hspace*{0.2cm}0.12\\
-5.7 & \hspace*{0.2cm}0.13 & \hspace*{0.3cm}-4.9 & \hspace*{0.2cm}0.08\\
-5.0 & \hspace*{0.2cm}0.11 & \hspace*{0.3cm}-4.0 & \multicolumn{1}{r}{0.06}%
\end{array}
$\\\hline
\end{tabular}
} }\caption{Coefficients of the polynomial expansion for the vector TDA. The
pion mass is expressed in MeV and $t$ is expressed in GeV$^{2}$. Notice that
the coefficients have to be multiplied by $10^{-3}$. }%
\label{tablepolyv}%
\end{table}
\end{landscape}
which is in agreement with those numerically obtained. The coefficients of the
polynomial expansions in the chiral limit are very close to those obtained for
the physical values of the pion mass. In Table~\ref{tablepolya}, the
coefficients $C_{n,i}^{\prime}(t)$ are therefore shown only for $m_{\pi}=140%
\operatorname{MeV}%
$. 
\\

The polynomiality property of the term containing the pion DA can also be
studied. The $t$-dependence only comes from the pion pole and it is divergent at the 
chiral limit when $t=0$.
We have numerically found
\begin{eqnarray}
\int_{-\xi}^{\xi} dx\,x^{n-1}\,\,\frac{4\,f^2_{\pi}}{m_{\pi}^2-t}\,\frac{1}{2\xi}\,\phi\left(\frac{x+\xi}{2\xi}\right)
&=& \frac{1}{2}\left( 1+(-1)^{n+1}\right)\,C^{\diamond}_{n-1}(t)\,\xi^{n-1}\qquad,
\label{dapol}
\end{eqnarray}
which is true for both $m_{\pi}=0$ and $140$ MeV. The even moments in $x$ disappear due to the pion
distribution amplitude is an even function in the interval $[-\xi,\xi]$.
\\

Calling the l.h.s. of  Eqs.~(\ref{oddevenax}-\ref{dapol}), respectively, $A_n(\xi,t)$ and $\Phi_n(\xi,t)$, and  using the {\it alternative} tensorial decomposition~(\ref{alter-decomp-tens}),  we tend to  write a general relation about the
polynomiality property of the whole axial bilocal matrix element, in the fashion of the {\it alternative} sum rule~(\ref{alt-sr-tda}). In the $\varepsilon^+=0$ gauge, it reads
\begin{eqnarray}
&&\int_{-1}^1 dx \, x^{n-1}\,\int \frac{dz^-}{2\pi}\,e^{ix\, p^+z^-}\,
\langle\gamma(p_{\gamma})|\bar q \left(-\frac{z}{2}\right)\, \thru n \gamma_5\,\tau^- q\left(\frac{z}{2}\right)|\pi(p_{\pi})\rangle\,\, |_{z^+=z^{\perp}=0}\nonumber\\
&=&\frac{1}{p^+}\,e\,\left(\vec{\varepsilon}^{\perp}\cdot\vec{\Delta}^{\perp}\right)\frac{1}{\sqrt{2} f_{\pi}}\,\Big[ 
A_n(\xi,t)-(2\xi)\,\Phi_n(\xi,t)\Big]\qquad,\nonumber\\
&=& \frac{1}{p^+}\,e\,\left(\vec{\varepsilon}^{\perp}\cdot\vec{\Delta}^{\perp}\right)\,\frac{1}{\sqrt{2} f_{\pi}}\,
\Big[ \sum_{i=0}^{n-1} \,C^{'}_{n,i}(t)\,\xi^{i}- \,2\, \frac{1}{2}\left( 1+(-1)^{n+1}\right)\, C^{\diamond}_{n-1}(t)\,\xi^{n}
\Big]\quad.
\label{polyaxialtot}
\end{eqnarray}

\vskip .2cm

 We can therefore
write a general relation about the polynomiality property of the whole axial
bilocal matrix element
\begin{eqnarray}
&&\int_{-1}^{1}dx\,x^{n-1}\,\int\frac{dz^{-}}{2\pi}\,e^{ix\,p^{+}z^{-}}%
\,\langle\gamma(p_{\gamma})|\bar{q}\left(  -\frac{z}{2}\right)  \,\thru n\gamma_{5}\,\tau^{-}q\left(  \frac{z}{2}\right)  |\pi(p_{\pi})\rangle
\,\,|_{z^{+}=z^{\perp}=0}\nonumber\\
&&=\frac{1}{p^{+}}\,e\,\left(\vec{\varepsilon}^{\perp}\cdot\vec{\Delta}^{\perp}\right)\,\frac
{1}{\sqrt{2}f_{\pi}}\,\sum_{i=0}^{n}\,C_{n,i}^{^{\star}}(t)\,\xi
^{i}\quad,
\end{eqnarray}
with the highest power in $\xi$ given by the pion pole contribution.
The latter result is the expected one from Lorentz invariance, e.g. for the GPDs~(\ref{poly-gpd-theo}).
\\
\\

\begin{table}[tb]
\centering
\begin{tabular}
[c]{|c|c|c|c|}\hline
$n$ & 1 & 2 & 3\\\hline
& $C^{\prime}_{1,0}\left(  t\right)  $ & $%
\begin{array}
[c]{rr}%
C^{\prime}_{2,0}\left(  t\right)  & C^{\prime}_{2,1}\left(  t\right)
\end{array}
$ & $%
\begin{array}
[c]{rrr}%
C^{\prime}_{3,0}\left(  t\right)  & C^{\prime}_{3,1}\left(  t\right)  &
C^{\prime}_{3,2}\left(  t\right)
\end{array}
$\\\hline\hline
$%
\begin{array}
[c]{c}%
m_{\pi}=140\,\,\,\,\,
\end{array}
$ &  &  & \\\hline
$%
\begin{array}
[c]{l}%
t=0\\
t=-0.5\\
t=-1.0
\end{array}
$ & $%
\begin{array}
[c]{l}%
22.4\\
\multicolumn{1}{r}{16.1}\\
\multicolumn{1}{r}{13.2}%
\end{array}
$ & $%
\begin{array}
[c]{ll}%
-3.77 & \hspace*{.2cm} 4.00\\
\multicolumn{1}{r}{-2.97} & \multicolumn{1}{r}{2.57}\\
\multicolumn{1}{r}{-2.53} & \multicolumn{1}{r}{1.97}%
\end{array}
$ & $%
\begin{array}
[c]{lll}%
6.8 & \hspace*{.2cm} -4.9 & \hspace*{.2cm} 2.3\\
\multicolumn{1}{r}{5.7} & \multicolumn{1}{r}{-3.5} & \multicolumn{1}{r}{1.7}\\
\multicolumn{1}{r}{5.0} & \multicolumn{1}{r}{-2.8} & \multicolumn{1}{r}{1.3}%
\end{array}
$\\\hline
\end{tabular}
\caption{Coefficients of the polynomial expansion for the axial TDA. The pion
mass is expressed in MeV and $t$ is expressed in GeV$^{2}$. Notice that the
coefficients have to be multiplied by $10^{-3}$. }%
\label{tablepolya}%
\end{table}

\section{Symmetries}
\label{sec:sym-tda}

In the previous Section we have defined the TDAs in the particular case of a transition
from a $\pi^+$ to a photon, parameterizing  processes  like hadron
annihilation into two photons and backward VCS in the kinematical
regime where the virtual photon is highly off-shell but with small
momentum transfer $t$ 
 \beq H\bar H\to
\gamma^{\ast}\gamma \to e^+ e^-\gamma \quad
\mbox{and}\quad\gamma^{\ast} H\to H\gamma\quad,
\label{hh}
\eeq with $H$ a hadron. Symmetries relate the latter distributions to TDAs
involved in other processes. For instance, we could wish to study the
$\gamma$-$\pi^-$ TDAs entering the factorized amplitude of the process
 \beq \gamma_L^{\ast}\gamma \to
M^{\pm}\pi^{\mp}\quad,\nonumber
\label{gg}
\eeq
 $M$ being either $\rho_L$ or $\pi$; in the same kinematical
regime.

By consistency with the processes that we are considering, we choose to change the definitions of the kinematical variables.
 The momentum transfer is $\Delta=p_{\pi}-p_{\gamma},$
therefore $p^{2}=m_{\pi}^{2}/2-t/4$ and $t=\Delta^{2}$. The skewness variable
describes the loss of plus momentum of the incident photon, i.e.
\beq
\xi&=&\left(
p_{\gamma}-p_{\pi}\right)  ^{+}/2p^{+}\quad,
\eeq
 and its value ranges between
 \beq
-1<\xi<{-t \over\left(  2m_{\pi}^{2}-t\right)  }\quad.
\eeq
Since the skewness variable defined for $\pi-\gamma$ transitions is minus the one defined for $\gamma-\pi$ transitions, we should hence consider the results corresponding to the results for negative $\xi$ is a previous Section. 
\\

Within these conventions, we define the $\gamma$-$\pi^{\pm}$ TDAs
 \beq
\int\frac{dz^{-}}{2\pi}e^{ixp^{+}z^{-}}&&\left.  \left\langle \pi^{\pm}(p_{\pi})
  \right\vert \bar{q}\left( -\frac{z}{2}\right) \gamma
  ^{+}\hspace*{-0.05cm}\tau^{\pm}q\left( \frac{z}{2}\right)
  \left\vert\gamma( p_{\gamma}\varepsilon)
  \right\rangle \right\vert _{z^{+}=z^{\bot}=0} \nonumber\\\nonumber\\
&=&\frac{1}{p^+} i\,e\,\varepsilon_{\nu}\,\epsilon^{+\nu\rho\sigma}\,P_{\rho}\,(p_{\pi}-p_{\gamma})_{\sigma
}\,\frac{V^{\gamma\to\pi^{\pm}}(  x, \xi,t)  }{\sqrt{2}f_{\pi}}\nonumber\quad,\label{tdarev-v}\\
\nonumber\\
\int\frac{dz^{-}}{2\pi}e^{ixp^{+}z^{-}}&&\left.  \left\langle
    \pi^{\pm}\left( p_{\pi}\right) \right\vert \bar{q}\left( -\frac{z}{2}\right)
  \gamma^+\hspace*{-0.05cm}\gamma_{5}\tau^{\pm}q\left(
    \frac{z}{2}\right) \left\vert\gamma\left(
      p_{\gamma}\varepsilon\right)   \right\rangle \right\vert _{z^{+}=z^{\bot}=0} \nonumber\\\nonumber\\
& =&\pm \frac{1}{p^+}\Big [- e\,\left( \vec{\varepsilon}^{\bot}\cdot(\vec{p}_{\pi}^{\bot
  }-\vec{p}_{\gamma}^{\bot})\right) \frac{A^{\gamma\to\pi^{\pm}}(x,\xi,t)
}{\sqrt{2}f_{\pi} } \nonumber\\
&&+ e\,\left( \varepsilon\cdot(p_{\pi}-p_{\gamma})\right)
\frac{2\sqrt {2}f_{\pi}}{m_{\pi}^{2}-t}~\epsilon(\xi)
~\phi\left( \frac{x+\xi}{2\xi}\right) \Big]\quad.
\label{tdarev}
\eeq %
\\
\\
With the help of  Appendix~\ref{sec:sym}, we find the following relations.
Time-reversal relates the $\pi^+$-$\gamma$ TDAs to $\gamma$-$\pi^+$
TDAs in the following way
 \beq
D^{\pi^+\to\gamma}(x,\xi,t)&=&D^{\gamma\to\pi^+}(x,-\xi,t)\quad,
 \eeq
where $D=V,A$. And $CPT$ relates the presently calculated TDAs to
their analog for a transition from a photon to a $\pi^-$
\beq
D^{\pi^{+}\rightarrow
\gamma}\left(  x,\xi,t\right)  &=&D^{\gamma\rightarrow\pi^{-}}\left(  -x,-\xi,t\right)  \quad.
\label{sym-gamma-pi-}
\eeq

\section{ Pion-Photon TDAs in the Nambu-Jona Lasinio Model}

\begin{figure}[bt]
\centering
 \mbox{\subfigure[]{\includegraphics [height=5.5cm]{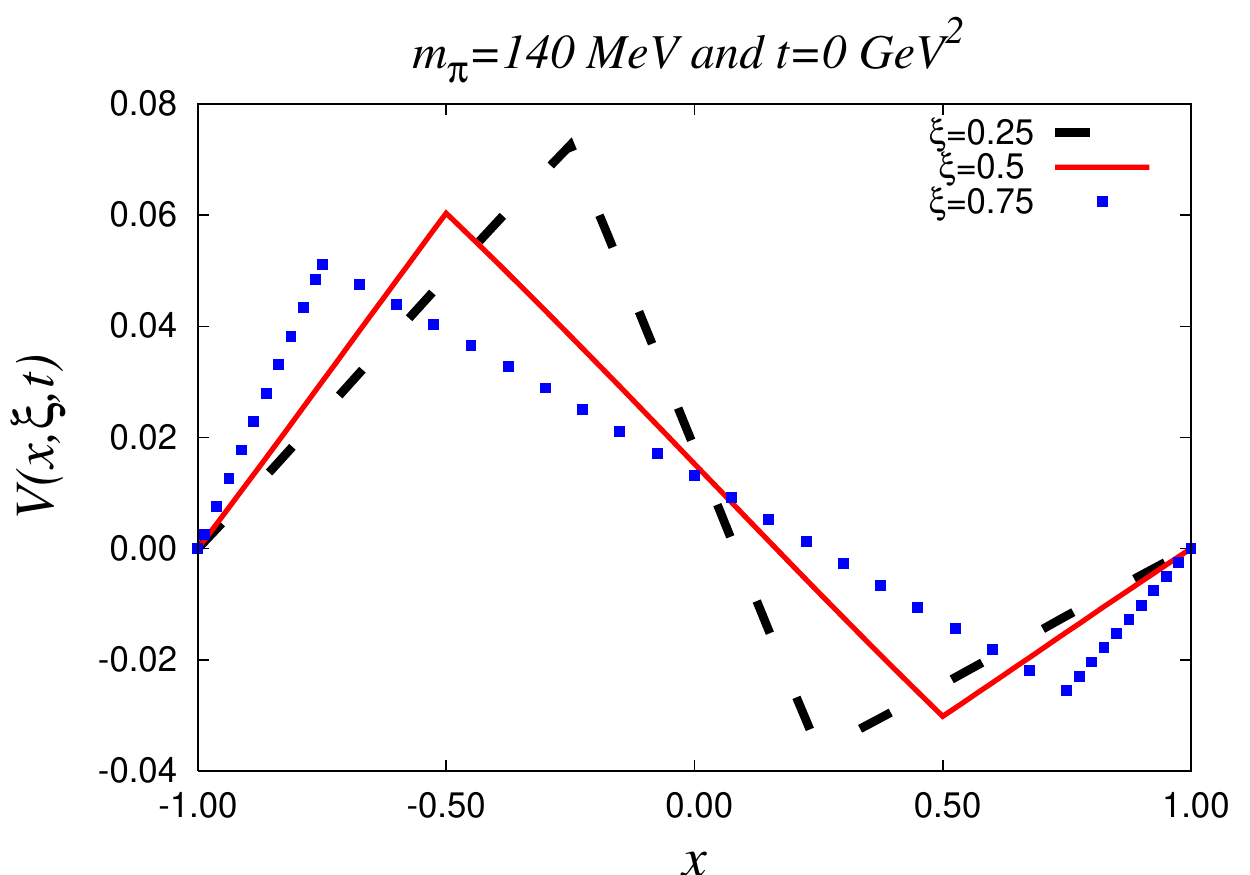}}
\subfigure[]{\includegraphics [height=5.5cm]{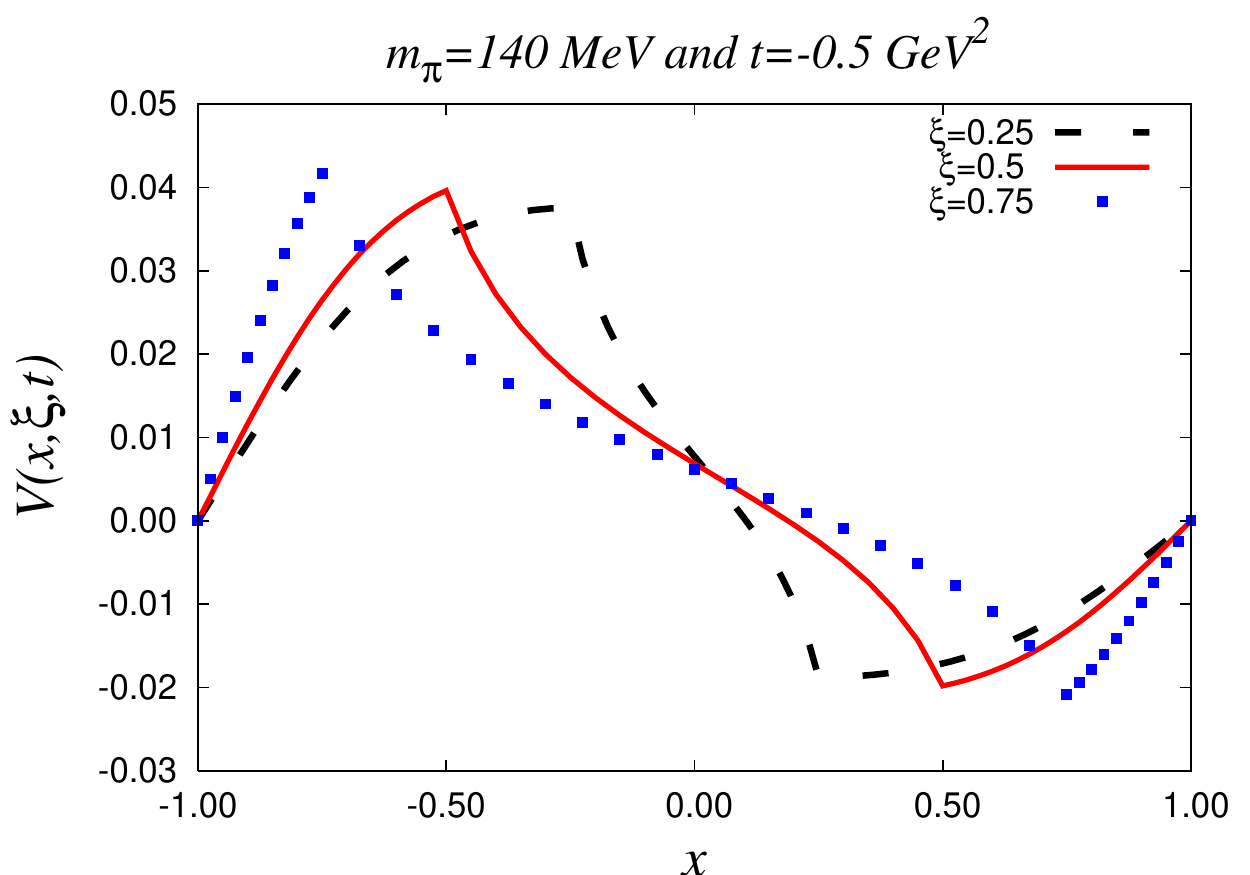}}}
\caption{ The vector TDA in the case
$m_{\pi}=140$ MeV and, respectively, $(a)$ $t=0$ and $(b)$ $t=-0.5$ GeV$^{2}$.}%
\label{TDAmpi140}%
\end{figure}

\begin{wrapfigure}[34]{l}{9cm}
\centering
\includegraphics [height=7cm]{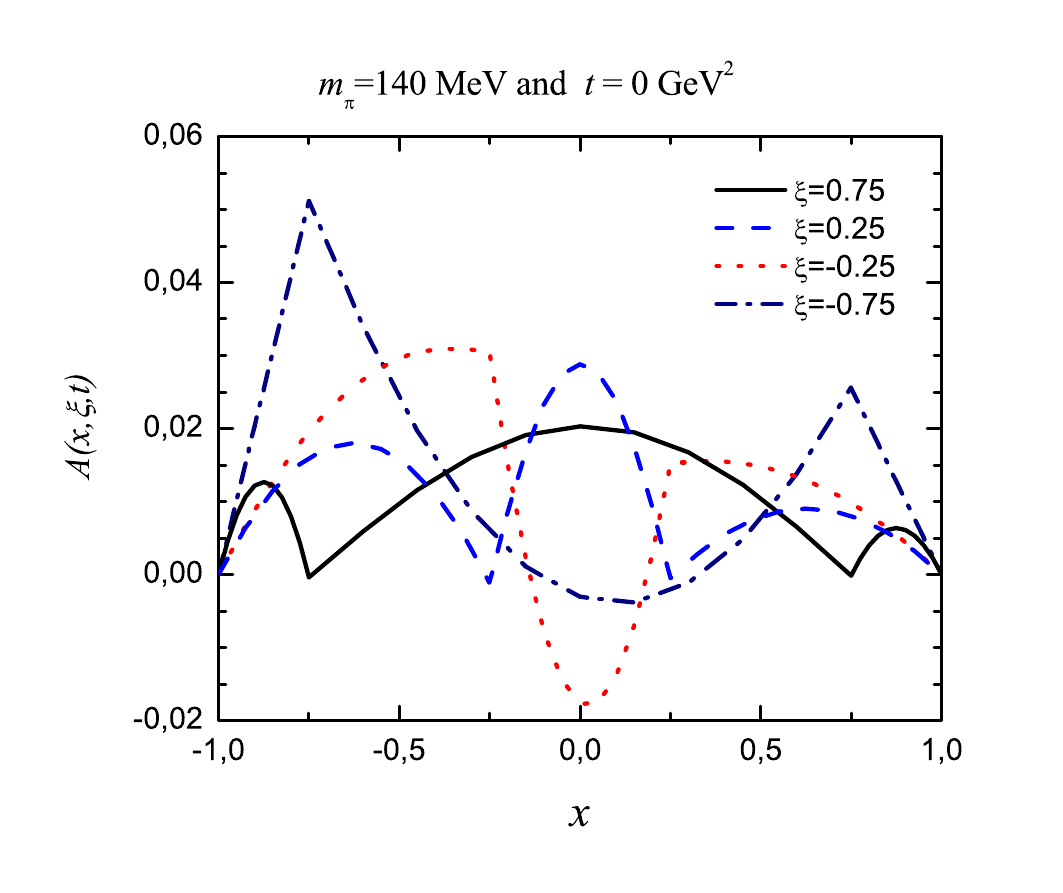}\\
%
\includegraphics [height=7cm]{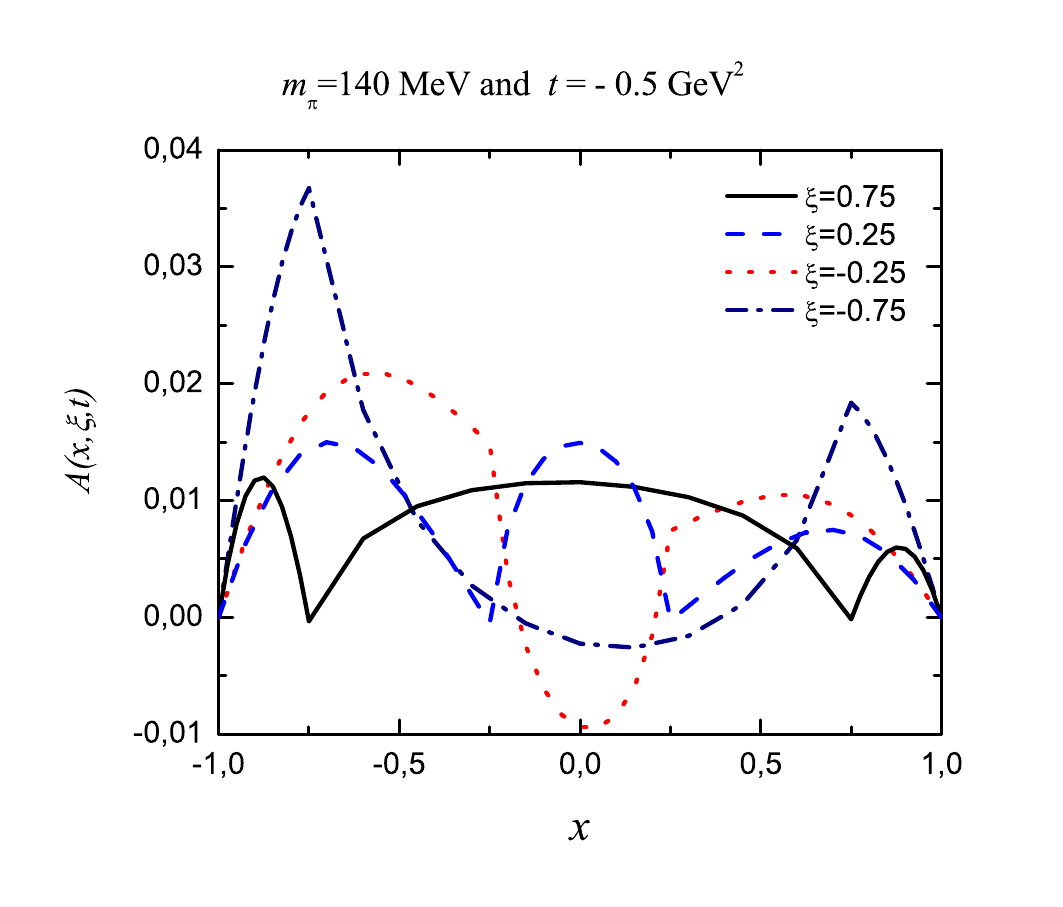}
\caption{ The axial TDA in the case
$m_{\pi}=140$ MeV and, respectively, $(a)$ $t=0$ and $(b)$ $t=-0.5$ GeV$^{2}$.}%
\label{axialTDAmpi140}%
\end{wrapfigure}

In Figs. \ref{TDAmpi140} and \ref{axialTDAmpi140}, the vector and axial $\pi^+\to\gamma$ TDAs obtained in the NJL model 
are plotted in function of $x$ for several values of $\xi$ and $t$. In these
figures, the $u$-quark contributes to the TDAs in the region of $x$ going from
$-\left\vert \xi\right\vert $ to $1$ and the $d$-antiquark going from $-1$ to
$\left\vert \xi\right\vert .$ Therefore, for $x\in\left[  \left\vert
\xi\right\vert ,1\right]  $ ($x\in\left[  -1,-\left\vert \xi\right\vert
\right]  $) only $u$-valence quarks ($d$-valence antiquarks) are present
(DGLAP regions). Besides, TDAs in the $-\left\vert \xi\right\vert
<x<\left\vert \xi\right\vert $ region (ERBL\ region) receive contributions
from both type of quarks. For the vector TDA, we observe in Fig.~\ref{TDAmpi140} that the position of the maxima is given by the $\xi$ value,
separating explicitly the ERBL region from the DGLAP regions. The vector TDA
is positive (negative) for negative (positive) values of $x$ with the change
of sign occurring around $x=0$. This change in the sign of the vector TDA is
originated in the presence of the electric charge of the quarks in
Eq.~(\ref{3.03}).

The process involved in the calculation of the TDAs allows negative values of
the skewness variable. In the chiral limit, the skewness variable goes from
$\xi=-1$ to $\xi=1,$ for any value of $t$. In the chiral limit, the vector TDA
for a negative value of $\xi$ is equal to the vector TDA for $\left\vert
\xi\right\vert $. This can be seen from the polynomiality expansion in this
limit, Eq. (\ref{even}), which has only even powers of $\xi.$ For the physical
value of the pion mass, negative values of $\xi$ are bounded by $t/\left(
2m_{\pi}^{2}-t\right)  <\xi.$ For each allowed value of $\xi$, we found that
the numerical value of $V\left(  x,-\left\vert \xi\right\vert ,t\right)  $ is
close to $V\left(  x,\left\vert \xi\right\vert ,t\right)  ,$ due to the
smallness of the coefficients of the odd powers of $\xi$ in the polynomial
expansion (\ref{oddeven}).
\\

Analyzing the axial TDAs, plotted in Fig.~\ref{axialTDAmpi140}, we observe two
different behaviors depending on the sign of $\xi$. For positive $\xi,$ the
position of the minima is given by the value of the skewness variable while
the position of the maxima is always $x\simeq0$ in the ERBL region and
$x\simeq\pm(1+\xi)/2$ in the DGLAP regions. 
For negative $\xi,$ the value of
the axial TDA at $x=\pm\xi$ is important and in some cases is a maximum and,
in the ERBL region, $A\left(  x,\xi,t\right)  $ presents a minimum near $x=0.$
As we have previously shown, the axial TDA in the ERBL region receives
contributions from two different diagrams, depicted in Fig.~\ref{Fig3}.
In the
second of these diagrams, a virtual quark-antiquark interacting pair in the
pion channel appears. The pion pole, contained in this diagram, has been
subtracted but the remaining non-resonant part contributes to the axial TDA.
We observe that the latter contribution is the dominant one and produces the
maxima around $x=0$ for positives $\xi$ (Fig.~\ref{comparisonxi}). Now, the
axial TDA does not change the sign when we go from negative to positive values
of $x$. In the axial TDA the change of sign originated in the presence of the
electric charge of the quarks in equation (\ref{3.10}) is compensated by the
change of sign between quark and antiquark contributions generated by the
$\gamma_{5}$ operator present in the axial current.

By comparing the plots~in Figures \ref{TDAmpi140} and \ref{axialTDAmpi140} for different values of the momentum transfer, it is
observed that the amplitudes are lower for higher $(-t)$ values, as it can be
inferred from the decreasing of the Form Factors with $\left(  -t\right)  $,
connected to the TDAs through the sum rules. In the case of the vector TDA, by increasing the $(-t)$ value,
not only the width and the curvature of the TDAs are changed but we also
observe that higher values of $\xi$ are preferred, i.e. the sign of the
derivative of the collection of maxima changes passing from a zero momentum
transfer to a non-zero one.

Isospin relates the value of the vector and axial TDAs in the DGLAP regions,%
\begin{equation}
V\left(  x,\xi,t\right)  =-\frac{1}{2}V\left(  -x,\xi,t\right)  ,~~~~A\left(
x,\xi,t\right)  =\frac{1}{2}A\left(  -x,\xi,t\right)  , ~~~~~~~~~~~\left\vert
\xi\right\vert <x<1~~, \label{5.01}%
\end{equation}
being the factor $1/2$ the ratio between the charge of the $u$ and $d$ quarks.
We observe in Figs.~\ref{TDAmpi140} and \ref{axialTDAmpi140} that our TDAs
satisfy these relations. It must be realized that the relation (\ref{5.01})
cannot be changed by evolution.
\begin{figure}[tb]
\centering
 \mbox{\subfigure[]{\includegraphics [height=5cm]{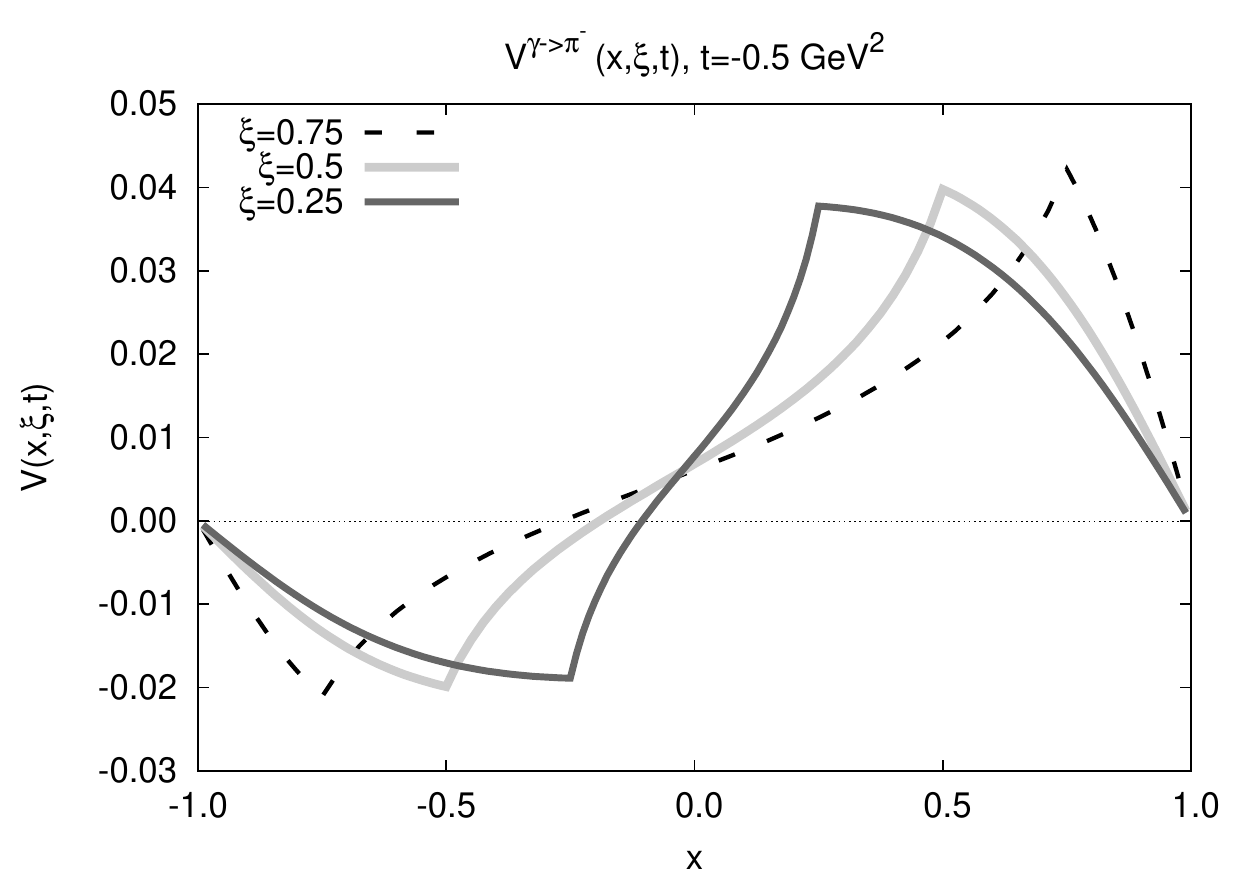}}
 \subfigure[]{\includegraphics [height=5cm]{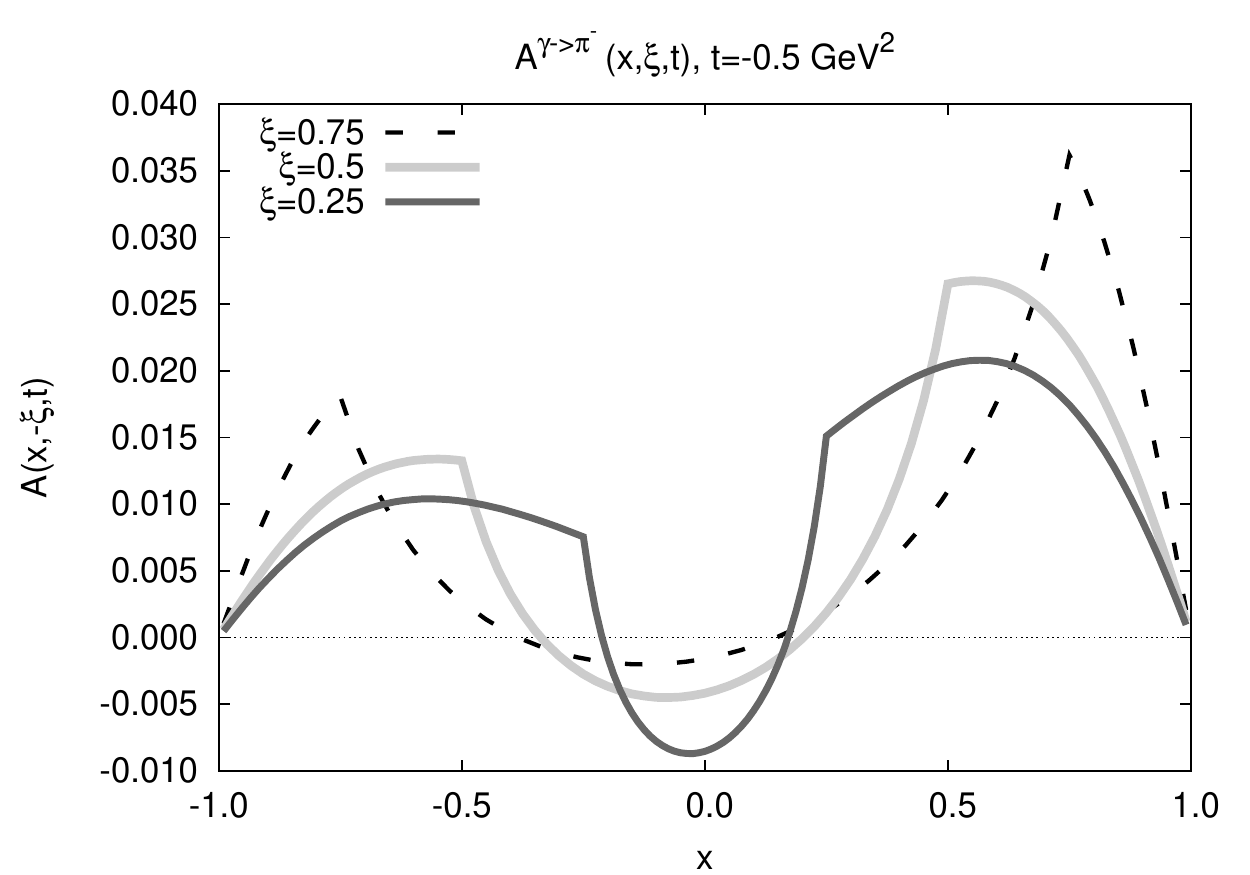}}}
 \caption{The functions (a) $V^{\gamma \to\pi^-}(x, \xi, t)$ and (b) $A^{\gamma \to\pi^-}(x, \xi, t)$ for different values of the skewness variable $\xi$ and  for $t=-0.5$ GeV$^2$.  }
\label{tdatot}
\end{figure}

Regarding the chiral limit, we observe that both the vector and
axial TDAs do not significantly change going from a non-zero pion mass to the
physical mass, except for the change in the lower bound of $\xi$.

In Fig.~\ref{tdatot} are illustrated the vector and axial TDA for the transition $\gamma \to\pi^-$, whose expressions have been given in~(\ref{sym-gamma-pi-}).

\section{Polynomiality and Double Distributions}
\label{sec:dd-tda}

In the previous Section, the definition of the pion-photon transition has been given from the tensorial structure of the pion radiative decay. This structure for the hadronic matrix element leads to the identification of the vector and axial TDAs as well as the pion pole contribution.

Generalized Parton Distributions can also be studied through their moments~\cite{Ji:1998pc}. In this Section, we build up the vector and axial-vector pion-photon transition distributions by considering matrix elements of towers of twist-two operators. This manifestly Lorentz-covariant approach leads us to define the Transition Distribution Amplitudes in terms of Double Distributions.

The Mellin moments in $x$ lead to derivative operators between the light-like separated  fields. Such  derivative operators correspond to the local twist-2 operators defined in Eq.~(\ref{vec-twist-2}) that, sandwiched between two hadronic states, give generalized Form Factors. 
\\

The pion-photon matrix elements of vector twist-2 operators can be decomposed in a fully Lorentz covariant fashion in terms of various twist-2 Form Factors $C_{nk} (t)$ similarly to Eq.~(\ref{towert2}) adapted to the tensorial structure~(\ref{2.01}). Namely~\cite{Tiburzi:2005nj}
\beq
&&\langle \gamma (p_{\gamma})|\bar \psi(0)\,\tau^-\, \gamma^{{\large\{}\mu}\,i\stackrel{\hspace{-.2cm}\leftrightarrow}{{\cal D}^{\mu_1}}\ldots 
i\stackrel{\hspace{-.5cm}\leftrightarrow}{{\cal D}^{\mu_{n-1}{\large\}}} }\psi(0) |\pi^+(p_{\pi})\rangle\nonumber\\
&&=-\frac{i\, e}{f_{\pi}}\,\varepsilon_{\nu}\,p_{\rho}\Delta_{\sigma}\epsilon^{\nu\rho\sigma\,\{\mu} \, \sum_{k=0}^{n-1}\,\frac{(n-1)!}{k!\, (k-n-1)!}\, C_{n k}(t)
\, p^{\mu_1}\ldots p^{\mu_{n-1-k}}\,\left(-\frac{\Delta}{2}\right)^{\mu_{n-k}}\ldots\left(-\frac{\Delta}{2}\right)^{\mu_{n-1}\}}\quad,\nonumber\\
\eeq
where the action of $\{\ldots\}$ on Lorentz indices produces the symmetric, traceless part of the tensor.  The polynomiality property immediately follows.
\\

The generalized moments can be written as a Double Distribution ${\cal W}(\beta,\alpha; t)$  which  is the generating function of the twist-2 Form Factors,
\beq
C_{n k} (t)&=& \int_{-1}^1d\beta\,\int_{-(1-|\beta|)}^{1-|\beta|}\, d\alpha \, \beta^{n-1 -k}\, \alpha^k\, {\cal W}(\beta,\alpha; t)\quad.
\label{gen-mmt-ff}
\eeq
It is exactly this step from the tower of twist-2 Form Factors as generated by the double distribution that ensures the fulfillment of the polynomiality property  when evaluating the distribution through this  parameterization.~\footnote{In Appendix~\ref{sec:dd-tda-kin}, the polynomiality property is derived from the expression of the Double Distribution in the $\alpha$-representation of the propagator.} By definition~\cite{Radyushkin:1996nd, Radyushkin:1997ki}, the matrix element of bilocal vector operator is written in terms of DD in the following way
\beq
&& \left\langle \gamma\left(p_{\gamma}\right)  \right\vert \bar{q}\left(  -\frac{z}{2}\right)
\rlap{$/$}z\hspace*{-0.05cm}\tau^{-}q\left(  \frac{z}{2}\right)  \left\vert\pi^+\left(  p_{\pi}\right)  \right\rangle\nonumber\\
&&=-i\, \frac{e}{\sqrt{2}\, f_{\pi}}\,\varepsilon_{\nu}\,p_{\rho}\Delta_{\sigma}\epsilon^{\nu\rho\sigma\,\mu} z_{\mu}\,\int_{-1}^1d\beta\,\int_{-(1-|\beta|)}^{1-|\beta|}\, d\alpha \,e^{-i\beta (p\cdot z)+i\alpha (\Delta\cdot z)/2} \,{\cal W}(\beta,\alpha; t)\quad.
\label{dd-tda-v}
\eeq
From the definition of the vector TDA~(\ref{2.04})  and the skewness variable~(\ref{skew}), we find the reduction formula
\beq
V^{\pi^+}(x,\xi,t) &=&   \int_{-1}^1  d\beta \int_{-(1-|\beta|)}^{1-|\beta|} d\alpha \, \delta(x-\beta-\xi \alpha) \,{\cal W}^{\pi^+}(\beta, \alpha; t)\quad.
\label{dd-v-tib}
\eeq
\\

The analysis of the axial matrix element is quite similar to the analysis of the vector one. Nonetheless, it is worth  reminding that the structure dependent tensorial form is not the only contribution. There is, actually, also the pion inner Bremsstrahlung to be separated out from the TDA.
As already mentioned in Section~\ref{sec:ffpiondecay},  we adopt  the  tensorial decomposition~(\ref{2.02}) given in Ref.~\cite{Bryman:1982et,Moreno:1977kx}, required  in order to ensure electromagnetic gauge invariance is pointed out. 

Thus, an appropriate tensorial decomposition  for the definition of the Double Distributions  would be
\beq
&&\langle \gamma (p_{\gamma})|\bar \psi(0)\,\tau^-\, \gamma^{{\large\{}\mu}\gamma_5\,i\stackrel{\hspace{-.2cm}\leftrightarrow}{{\cal D}^{\mu_1}}\ldots 
i\stackrel{\hspace{-.5cm}\leftrightarrow}{{\cal D}^{\mu_{n-1}{\large\}}} }\psi(0) |\pi^+(p_{\pi})\rangle\nonumber\\
&&=e\, \left( \varepsilon\cdot \Delta\right)
\,\sqrt{2} \, f_{\pi}\,\frac{\,\Delta^{{\large\{ }\mu}}{t-m_{\pi}^2}\, \phi^{(n-1)}\, \left(-\Delta\right)^{\mu_1}\ldots \left(-\Delta\right)^{\mu_{n-1} {\large \}}} 
\nonumber\\
&&+
\frac{e}{2}\, \left(\vec{\varepsilon}^{\perp}\cdot \vec{\Delta}^{\perp}\right)
\left[
\frac{2\,p^{ \{\mu}}{\sqrt{2} \, f_{\pi}}\, \, \, \sum_{k=0}^{n-1}\,\frac{(n-1)!}{k!\, (k-n-1)!}\, C'_{nk}(t)
\, p^{\mu_1}\ldots  p^{\mu_{n-1-k}}\,\left(-\frac{\Delta}{2}\right)^{\mu_{n-k}}\ldots\left(-\frac{\Delta}{2}\right)^{\mu_{n-1}\}}
\right . \nonumber\\
&&\hspace{2.7cm}+\left .
\frac{\Delta^{ \{\mu}}{\sqrt{2} \, f_{\pi}}\,  \,\, B_{n}(t)\left(-\frac{\Delta}{2}\right)^{\mu_{1}}\ldots\left(-\frac{\Delta}{2}\right)^{\mu_{n-1}\}}
\right]\quad,
\label{matr-elmt-a}
\eeq
where we have chosen to decompose the second term on the r.h.s. in such a way that it does explicitly not depend on the gauge choice. Also, since there is a large freedom in the way of writing down this decomposition, we have chosen to
 isolate the highest power of $\xi$, in the third line. However, the generalized Form Factor $B_{n}(t)$ could perfectly be included into $C'_{nk}(t)$. This is more  {\it naturally} seen  if we adopt  the tensorial decomposition $\varepsilon_{\nu} \, p_{\gamma}^{\mu} \, p_{\pi}^{\nu}$.

Time-reversal invariance does not imply any symmetry on either term. The polynomiality property immediately follows. %

The  decomposition of the axial twist-2  matrix element~(\ref{matr-elmt-a}) leads to a relation in agreement with the general relation~\ce{polyaxialtot}.  From  the polynomiality property  in the NJL calculation, we know that  the pion pole term yields the highest power of $\xi$ for odd moments in $x$~(\ref{dapol}). In particular, in the NJL model calculation, no $B$-like term is present.
\\

The moments of the pion DA are expressed  in terms of 
\beq
\phi^{(n)}&=&\int_0^1\, dx\, \left( x-\frac{1}{2}\right)^n\, \phi(x)
\quad,
\nonumber
\eeq
with $n$ even;
what corresponds to the expression~(\ref{dapol}) to which we have extracted  the $t$-dependence to keep trace of the pion pole.

The generalized moments can be written as  Double Distributions  following~(\ref{ddpion}). These DD's are the generating functions of the twist-2 Form Factors. The term depending on both $p$ and $\Delta$ is expressed as ${\cal W}'(\beta,\alpha; t)$,
\beq
C'_{nk} (t)&=& \int_{-1}^1d\beta\,\int_{-(1-|\beta|)}^{1-|\beta|}\, d\alpha \, \beta^{n-1-k}\, \alpha^k\, {\cal W}'(\beta,\alpha; t)\quad.
\label{gen-mmt-ff-a}
\eeq
On the other hand, we have  extracted the pure $\Delta$-dependent generalized moments whose expression is therefore
\beq
B_{n} (t)&=& \int_{-1}^1d\alpha\,\alpha^{n-1}\, {\cal B}(\alpha,t)\quad.
\eeq

After all these considerations, we can write down  the light-like separated matrix element 
\beq
&& \left\langle \gamma\left(p_{\gamma}\right)  \right\vert \bar{q}\left(  -\frac{z}{2}\right)
\rlap{$/$}z\, \gamma_5\,\tau^{-}q\left(  \frac{z}{2}\right)  \left\vert\pi^+\left(  p_{\pi}\right)  \right\rangle
\nonumber\\
&&=e\, \left(\varepsilon\cdot \Delta\right)
\,\sqrt{2} \, f_{\pi}\,\frac{\,\Delta \cdot z}{t-m_{\pi}^2}\,\,\int_{-1}^{1} d\alpha\, e^{i\alpha\, (\Delta\cdot z)/2}\, \phi\left(\frac{\alpha+1}{2}\right)\nonumber\\
&&+\frac{e}{2}\, \left(\vec{\varepsilon}^{\perp}\cdot \vec{\Delta}^{\perp}\right)
\left[
\frac{1}{\sqrt{2} \,f_{\pi}}\, \,2\,p\cdot z\, \,\int_{-1}^1d\beta\,\int_{-(1-|\beta|)}^{1-|\beta|}\, d\alpha \,e^{-i\beta (p\cdot z)+i\alpha (\Delta\cdot z)/2} \,{\cal W}'(\beta,\alpha; t)\right.\nonumber\\
&&\left .  \hspace{2.7cm}+
\frac{1}{\sqrt{2} \,f_{\pi}}\, \Delta \cdot z\, \int_{-1}^{1}\, d\alpha \,e^{i\alpha\, (\Delta\cdot z)/2}\, {\cal B}(\alpha,t)
 \right]
 \quad,
\label{dd-tda-a}
\eeq
where the pion resonance   structure has been subtracted.
\\

From the definition of the axial TDA~(\ref{2.05}), we find the reduction formula
\beq
A^{\pi^+}(x,\xi,t) &=&   \int_{-1}^1  d\beta \int_{-(1-|\beta|)}^{1-|\beta|} d\alpha \, \delta(x-\beta-\xi \alpha) \,{\cal W}^{' \pi^+}(\beta, \alpha, t)
+
\mbox{sgn}(\xi)\, B\left(\frac{x}{\xi}, t\right)
\quad.
\label{red-dd-tda-a}
\eeq
The Fourier transform of the first term on the r.h.s. of Eq.~(\ref{dd-tda-a}) manifestly reduces to the pion pole term parameterization of Eq.~(\ref{2.05}) as expected.
\\

A D-term was introduced in the matrix elements defining the GPDs. This new term is necessary in order to modify the tensorial structure to reproduce the
 correct behavior with the two independent variables $p \cdot z, \Delta \cdot z$.  We therefore interpret the $D$-term in the context of TDAs as the minimal contribution that we are compelled to introduce in order to restore gauge invariance in the tensorial structure  adapted from the axial vector hadronic  current involved in the pion radiative decay~(\ref{2.02}).
In other words, the equivalent of the $D$-term is  the pion pole contribution in the  language of Double Distribution parameterization.

\section{ Pion-Photon TDAs in Other Approaches}
\label{sec:compa-tda}

Other studies of pion-photon TDAs have already been undertaken~\cite{Tiburzi:2005nj, Broniowski:2007fs, Kotko:2008gy}.  We also mention the phenomenological approach, coming from the GPD analysis, used in Ref.~\cite{Lansberg:2006fv}. All these calculations have been performed using the Double Distributions described in Section~\ref{sec:dd-dt}.

In this Section, we  aim to compare the $(x, \xi, t)$ dependencies as obtained by different methods and models in order to give a critical point of view of our results.

\subsection{The $\alpha$-Representation of the DD for the  TDAs}

In this Section, we consider the $\alpha$-representation of the propagators in the kinematics of the pion-photon transition described in Section~\ref{sec:kin-tda}. This technique leads to the identification of the Double Distributions defined here above. Nevertheless and as it is been highlighted before, the model imposes the integral decomposition as well as the relevant diagrams to be considered. The forthcoming considerations aim to be general and applicable to the model calculations whose results will be given in a further Section.
\\

In the vector current case, the trace over the Dirac indices  yields to the exact tensorial structure~(\ref{2.04}). As a consequence, the 3-point function corresponds exactly to the vector TDA  for the $u$-quark up to an overall factor $g$ coming from the trace over isospin, flavor, color and Dirac spaces.~\footnote{In the very particular case of the NJL model calculation, the factor is $g=8 N_c\, f_{\pi}\, g_{\pi qq}$, e.g. Eq.~(\ref{3.04}).} 
In terms of Double Distribution,  it reads
\beq
v(x, \xi, t)&=&g  \,\tilde I_3(x, \xi,t)\quad,\nonumber\\
&=&\frac{1}{(4\,\pi)^2}\,\int_0^1 d\beta \int_{-1+|\beta|}^{1-|\beta|} d\alpha\, \delta\left(x-\beta-\alpha \xi\right)\, {\cal I}_3(\beta,\alpha,t)\quad,
\eeq
with the Double Distribution in the same  kinematics being
\beq
{\cal I}_3(\beta,\alpha,t)&=&g\,\frac{\theta \left(  1-\alpha-\beta\right)}{m^2- m_{\pi}^2 \,\frac{1}{2}\,\beta(1-\beta-\alpha) - \frac{t}{4}\,(1+\alpha-\beta)(1-\alpha-\beta) }\quad.
\label{dd-3prop-ab}
\eeq
A detailed derivation of this result is given in Appendix~\ref{sec:dd-tda-kin}. 
The total DD for the vector TDA ${\cal W}(\beta,\alpha,t)$ is obtained applying the previous result to the isospin decomposition~(\ref{3.03}).
The support property follows  from the $\beta$-integration~(\ref{supp-xi-pos}, \ref{supp-xi-neg}). 

From the DD~(\ref{dd-3prop-ab}), one can easily see that the limit $m_{\pi}^2\to 0$ implies that
\beq
{\cal I}_3(\beta, \alpha, t)&=&{\cal I}_3(\beta, -\alpha, t)\quad.
\nonumber
\eeq
 This so-called $\alpha$-symmetry reflects the appearance of only even powers of $\xi$ in the polynomiality expansion of the vector TDA; a result that has been obtained in the calculation of {\it skewed} PDs in the NJL model~(\ref{even}).
\\

The trace over the Dirac space for the  axial current  involves  2- and 3-propagator integrals. Besides the terms proportional to the external momenta, the  trace evaluation of the diagram of the type $(a)$ of Fig.~\ref{Fig3}  yields a factor $-2\varepsilon\cdot k\, p_{\pi}^{\mu}$. This renders the calculation  slightly more complicated but the $k^{\nu}$-dependent integral can be reduced to the same 3-propagator integral as for the vector one.

 The evaluation of the Double Distributions is similar to the one undertaken for the 3-propagator integral of the vector TDA.
 The subsequent steps are as described in Appendix~\ref{sec:dd-tda-kin}, and the result for the $k^{\nu}$-dependent integral is
\beq
\varepsilon \cdot \tilde I_3(x,\xi, t)&=&\frac{\varepsilon \cdot \Delta}{(4\,\pi)^2}\,\int_0^1 d\beta \int_{-1+|\beta|}^{1-|\beta|} d\alpha\, \delta\left(x-\beta-\alpha \xi\right)\,  \left(1+\alpha-\beta\right)\,{\cal I}_3(\beta,\alpha,t)
\quad,
\eeq
with the same double distribution ${\cal I}_3(\beta,\alpha,t)$ defined by Eq.~(\ref{dd-3prop-ab}).
\\

\subsection{Phenomenological Parameterizations of TDAs}

Phenomenological parameterizations of the $(x, \xi, t)$ dependence of the parton distributions constitute a good alternative and/or complementary approach of the diagrammatic analysis of the Double Distributions.
In what follows, we report  a fully phenomenological $t$-independent parameterization and a semi-kinematical approach improved by phenomenological considerations.
\\

It is easier to start describing the $(x, \xi)$ dependency by setting $t=0$. 
Hence, we will, in a first time, consider the
  $t$-independent Double Distributions, entering the reduction formulae~(\ref{dd-v-tib}, \ref{red-dd-tda-a}). 
  
  In the DVCS framework, the suggestion has been made by Radyushkin~\cite{Radyushkin:1998bz,Radyushkin:1998es}  to parameterize the GPDs in such a way that the forward limit, i.e. 
\beq
\int_{-1+|\beta|}^{1-|\beta|} d\alpha\, {\cal F}^q (\beta,\alpha)&=& q(\beta)\quad,
\nonumber
\eeq  
is ensured. 
The function $q(\beta)$ is hence related to the parton distribution for the flavor $q$. %
  Basically, the phenomenological forward quark distribution as measured from DIS can be used as an input in the parameterization. 
  
  The TDA do not obey any forward limit, but it has nevertheless been proposed to apply the DIS phenomenology as a first approach~\cite{Lansberg:2006fv}.
 Hence, for the TDAs, we will use the same form as for the GPDs, i.e.
\beq
{\cal F}^q (\beta,\alpha)&=& h (\beta,\alpha) \,q(\beta)\quad,
\label{t-ind-dd}
\eeq
for ${\cal F}^q$ corresponding either  to the vector or axial DD, namely ${\cal W}$ or ${\cal W}'$ respectively. 
The function $h (\beta,\alpha) $ denotes a profile function, that  can be parameterized through a one-parameter ansatz, following Refs.~\cite{Radyushkin:1998bz,Radyushkin:1998es,Radyushkin:2000ap},
\beq
h (\beta,\alpha) &=& \frac{\Gamma(2b+2)}{2^{2b+1} \Gamma^2(b+1)}\,\frac{\left[(1-|\beta|)^2-\alpha^2\right]^b}{(1-|\beta|)^{2b+1}}\quad.
\label{profile}
\eeq
The profile function is normalized to $1$.
\\
\begin{wrapfigure}[19]{l}{9.5cm}
\centering
\includegraphics[height=5.3cm]{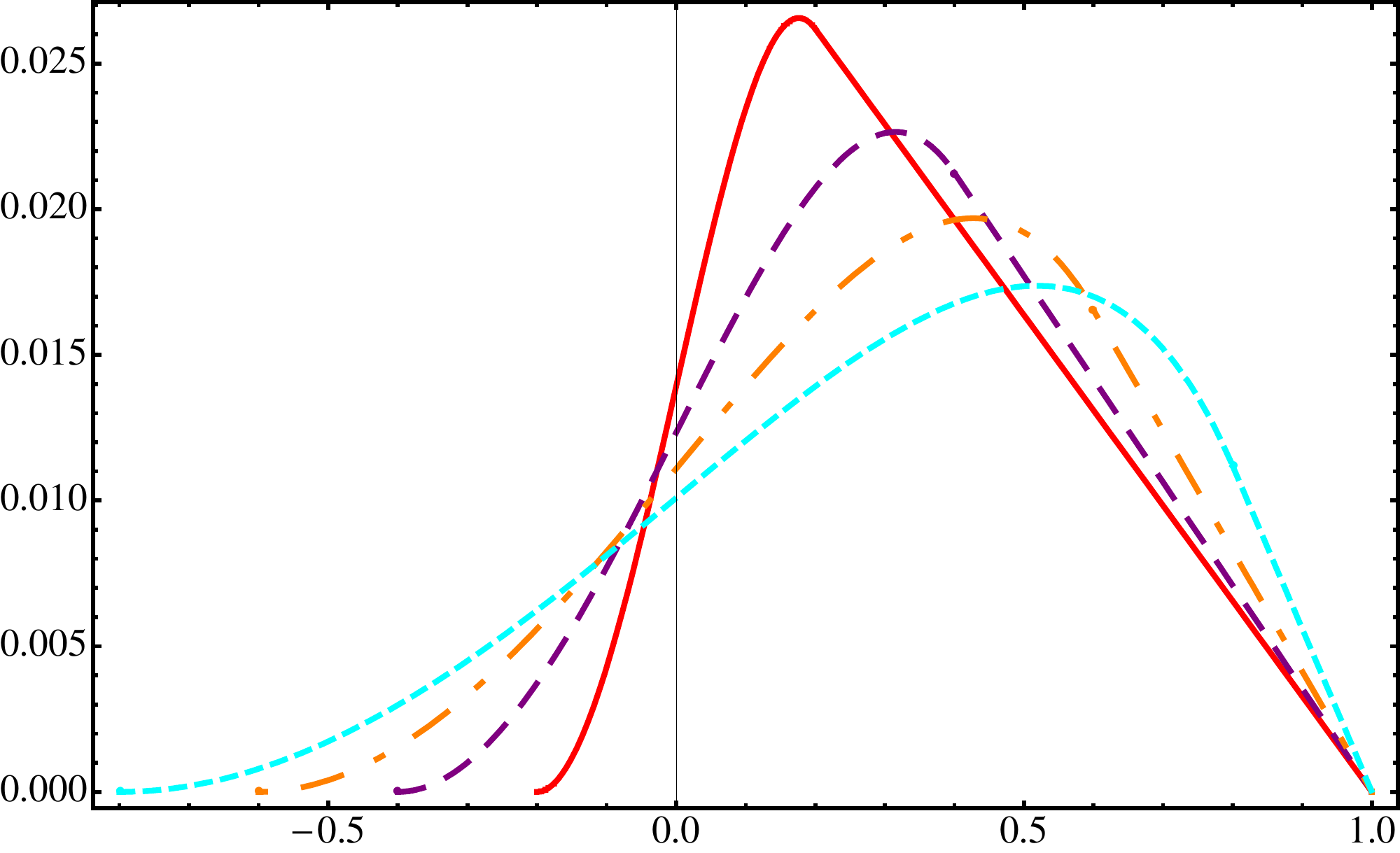}
\caption{The  TDA $v^{\pi}$ as described in Ref.~\cite{Lansberg:2006fv} through the ansatz for the profile function (\ref{profile}); for values of the $\xi$ variable $0.2, 0.4, 0.6 \,\&\, 0.8$, respectively, the plain red, the dashed purple, the dashed-dotted orange and the short-dashed cyan curves. }%
\label{t-ind-param-rad}%
\end{wrapfigure}
The parameter $b$ in~(\ref{profile}) characterizes the strength of the $\xi$-dependence of the resulting GPD. The limiting case $b\to\infty$ corresponds to the $\xi$-independent ansatz for the GPD, namely $H(x,\xi)=q(x)$.
The power in $b$ is a free parameter for both the valence and sea distributions. They can be used as fit parameters in the extraction of GPDs from the experiments. For instance, the twist-2 DVCS predictions of Ref.~\cite{Vanderhaeghen:1999xj} corresponds to the choice
 \beq
b=b_{val}=b_{sea}=1\quad.
\nonumber
\eeq
 This mild dependence on $\xi$ is followed by the authors of Ref.~\cite{Lansberg:2006fv} for the parameterization of the pion-photon TDA.
\\

Also, Lansberg {\it et al.}~\cite{Lansberg:2006fv} assume, as a first guess, that  the $\beta$-dependence of $q$  for the $\gamma\to\pi^-$ transition\footnote{See in Section~\ref{sec:sym-tda} how to go link this transition to the $\pi^+\to\gamma$. } is given by a simple linear law~\footnote{The parameterization is here given for the quark distribution, the antiquark would be $\bar q(\beta)=-q(-\beta)$.}
\beq
q(\beta)&=& 2(1-\beta)\,\theta(\beta)\quad.
\nonumber
\eeq
The $t$-dependence comes solely from the Form Factors through the sum rules. Therefore, the resulting distribution $d(x, \xi)$, with $d=v,a$, must be normalized to $1$ when integrated over $x$. This leads to a TDA whose dependence on the three kinematical variables is given by
\beq
D(x, \xi, t)&=&\left(Q_d\, d^u(x, \xi)+ Q_u \,d^d(x, \xi)\right)\,\frac{\sqrt{2}f_{\pi}}{m_{\pi}}F_{D}\left(  t\right) \quad,
\label{ans-jp}
\eeq
$D$ denoting either the vector or axial quantities. %

In the same Reference  the $t$-dependence coming from the Form Factors is considered negligible as we are concerned with small $t$ region.  In this way, we can use the values given at zero momentum transfer, see Section~\ref{sec:sr-pol}.

As explained in Section~\ref{sec:dd-dt}, the reduction from the DD to {\it skewed} distributions requires a particular dependence on the skewness variable, imposed in the $\delta$-function of the reduction formula. This choice leads to the identification of two kinematically different regions.
The result obtained through the above-given ansatz is shown, for the $u$-quark and for a vector current, on Fig.~\ref{t-ind-param-rad}. The shape of the vector $u$-quark TDA reasonably follows what we obtained in the NJL model, Fig.~\ref{TDAmpi140}. The same ansatz is used for the axial TDA with the simple replacement $D=A$ in~(\ref{ans-jp}). Therefore the difference between the vector and axial TDA resulting from this ansatz resides in the ratio $F_A^{\pi}/F_V^{\pi}$ and the charge combination. The total $\gamma$-$\pi^-$ distribution does, in turn, change according to the charge combinations involved in the transition, see Section~\ref{sec:sym-tda}. Also, this parameterization from the phenomenology of GPDs does not include the term compensating for the gauge invariance in~(\ref{dd-tda-a}). However, this contribution is closely related to the pion pole whose importance in fulfilling the sum rule for the axial current has been highly emphasized in the previous Sections. We conclude that the axial TDA proposed by~\cite{Lansberg:2006fv} considerably and conceptually  differs  from the one obtained in the NJL model.
\\

Another (semi-)phenomenological approach including, this time, the phenomenology of photon distributions has been proposed in  Ref.~\cite{Tiburzi:2005nj}.
The $t$-dependent double distributions defined in Eqs.~(\ref{dd-tda-v}, \ref{dd-tda-a}) are calculated in a simple quark model. In doing so, Tiburzi determines the $t$-dependence of the distributions and reproduces the Form Factors. His intermediate  results, in agreement with the results of Ref.~\cite{Broniowski:2007fs},  are  $F_A(0)/F_V(0)=0.98$ and $ F_A(0)\sim 0.026$.
 However the partonic content of the model has been modified to give reasonable phenomenology while evolved. But
positivity bounds for the pion-photon Transition Distribution Amplitudes are investigated instead.
The positivity constraints are an application of the Schwartz inequality in the Hilbert space~\cite{Diehl:2000xz, Pire:1998nw, Pobylitsa:2001nt, Pobylitsa:2002gw, Pobylitsa:2002iu, Radyushkin:1998es}. Basically, in the $[-1,-\xi]$ and $[\xi,1]$
regions, we can find a bound of the type, for the pion GPD,
\begin{eqnarray}
\theta(1-x)\theta(x-\xi)\Big| H_q^{\pi^+}(x,\xi, t)\Big| \leq \sqrt{q^{\pi^+}\left(x_{\mbox{\small in}}\right)\,q^{\pi^+}\left(x_{\mbox{\small out}}\right)}\quad,
\label{pos-bound-gpd}
\end{eqnarray}
where the momentum fractions of the quark before and after the scattering are
\beq
x_{\mbox{\small in}}=\frac{x+\xi}{1+\xi}\quad,&&\quad x_{\mbox{\small out}}=\frac{x-\xi}{1-\xi}\quad.
\nonumber
\eeq
\\

The author of  Ref.~\cite{Tiburzi:2005nj} provides for some estimates of the TDAs on the basis of positivity bounds. The latters, determined using the technique described in Ref.~\cite{Pire:1998nw}, involve the $u$-quark and $d$-antiquark distributions, the first one related to the pion and the second to the photon.
\\
\begin{figure}[ptb]
\centering
 \includegraphics [height=4.3cm]{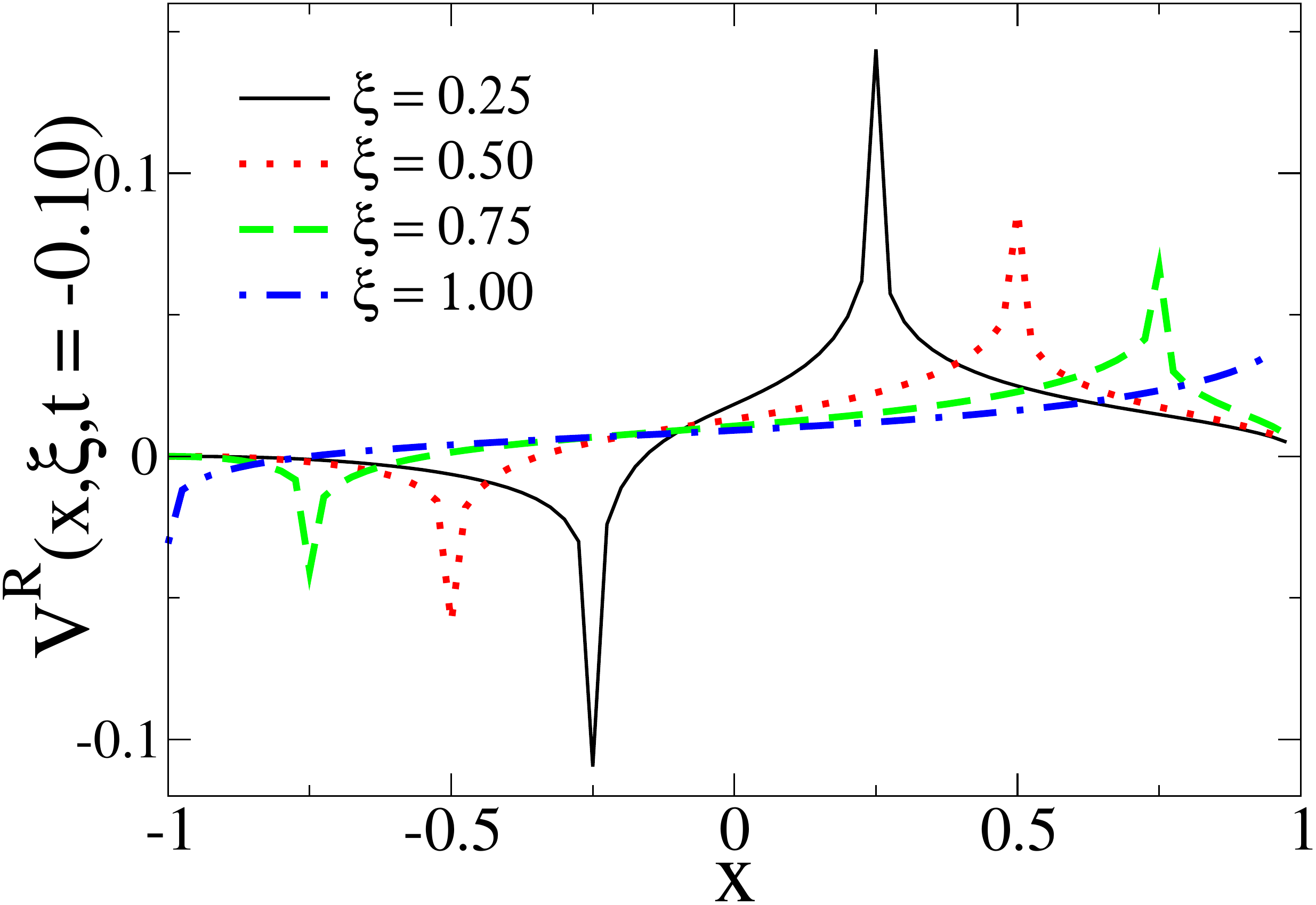}
 \hspace{.5cm}
 \includegraphics [height=4.3cm]{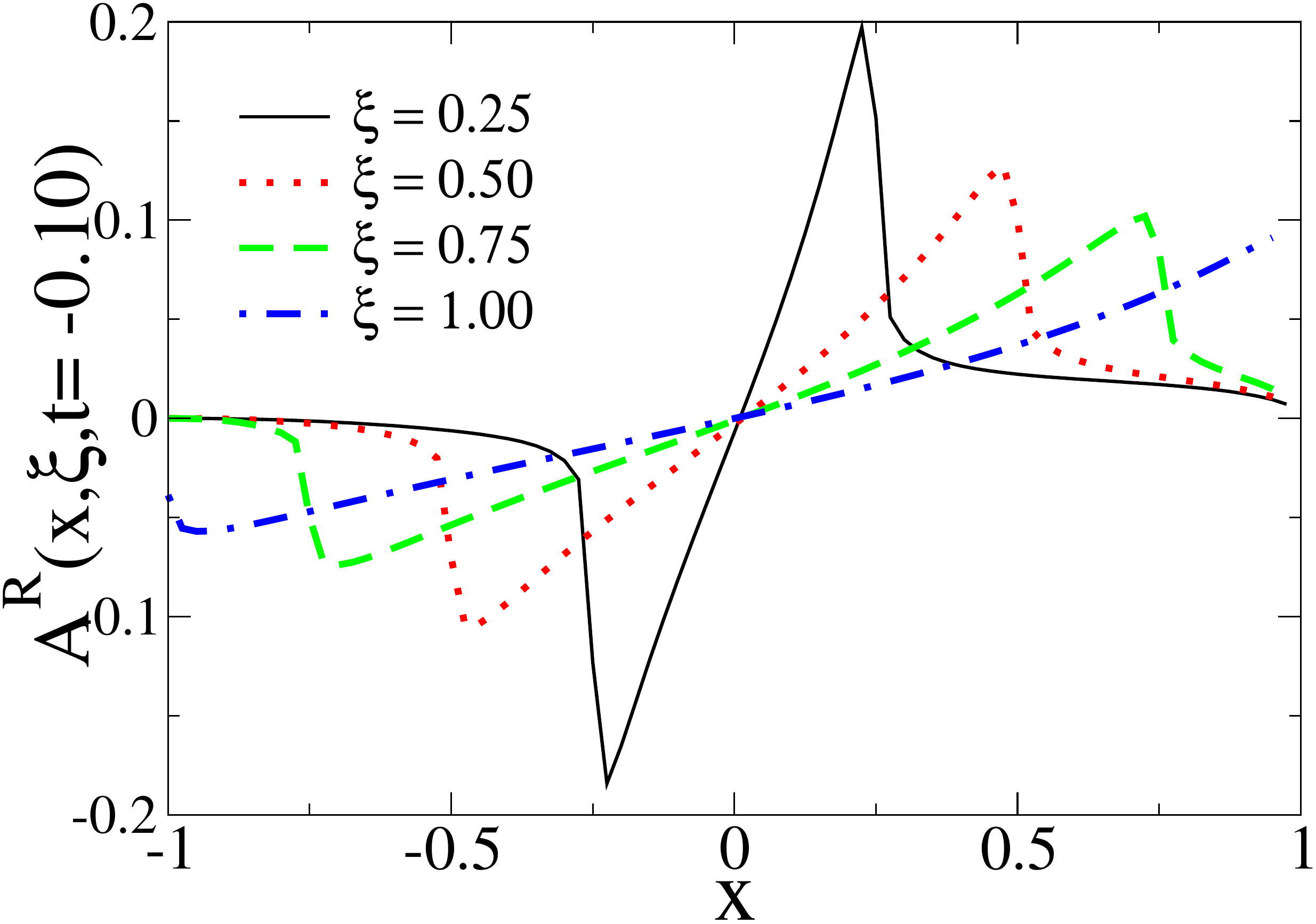}
\caption{The vector and axial
 TDAs in the approach of Ref.~\cite{Tiburzi:2005nj}.}
\label{tibur}
\end{figure}
The Double Distributions obtained in the simple quark model are modified in order to saturate those constraints.
The modifications of the model are motivated by the apparition of pion and photon parton distributions  in the new realistic double distributions. 
Since the evolved PDFs are well-known, this allows one to establish the renormalization scale of the model avoiding evolution.
 For the pion, a simple analytic parameterization of the valence and sea quark
has been chosen following Ref.~\cite{Gluck:1999xe}. The real photon distribution is parameterized in a vector-meson dominance fashion~\cite{Gluck:1999ub}.
 From the perspective of model building, this comes out to be an attempt to satisfy the known constraints and generate an input distribution at a moderately low scale.
\begin{wrapfigure}[16]{l}{9cm}
\centering
 \includegraphics [height=5.7cm]{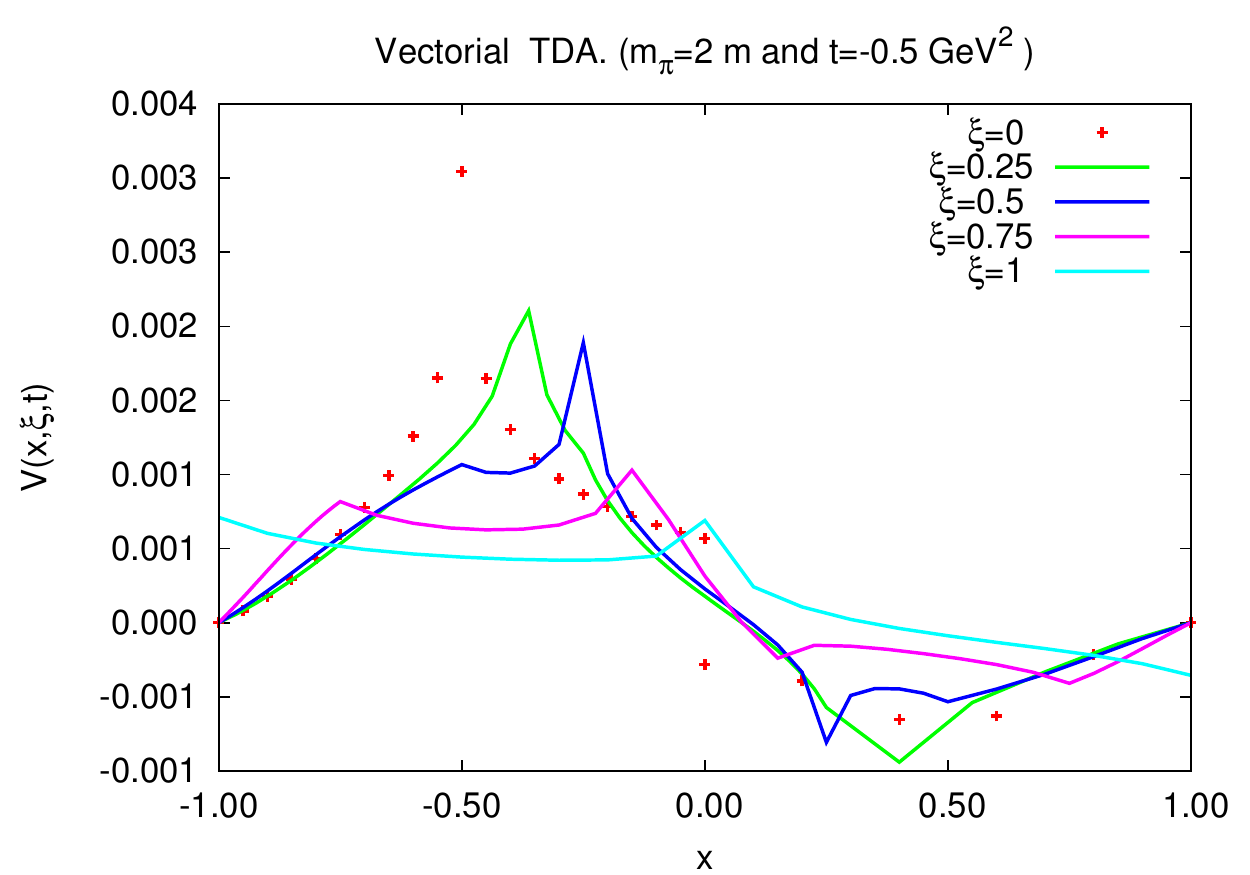}
\caption{The vector
 TDA for $m_{\pi}=2 m$ and for $t=-0.5$ GeV$^2$.}
\label{mpi2}
\end{wrapfigure}

Thus the values of the vector and axial Form Factor, as given by~\cite{Amsler:2008zz}, are used in order to determine the empirical constants of the parametrization. In the same way, the constituent quark mass parameter $m$ is obtained (fitted) requiring that the model pion-photon transition Form Factor comes close to the experimentally parameterized form. The value $m=0.2$ GeV is obtained.
The resulting TDAs, illustrated on Fig.~\ref{tibur},
do not satisfy the isospin relation (\ref{5.01}) due to the different choice
of the $u$-quark and $d$-antiquark distributions used in the saturation of the positivity
bounds. We have  studied the positivity bounds~(\ref{pos-bound-gpd}) for GPDs in the NJL model  and noticed that
it is actually an upper bound that is sometimes very higher than the value of
the GPD itself. 

Moreover the TDAs shown by Tiburzi are rather peaked at $x=\xi$, what is not reproduced in the NJL model calculation. Rather, at weak binding when  $m_{\pi}=2m$,
the latter presents peaks at $x=\frac{1}{2}(1-\xi)$ and $x=\frac{1}{2}(1+\xi)$ for the vector TDA \cf{mpi2}. This is what one expects from a free quark picture. This has already been observed for the GPDs in the NJL model~\cite{Theussl:2002xp} and in a previous work of the same Tiburzi~\cite{Tiburzi:2001ta}. Given the previous argument, the position of the peaks obtained by Tiburzi is not understandable in the NJL model.

\subsection{Model Calculations of the TDAs}

Other model calculations of  the pion-photon TDAs have already been done, respectively, in the Spectral Quark Model (SQM)~\cite{Broniowski:2007fs} and a non-local Chiral Quark Model ($\chi$QM)~\cite{Kotko:2008gy}.  The Double Distributions defined through the $\alpha$-representation of Section~\ref{sec:dd-tda}
have been used in both calculations. We here after comment and compare their results.

 The vector and axial TDAs calculated in the SQM, NJL model and non-local $\chi$QM  are compared in Fig.~\ref{comp}.
 \\
 
The authors of the first reference use the asymmetric notation~(\ref{asym-not}, \ref{kin-asym-tda}).  The comparison is here awkward since the authors define, by turns,  
\beq
\zeta=(p_{\gamma}-p_{\pi})^+/p_{\pi}^+ &;& -\zeta=(p_{\gamma}-p_{\pi}) \cdot n
\nonumber
\eeq
 We nevertheless decide to use  the standard relation between their asymmetric notations and the symmetric ones~(\ref{asym-not})~\cite{Diehl:2003ny}.
Their functions $V_{\mbox{\tiny SQM}}$ and $A_{\mbox{\tiny SQM}}$ correspond, respectively,  to the $v$ and $a$   given in Eqs.~(\ref{3.03}, \ref{3.10}).
They  are obtained through the reduction formulae, e.g.~(\ref{dd-v-tib}) for the vector TDA. As for the axial TDA, the decomposition~(\ref{dd-tda-a}) is not exactly adopted: Broniowski {\it et al.} {\it structurally} isolate the pion pole by neglecting the whole $\Delta\cdot z$ structure instead.
 In so doing they select from the trace only the piece proportional to $(\vec{\varepsilon}^{\perp}\cdot \vec{\Delta}^{\perp}) \,p^{ \mu}$ of the tensorial  decomposition
\beq
&&\hspace{-1cm}\int\frac{dz^{-}}{2\pi}e^{ixp^{+}z^{-}}\left.  \left\langle \gamma\left(
p_{\gamma}\right)  \right\vert \bar{q}\left(  -\frac{z}{2}\right)
\rlap{$/$}n\hspace*{-0.05cm}\gamma_{5}\tau^{-}q\left(  \frac{z}{2}\right)
\left\vert \pi\left(  p_{\pi}\right)  \right\rangle \right\vert _{z^{+}=z^{\bot}=0}\nonumber\\
& & =\frac{1}{p^{+}}e\, \left(\vec{\varepsilon}^{\perp}\cdot \vec{\Delta}^{\perp}\right) \,  \frac{ A^{\pi^{+}}\left(  x,\xi,t\right)  }{\sqrt{2}f_{\pi}}
+\ldots\quad, \nonumber
\eeq
where, here, $A$ is seen as $\bar A$~(\ref{Abarra-A}) to which the pion resonance has been {\it structurally} subtracted. Therefore the piece proportional to $(\varepsilon\cdot \Delta) \Delta^{\mu}$ leads to both $D$-term like structures, one being the resonant pion pole~(\ref{2.05}), and is represented by the ellipses. 
\\

The normalization condition is different from the one used in the NJL model calculation and the results quoted by these authors must be multiplied, for the vector TDA, by a factor  
\beq
48\pi^2\,\frac{\sqrt 2 f_{\pi} }{m_{\pi}}\,F_V(0)\sim 10
\nonumber
\eeq
 before comparison. From Fig.~\ref{comp} we conclude that there is a
qualitative and quantitative agreement, for the vector TDA, between the results of
Ref.~\cite{Broniowski:2007fs} and those obtained in the NJL model.

Regarding the axial TDA we observe, in addition to the normalization factor 
\beq
48\pi^2\,\frac{\sqrt 2 f_{\pi}}{m_{\pi}} \,F_A(0)&\sim& 10\quad,
\nonumber
\eeq
  a change in the global sign due to different definitions.
In the first version of \cite{Broniowski:2007fs} the axial TDA for  positive values of $\xi$ is given\footnote{There is a typographic error in Eq.~(23) of the first version of this reference, where a factor $M_V^2/6$ must be dropped.}. It coincides with the results in the NJL model,  as observed in Fig.~\ref{comp}.
Surprisingly, the result presented in Ref.~\cite{Broniowski:2007fs} coincides with our result for negative $\xi$. It is perhaps due to the change in the definition of $\xi$  mentioned above.  Rather, one might call into question the selection procedure of  the structure dependent quantities.
\\

In
Ref.~\cite{Kotko:2008gy} the TDAs are calculated in three different
models. The first one is a local model whose pole structure has some similarity with   the one of the  NJL model. The two other models, i.e. semi-local and fully non-local,
follow the results of the local one.  The same structure decomposition as in the SQM is followed.
The most prominent difference
between the results obtained in the non-local and the NJL models is
the appearence of important odd powers in $\xi$ in the polynomial
expansion of the vector TDA.
We know from Ref.~\cite{Noguera:2005ej}
that, for non-local models, there are additional contributions to
those calculated in Ref.~\cite{Kotko:2008gy}. In the case of PDFs,
disregarding these contributions can produce small isospin violations
\cite{Noguera:2005cc}.  It can therefore be considered that, on that point, the results of Ref.~\cite{Kotko:2008gy} must be confirmed.
 
 For numerical comparison, the results obtained in \cite{Kotko:2008gy}  must be corrected
by a factor 
\beq
2\pi^2\,\frac{\sqrt 2 f_{\pi}}{m_{\pi}} \,F_V(0)&\sim& 0.45\quad,
\nonumber
\eeq
 due to the use of a different normalization
condition. 
We observe, see  Fig.~\ref{comp},  that our results in the NJL model coincide with the results of the latter reference.~\footnote{ For the axial TDA, there is a change in the definition of the skewness variable between the caption of Figs.~3, 5 and Fig.~9 of Ref.~\cite{Kotko:2008gy}. The agreement is valid if the convention of the caption of their  Fig.~9 is chosen.}
\\

\begin{wrapfigure}[24]{l}{100mm}
\centering
\includegraphics [height=7.5cm]{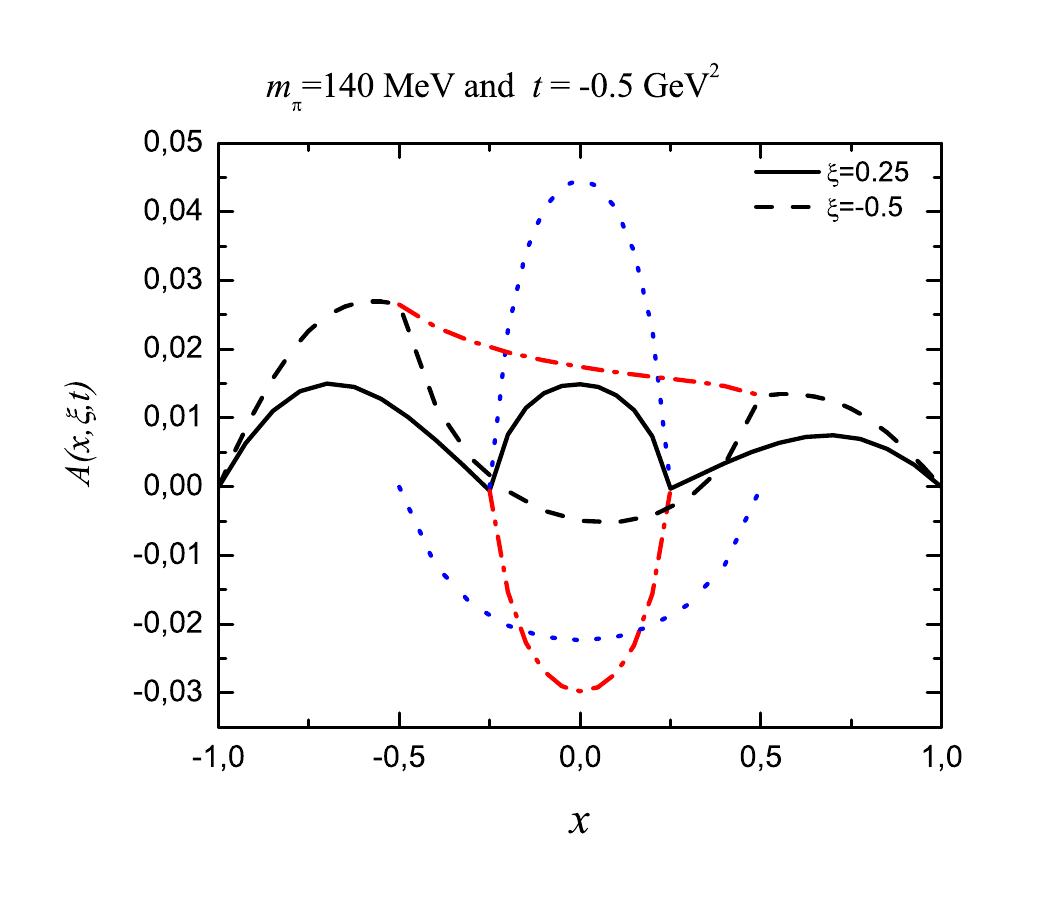}\caption{Contributions to the axial $\pi^+\to\gamma$ TDA
for both positive ($\xi=0.25,$ plain line) and negative ($\xi=-0.5,$ dashed
line) values of the skewness variable and for $m_{\pi}=140$ MeV and $t=-0.5$
GeV$^{2}$. In each case, and in the ERBL region, the contribution coming from
the first diagram of Fig.~\ref{Fig3} is represented by the dashed-dotted lines
and the non-resonant part of the second diagram of Fig.~\ref{Fig3} is
represented by the dotted lines.}%
\label{comparisonxi}%
\end{wrapfigure}

Regarding our axial TDA, it is worth noticing that it differs from the previous calculations
\cite{Tiburzi:2005nj, Broniowski:2007fs,Kotko:2008gy} due to the effect of the non-resonant
part of the second diagram of Fig. \ref{Fig3}. This contribution,
corresponding to the last term of Eq.~(\ref{3.11}), is proportional to
$\left(  t-m_{\pi}^{2}\right)  ^{-1}$ but with zero value for the residue. The
presence of this term is crucial in order to obtain the axial Form Factor
using the sum rule as shown by the reduction formula~(\ref{red-dd-tda-a}). Furthermore this term is dominant in the ERBL region as
we can infer from Fig. \ref{comparisonxi}.
\\
\\

It can eventually be concluded that there is no disagreement
between the different studies concerning the $\pi$-$\gamma$ TDAs besides the ambiguity in the definition of the skewness variable.
Calculations of Refs.~\cite{Broniowski:2007fs, Kotko:2008gy} are performed in the chiral limit, where $\xi$ runs from $-1$ to $1$. 
 The symmetric nature of this interval makes difficult to check the sign of $\xi$. On the other hand, the NJL calculation  is given for the physical pion mass. In this case, the kinematics of the process imposes $t/(2m_{\pi}^2-t)<\xi<1$. 
From Eqs.~(\ref{a-dglap}, \ref{3.11})  we observe that there is a pole in the axial TDA for the limit value $\xi=t/(2m_{\pi}^2-t)$, preventing us from going through unphysical values of $\xi$. 

Moreover, the sum rules~(\ref{2.06}, \ref{2.07}) for both the vector and axial TDAs are here satisfied for physical values of $\xi$, and broken in the unphysical regions $\xi<t/(2m_{\pi}^2-t)$ and $\xi>1$. We therefore conclude that the choice of sign in our calculation in the NJL model is consistent and gives a guideline for  comparing with other models.

\begin{figure}
\begin{minipage}{7.5cm}
\hspace{-.5cm}
\includegraphics[height=.30\textheight]{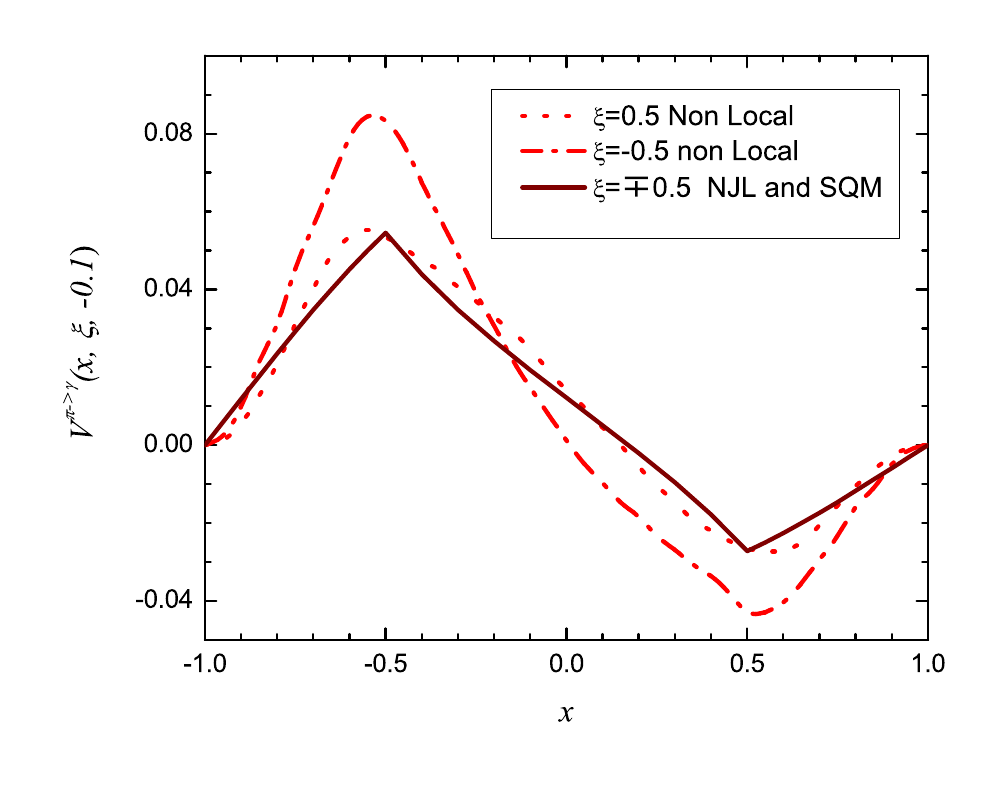}
\end{minipage}
\begin{minipage}{7.5cm}
\hspace{-.5cm}
\includegraphics[height=.31\textheight]{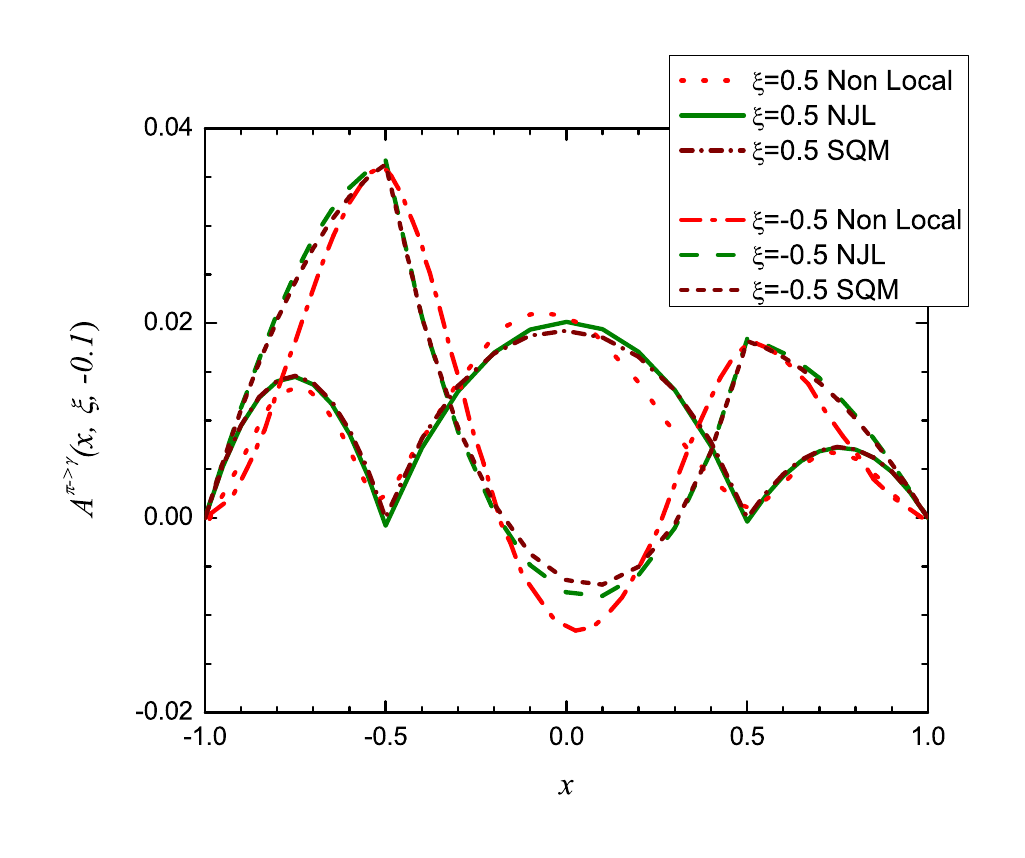}
\end{minipage}
  \caption{ Comparison of the $\pi^+\to\gamma$ TDAs for $t=-0.1$ GeV$^2$ of Refs.~\cite{Broniowski:2007fs, Kotko:2008gy} for $m_{\pi}=0$ MeV  and  Ref.~\cite{Courtoy:2007vy} for $m_{\pi}=140$ MeV.
  On the left, we have: the vector  TDA  for $\xi=\pm 0.5$ as a single (solid) curve  for the results of both  the NJL model and SQM (these four curves are indistinguishable);
    the result of the non-local $\chi$QM calculation  for $\xi=0.5$ (dotted line) and for $\xi=-0.5$ (dashed-dotted line). On the right, we have the results of the axial TDA 
  with the choice for the sign of $\xi$ as discussed in the text.}

\label{comp}
\end{figure}

\section{Discussion about the Results}

In this Chapter we have defined the pion-photon vector and axial Transition
Distribution Amplitudes using the Bethe-Salpeter amplitude for the pion. In
order to make numerical predictions we have used the Nambu-Jona Lasinio model.
The Pauli-Villars regularization procedure is applied in order to preserve
gauge invariance.

We know from PCAC that the axial current couples to the pion. Therefore, in
order to properly define the axial TDA, i.e. with all the structure of the incoming
hadron being included in $A\left(  x,\xi,t\right) $, we need to extract the
pion pole contribution. In so doing, we found that the axial TDA had two
different contributions, the first one related to a direct coupling of the
axial current to a quark of the incoming pion and to a quark coupled to the
outcoming photon and a second related to the non-resonant part of a
quark-antiquark pair coupled with the quantum numbers of the pion.

The use of a fully covariant and gauge invariant approach guaranties that we
will recover all fundamental properties of the TDAs. In this way, we have the
right support, $x\in\left[  -1,1\right]  $, and the sum rules and the
polynomiality expansions are recovered. We want to stress that these three
properties are not inputs, but results in our calculation. The value we found
for the vector Form Factor in the NJL model is in agreement with the
experimental result \cite{Amsler:2008zz} whereas the value found for the axial
Form Factor is two times larger than in \cite{Amsler:2008zz}. This discrepancy is
a common feature of quark models \cite{Broniowski:2007fs}. Also the neutral
pion vector Form Factor $F_{\pi\gamma^{\ast}\gamma}(t)$ is well described.
These results allow us to assume that the NJL model gives a reasonable
description of the physics of those processes at this energy regime.

Turning our attention to the polynomial expansion of the TDAs, we have seen
that, in the chiral limit, only the coefficients of even powers in $\xi$ were
non-null for the vector TDA. This result is quoted as the recovery of the so-called $\alpha$-symmetry of the Double Distributions.  No constraint is obtained for the axial TDA.
Nevertheless, the NJL model provides simple expressions for the coefficients
of the polynomial expansions in the chiral limit, Eqs. (\ref{4.03}) and
(\ref{coefaxial}).

We have obtained quite different shapes for the vector and axial TDAs. This is
in part, at least for the DGLAP regions, imposed by the isopin relation
(\ref{5.01}). We have pointed out the importance of the non-resonant part of
the $qq$ interacting pair diagram for the axial TDA in the ERBL region.

It is interesting to inquire about the domain of validity of the isospin relations
(\ref{5.01}). These relations are obtained from the isospin trace calculation
involved in the diagram $(a)$ of Fig.~\ref{Fig3}. Due to the simplicity of the
isospin wave function of the pion these results are more general than the NJL
model and could be considered as a result of the diagrams under consideration.
These diagrams are the simplest contribution of handbag type.

Also, in the last Section, we have presented the different approaches proposed to gather information about the pion-photon transition distribution amplitudes. The model calculations are in good agreement one to others. On the other hand, the phenomenological approaches are either based on GPD's phenomenology or pion and photon PDF's. It is unfortunately the best we can do for now.
\\

The Transition Distribution Amplitudes proposed by the authors of
\cite{Pire:2004ie} open the possibility of enlarging the present knowledge of
hadron structure for they generalize the concept of GPDs for non-diagonal
transitions. Calculated here, as a first step, for pion-photon transitions,
these new observables  lead to interesting estimates of cross section
for exclusive meson pair production in $\gamma^{\ast}\gamma$ scattering
\cite{Lansberg:2006fv} as we will see in the next Chapter. %
\\
\\
\begin{wrapfigure}[13]{l}{10cm}
\centering
\includegraphics[height=3.3cm]{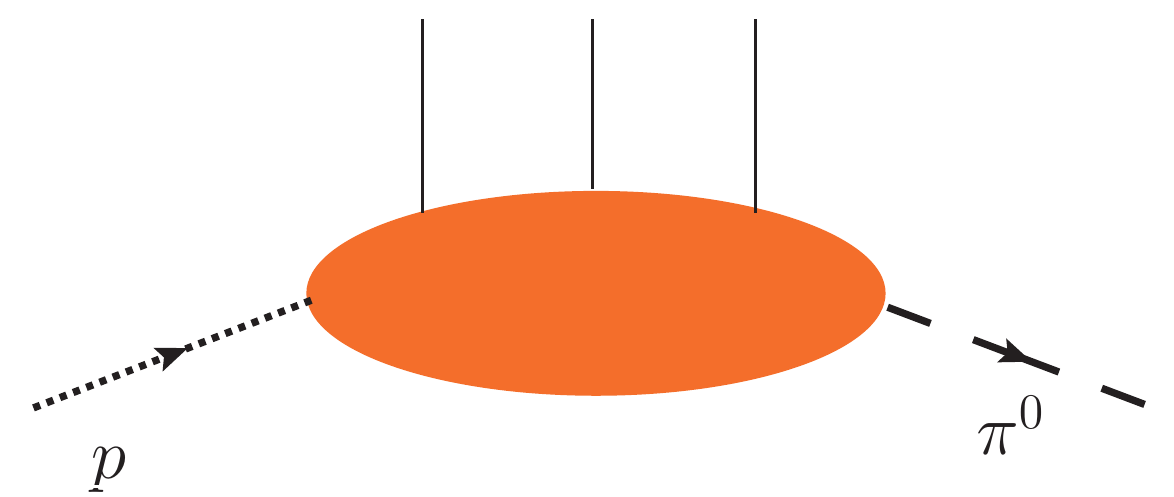}
\caption{Baryonic $p\to\pi^0$ TDA.}%
\label{btda}%
\end{wrapfigure}

Transition Distribution Amplitudes have also been extended to baryonic transitions~\cite{ Lansberg:2007bu, Lansberg:2007se, Lansberg:2007ec, Pire:2004ie}. For instance, the nucleon-to-pion TDAs describe  exactly how a baryon can turn into a meson, namely the ÒtransitionÓ from a baryonic to a mesonic state. 
The leading twist TDAs for the $p\to\pi^0$  transition, depicted in Fig.~\ref{btda}, are defined from the correlator 
\beq
\langle \pi^0(p_{\pi})| \epsilon^{ijk}\, u_i^{\alpha}(z_1 n)\, u_j^{\beta}(z_2 n)\, u_k^{\gamma}(z_3 n)|p(p_1, s_1)\rangle\quad,
\nonumber
\eeq
the latter matrix element being also related to the baryonic DAs.
\\

There are 8 leading-twist TDAs for the $p \to \pi^0$ transition, namely, two vector $ V^{p\pi^0}_{i}\!\!(x_i,\xi, \Delta^2)$, two axial
$A^{p\pi^0}_{i}\!\!(x_i,\xi, \Delta^2)$ and four tensor
$T^{p\pi^0}_{i}\!\!(x_i,\xi, \Delta^2)$ ones. 
The peculiarity of the baryonic TDAs is that, for both the meson and photon cases,  3 quarks are exchanged in the $t$-channel, 
satisfying the relation $x_1 + x_2 + x_3 = 2\xi$  ($\xi \geq 0$). The interplay of the variables implies 3 kinematical regions;  one  DGLAP region, $x_i \geq 0$, and two ERBL  regions, $x_1 \geq 0; x_2 \geq 0;  x_3\leq 0$ and $x_1 \geq 0; x_2 \leq 0;  x_3\leq 0$.
\\

A calculation in the Meson-Cloud model of Ref.~\cite{Pasquini:2006dv}  has been performed in the ERBL region~\cite{Pasquini:2009ki, Pincetti:2008fh, tesimanuel}, where the pion cloud contribution -as coming from a higher order in the Fock state expansion- plays an important role and could therefore be tested.
It is so far the only non-perturbative approach to these new distributions. 
\\
The experimental importance of baryonic TDAs will be shortly overviewed in the next Chapter.

\chapter{
Exclusive meson  production in
$\gamma^{\ast}\gamma$ scattering}
\label{sec:cross}
This Chapter is devoted to the applications of the results for the photon-pion Transition Distribution Amplitudes displayed in the previous Chapter. In particular, we discuss the results of Ref.~\cite{Courtoy:2008nf}.
\\

Cross section estimates for the processes
\begin{equation}
\gamma_{L}^{\ast}\gamma\rightarrow\pi^{+}\pi^{-}\ \ ,\quad\gamma_{L}^{\ast}
\gamma\rightarrow\rho^{+}\pi^{-} \quad,
\label{processes}%
\end{equation}
factorized according to Fig.~\ref{facto},
have been proposed in Ref.~\cite{Lansberg:2006fv} using for the TDA the
$t$-independent double distributions~(\ref{t-ind-dd}), in a first approach, and, in a second,
the $t$-dependent Double Distribution~(\ref{dd-3prop-ab}) obtained through  diagrammatic analysis. %
\\

To implement these analyses,  in this Chapter, we  display the cross section for the processes~(\ref{processes}) with the large-distance part calculated in models.
In a previous Section we have concluded that there is a
clear agreement between the different model calculations of the pion-photon TDAs,
we are allowed to analyze the result of a single model, e.g. the NJL model. 
Furthermore, this choice allows us to keep trace of the pion pole contribution in order to include it in our analysis.

\section{Processes and Kinematics}

The $\gamma^{\ast}\gamma\rightarrow M^{+}\pi^{-}$ process$,$ with $M^{+}%
=\rho_{L}^{+}$ or $\pi^{+},$ is a subprocess of the 
\beq
e\left(  p_{e}\right)
+\gamma\left(  p_{\gamma}\right)  \rightarrow e\left(  p_{e}^{\prime}\right)
+M^{+}\left(  p_{M}\right)  +\pi^{-}\left(  p_{\pi}\right)  \quad,
\eeq
process. We
follow all the definitions of the kinematics given in Ref.~\cite{Lansberg:2006fv},~\footnote{In  Section III.A and Fig. 3.} with the exception that our $n^{\mu}$ vector is defined following~(\ref{light-cone-vector}), which is twice the $n^{\mu}$ vector used in Ref.~\cite{Lansberg:2006fv}. 
In particular, for massless pions,\footnote{The pion mass is set to $m_{\pi}=0$ MeV throughout the Chapter.} we have, 
\\
\begin{wrapfigure}[10]{l}{9cm}
\begin{center}
\includegraphics[height=5cm]{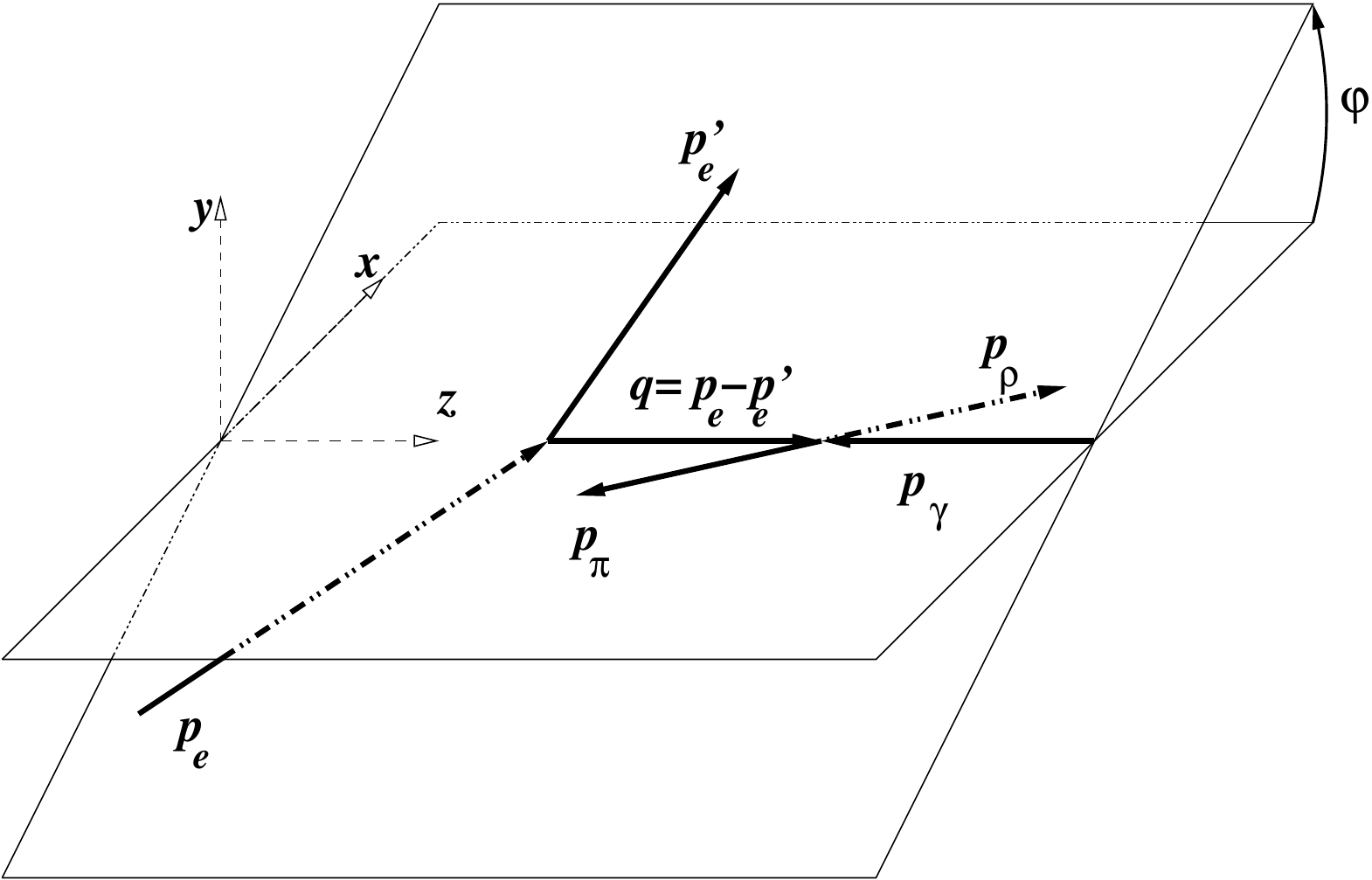}
\caption{Kinematics for the processes~(\ref{processes})~\cite{Lansberg:2006fv}.}
\label{kin-jp}
\end{center}
\end{wrapfigure}
\beq
&&Q^{2}=-q^{2}=-\left(  p_{e}-p_{e}^{\prime}\right)  ^{2}\quad, \nonumber\\
&&W^{2}=\left(  q+p_{\gamma} \right)  ^{2}\quad, \nonumber\\
&&s_{e\gamma}=\left(  p_{e}+p_{\gamma}\right)  ^{2}\quad,
\eeq
with $p_e$ and $p'_e$, respectively, the momentum of the incoming and outgoing electron;
\\
\begin{align}
&&p_{\gamma}=(1+\xi)\bar{p}\quad ,\nonumber\\
&&\quad  p_{\pi}=(1-\xi)\bar{p}+\frac{\vec{\Delta}^{\perp2}}{2\left(  1-\xi\right)  }n+\vec{\Delta}^{\perp},\nonumber\\
&&q  =- \frac{Q^2}{Q^2+W^2}\,(1+\xi)\, \bar p+\frac{Q^2+W^2}{1 +\xi} n\quad,
\label{kin-mesonprod}
\end{align}
where 
$
\Delta_{T}=(0,\vec{\Delta}^{\perp},0)$,
and therefore 
$
\Delta_{T}^{2}=-\vec{\Delta}^{\perp2}$. Notice that 
\beq
\vec{\Delta}^{\perp2}=(-t)(1-\xi)/(1+\xi)\quad,
\nonumber
\eeq
 with $t<0$. 
 \\
 
 It is worth noticing that, even if the skewness variable is not constrained 
to be positive but rather to belong to $\xi \in [-1,1]$, the  external kinematics of the process (\ref{processes}) restrict $\xi$. 
Effectively, the kinematics~(\ref{kin-mesonprod}) have been given with the photon momentum expressed in terms of the momentum transfer squared $Q^2$ and the center of mass energy squared of the $\gamma^{\ast} \gamma$ system $W^2$, through which we can determine the skewness variable, i.e. 
\begin{align}
&y=\frac{q\cdot p_{\gamma}}{p_{\gamma}\cdot p_e}= \frac{1+\xi}{2\xi} \frac{Q^2}{s_{e\gamma}}=\frac{Q^2+W^2}{s_{e\gamma}}\nonumber\\
\Rightarrow \,&
\xi= \frac{Q^2}{Q^2+2W^2}\quad.
\nonumber
\end{align}
%
The momentum transfer therefore reads
\beq
q  &=&-2\xi\bar{p}+\frac{Q^{2}}{4\xi}n\quad,
\label{q-xi}
\eeq
\\
The terms contributing to the cross section are selected by multiplying  the amplitudes  by the longitudinal polarization of the 
incoming photon $\varepsilon_{L \mu}$.
An expression for $\varepsilon_{L \mu}$ is found knowing that $\varepsilon_L^2=1$ and that $\varepsilon_L.q=0$.
For $q$ as given by~(\ref{q-xi}), we find  
\begin{eqnarray}
\varepsilon_{L }= \left(\frac{2\xi}{Q}\bar p +\frac{Q}{2\xi}n\right)\quad,
\label{epsl}
\end{eqnarray}
following Ref.~\cite{Lansberg:2006fv} for the convention for the overall sign.
The real photon polarization is defined by the condition
 $\varepsilon^{-}=0$ together with the gauge
condition $\varepsilon^{+}=0$. The kinematics are illustrated on Fig.~\ref{kin-jp}.

\section{Cross Sections}

The differential cross section for the kinematical variables $Q^2,  t,  \xi$ and the angle $\varphi$ between the hadronic, where takes place the subprocess
$\gamma_L^{\ast}\gamma\rightarrow M^+\pi^-$ with $M$ being either $\rho_L$ or $\pi$, and leptonic planes  reads
\beq
\frac{d\sigma^{e\gamma\to e M^+\pi^-}}{dQ^2 dt d\xi d\varphi}&=& \frac{1}{32\,(2\pi)^4\,s^2_{e\gamma}}\,\frac{1}{\xi(1+\xi)}\, 
 | {\cal M}^{e\gamma\to eM^+\pi^-} |^2\quad.
\label{totcross}
\eeq
The square amplitude for the process $e\gamma\to eM^+\pi^-$ is obtained through  the square amplitude for the
subprocess $\gamma_L^{\ast}\gamma\rightarrow M^+\pi^-$ and the contribution of the fermionic line from which the 
highly virtual photon is emitted
\beq
| {\cal M}^{e\gamma\to eM^+\pi^-}|^2&=& \frac{4\pi\,\alpha_{elm}}{2Q^4}\,\alltrace \left( \thru p'_e \thru \epsilon_L (q)
\thru p_e \thru \epsilon_L^{\ast} (q)\right) | {\cal M}^{M^+\pi^-} |^2\quad.
\eeq
Using Eq.~(\ref{epsl}), we find that the trace yields
\beq
16\xi\,s_{e\gamma}/(Q^2\,(1+\xi)^2)\{2\xi\,s_{e\gamma}\,-\,(1+\xi)Q^2\} \quad.
\nonumber
\eeq
And the  amplitude for the subprocess is the factorized amplitude of Fig.~\ref{facto}. It reads 
\beq
{\cal M}^{\rho_L^+ \pi^-}(Q^2, \xi,t)&=&-\int_{-1}^1 dx\,\int_0^1 dz\,\frac{f_{\rho}}{\sqrt{2}\, f_{\pi}}\,\phi_{\rho}(z) \,M^V_h(z,x,\xi)\, V(x,\xi,t)\quad,
\label{ampl-rho}\\
{\cal M}^{\pi^+ \pi^-}(Q^2, \xi,t)&=&\int_{-1}^1 dx\,\int_0^1 dz\,\phi_{\pi}(z) \,M^A_h(z,x,\xi)\nonumber\\
&& \left[A^{\gamma\rightarrow \pi^{-}}(x,\xi,t)-\frac{4f_{\pi}^{2}}{m_{\pi}^{2}-t}~\epsilon(\xi)~\phi^{\pi}\left(
\frac{x+\xi}{2\xi}\right) \right]\quad
\label{ampl-pi}
\eeq
where $z$ is the light-cone momentum fraction carried by the quark entering the $M^+$ and  $f_{\rho}=0.216$ GeV. The hard parts  are defined as in \cite{Lansberg:2006fv}
\beq
M^V_h(z,x,\xi)
&=&8\pi^2\alpha_{elm}\alpha_s\,\frac{C_F}{N_c\,Q}\,\epsilon^{+\nu\rho\sigma}\,P_{\rho}\,(p_{\pi}-p_{\gamma})_{\sigma}
\frac{1}{z (1-z)}\left(\frac{Q_u}{x-\xi+i\epsilon}+\frac{Q_d}{x+\xi-i\epsilon}\right)\quad, \nonumber\\
 \nonumber\\
M^A_h(z,x,\xi)
&=&4\pi^2\alpha_{elm}\alpha_s\,\frac{C_F}{N_c}\,\frac{(\vec{\varepsilon}^{\perp}.\vec{\Delta}^{\perp})}{Q}\,
\frac{1}{z (1-z)}\left(\frac{Q_u}{x-\xi+i\epsilon}+\frac{Q_d}{x+\xi-i\epsilon}\right)\quad.\nonumber\\
\label{factdapi}
\eeq
The pion DA is chosen to be the usual asymptotic normalized meson 
DA, i.e. $\phi_{\pi}(z)=6 z(1-z)$, what cancels the $z$-dependence of the hard amplitude. The  integral over $x$ remains which contains the model-dependent TDA.
Separating the real and imaginary parts of the amplitude, but only for the structure dependent part, this integral is defined
\beq
I_x^D&=&Q_u\, \int^1_{-1} dx \frac{D(x,\xi,t)-D(\xi,\xi,t)}{x-\xi}+Q_d\, \int^1_{-1} dx \frac{D(x,\xi,t)-D(-\xi,\xi,t)}{x+\xi}\nonumber\\
&&+ Q_u\, D(\xi,\xi,t) \,\left( \mbox{ln}\frac{1-\xi}{1+\xi} -i\pi\right)+ Q_d\, D(-\xi,\xi,t) \,\left( \mbox{ln}\frac{1+\xi}{1-\xi} +i\pi\right)\quad,
\label{intx}
\eeq
where $D$ stands for $V,  A$.
This non-perturbative part will be given by the TDAs evaluated in the NJL model of Section~\ref{sec:tda-njl}.
\\

\begin{figure}[bt]
\centering
\mbox{\subfigure[] {\includegraphics [height=3cm]{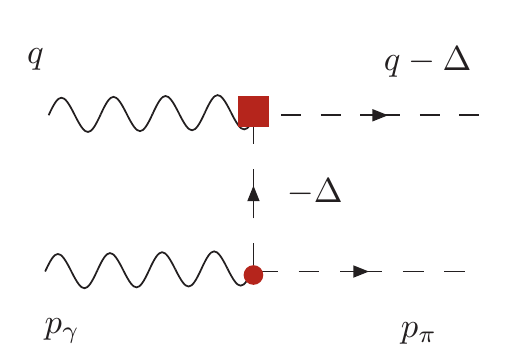}}
\subfigure[] {\includegraphics [height=3cm]{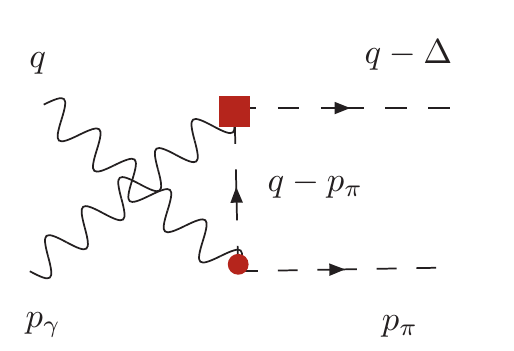}}
\subfigure[] {\includegraphics [height=3cm]{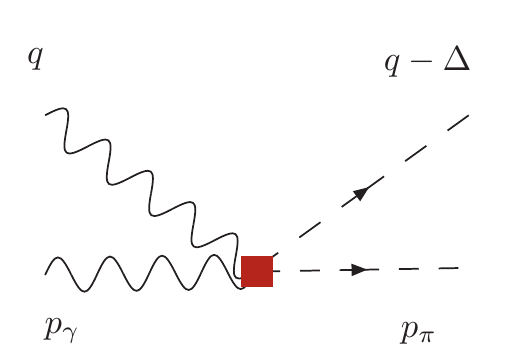}}}
\caption{Pion Compton Scattering. The direct and the cross contributions, respectively $(a)$ and $(b)$, as well as the seagull contribution
$(c)$. The dot vertex is the usual $\pi\pi\gamma$-vertex and the square vertex is the Brodsky-Lepage vertex. }
\label{cs}
\end{figure}

As for the pion pole contribution,   the expression for the amplitude~(\ref{ampl-pi}) suggests that the pion amplitude is 
defined by performing the entire integral procedure just as for the TDAs.  The pion
pole term  becomes proportional to the electromagnetic pion Form Factor studied in Section~\ref{sec:pion-ff}
\begin{equation}
I^{\pi}=
-\frac{1}{\alpha_{s}}\,\frac{3}{4\pi}\frac{Q^{2}\,F_{\pi}\left(
Q^{2}\right)  }{t-m_{\pi}^{2}}
\quad.
\label{pi-FF}
\end{equation}
In this case we would use the asymptotic form for the pion DA by consistency.
If we use $\phi_{\pi}\left(  z\right)  =6z\,(1-z)$ with $z=\left(  x+\xi\right)  /2\xi,$  we obtain the
Brodsky-Lepage pion Form Factor~(\ref{bl-ff})
\begin{equation}
Q^{2}\,F_{\pi}\left(  Q^{2}\right)  =16\pi\ \alpha_{s}\ f_{\pi}^{\,2}\quad.
\end{equation}
Therefore the pion pole contribution to the amplitude~(\ref{ampl-pi}) becomes
\beq
{\cal M}_{\pi}&=& -4\pi\, \alpha_{elm}\,2\,  \frac{Q^2 \,F_{\pi}(Q^2)}{t-m_{\pi}^2}\, \frac{\vec{\varepsilon}^{\perp}.\vec{\Delta}^{\perp}} {Q} \quad.
\label{ampl-pion-comp}
\eeq
\\
In an alternative way, the analyses of the pion Compton scattering yields the same result. 
For the sake of generality, we choose, for the purpose of that calculation, the general expression
for $\varepsilon$ given by $\varepsilon=\vec{\varepsilon}^{\perp}+\alpha \bar p$.
 Since it  does not depend on $\alpha$, we can state that the  result obtained is independent of the gauge's choice. Therefore, in the $\alpha=0$ gauges,  only the direct diagram $(a)$ of Fig.~\ref{cs} does contribute to the cross section. 
Both the crossed $(b)$ and the seagull $(c)$ diagrams are higher-twist.
\\
Contrarily to the amplitudes involving the TDAs, the pion amplitude is model-independent thanks to the asymptotic expression of the pion DA that will be
henceforth adopted.
\\

An analysis of Eq.~(\ref{factdapi}) including the isospin decomposition (\ref{3.03}, \ref{3.10}) shows that only the direct terms of the charge combinations, i.e. $\propto Q_q^2$,
 contribute to the amplitude ${\cal M}^{\rho_L^+ \pi^-}$.  A similar analysis for the pion  production but at zero momentum transfer indicates that the cross terms, i.e. $\propto Q_u Q_d$, cancel out.

Going back to the differential cross section Eq.~(\ref{totcross}), we average over the real photon polarization, which give an overall $ \vec{\Delta}_{\perp}^2/2$ factor.
The integration over $\varphi$ being straightforward, we are left with\footnote{A
factor $1/4$ is missing in Eq.~(23) of Ref.~\cite{Lansberg:2006fv}. This
typo does not affect to the numerical results reported in that reference \cite{jp}.} 
\beq
\frac{d\sigma^{e\gamma\to e\rho_L^+\pi^-}}{dQ^2 dt d\xi }&=& 2\pi\,\frac{32\pi}{9}\,\frac{\alpha_{elm}^3\alpha_s^2}{s_{e\gamma}\,Q^8}\,\left (\frac{f_{\rho}}{\sqrt{2}\,f_{\pi}}\right)^2\,(-t)\,\frac{1-\xi}{(1+\xi)^4}\nonumber\\
&&
\,\left(2\xi\,s_{e\gamma}\,-\,(1+\xi)Q^2 \right)\,\left \{\left(  \Re \,I_x^V\right)^2+ \left( \Im\, I_x^V\right)^2\right\}\quad,
\label{rho}\\
\nonumber\\
\frac{d\sigma^{e\gamma\to e\pi^+\pi^-}}{dQ^2 dt d\xi }&=& 2\pi\,\frac{32\pi}{9}\,\frac{\alpha_{elm}^3\alpha_s^2}{s_{e\gamma}\,Q^8}\,(-t)\,\frac{1-\xi}{(1+\xi)^4}
\,\nonumber\\
&&\left(2\xi\,s_{e\gamma}\,-\,(1+\xi)Q^2 \right)\,\left \{\left(\Re\, I_x^A-\,72\, \frac{f_{\pi}^2\,}{t-m_{\pi}^2}\right)^2+ \left( \Im\,  I_x^A\right)^2\right\}\quad.
\label{fincross}
\label{CroosSecc}
\eeq
 From Eq.~(\ref{CroosSecc}) it can be
observed  that 
\beq
\xi\geq Q^{2}/\left( 2\, s_{e\gamma}-Q^{2}\right)\quad.
\label{xi-lim-kin}
\eeq
 In other words, there is a (positive) lower limit on the value of $\xi$.
Let us now proceed  to evaluate these integrals.

\section{Results from the Phenomenological Parameterization}

The first estimates for the cross sections of meson production  have been given for the phenomenological parameterization of the Double Distribution~(\ref{t-ind-dd}), what  is referred to as Model 1. The $t$-dependent Double Distribution~(\ref{dd-3prop-ab}) are considered for the vector TDA and is called Model 2.  In none of these approaches the  pion pole contribution to the cross section for the pion  production is taken into account. It means that the term proportional to the pion decay constant is not present in the expression for the cross section~(\ref{CroosSecc}).

Also, no QCD evolution is taken into account, the effects of which are supposedly less important than the uncertainty of the modeling of the TDAs. 

For the vector TDA both the real and imaginary part of the amplitude contribute significantly to the cross section.
Since we are dealing with $t$-independent DDs, the $t$-dependence comes solely  from the overall $-t$ factor in the trace.
\\

\begin{figure}
[ptb]
\begin{center}
\includegraphics[height=6cm, angle=-90]{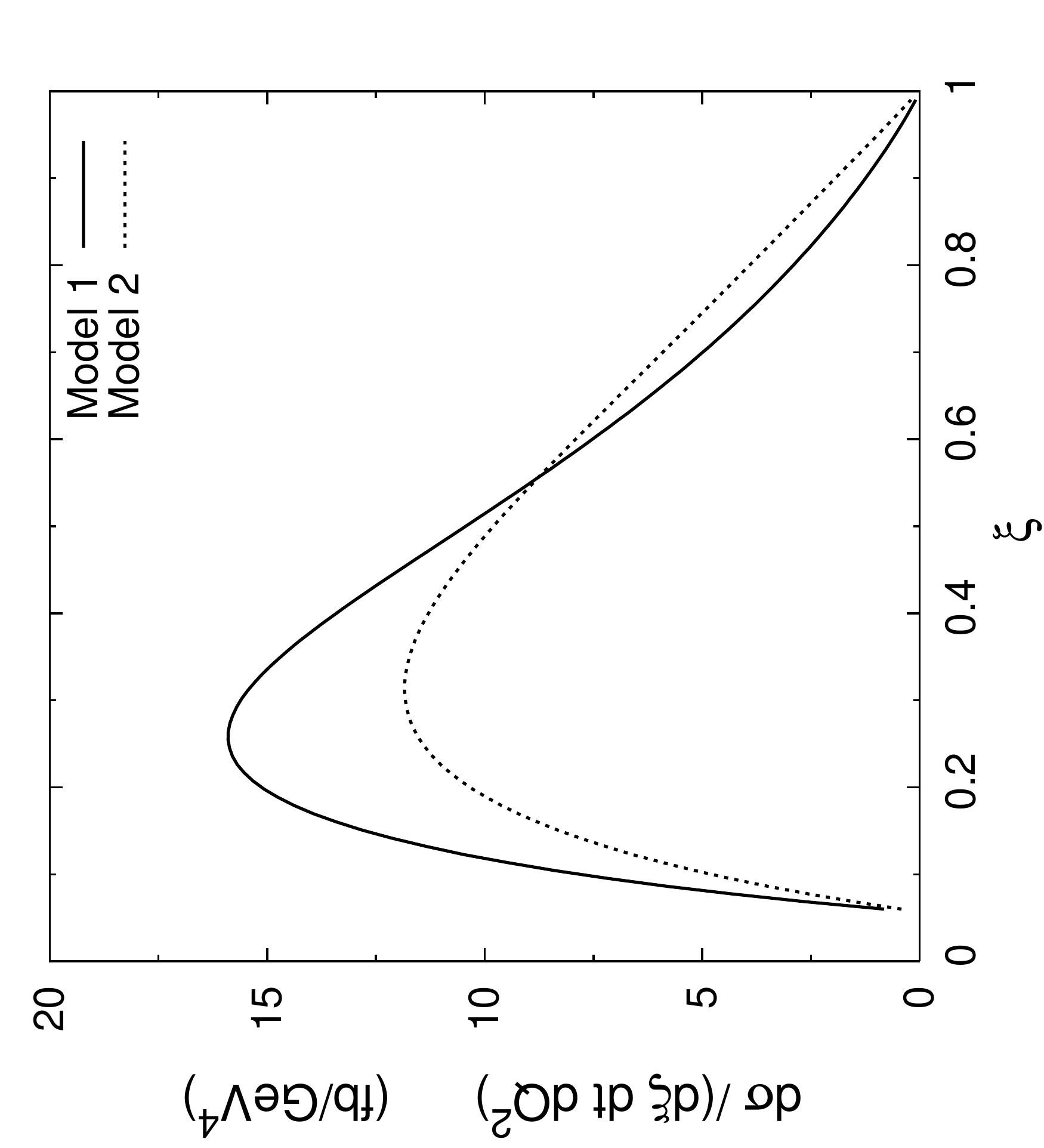}
\includegraphics[height=6cm, angle=-90]{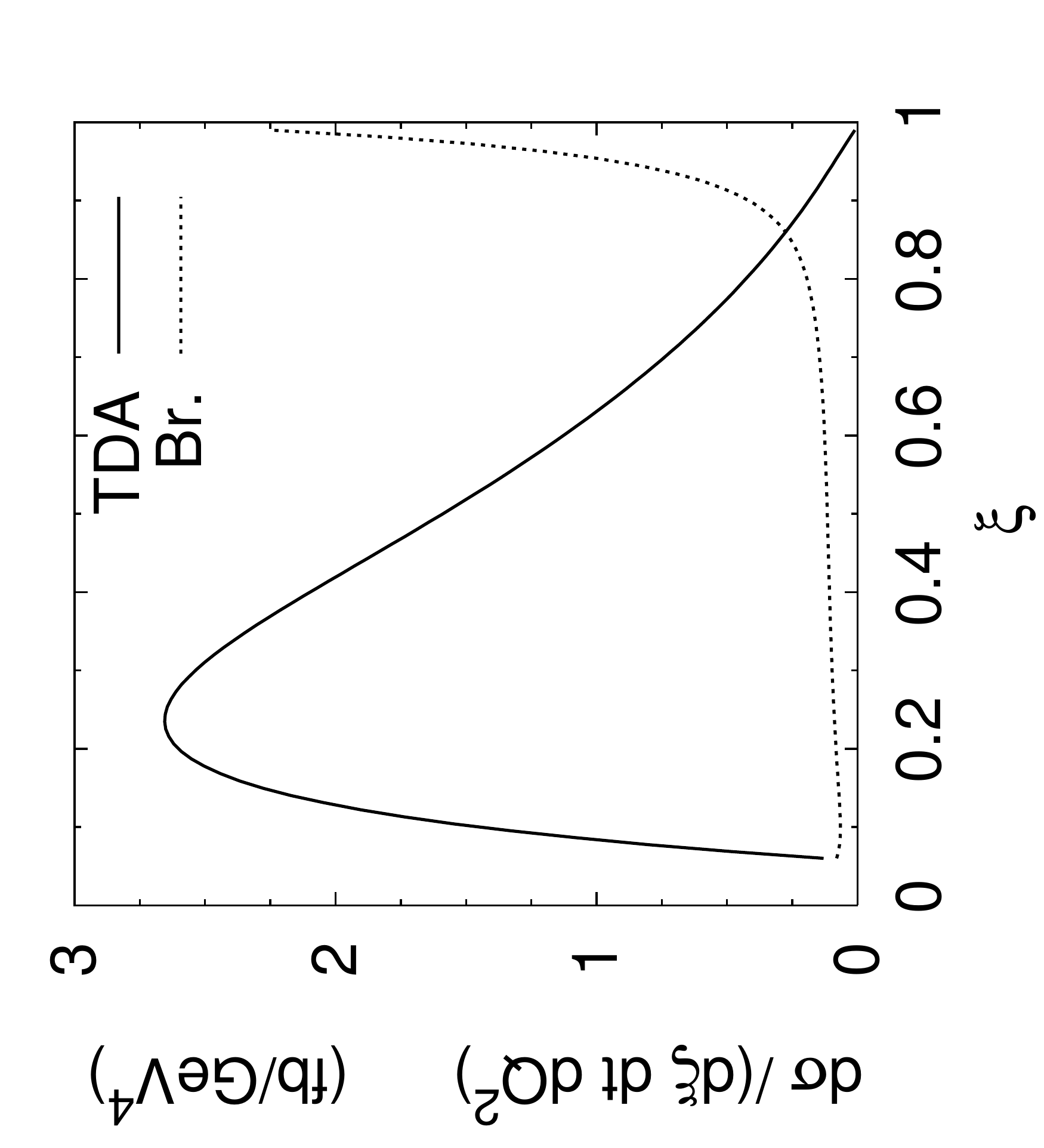}
\caption{The cross section for $\pi\rho$ production (on the left) in a $t$-independent parameterization (Model 1) and a $t$-dependent one (Model 2) ; and for $\pi\pi$ on the right~\cite{Lansberg:2006fv}.}
\end{center}
\label{cross-jp}
\end{figure}

 In order to  numerically estimate the cross sections,  the strong coupling constant is fixed to a 
 value of $\alpha_{s}\simeq1$.  This large value of $\alpha_{s}$
 is indicated from the analyses of the electromagnetic Form Factor~\cite{Braun:2005be}.
On the left panel of Fig.~\ref{cross-jp}, the cross section for the $\pi\rho$ production is plotted for $Q^2=4$ GeV$^2$ and $s_{e\gamma}=40$  GeV$^2$. 
 The high-$\xi$ behavior is similar for both model while the intermediate-$\xi$ region is sensitive to specific models for the TDAs. On the other hand, as it is shown in Ref.~\cite{Lansberg:2006fv}, the $Q^2$ behavior at fixed $\xi$ is model independent, what constitutes a crucial test for the validity of the approach.
\\

Turning our attention to the $\pi\pi$ production, we should compare the estimates with the cross section for the Bremsstrahlung process in the same kinematical regime. It is shown in Ref.~\cite{Lansberg:2006fv} that the process where the $\pi^+\pi^-$  is produced by a photon radiated from the leptonic line does not interfere much with the $\pi\pi$ production from $\gamma^{\ast}\gamma$ for reasonable values of $\xi$ and with the exception of small values of $s_{e\gamma}$. In other words, the hard hadronic process dominates the Bremsstrahlung contribution in the kinematical region under scrutiny.
This statement is illustrated on the right panel of Fig.~\ref{cross-jp}. The axial TDA is given for the Model 1 and the dotted line represents the contribution of the Bremsstrahlung.
\\

A general observation is that both the cross sections for $\rho_L^+\pi^-$ and $\pi^+\pi^-$ are rather small.
In the next Section, we will see how these results are enhanced by the pion pole amplitude  for the pion  production as well as by QCD evolution for the rho production.

\section{Results from the TDAs in the NJL Model}
\label{sec:cross-njl}

For the nonperturbative part of the process we use the TDAs evaluated in the
NJL model. \\

The kinematics of the process restrict the domain of relevance  of the variables $(\xi, t)$, which, in turn, define the shape and the magnitude of the distributions.
The restriction~(\ref{xi-lim-kin}) on the value of the skewness variable is indeed  particularly interesting  because  the value of $\xi$ defines  the shape of the TDAs. 
In particular, the shape of the axial TDA calculated in models radically changes according to the sign of the skewness variable; $A(x,\xi,t)$ has, at $x=\pm\xi$,
 its maximum values for $\xi>0$ (see right panel of Fig.~\ref{tdatot}) while it has its minimum values for $\xi<0$.
 However the vector TDA has its maximum and minimum values at $x=\pm\xi$ (see left panel of  Fig.~\ref{tdatot}) 
independently of the sign of the skewness variable. 
\\
 On the other hand, the magnitude of the distributions is controlled by the $t$-dependence. This can be
easily understood because the TDAs, that  must satisfy the sum rules~(\ref{2.06}, \ref{2.07}),  are 
expected to decrease at least as $t^{-1}.$ The $t$-behavior of the cross sections will therefore be dictated by both the overall $(-t)$ factor coming from the trace and the TDAs.

\begin{figure}
[ptb]
\begin{center}
\includegraphics[
trim=0.697746cm 0.698118cm 0.699478cm 0.699060cm,
height=8.1385cm,
width=16.0705cm
]%
{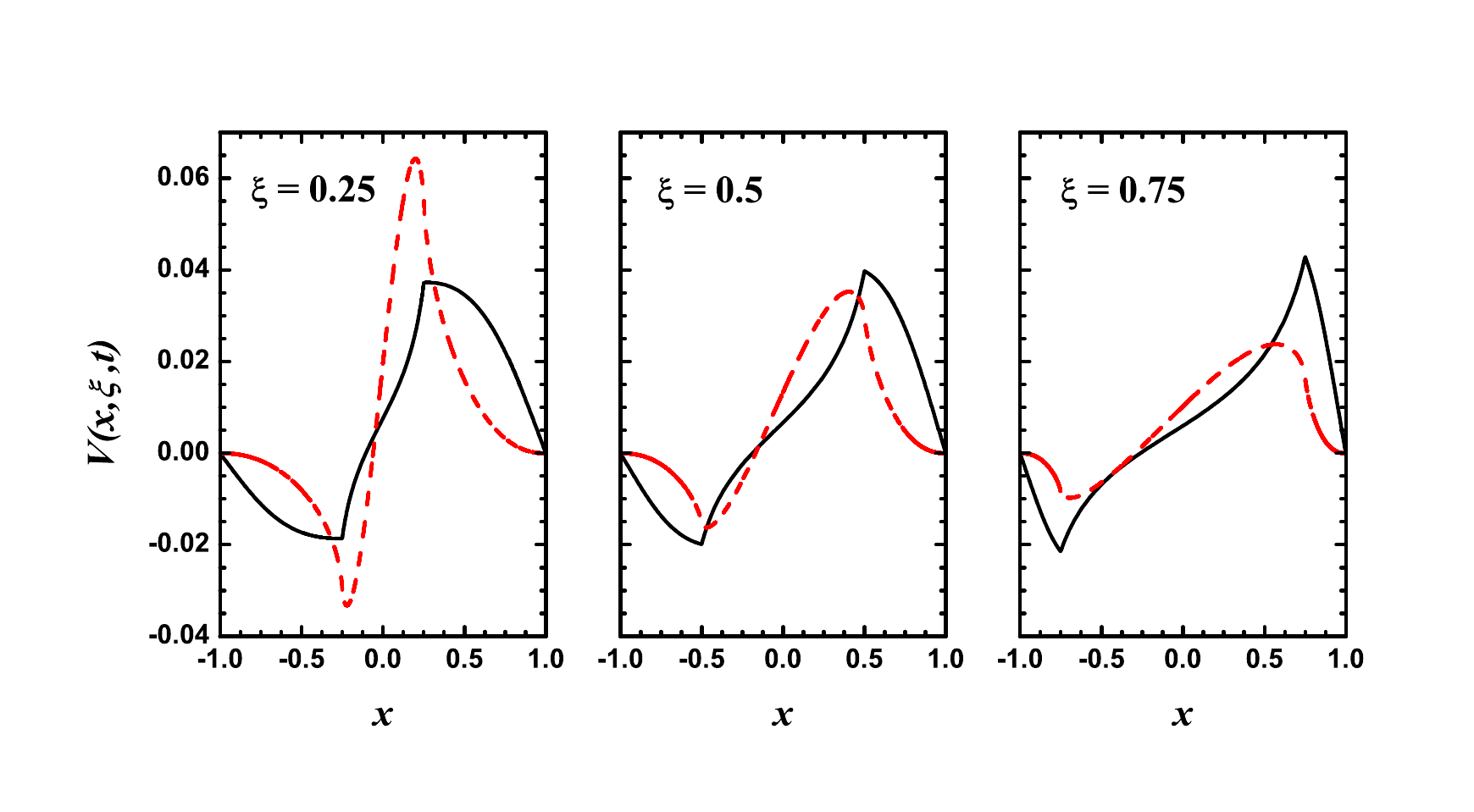}%
\caption{The functions $V^{\gamma\rightarrow\pi^{-}}(x,\xi,t)$ and for
$\xi=0.25,$ $0.5,$ $0.75$ and for $t=-0.5$ GeV$^{2}$. \ In each figure, the
solid line corresponds to the NJL model prediction and the dashed line to its
LO evolution.}%
\label{vTDA}%
\end{center}
\end{figure}

The value $\alpha_{s}\simeq1$ chosen in Ref.~\cite{Lansberg:2006fv} is also adopted here.
 Using
this value for $\alpha_{s}$\ we have evaluated the cross section for $\rho$
production. In Fig.~\ref{pi-rho-4} we plot this cross section as a function of
$\xi$. As we observe, the cross section is largely dominated by the imaginary
part of the integral~(\ref{rho}). 
Comparing
with the previous results shown on Fig.~\ref{cross-jp}, we
observe that our predictions are higher by a factor 2 or 3.%
\\
\begin{wrapfigure}[25]{l}{10cm}
\centering
\includegraphics[height=7cm]{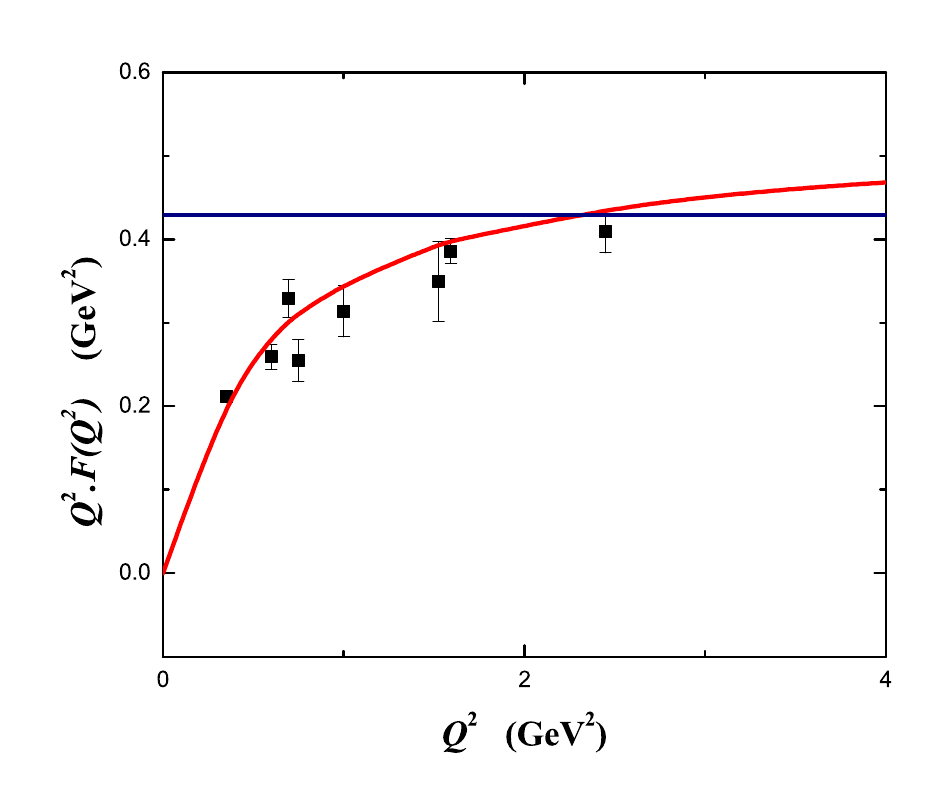}
\caption{Electromagnetic Pion FF. The Brodsky-Lepage asymptotic FF for $\alpha_s=1$ and $Q^2=2.45$ GeV$^2$ (blue line) and the experimental data given in Table~\ref{data-elmff}. The red curve represents a fit in the monopole form $Q^2/(1+0.44 \,Q^2/0.2336) $.}%
\label{epiff}%
\end{wrapfigure}
\begin{figure}
[ptb]
\begin{center}
\includegraphics[
trim=0.553911cm 0.555731cm 0.557113cm 0.556600cm,
height=7.5cm
]%
{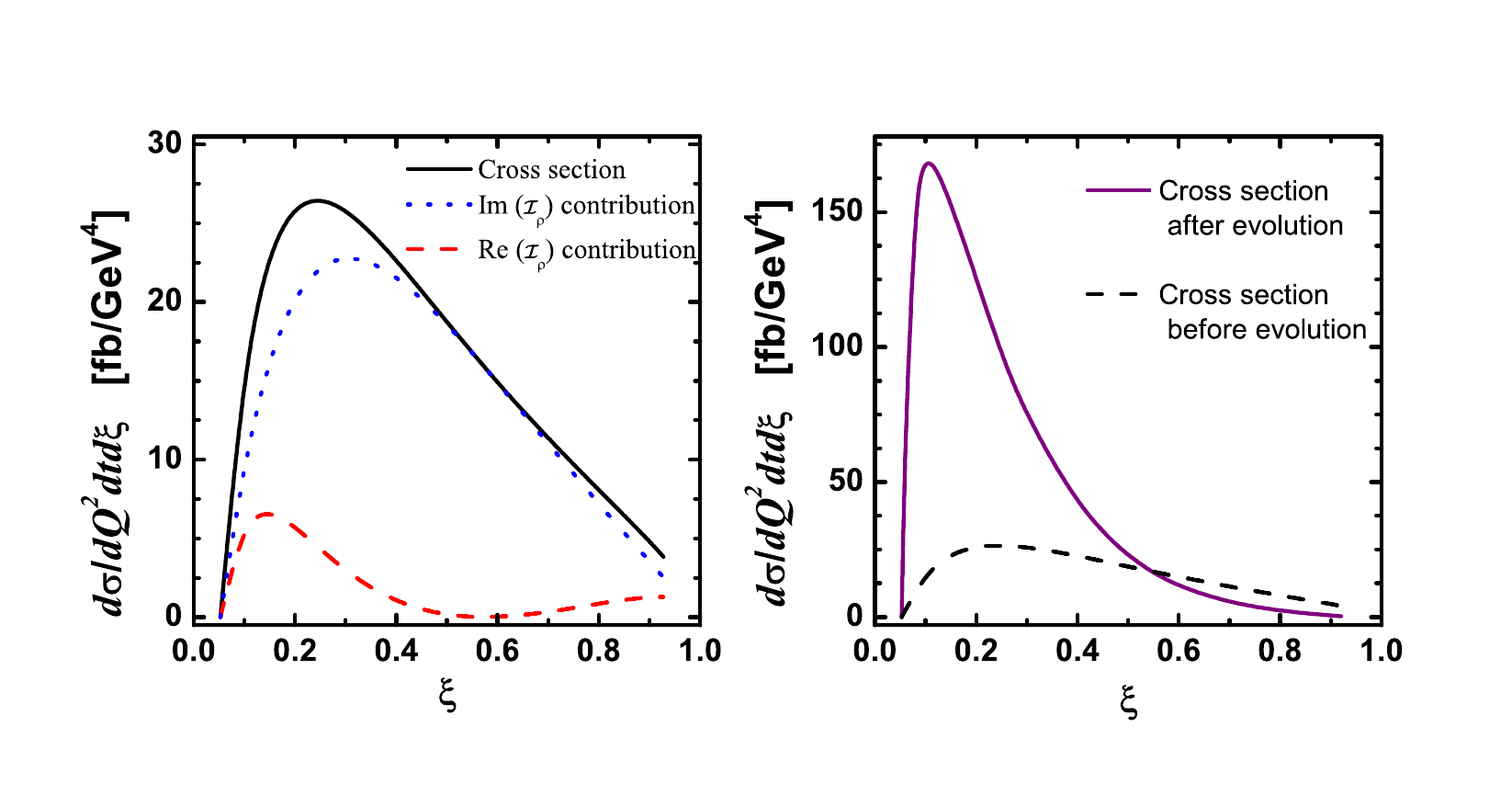}%
\caption{$e\gamma\rightarrow e^{\prime}\rho_{L}^{+}\pi^{-}$ differential cross
section plotted as a function of $\xi$ for $Q^{2}=4\operatorname{GeV}^{2},$
$s_{e\gamma}=40\operatorname{GeV}^{2},$ $t=-0.5\operatorname{GeV}^{2}.$ In the
first layer, the dotted\ (dashed) line is the contribution to the cross
section coming from the imaginary (real) part of the integral given in Eq.
(\ref{rho}). In the second layer\ we give the cross sections before and after
evolution.}%
\label{pi-rho-4}%
\end{center}
\end{figure}
The cross section for pion production at $Q^{2}=4$ GeV$^2$ as a function of $\xi$ is given in
Fig.~\ref{pi-pi-4} (left). This cross section is dominated by the pion pole
contribution which is determined by the Brodsky-Lepage pion Form Factor. Alternatively,
if the pion Form Factor is experimentally known, we can infer phenomenologically this
contribution. And hence the axial TDA could be extracted from the interference term. The latter shows already an impressive enhancement with respect to the cross section shown in the right panel of Fig.~\ref{cross-jp}.

From Ref.~\cite{Horn:2006tm} we know that the pion Form Factor at
$Q^{2}=2.45$ GeV$^{2}$ is $0.167\pm0.010$. The pion Form Factor  is shown on Fig.~\ref{epiff} for the different experimental results, see Table~\ref{data-elmff}. In Fig.~\ref{pi-pi-4} (right) we depict, for each contribution, the prediction
 using the interval defined by the experimental value of the Form Factor  (filled areas), including also the  theoretical
prediction using the Brodsky-Lepage pion Form Factor (lines).
We observe an unexpectedly huge enhancement of the total cross section of a factor of about 60.
\\

\begin{figure}
[ptb]
\begin{center}
\includegraphics[
trim=0.624743cm 0.625582cm 0.626528cm 0.626517cm,
height=7.5cm
]%
{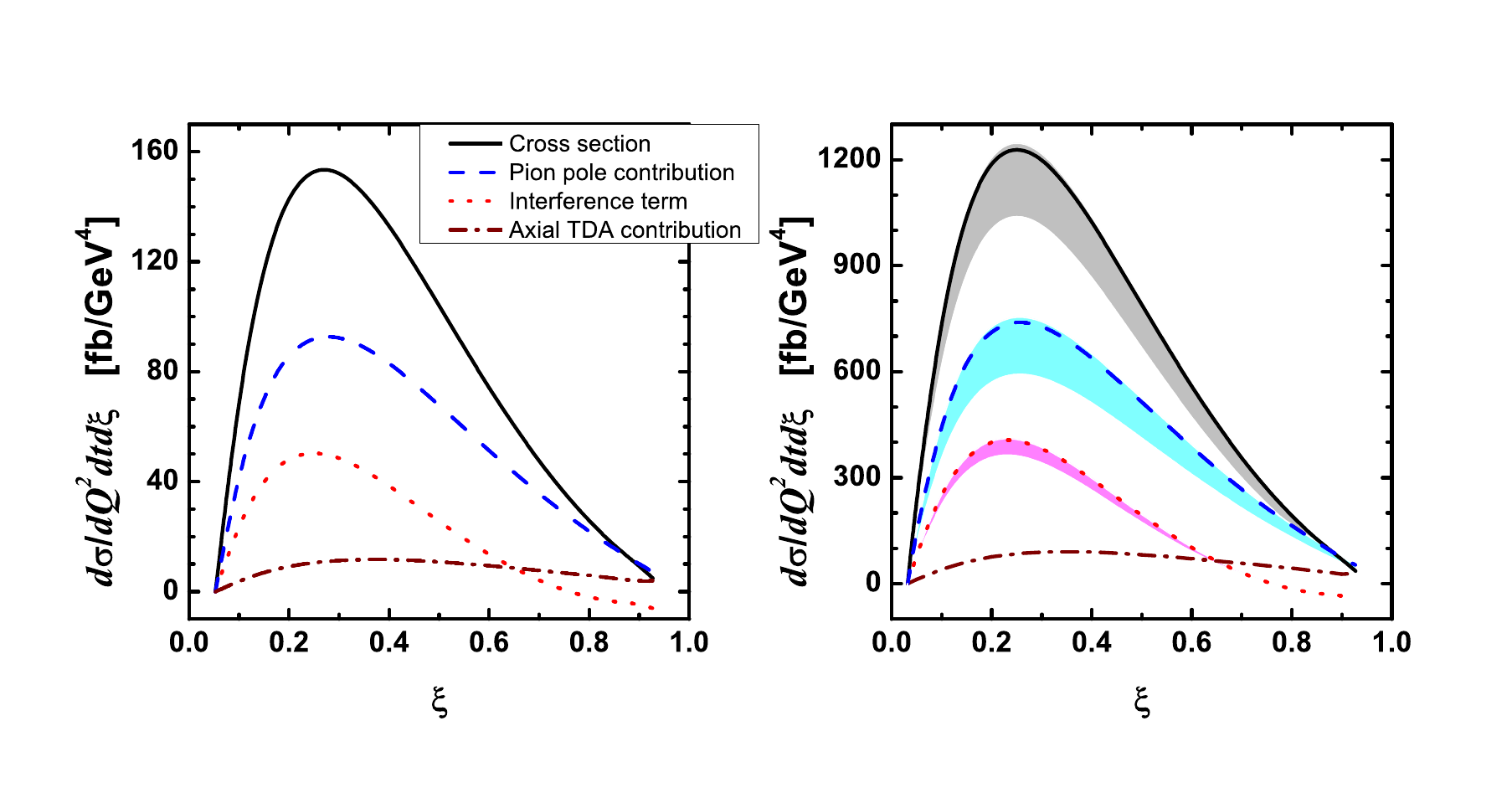}%
\caption{$e\gamma\rightarrow e^{\prime}\pi^{+}\pi^{-}$ differential cross
section plotted as a function of $\xi$ for $s_{e\gamma}=40\operatorname{GeV}%
^{2},$ $t=-0.5\operatorname{GeV}^{2},$ $Q^{2}=4\operatorname{GeV}^{2}$ (left)
and $Q^{2}=2.45\operatorname{GeV}^{2}$ (right)$.$ The
dashed\ (dashed-dotted)[dotted] line is the contribution to the cross section
coming from the pion fom factor (axial TDA) [interference term between the
pion Form Factor and A]. The pion pole contribution is calculated using the
Brodsky-Lepage pion Form Factor. The filled areas in the right layer correspond to the
same contributions but with the experimental value for the pion Form Factor $F_{\pi
}=0.167\pm0.010$ \cite{Horn:2006tm}.}
\label{pi-pi-4}
\end{center}
\end{figure}

The $t$-dependence of the cross section for pion production includes a strong
dependence on $t$ coming from the pion pole. Neglecting the pion mass in
(\ref{fincross}), we observe that the pion pole contribution to $I_{\pi}$
is proportional to $t^{-1}.$ Therefore, the cross section grows as $t^{-1}$
for small $t$ values. For large $t$ values we expect, as in the $\rho$ production
case, a decreasing as $t^{-1}.$%

\begin{table}[ptb]
\centering
\begin{tabular}
[c]{|c|c|c|c|c|}\hline
&$Q^2$ [GeV$^2$] & $F^{\pi}(Q^2)$ & $F^{\pi}(Q^2)$+ error & $F^{\pi}(Q^2)$- error\\\hline\hline
DESY~\cite{Ackermann:1977rp} & 0.351405 & 0.211314 &&\\\hline
DESY~\cite{Brauel:1977ra}& 0.695093 & 0.329482 &0.353265 &0.306529 \\\hline
JLab~\cite{Tadevosyan:2007yd}&
0.599788   &  0.259397   &   0.274914   &      0.246048 \\
&0.748309  &   0.254932  &    0.280412    &     0.230928 \\
& 1.00094  &    0.313972  &   0.345017      &   0.284536\\
&1.52456    & 0.34985    &   0.397251       &  0.309278 \\\hline
JLab~\cite{Horn:2006tm}&
1.59404   &   0.385976     & 0.408247 &        0.368385\\
&2.44972   &   0.405233      &0.434364      &   0.38488\\\hline
\end{tabular}
\caption{The experimental results for the pion Form Factor. The last two columns represent the upper and the lower limit of the error bars.}%
\label{data-elmff}%
\end{table}

\subsection{QCD Evolution}

We have also studied the effect of the QCD evolution on our estimates, using for this purpose the code of Freund
 and McDermott \cite{FreundMcDermott}. The same procedure as for the Distribution Amplitude and the GPDs is followed.
 The value of $Q_0$ for the LO evolution in the NJL model is given by Eq.~(\ref{modelscale}).
 \\

Turning our attention to the vector TDA evolved at LO  to a scale of $4$ GeV$^2$ (Fig. \ref{vTDA}), we observe that
the value of $V\left(  x,\xi,t\right)  $ at $x=\pm\xi$ grows for small values
of $\xi$ and decreases for large $\xi$ values in comparison with the TDA at the scale of the model.
This implies that the cross
section for $\rho$ production, which is largely dominated by the imaginary part,
will grow appreciably in the small $\xi$ region. In Fig.~\ref{pi-rho-4} we
compare the cross section after evolution, calculated only through
contribution of the imaginary part of $I^V$ with the one
obtained before evolution. We observe that this cross section is multiplied by
a factor about 5 in the $\xi\sim0.2$ region. In the case of the axial TDA, the
cross section is dominated by the pion Form Factor contribution, therefore the effect
of the evolution is expected to be small.
\\

\section{The Experimental Situation}

In the previous Section we have looked at the expected cross section for $\pi$-$\pi$
and $\pi$-$\rho$ production in exclusive $\gamma^{\ast}\gamma$ scattering in
the forward kinematical region using realistic models for the description of
the pion. First
we confirm the
previous estimates for $\rho$ production,
even if our results for the cross section are a factor 2 larger than the one
obtained in Ref.~\cite{Lansberg:2006fv}.
Second, in comparison with this previous evaluation of the cross
sections, we have improved in considering the effect of evolution on the vector current.
 In doing so,
 an additional factor 5 appears  in the small $\xi$ region,
leading to a cross section for $\rho$ production of one order of magnitude larger than the previous
calculation. We have also improved in including the pion pole term in the tensor decomposition of  the axial current.
 Then an even larger
enhancement factor, of about 60 in this case, is found in the cross section for pion production.
 The
interference term becomes a factor 15 larger than the pure TDA contribution,
making the axial TDA more accessible experimentally. 
\\

Data have been collected at LEP and CLEO on these reactions with some phenomenological success, even if mainly on  $\rho\rho$ final states in a different factorization regime~\cite{Anikin:2003fr, Anikin:2005ur}. 
More data are obviously needed and are eagerly waited for in the TDA region, and much hope comes from the high luminosity electron colliders. 

\section{Backward Pion Electro-production}

Recently Pire and collaborators have shown in a series of articles~\cite{Pire:2004ie,Lansberg:2007bu, Lansberg:2007se, Lansberg:2007ec} that the TDAs represent the ideal framework 
through which it is possible to describe the backward pion electro-production, e.g., $ep\to  e' p' \pi^0$ , as well as the meson production via a nucleon-antinucleon Drell-Yan process, e.g., $\bar N N\to \gamma^{\ast}\pi$, in the forward kinematics. 

These processes are experimentally more accessible then the ones previously analyzed. 
For instance, the $\bar p p$ annihilation, e.g. $\bar p p\to \gamma\gamma$, 
the meson production channel $\bar p p\to \gamma\pi^0$,
that will be intensively studied at  GSI-HESR antiproton program by the planned FAIR project,~\footnote{A great wealth of up-to-date informations about the experiments that will be performed at GSI-HESR can be found at this web-page: 
http://www.gsi.de/zukunftspro jekt/indexe.html.} and by 
the meson photo-production process, $p\gamma\to \pi^0 p$, with forthcoming data from 
JLab.

\begin{figure}
[ptb]
\begin{center}
\includegraphics[height=5cm]{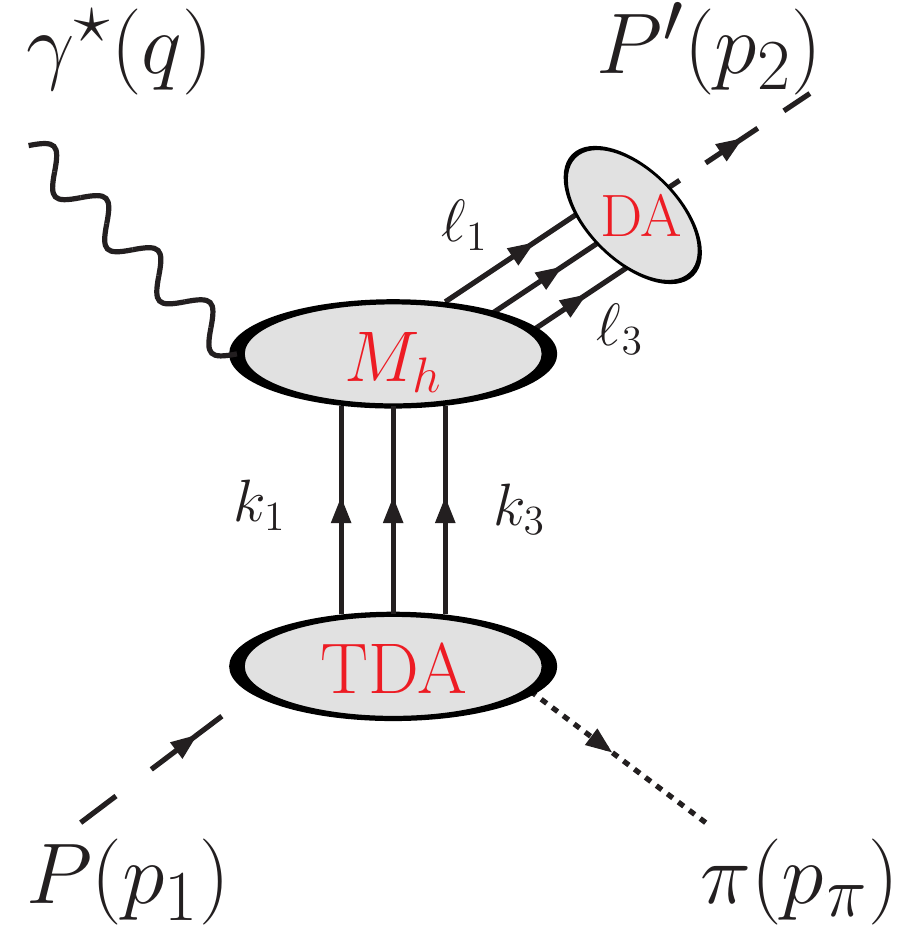}
\end{center}
\caption{The factorization of the process $\gamma^{\ast} P\to  P' \pi^0$ into proton DA, hard subprocess amplitude $M_h$ and proton-pion TDA. Figure from~\cite{Lansberg:2007ec}.}
\label{p-pi0-diag}
\end{figure}

In Ref.~\cite{Lansberg:2007ec}, a soft-pion limit for the baryonic TDA is used to give an estimate of the differential cross section
for $  \gamma^\ast P \to P'\pi^0 $.
However, the  lack of knowledge of the TDAs unfortunately prevents  from giving an accurate estimate of the latter quantity.

Data should be available, at higher energies, 
as for instance at HERMES and with CEBAF at 
12 GeV ~\cite{JLAB12}. The  calculation of the cross section~\cite{Lansberg:2007ec}, in an admittedly quite narrow 
range of the parameters, can thus serve as a  reasonable input to the feasibility  study of  
backward pion electroproduction at CEBAF at 12 GeV, in the hope to reach the scaling regime, 
in which we are interested.

\chapter{Peeping through Spin Physics: the Sivers  Function}
\label{sec:sivers}
\pagestyle{fancy}

\begin{wrapfigure}[16]{l}{85mm}
\centering
\includegraphics [height=5cm]{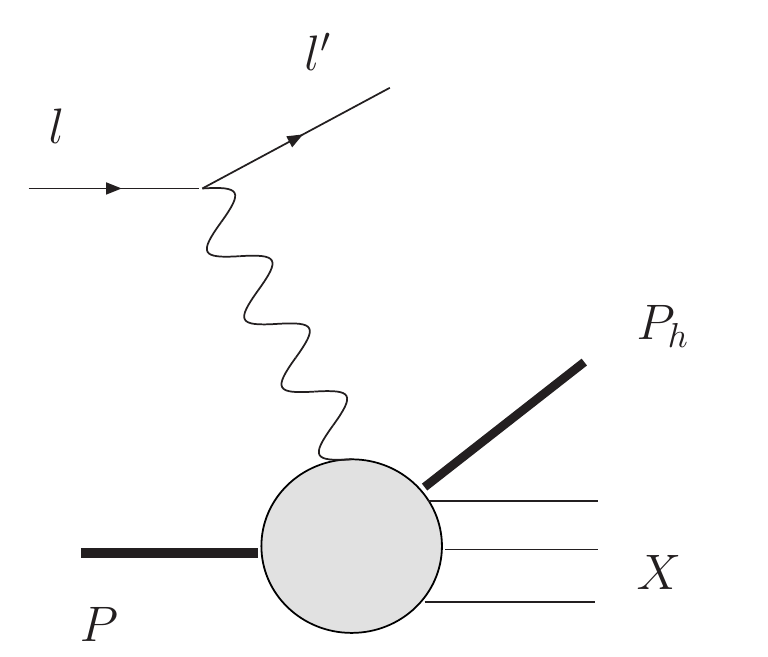}
\caption{Semi-Inclusive Deeply Inelastic Scattering. }
\label{sidis}%
\end{wrapfigure}
\begin{wrapfigure}[13]{r}{85mm}
\centering
\includegraphics [height=3cm]{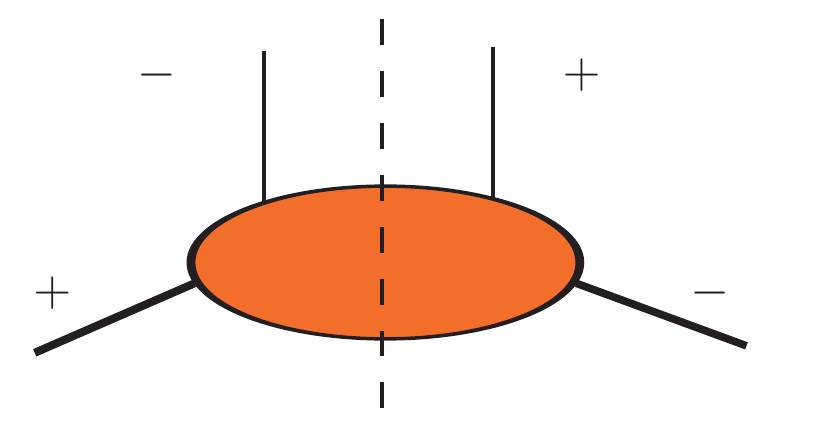}
\caption{Representation of a chirally odd distribution. The corresponding handbag diagram is forbidden by conservation of chirallity. }
\label{heli-flip}%
\end{wrapfigure}

The partonic structure of transversely polarized hadrons is  one of their less known features. A review~\cite{Barone:2001sp} on the  transverse polarization of quarks  in hadrons gathered all the knowledge which is, nevertheless, appreciably growing on both the theoretical and experimental sides.
Experiments for the determination of the transversity properties are progressing very fast
and the relevant experimental effort  has motivated a strong theoretical
activity.
\\
When extending our analyses to {\it unintegrated} parton distributions~(\ref{decom-bm}), we have to deal with the intrinsic transverse motion of quarks. This supplementary degree of freedom implies the existence of new parton distribution functions, i.e. ~(\ref{decom-bm}). Besides their dependence on $k_T$, the additional parton distributions can either be chiral-even or odd. In the former case, the distributions can be probed in fully inclusive DIS. 
The conservation of chirality  does however not allow the identification, through the optical theorem,  of chiral-odd distributions, for helicity must be flipped twice as shown on Fig.~\ref{heli-flip}. The {\it Transverse Momentum Dependent} Parton Distributions (TMD)  studied in the context of transversely polarized nucleons do flip helicity  and cannot be accessed through  DIS.
\\

Semi-inclusive deep inelastic scattering (SIDIS), i.e. the process depicted on Fig.~\ref{sidis}
\beq
e({\it l})\, N({\cal P}) \to e'({\it l}')\, h({\cal P}_h) \, X ( {\cal P}_x)\quad,
\eeq
\begin{wrapfigure}[17]{l}{100mm}
\centering
\includegraphics [height=6cm]{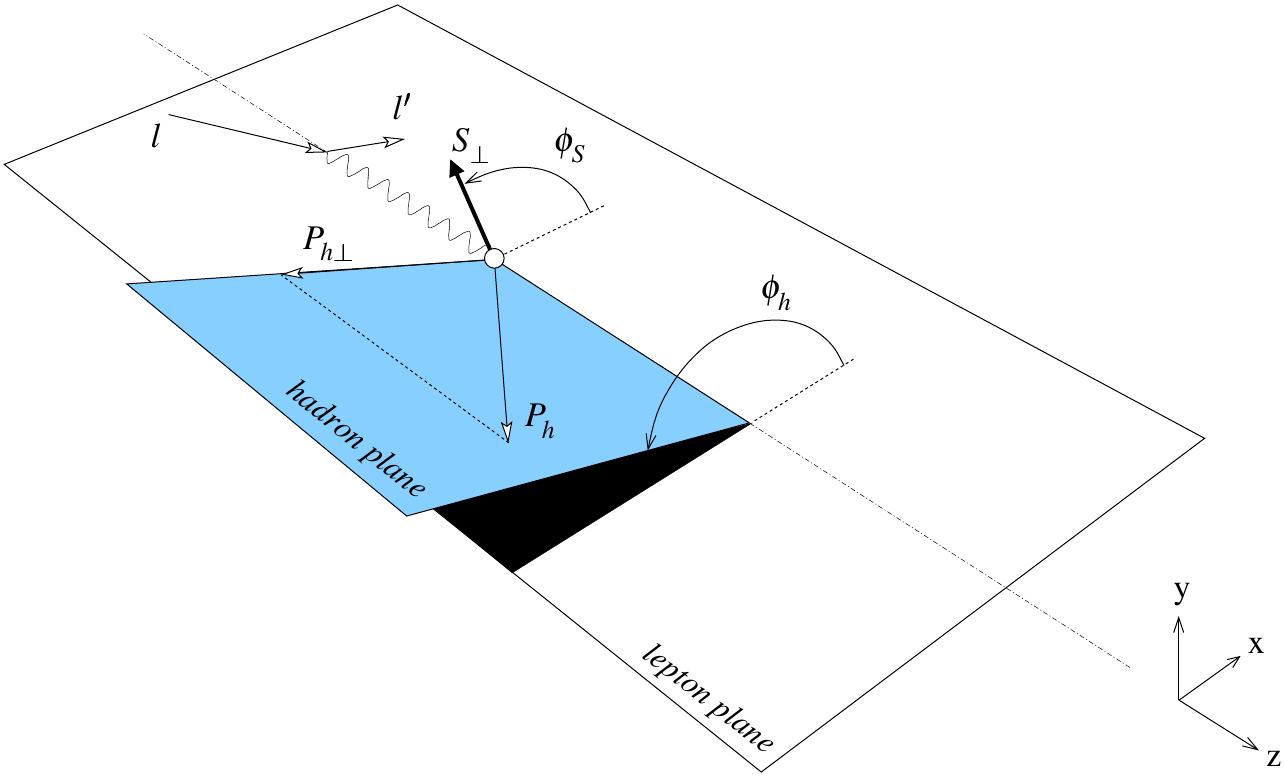}
\caption{SIDIS kinematics in  the $\gamma^{\ast} p$ frame as defined in the Trento conventions~\cite{trento}. }
\label{anglestrento}%
\end{wrapfigure}
with the detection in the final state of a produced hadron $h$ in coincidence with the scattered electron $e'$, is one of the proposed processes to access the parton distributions of transversely polarized hadrons.~\footnote{In this case, we can define two different {\it transverse directions}. In general, a quantity which is transverse
in a frame where $P$ and $h$ have no transverse components (thus ${\vec P_T}={\vec h_{T}}=0$), 
is indicated with a subscript $T$,  so that $T$ means transverse
with respect to ${\vec h}$, while those with a subscript $\perp$  are defined in a frame where $q$ and $P$ have no transverse components
(standard DIS frame), so that $\perp$ means transverse with respect to
$\vec q$ (thus ${\vec q_\perp} = {\vec P_\perp=0}$)~\cite{mu-ta}. There is no such distinction in DIS, so that $T=\perp$ and they can be used equivalently.}
For several years it has been known that  SIDIS off a transversely polarized target shows azimuthal asymmetries,
the so called {\it single spin asymmetries} (SSAs) \cite{Collins,sivers, sivers2}.%
\\
As a matter of fact, it is predicted that the number of produced hadrons in a given direction or in the opposite one in the hadronic plane, with respect to the reaction plane,  depends on the orientation of the transverse spin  of a polarized target with respect to the direction of the unpolarized beam,~\footnote{ Here $U$ means Unpolarized beam, $T$ means Transversely polarized target and it is assumed that the produced hadron is spinless, or that its polarization is not detected.}
\beq
A_{UT}(\phi_h,\phi_S)& \equiv&  
{ d \sigma (\phi_h,\phi_S) - d \sigma (\phi_h,\phi_S + \pi)
\over
d \sigma (\phi_h,\phi_S) + d \sigma (\phi_h,\phi_S + \pi)}
\neq 0\quad.
\label{main}
\eeq
\\
 It can be shown that the SSA in SIDIS off transverse polarized targets is essentially due to two different physical mechanisms, whose contributions can be technically distinguished~\cite{Bacchetta:2006tn, Boer:1997nt, ko-mu, mu-ta}.
\\

One of the two different mechanism has been proposed by Collins \cite{Collins} and is therefore referred to after his name. It is due to parton final state interactions in the production of a spinless hadron  by a transversely polarized quark, and will not be discussed here.

The other is the Sivers mechanism, producing a term in the SSA which is given by the product of the unpolarized fragmentation function with the Sivers Parton Distribution~\cite{sivers, sivers2},  describing the  modulation of the number density 
of unpolarized quarks in a transversely polarized target due to the correlation between the transverse spin of the target and the intrinsic transverse parton momentum.

The Sivers function, $ f_{1T}^{\perp q} (x, {k_T} )$,  is  the quantity of interest of this Chapter. It is formally defined through the $\Phi^{[\gamma^+]}$ of the decomposition~(\ref{decom-bm}). Namely,~\cite{trento,goeke},
\begin{eqnarray}
\Phi^{[\gamma^+] \,q}(x,\vec k_T,S) & = & f_1^q(x, k_T) - {\epsilon_T^{ij}k_{Ti}S_{Tj} \over M} f_{1T}^{\perp q} (x, k_T)
\nonumber\\
& = & {1 \over 2} \int { d \xi^- d^2 \vec \xi_T\over (2 \pi )^3} e^{-i ( x \xi^- P^+  - \vec{\xi}_T \cdot {\vec k_T} )}
\langle P,S | \hat O_q | P, S \rangle \quad,
\label{def0}
\end{eqnarray}
with the operator defined by~\cite{jiyu,bjy}
\begin{eqnarray}
\hat O_q = 
\bar \psi_q(0,\xi^-,\vec{\xi}_T)
{\cal L}^\dagger_{\vec{\xi}_T}(\infty,\xi^-)
\gamma^+ 
{\cal L}_0(\infty,0)
\psi_q(0,0,0)~,
\label{o_siv}
\end{eqnarray}
where $\psi_q(\xi)$ is the quark field and  
the gauge link has been defined as follows,
\begin{eqnarray}
{\cal L}_{\vec{\xi}_T}(\infty,\xi^-) =
P \exp \left (-ig \int_{\xi^-}^\infty A^+(\eta^-,\vec{\xi}_T) d \eta^- 
\right )\quad.
\label{link}
\end{eqnarray}

The number density of unpolarized quarks in a transversely polarized target is effectively modulated by the quantity
\beq
&&\Phi^{[\gamma^+] \,q}(x,\vec k_T,S_j=\uparrow)- \Phi^{[\gamma^+] \,q}(x,\vec k_T,S_j=\downarrow)\nonumber\\
&&=-2\,\epsilon^{ij}\,\frac{ k_i}{M}\,\, f_{1T}^{\perp q} (x, k_T)\quad,
\label{iden-siv}
\eeq
with  $i,j$ being either $x,y$  and $S=\uparrow /\downarrow$ standing for $S$ in a transverse direction, in opposition with the helicity $S=\pm$.
\\
\begin{wrapfigure}[15]{r}{80mm}
\centering
\includegraphics [height=4.7cm]{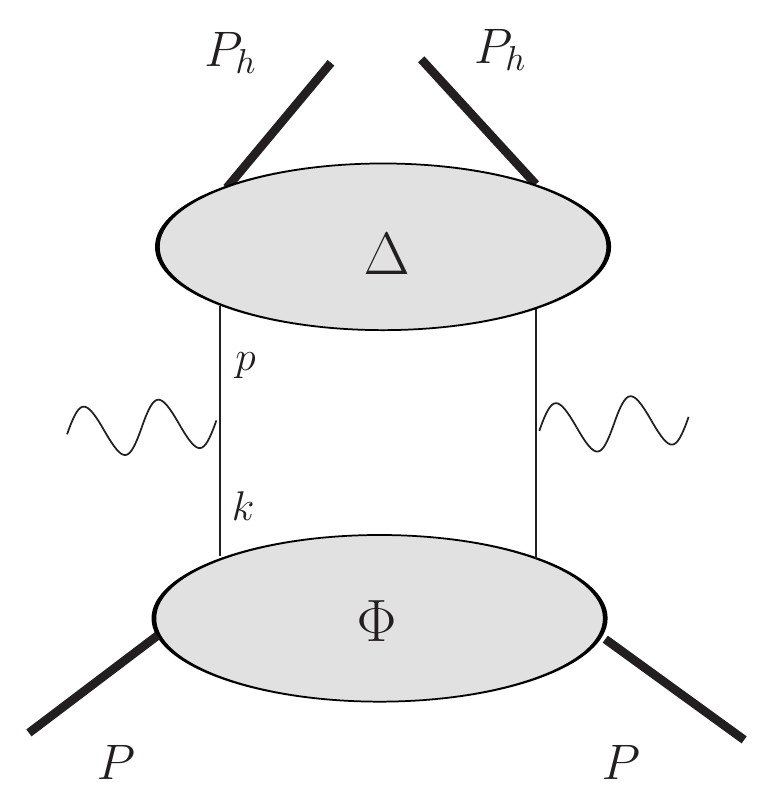}
\caption{Factorization in SIDIS. }
\label{facto-sidis}%
\end{wrapfigure}
The Sivers function contributes with a $\sin(\phi_h-\phi_S)$ weighting function to the  cross section of SIDIS with unpolarized beam off  a transversely polarized proton~(\ref{main}), with the angles $\phi_h, \phi_S$ defined in Fig.~\ref{anglestrento}. 
The Sivers asymmetry, according to the ``Trento convention'' \cite{trento}, is defined in terms
of the experimental cross sections, 
\beq
A_{UT}^{Sivers} = 
{\int d \phi_S d \phi_h \sin ( \phi_h - \phi_S )d^6 \sigma_{UT}
\over
\int d \phi_S d \phi_h  
d^6 \sigma_{UU}}\quad.
\label{siv_ex}
\eeq
The cross sections themselves can be
written, through the factorization theorems, as a convolution  of $ k_T$-dependent
distribution and fragmentation functions \cite{mu-ta}  as shown on Fig.~\ref{facto-sidis}; the former being here  the Sivers function.
\\

Recently, the first data of SIDIS off transversely polarized targets have been published, by the HERMES Collaboration~ \cite{hermes} for the proton and by the COMPASS Collaboration~\cite{compass}  for the deuteron. It has been found that, while the Sivers effect is sizable for the proton, it becomes negligible for the deuteron, so that apparently the neutron contribution cancels the proton one, showing a strong flavor dependence of the mechanism.

Different parameterizations of the available SIDIS data have been published \cite{ans,coll3, Anselmino:2008sga}, still with large uncertainties. 
Deeper analyses are in progress, e.g.,  \cite{bog, bog-2}. New data will be soon available, e.g., thanks to the improvement of the CEBAF accelerator at 12 GeV at JLab~\cite{prop-jlab} or the improvement of the experiments at HERMES~\cite{prop-hermes}, that will  reduce the errors on the extracted Sivers function and will help  discriminate between different theoretical predictions. 

\section{The Gauge Link}

The Sivers function is a Transverse Momentum Dependent PD and vanishes if integrated over the transverse momentum.
It is a time-reversal odd object~\cite{Barone:2001sp} and for this reason, for several years, it was believed to vanish due to time reversal invariance. 
However, this argument was invalidated by a calculation in a spectator model \cite{brohs}, following the observation
of the existence of leading-twist  Final State Interactions (FSI) \cite{brodhoy}. 
The current wisdom is that a non-vanishing Sivers function is generated by the gauge link in the definition of TMD
parton distributions \cite{coll2,jiyu,bjy}, whose contribution does not vanish in the light-cone gauge,~\footnote{ The gauge link is usually taken in a non-singular gauge, i.e. not in the light-cone gauge, for further details we refer to Ref.~\cite{bjy}.} 
as  happens for the standard parton distribution functions.
\\

\begin{figure}[ptb]
\begin{center}
\mbox{
\subfigure[]{
\includegraphics[height=2.7cm]{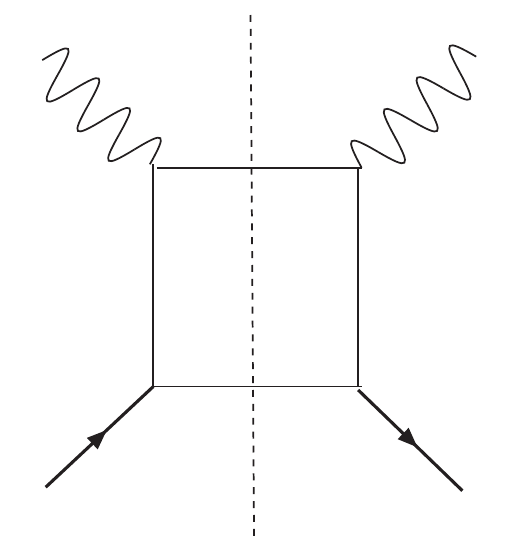}}
\subfigure[]{
\includegraphics[height=2.7cm]{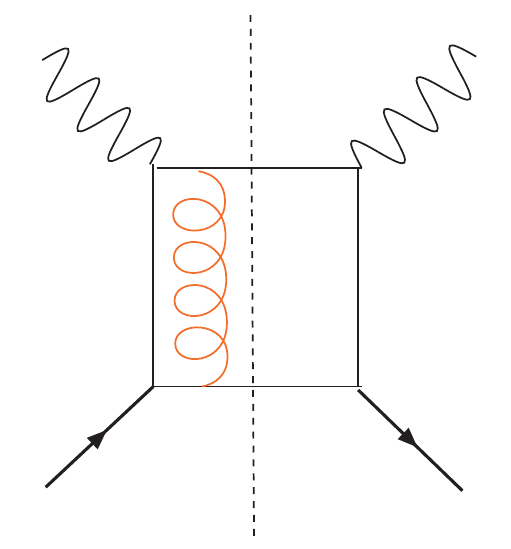}}
\subfigure[]{
\includegraphics[height=2.7cm]{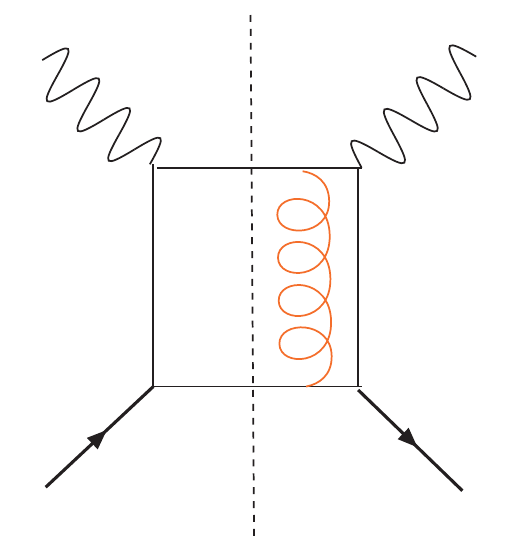}}
\subfigure[]{
\includegraphics[height=2.7cm]{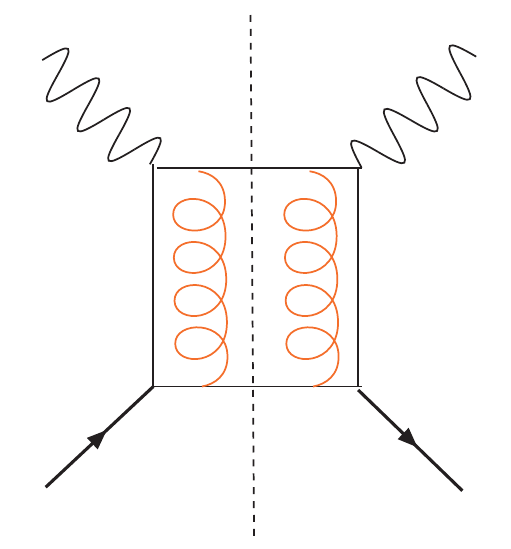}}
}
\caption{The expansion of the gauge link: $(a)+(b)+(c)+(d)+\ldots$.}
\label{gauge link}
\end{center}
\end{figure}
Technically,  Final State Interactions are reflected in the expansion of the gauge link~(\ref{link})  as shown on Fig.~\ref{gauge link}.
It is hence clear by  the identification of the Sivers function in~(\ref{iden-siv}) that,  if the Sivers function were defined without the gauge link, it would be identically zero.

To the first non trivial order, i.e.  $(b)-(c)$ of Fig.~\ref{gauge link}, giving a contribution to the asymmetry,
the Sivers function is obtained
for the flavor $\alpha$, as follows,
\begin{eqnarray}
f_{1T}^{\perp \alpha} (x, {k_T} ) = \Im \left \{
{M \over 2 k_x} \int { d \xi^- d^2 \vec{\xi}_T \over (2 \pi)^3 }e^{-i ( x \xi^- P^+  - \vec{\xi}_T \cdot {{\vec k_T}} )}
\langle \hat O^\alpha \rangle \right \}\quad,
\label{sivers-vers-vers}
\end{eqnarray}

where, in a helicity basis, the operator reads
\begin{eqnarray}
\langle \hat O^\alpha \rangle & = & 
\langle P S_z = + | \bar \psi_{\alpha i}(\xi^-,\vec{\xi}_T)(ig) \left \{ \int_{\xi^-}^\infty A_a^+(0,\eta^-,\vec{\xi}_T) d \eta^- T^a_{ij}  \right \}
\gamma^+ \psi_{\alpha j }(0) | P S_z = -  \rangle + \mbox{h.c.}\quad.\nonumber \\
\end{eqnarray}

\section{Final State Interactions at the Quark Level}

In this theoretical and experimental scenario, due to the lack of correct QCD calculations of Parton Distributions, it becomes important to perform model calculations of the Sivers function. Several estimates exist, in a quark-diquark model \cite{brohs,jiyu,bacch, Bacchetta:2008af};
in the MIT bag model, in its simplest version~\cite{Yuan:2003wk} and introducing an instanton contribution \cite{d'a}; in a light-cone model \cite{luma}; in a nuclear-style framework, relevant to establish the arising of the Sivers function
in proton-proton collisions \cite{bianc}; in the Constituent Quark Model (CQM) of Isgur and Karl (IK) \cite{Courtoy:2008vi}.
\\

Constraints on the Sivers function are obtained from {\it first principles}, following the example of, e.g.,  sum rules in GPDs.
A model independent constraint on calculations of $f_{1T}^{\perp \cal{Q}} (x, {k_T} )$ 
is the Burkardt sum rule  \cite{Burkardt:2004ur, Burkardt:2004ur-2}. 
It states that the total average transverse momentum of the partons in a hadron, $\langle \vec k_T \rangle$, which can be defined through the first moments of  $f_{1T}^{\perp \cal{Q}} (x, {k_T} )$, for all the partons in the target, has to vanish.
If the proton is polarized in the positive $x$ direction,
the Burkardt sum rule reads,
\begin{equation}
\sum_{{\cal Q}=u,d,s,g..} \langle k_y^{\cal{Q}} \rangle = - \int_0^1 d x \int d \vec k_T {k_y^2 \over M}  f_{1T}^{\perp \cal{Q}} (x, {k_T} )=0~\quad.
\label{burs}
\end{equation}
The Burkardt sum rule remains unchanged under evolution.
\\
To distinguish between the model estimates, data
and model independent relations, such as this one, can be used.

\begin{figure}[ptb]
\begin{center}
\includegraphics[height=3cm]{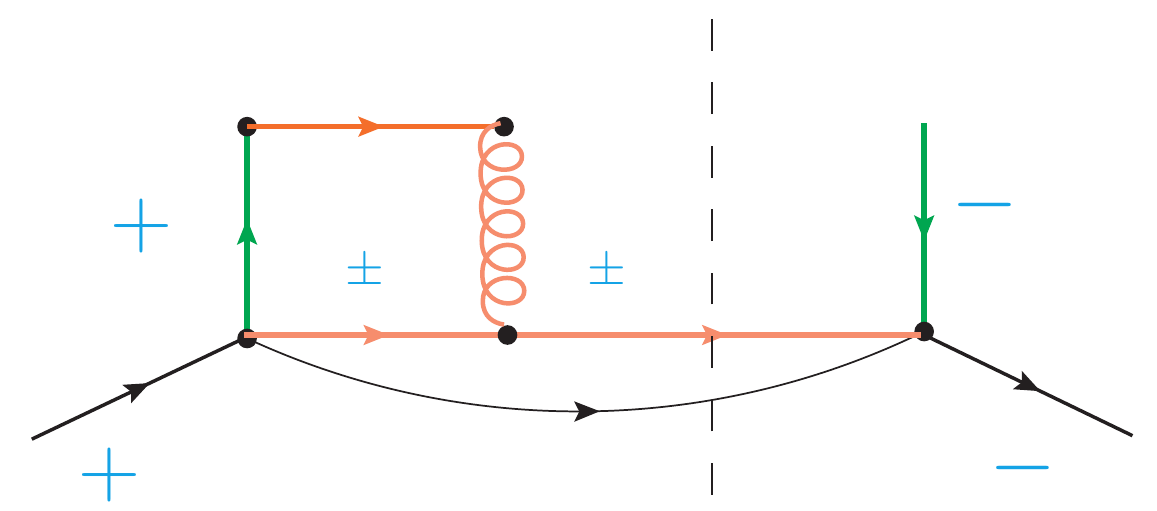}
\includegraphics[height=3cm]{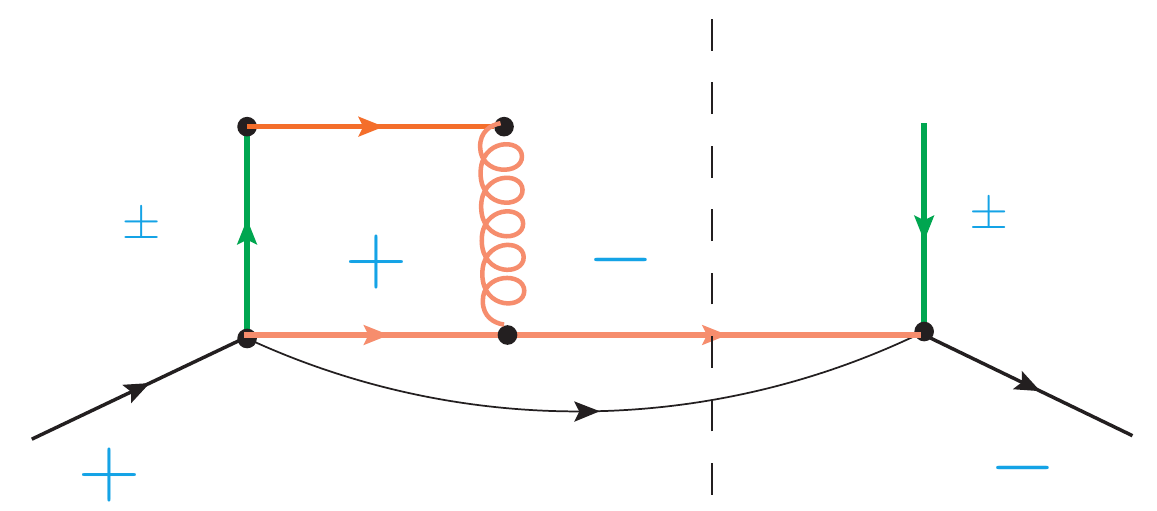}
\caption{Contributions to the Sivers function in quark models. The $+, -$ signs  represent the helicities.}
\label{flip-quark}
\end{center}
\end{figure}
It is also difficult to relate the Sivers Function to the target helicity-flip, impact parameter dependent (IPD), generalized parton  distribution $E$~(\ref{GPD}). Although simple relations or ansatze between the two quantities are found in models \cite{burk,sch}, a clear model independent formal relation is still to be proven, as shown in Ref. \cite{mgm}. However some {\it gross} features, such as the sign of the Sivers function, can be deduced from the results for the IPD parton distributions.
\\

In the context of  Constituent Quark Model for the proton, one  expects, as a consequence of the Burkardt sum rule,  the $u$ and the $d$ Sivers functions to contribute with a similar magnitude but opposite sign.
\\

In this Section, we  analyze  the Sivers functions in two such models; the MIT bag model and the CQM of Isgur and Karl. The gluon exchange is here treated in a perturbative way; both models moreover assume an $SU(6)$ proton wave function, even if only in the first order of the IK wave expansion

If an $SU(6)$ state is assumed, any  model calculation of Parton Distributions gives $u$ and $d$ parton distributions which are proportional to each other~\cite{Jaffe:1974nj, Jaffe:1991kp}. In the case of  $f_{1T}^{\perp \cal{Q}}$  one is not investigating single particle properties but two-body ones, due to the two-body Final State Interaction operator generated by the gauge-link. There 
is therefore in principle no reason to have proportionality, which is a good sign that the results in such models might a priori fulfill the Burkardt sum rule~(\ref{burs}).~\footnote{A proportionality of $-1$ would nevertheless be more than acceptable for the fulfillment of the sum rule. }

The diagrammatic representation of the leading-twist Final State Interactions contributing to a non-zero Sivers function~(\ref{sivers-vers-vers}) in those models is depicted in Fig.~\ref{flip-quark}. As it has been mentioned above, helicity has to flip twice: both at the nucleon state and at the quark level. At the quark level, it is the presence of a 2-body {\it light-like} gluon interaction that enables a spin-flip. By symmetry among the active quarks, the quark interacting with the virtual photon (green quark line) can either flip its helicity or not. The first case is represented on the left panel of Fig.~\ref{flip-quark}, while the second case implies that the "active-but-non-interacting-with-the-photon" quark flips in turn as shown on the right panel. 
\\

Let us now concentrate on the results of these two model calculations. Historically the first calculation performed in the MIT bag model has been undertaken by Yuan~\cite{Yuan:2003wk} whose conclusions were followed by the whole Community. Our  CQM calculation has led us to satisfactory Sivers functions~\cite{Courtoy:2008vi} even if non complying with the  conclusions of Ref.~\cite{Yuan:2003wk}. This is why we "peeped through" the first calculation in the bag model.  A more  physical picture of the Sivers function as allowing symmetry between the two active quarks~\ref{flip-quark} 	was revealed~\cite{Courtoy:2008dn}.
Therefore, the results of Refs.~\cite{Courtoy:2008vi, Courtoy:2008dn} are presented here as illustrations of $1$ and $3$-body approaches of the Sivers function.

\subsection{$1$-body Model}

\begin{wrapfigure}[11]{r}{90mm}
\centering
\includegraphics [height=3.3cm]{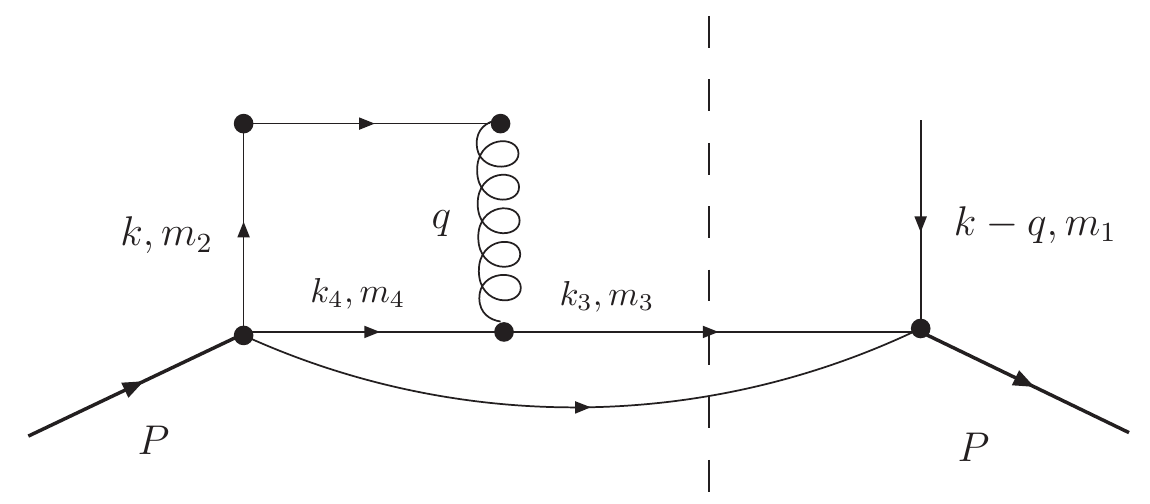}
\caption{The quark helicity and momentum label for the bag calculation. }
\label{sivyuan}%
\end{wrapfigure}
Here we revise the first formal calculation of the Sivers function that was performed in the MIT bag model~\cite{Yuan:2003wk}. The details of the calculation are given in Ref.~\cite{Courtoy:2008dn} with the formalism developed in Ref.~\cite{Yuan:2003wk}.

The expression~(\ref{sivers-vers-vers}) is written inserting the bag model wave function in momentum space, $\varphi(k)$.~\footnote{ The expression for the wave function  of the MIT bag model in momentum space is
\begin{eqnarray}
\varphi_m(\vec{k})&=&i\, \sqrt{4\pi}\, N\, R_0^3 
 \begin{pmatrix}
 t_0(|\vec{k}|) \chi_m\\ 
   {\vec{\sigma} \cdot \hat{k}}\,t_1(|\vec{k}|)\, \chi_m
  \end{pmatrix} \quad,
  \label{bagwf}
\end{eqnarray}
with the normalization factor $N=\left( \frac{\omega^3}{2R_0^3\, (\omega-1)\sin^2\omega}\right)^{1/2}$ 
where $\omega=2.04$ for the lowest mode and $R_0$ is the bag radius. The two last quantities are related through the relation $R_0M_P=4\omega$.
The functions $t_i(k)$ are defined as
$ t_i(k) =\int_0^1 u^2\, du\, j_i	(ukR_0)j_i(u\omega)$. 
}
Using Fig.~\ref{sivyuan} for the definition of the quark helicity and momentum labels,
the Sivers function in the bag model can be written~\cite{Yuan:2003wk}
\begin{eqnarray}
f_{1T}^{\perp\alpha}(x,k_\perp)&=&-2g^2\frac{ME_P}{k^y} \int
\frac{d^2q_\perp}{(2\pi)^5}\frac{i}{q^2}
\sum\limits_{\beta,m_1,m_2,m_3,m_4}T^a_{ij}T^a_{kl}
    \langle PS_x |b_{\alpha m_1}^{i\dagger}b_{\alpha m_2}^{j}
    b_{\beta m_3}^{k\dagger}b_{\beta m_4}^{l} |PS_x\rangle\nonumber\\
&&~~\times\varphi_{m_1}^\dagger(\vec{k}-\vec{q}_\perp)\gamma^0\gamma^+
\varphi_{m_2}(\vec{k})
\int \frac{d^3k_3}{(2\pi)^3}
\varphi_{m_3}^\dagger(\vec{k_3})\gamma^0\gamma^+
\varphi_{m_4}(\vec{k_3}-\vec{q}_\perp) \quad, 
\label{sivers1}
\end{eqnarray}
where 
$M$ is the proton mass, $E_p$ its energy, $b_{{\cal Q},m}^i$
is the annihilation operator for a quark with flavor
${\cal Q}$, helicity $m$, and color index $i$, 
and
$T_{ij}^a$ is a Gell-Mann matrix.
\\

\begin{figure}[tb]
\begin{center}
\mbox{
\subfigure[]{
\includegraphics[height=9cm]{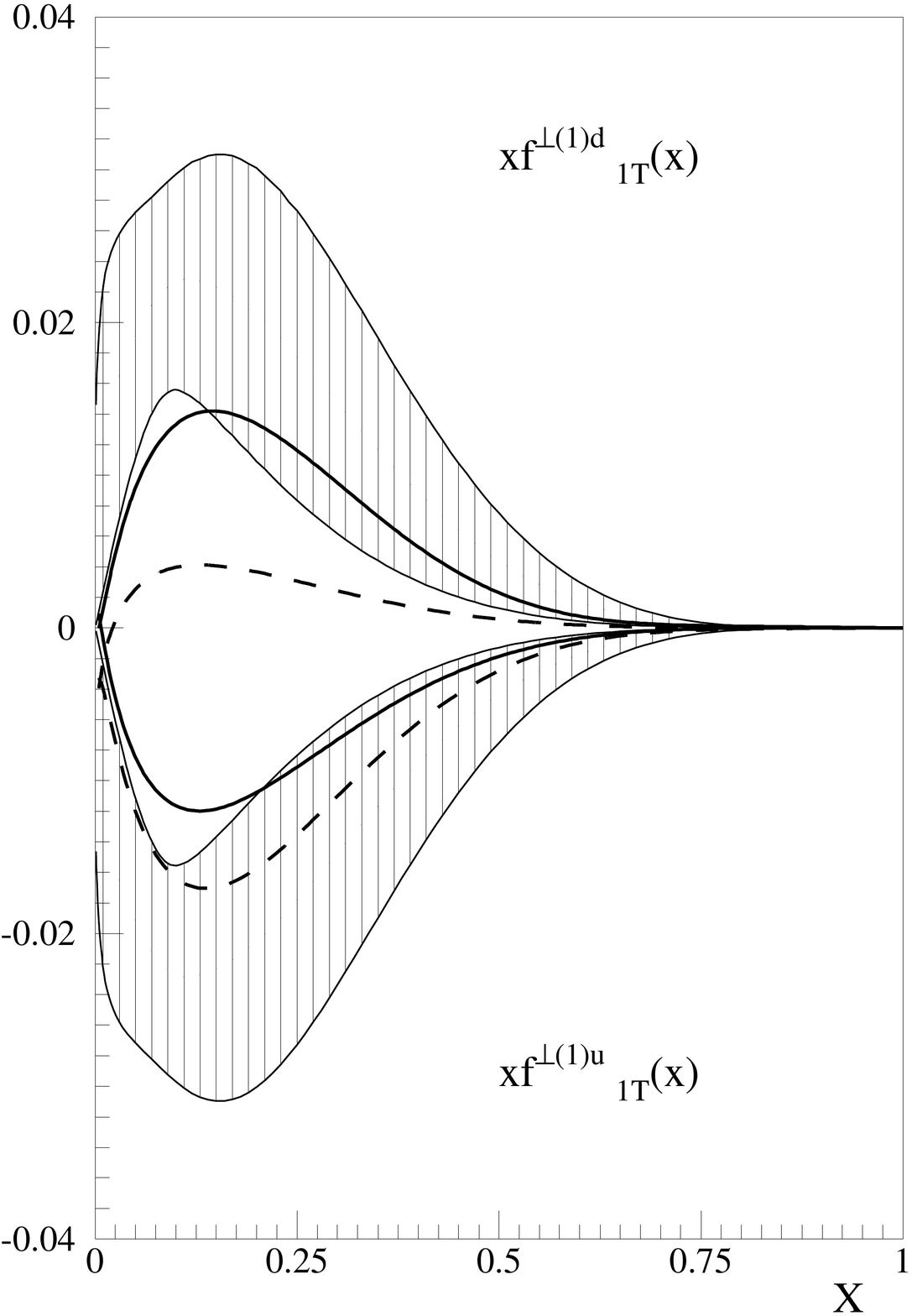}}
\subfigure[]{
\includegraphics[height=9cm]{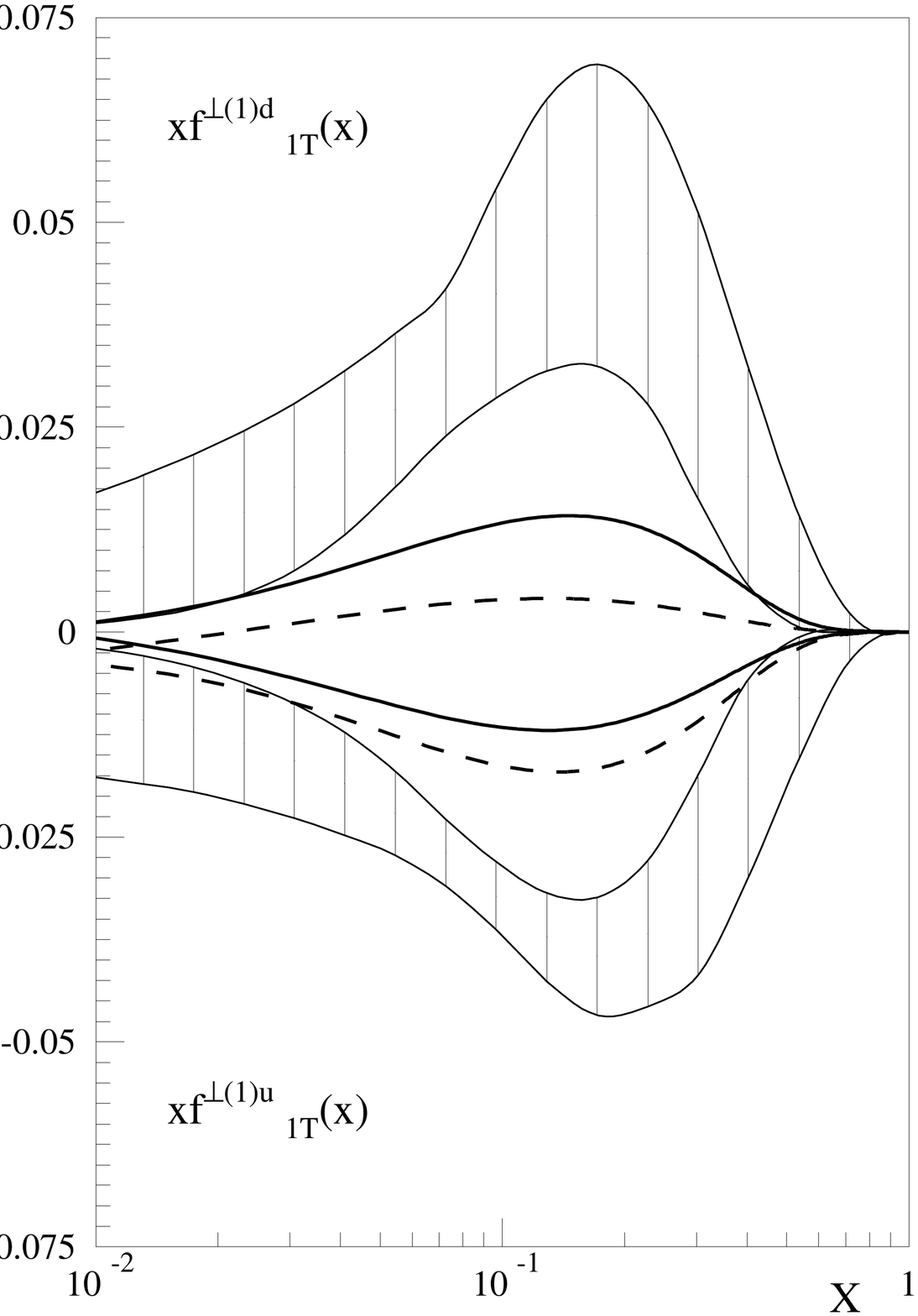}}}
\caption{The (evolved) $u$ and $d$ Sivers functions obtained in the MIT Bag model and compared with the extraction from the data; respectively from  $(a)$ \cite{coll3} and $(b)$ \cite{Anselmino:2008sga}. The plain curves represent the results obtained through Eq.~(\ref{sivers1})~\cite{Courtoy:2008dn}, whereas the dotted curves represent the results of Ref.~\cite{Yuan:2003wk}.}
\label{siv-1}
\end{center}
\end{figure}

Performing the integral over $k_3$ and then $q_{\perp}$ and calculating the appropriate flavor-color factors, we find that there is actually no proportionality between the $u$ and the $d$ quark distributions.
\\
Numerical results are shown in Fig.~\ref{siv-1} for the first moment of the Sivers function (plain curves),
\beq
f_{1T}^{\perp (1) {\cal Q} } (x)&=& \int {d^2 \vec k_T}  { k_T^2 \over 2 M^2}
f_{1T}^{\perp {\cal Q}} (x, {k_T} )\quad.
\label{momf}
\eeq
The obtained Burkardt sum rule is 
\beq
\sum_{{\cal Q}=u,d,s,g..} \langle k_y^{\cal{Q}} \rangle &=&-0.78\, \mbox{MeV}\quad.
\nonumber
\eeq
To have an estimate of the quality of the agreement of this result with this sum rule, we consider the ratio
\beq
r&=& \frac{\langle k_x^{d} \rangle+\langle k_x^{u}\rangle }{\langle k_x^{d} \rangle-
\langle k_x^{u} \rangle}
\simeq 0.05\quad,\nonumber
\eeq
i.e., the Burkardt sum rule seems to be violated by  5 \%.
\\

In order to compare the results with the data,  one should realize that one step of the analysis is still missing. The scale of the model is much lower than the one of the HERMES data, which is $Q^2 =2.5$ GeV$^2$.
For a proper comparison, the QCD evolution from the model scale to the experimental one would be necessary \cite{jaro}. Unfortunately, the evolution of TMDs is still to be understood. In order to have an indication of the effect of the
evolution, we perform a NLO evolution of the model results assuming, for the moments of the Sivers function, Eq. (\ref{momf}), the same anomalous dimensions of the unpolarized PDFs.~\footnote{Important progress concerning the evolution of the first moment of the Sivers function have been made recently, see \cite{Kang:2008ey,Vogelsang:2009pj}.}
The parameters of the evolution have been fixed in order to have a fraction $\simeq 0.55$ of the momentum
carried by the valence quarks at 0.34 GeV$^2$, as in typical parameterizations of PDFs , starting from
a scale of $\mu_0^2 \simeq 0.1$ GeV$^2$ with only valence quarks.

The results  show an impressive improvement of the agreement with data once the results  of Ref.~\cite{Yuan:2003wk} are evolved in the way just described. The data, represented by the dashed bands, are perfectly described for both flavors. 
Although one should not forget that the performed evolution is not really correct,
the analysis shows that, after evolution, model calculations can be
consistent with the available data.

The slight violation of the Burkardt sum rule, which  is a momentum property, is probably due to the fact that  the proton state is not a momentum eigenstate in the MIT bag model.
\\
\\

\subsection{$3$-body Model}

\begin{wrapfigure}[13]{r}{90mm}
\centering
\includegraphics [height=3.3cm]{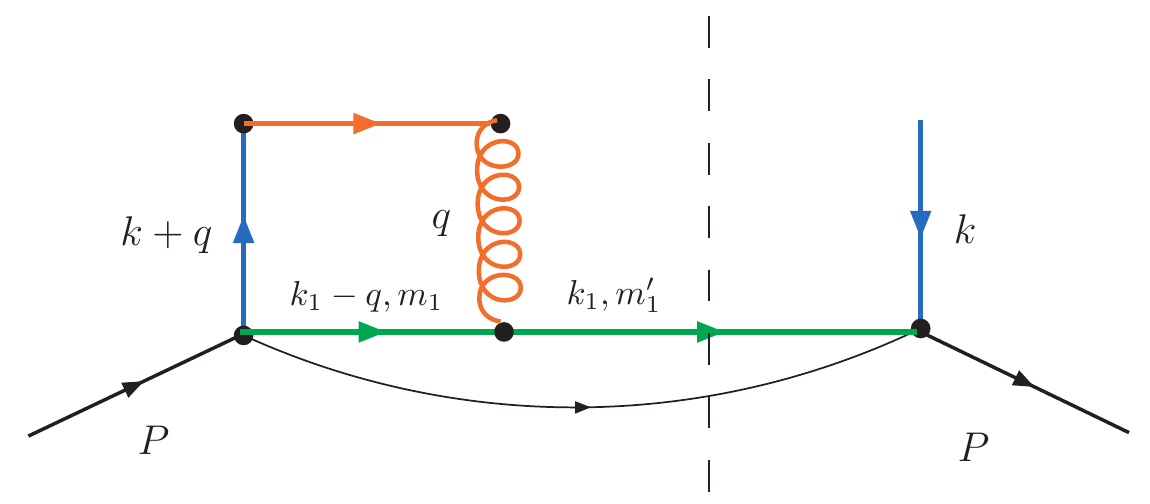}
\caption{The quark helicity and momentum label for the CQM calculation. }
\label{sivmayo}%
\end{wrapfigure}
The Constituent Quark Model (CQM) has a long story of successful predictions in low energy studies of the electromagnetic
structure of the nucleon. 
The Isgur-Karl model \cite{ik, ik2} is chosen to perform a calculation, in order to describe the performances of the approach developed in the context of CQM in Ref.~\cite{Courtoy:2008vi}.
A difference with respect to calculations of PDFs and GPDs is that,
in the calculation of TMDs, the leading-twist contribution to the one-gluon-exchange (OGE)
Final State Interaction has to be evaluated. This is done through a non-relativistic reduction of the relevant operator, according to the philosophy of
Constituent Quark Models \cite{ruju}.~\footnote{
In this model
the proton wave function is obtained in a
OGE potential
added to a confining harmonic oscillator;
including contributions up to the $2 \hbar \omega$ shell,
 the proton state is given by the
following admixture of states
\begin{eqnarray}
|N \rangle =
a | ^2 S_{1/2} \rangle_S +
b | ^2 S'_{1/2} \rangle_{S} +
c | ^2 S_{1/2} \rangle_M +
d | ^4 D_{1/2} \rangle_M~,
\label{ikwf}
\end{eqnarray}
where the spectroscopic notation $|^{2S+1}X_J \rangle_t$,
with $t=A,M,S$ being the symmetry type, has been used.
The coefficients were determined by spectroscopic properties to be
$a = 0.931$,
$b = -0.274$,
$c = -0.233$, $d = -0.067$
\cite{mmg}.}
\\

The expression~(\ref{sivers-vers-vers}) is developed by inserting  proper complete sets of intermediate one-quark states, using translational invariance and the definition 
of the gluon field as explained in Ref.~\cite{Courtoy:2008vi} plus the definition of the labels of Fig.~\ref{sivmayo}, one is left with the following expression for the Sivers
function,
\begin{eqnarray}
&&f_{1T}^{\perp \alpha} (x, {k_T} ) \nonumber\\
& &=  \Im\left \{i g^2 {M \over 2 k_x }\int 
d \tilde k_1
d \tilde k_3
{d^4 q \over (2 \pi)^3} \delta(q^+) 
\delta(k_3^+  + q^+ - xP^+)
\delta(\vec k_{3 T} + \vec q_T - \vec k_T)
{ (2 \pi) \delta(q_0)}
\right.
\nonumber \\
& &   \hspace{-.2cm}
\left.
\sum_{\beta, m_1,m_1',m_3,m_3'} 
\langle P S_z = + | T^a_{ij}  T^a_{kl}\,\,
b^{\alpha i \dagger}_{m_1} (\tilde k_1) 
b^{\beta k \dagger}_{m_3 } (\tilde k_3) 
b^{\alpha j}_{m_1'}(\tilde k_1 - \tilde q) 
b^{\beta l}_{m_3' }(\tilde k_3+\tilde q) 
\,\,
V(\vec k_1, \vec k_3, \vec q)
| P S_z = - \rangle \right \}\,,\nonumber\\
\label{start}
\end{eqnarray}
with the interaction determined by
\begin{eqnarray}
V(\vec k_1, \vec k_3, \vec q)=
{ 1 \over q^2} 
\bar u_{m_3}(\vec k_3) {\gamma^+ }
u_{m_3'}(\vec k_3 + \vec q)
\bar u_{m_1}(\vec k_1) \gamma^+  
u_{m_1'}(\vec k_1 - \vec q)~.
\label{pot}
\end{eqnarray}

\begin{figure}[ptb]
\begin{center}
\mbox{
\subfigure[]{
\includegraphics[height=6cm]{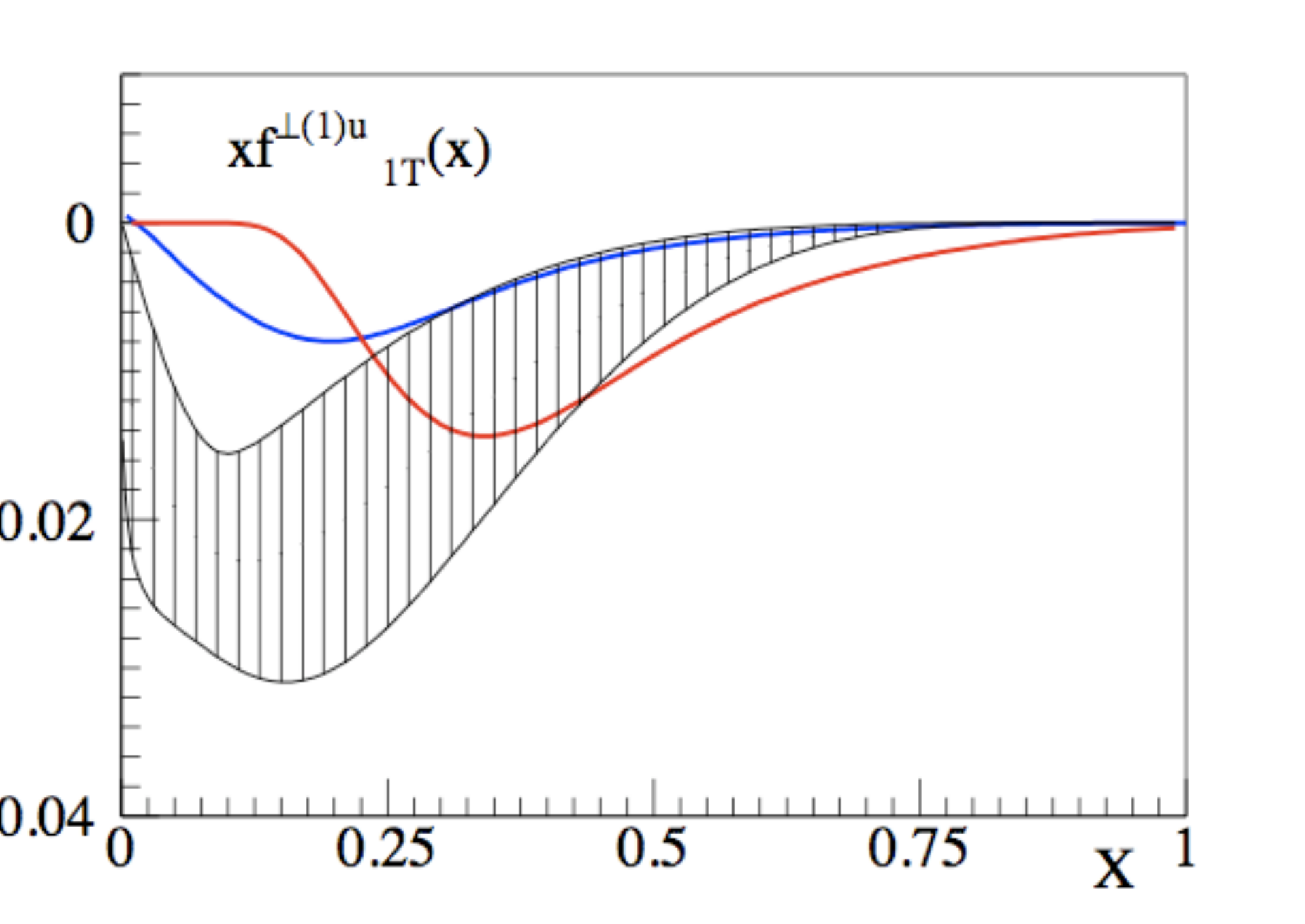}}
\subfigure[]{
\includegraphics[height=6cm]{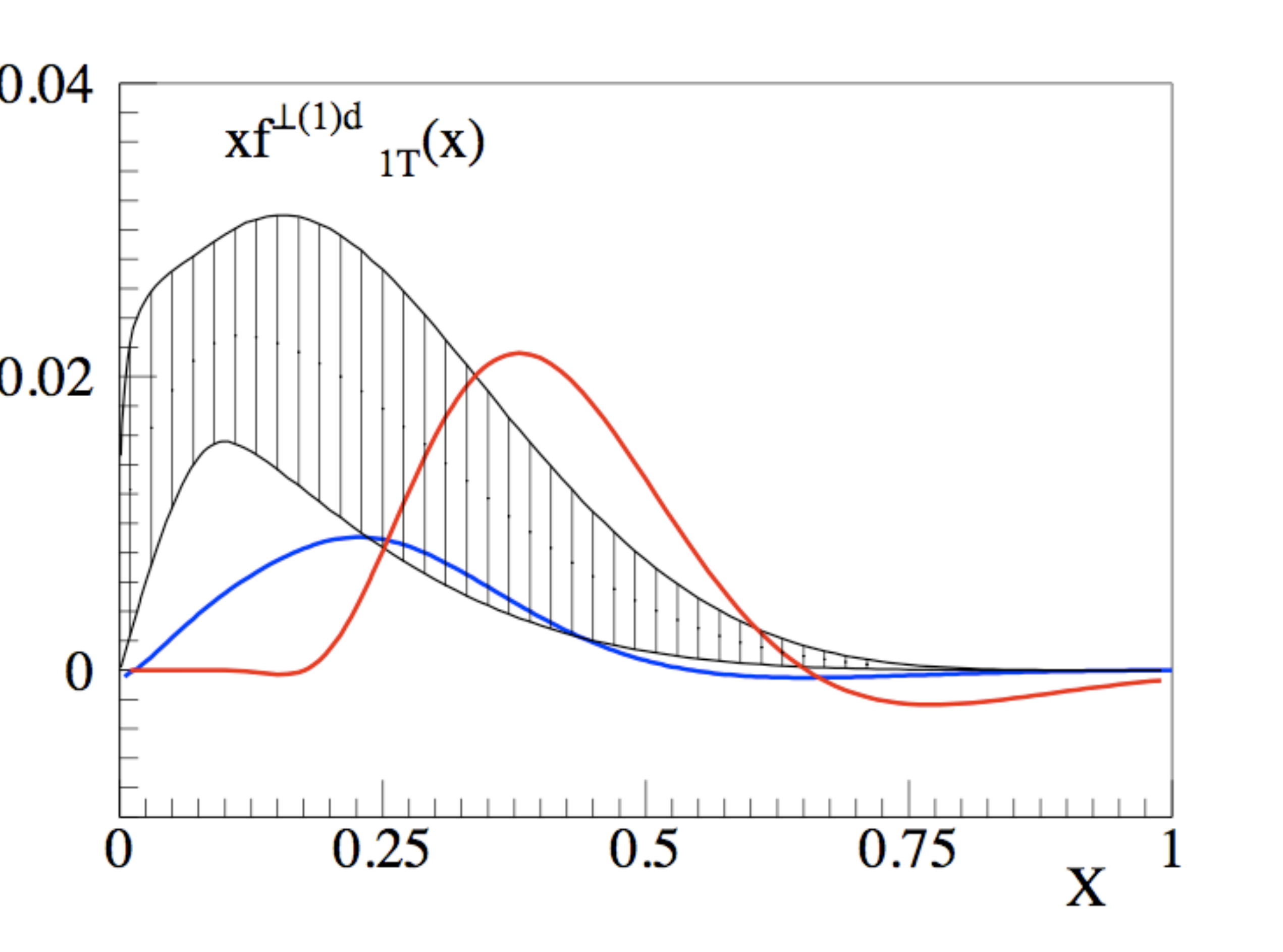}}}
\caption{
The Sivers functions obtained in the IK model and compared with the extraction from the data~\cite{coll3}.The red curves represent the results before evolution, whereas the blue curves are after evolution to the scale of the data.}
\label{siv-3-b}
\end{center}
\end{figure}

For an easy presentation,
the quantity which is usually shown for the results of calculations
or for data of the Sivers function is its first moment~(\ref{momf}).
 The results of the CQM  approach are given by the red curves in Fig.~\ref{siv-3-b} for
the $u$ and the $d$ flavor following the example of Fig.~\ref{siv-1} in the bag calculation. They are compared with a parameterization of the HERMES data,
corresponding to an experimental scale of $Q^2=2.5$ GeV$^2$
\cite{coll3}.
The two dashed curves represent the best fit proposed in Ref. \cite{coll3} plus 30 $\% $
and minus 30 $\% $, respectively.
Such an interval gives a rough estimate of the experimental errors and theoretical uncertainties in the extraction of the Sivers function from data. \\
The evolved results are given by the blue  curves in Fig.~\ref{siv-3-b} for the $u$ and the $d$  flavors.
As it is clearly seen, the agreement with data improves dramatically and their
trend is reasonably reproduced at least for $x \ge 0.2$.
\\

Within our scheme, at the scale of the model, the  Burkardt sum rule is 
\beq
\langle k_x^{u} \rangle + \langle k_x^{d} \rangle &=& -0.40 \, \mbox{MeV}\quad, 
\nonumber
\eeq
obtaining $r \simeq 0.02$, so that we can say that our calculation
fulfills the Burkardt sum rule to a precision of a few percent.

\section{The Way to a Conclusion}

With  respect to Ref.~\cite{Yuan:2003wk}, the results in the MIT bag model (Fig.~\ref{siv-1} ) show an impressive improvement of the agreement with data once the full contribution, i.e. coming from both diagrams~\ref{flip-quark}, is taken into account. The data are described rather well for both flavors.
Comparing this encouraging outcome with that of the IK model, one can notice that the Burkardt SR is better fulfilled in the CQM.

In closing our analyses of the Sivers functions in a $1$ versus $3$-body model, we can say that, for the first time,
it has been established that correct model calculations provide phenomenological successful interpretations of the Sivers function, which are consistent with  each other.
\\

Let us compare our approach with other calculations mentioned above.
\\
The first version of the MIT bag model calculation~\cite{Yuan:2003wk} has non-vanishing $u$ and $d$-quarks contribution of opposite sign which are proportional in magnitude. The $d$-quark contribution is much smaller than the one obtained in the revised version of the bag model as well as the one of the CQM and therefore does not satisfy the Burkardt sum rule.
The MIT bag model modified by instanton effects~\cite{d'a} has $u$ and $d$-quark contributions of the same sign and therefore does not satisfy the Burkardt sum rule. 
\\
The diquark model with scalar diquarks~\cite{bacch} has no contribution for the $d$-quark   and therefore does not satisfy the Burkardt sum rule, Eq.~(\ref{burs}).
The diquark model with axial-vector diquarks has contributions to both $u$ and $d$-quarks and with opposite sign, but with the magnitude of the $d$ 10 times smaller than that of the $u$. The Burkardt sum rule  is not satisfied. The results are improved in the version of Ref.~\cite{Bacchetta:2008af}.

As a summary, we can say that the calculations presented here~\cite{Courtoy:2008vi, Courtoy:2008dn}, despite the naive wave functions used, are in better agreement with the data with respect to the other approaches, and fulfill the Burkardt sum rule. The model calculations have still to be improved and their evolution properly undertaken. So have to be the extraction of the data as well as the experiments. Much efforts in that direction is being put from the whole {\it Transversity} community.~\footnote{ We cite, e.g., the latest workshop "Transversity 2008: 2nd International Workshop On Transverse Polarization Phenomena In Hard Processes",
28-31 May 2008, Ferrara, Italy.}

\chapter{Conclusion and Future Perspectives}
\pagestyle{fancy}
\label{sec:concl}
\label{sec:out-tda}
\label{sec:out-gpd}
\label{sec:spinpion}

Our main source of information about the internal pion and nucleon structure is provided by Deep Inelastic Scattering. Unpolarized DIS has revealed the partonic distributions of the hadrons. It has shown that the quarks carry only about half of the total hadron's momentum.  Polarized DIS has provided information about spin distributions, telling us that only a small fraction of the spin of the proton is carried by the intrinsic spin of the quarks.
\\

In the last decade, a theoretical treatment of  semi-inclusive as well as  exclusive deep reactions has been developed, increasing our knowledge on the structure of hadrons. 
For instance, Semi-Inclusive Deep Inelastic Scattering allows to obtain information also on chiral-odd Parton Distributions. In a transversely polarized hadron,  the transversity distribution tells us what is the number density of quarks with polarization parallel to that of the hadron, minus the number density of quarks with antiparallel polarization. Moreover, the complete study of the features of the produced hadron in SIDIS requires that we take into account the transverse motion of quarks. The theoretical tools for the description of this process are therefore the {\it Transverse Momentum Dependent} PDs. Among this class of distributions, the Sivers function is  the object of important theoretical studies as it has been proposed to explain single-spin asymmetries in SIDIS.

On the other hand, exclusive electroproduction of photons (Deep Virtual Compton Scattering) have been the most studied processes. Under the conditions of large virtuality $Q^2$ of the exchanged photon and low momentum transfer $t$, i.e. near the forward direction, the amplitude of the reaction factorizes into a hard and a soft part. This soft part have opened new windows on the structure of hadrons.

The theoretical tools for the understanding of theses processes are the {\it Generalized Parton Distributions}. The GPDs connect in different limits to  other well known physical quantities: in the limit of low momentum transfer, $t\to 0$, they connect to the usual Parton Distributions. Integrated over the momentum fraction of the quark, GPDs lead to the electromagnetic hadron's Form Factors. Moreover, through  Ji's sum rule,  a particular combination of moments of GPDs can be related to the spin of the proton. 

Exclusive hadron production in $\gamma^{\ast}\gamma$ scattering at small momentum transfer and large invariant mass is also assumed to factorize into hard and soft parts. The soft part provides for additional information on hadron structure through the {\it Transition Distribution Amplitudes}.
\\
\\

In this thesis, we have considered some of the distributions that we have just described. 
Throughout the Chapters~\ref{sec:pionda}, \ref{sec:gpd}, \ref{sec:tda} $\&$ \ref{sec:cross}, the partonic structure of the pion has been scrutinized under the Nambu~-~Jona-Lasinio description of the pion.
In Chapters~\ref{sec:pionda} $\&$ \ref{sec:gpd}, we have reviewed the previous situation; whereas, in Chapters~\ref{sec:tda} $\&$ \ref{sec:cross}, we have presented our results on the Transition Distribution Amplitudes.
\\

In Chapter~\ref{sec:pionda}, the {\it Distribution Amplitude} and the  {\it Parton Distribution Function} of the pion have been presented together with their relation through the Soft Pion Theorems. The study of these well-known quantities have provided for a way to set  the scene for the subsequent and original part of the manuscript. 
%
%
To be specific, more than a simple illustration of the formalism in which we were to be working, the calculation of the pion DA has allowed for a first criticism of the choice of the NJL model. %
%
%
%
%
%
This has been done, among other things, by determining the scale of validity of this quark model $Q_0$ by evolving according to the QCD evolution equations. In particular, the Parton Distribution Functions of the pion have been found to be in a surprisingly good agreement with the data, once evolved to Leading-Order. 
We have also considered the Next-to-Leading-Order evolution, and we have reached the conclusion  that the effect of the  NLO evolution has been compensated at LO in the adaptation of the low model's scale $Q_0$. 
On the other hand, the Distribution Amplitude of the pion calculated in the NJL model have been found to be in a good although less impressive agreement with the data. 

The conclusions that we can draw from the analysis of the pion PDF and DA also follow from  the questions of the support property as well as the end-point behavior. 
First, we conclude that  the conservation of covariance on the light-cone is a must-have feature of the model that one would have chosen for the analyses of distribution functions. We have actually learned that the support property is closely related to this symmetry, which is respected in the NJL model.
Second, we emphasize the importance of perturbative QCD when going to higher energies in which the experiments are performed. We therefore advocate discarding some models only after having evolved their results to a perturbative scale.

 Given the symmetries it respects, and the good results it has led to, the NJL model, with all its merits and faults,  can be considered as a realistic model for the pion not only for low energy observables, but also for the determination of the Parton Distributions. The emergent image of the pion 
  is that it is basically determined by the chiral symmetry.
\\

Chapter~\ref{sec:gpd} has been devoted to the {\it Generalized Parton Distributions} of the pion. In particular, the results of Ref.~\cite{Theussl:2002xp} have been displayed together with the  parameterization of GPDs  via Double Distributions~\cite{Polyakov:1999gs}. The expected support and the polynomiality for the GPD of the pion  have been obtained.  There is no relevant discrepancy between the results obtained in the NJL model and the other calculations, e.g.~\cite{Broniowski:2007si, Polyakov:1999gs}, before and after QCD evolution.  The agreement after evolution gives us confidence on the evolution code~\cite{FreundMcDermott} that we have used.
\\

In Chapter~\ref{sec:tda}, we have principally presented  our results for the pion-photon vector and axial {\it Transition
Distribution Amplitudes}, respectively $V(x,\xi,t)$ and $A(x,\xi,t)$~\cite{Courtoy:2007vy, Courtoy:2008af, Courtoy:2008ij}. %
We have used the formalism developed in Chapter~\ref{sec:pionda}:  the Bethe-Salpeter amplitude for the pion has been constructed with the quark-pion vertex function coming from the Nambu-Jona Lasinio model. This technique  allows for  numerical predictions. Also,  the Pauli-Villars regularization procedure has been applied in order to preserve gauge invariance.

The definition of these pion-photon TDAs have been displayed in Section~\ref{sec:kin-tda}.
The Partial Conservation of the Axial Current tells us that the axial current couples to the pion. Therefore, in
order to properly define the axial TDA, namely with all the structure of the incoming hadron included in $A\left(  x,\xi,t\right)  $, the pion pole contribution must be extracted. By doing so, we have found that the axial TDA had, in the NJL model and  the handbag approximation, two different contributions: the first one is related to a direct coupling of the axial current to a quark of the incoming pion and to a quark coupled to the outgoing photon. The second one is related to the non-resonant part of a quark-antiquark pair coupled with the quantum numbers of the pion. The presence of the latter contribution guarantees the gauge invariance of the axial TDA. 

The use of a fully covariant and gauge invariant approach guarantees  that all the fundamental properties of the TDAs   will be  recovered. In this way, the right support, $x\in\left[  -1,1\right]  $, has been obtained and the sum rules, together with the polynomiality expansions, have been recovered. We want to stress that these three properties are not inputs, but they are results springing from our calculation~\cite{Courtoy:2007vy}. The value obtained in the  NJL model
for the vector Form Factor is in agreement with the experimental result provided in the Particle Data Group~\cite{Amsler:2008zz}, whereas the value found for the axial Form Factor is two times larger than the ones found in the PDG~\cite{Amsler:2008zz}. This discrepancy is a common feature of quark models as it has been obtained in different calculations~\cite{Broniowski:2007fs, Kotko:2008gy}. Also, the neutral pion vector Form Factor $F_{\pi\gamma^{\ast}\gamma}(t)$ is well described, reproducing the value given by the anomaly~\cite{Brodsky:1981rp}. These results allow us to assume that the NJL model gives a reasonable
description of the physics of those processes at this energy regime.

Turning our attention to the polynomial expansion of the TDAs, it has been shown that, in the chiral limit, only the coefficients of even powers in $\xi$ were non-null for the vector TDA~\cite{Courtoy:2007vy}. No constrain has been obtained for the axial TDA. Nevertheless, the NJL model have provided for simple expressions for the coefficients of the polynomial expansions in the chiral limit (\ref{coefaxial}).

We have obtained quite different shapes for  the vector and axial TDAs.   At least for the DGLAP regions, this is in part due to the isospin relation 
\begin{equation}
V\left(  x,\xi,t\right)  =-\frac{1}{2}V\left(  -x,\xi,t\right)  ,~~~~A\left(
x,\xi,t\right)  =\frac{1}{2}A\left(  -x,\xi,t\right)  , ~~~~~~~~~~~\left\vert
\xi\right\vert <x<1\quad.
\nonumber
\end{equation}
The difference in the shapes is also due to  the non-resonant part of the $qq$ interacting pair diagram for the axial TDA in the ERBL region.	
It is interesting to inquire about the domain of validity of the previous relations. These relations are obtained from the isospin trace calculation involved in the central diagram of Fig.~\ref{Fig3}. Due to the simplicity of the
isospin wave function of the pion, these results are more general than the NJL model, and therefore more general than a result of the diagrams under consideration, which  are the simplest contributions of the handbag type.

Since there exist other model calculations as well as phenomenological approaches of the pion-photon TDAs~\cite{Broniowski:2007fs, Kotko:2008gy, Lansberg:2006fv, Tiburzi:2005nj}, a qualitative comparison and/or criticism of our approach was in order.  This has been done in Section~\ref{sec:compa-tda}, where the Double Distributions for the TDAs have been defined.
We have concluded that there is no disagreement between the different studies concerning the $\pi$-$\gamma$ TDAs, besides some  ambiguities in the definition of the skewness variable.
\\

The TDAs, calculated, as a first step, for pion-photon transitions, have led to interesting estimates of cross section
for exclusive meson pair production in $\gamma^{\ast}\gamma$ scattering. These processes involve photon-pion transitions. Also, the symmetries relating the different transitions, e.g., $\pi^+\to\gamma$ to $\gamma\to\pi^-$ have been displayed.
 In Chapter~\ref{sec:cross} we have presented our results for a phenomenological application of the TDAs~\cite{Courtoy:2008nf}. We have examined the expected cross section for $\pi$-$\pi$ and $\pi$-$\rho$ production in exclusive $\gamma^{\ast}\gamma$ scattering in the forward kinematical region, by using  realistic models for the description of the pion. First, we have confirmed the previous estimates for $\rho$ production obtained in Ref.~\cite{Lansberg:2006fv}, even if our results for the cross section are  larger of a factor 2. Second, in comparison with this previous evaluation of the cross
sections, we have improved our findings by considering the effect of QCD  evolution on the vector current.  By so doing,
 an additional factor 5 has appeared  in the small $\xi$ region, leading to a cross section for $\rho$ production of one order of magnitude larger than the previous calculation. We have also ameliorated  by  including the pion pole term in the tensorial decomposition of  the axial current.
 Then, an even larger enhancement factor, of about 60 in this case, has been found in the cross section for pion production.
 The interference term has become  larger by a factor 15 with respect to the pure TDA contribution, making the axial TDA more accessible experimentally. 
\\
\\

Besides the structure of the pion, we have also been interested in semi-inclusive processes involving Parton Distributions of the  proton.  Since the features of the proton are rather different than the pion's one, the description of the proton requires the use of models whose properties are adequate for that problem.  In particular, we have used two models for the proton which do not include the realization of the chiral symmetry but rather  confinement: the NR Constituent Quark Model of Isgur-Karl as well as the MIT bag model. 

We now draw some conclusions on this  part of the thesis, which  discusses  the Sivers function. 
In Chapter~\ref{sec:sivers} we have presented our results for the calculation of the Sivers function in the two  models described above.
The formalism that we have developed in  Ref.~\cite{Courtoy:2008vi} for the evaluation of the Sivers function  is  rather general as it  can  be used in any Constituent Quark Model.
The crucial ingredient of this calculation has been  the Non Relativistic  reduction of the leading twist part of the One-Gluon Exchange diagram in the final state.
It has been shown that the Isgur-Karl model, based also on a OGE contribution to the Hamiltonian, is a proper framework for the estimate  of the Sivers function. The obtained results show a sizable effect, with an opposite sign for the $u$ and $d$ flavors.
This is in agreement with the pattern found from an analysis of impact parameter dependent
GPDs in the IK model~\cite{fratinif} and allows for the fulfillment of the Burkardt sum rule. More precisely, this sum rule is verified with a precision of a few percents. 
The connection of the Sivers function with Impact Parameter Dependent PDs deserves a careful analysis and will be discussed elsewhere. In particular, some future directions of work are given in a next Section.

The Sivers function calculated in the MIT bag model has been revised~\cite{Courtoy:2008dn}, and now it fulfills the Burkardt sum rule to a level of 5$\%$. Comparing this encouraging outcome with that of the CQM,  one can say that the Burkardt sum rule is better fulfilled  in the CQM. Most probably this has to do with the fact 
that the Burkardt sum rule is associated with transverse momentum conservation.
The evolved Sivers functions, as obtained in both models, show an impressive  agreement with data. We would like to remark that the data are described rather well for both flavors.

In conclusion, we want to stress that we have established, for the first time, that correct model calculations provide phenomenological
successful interpretations of the Sivers function,
which are consistent with
each other.
\\

In what follows, we present the possible future directions for both lines of work, trying to merge Spin Physics together with  pion structure in the very last Section.

\subsubsection{The Future of Transition Distribution Amplitudes}

In  Chapters~\ref{sec:tda} and \ref{sec:cross}, we have introduced the concept of {\it Transition Distribution Amplitudes} as an extension of Distribution Amplitudes as well as of Generalized Parton Distributions. We have presented the results in a model calculation, namely in the NJL model, and compared with different model calculations carried out until now. The mesonic TDAs have been applied in the estimation of the cross section for meson pair production in two-photon processes, awaiting experimental data.

A glimpse of  distribution parameterizations have been given including phenomenological modeling. In this short Section, we mention some other approaches and applicability frameworks for TDAs.
\\

Besides the parameterization through the  Radyushkin's Double Distribution, there exists the so-called Dual Representation~\cite{Polyakov:1998ze, Polyakov:2002wz}. Duality comes by  representing  the parton distributions
as an infinite series of distribution amplitudes in the $t$-channel, obtained by decomposing the GPDs in $t$-channel partial waves. This representation leads to an alternative parameterization of GPDs with respect to~(\ref{t-ind-dd}). 

A further study of the TDAs as parameterizing the processes~(\ref{processes}) through this representation should lead to interesting, model-independent results. The latter could, in turn,  be compared with the model-dependent estimations of the cross sections given in Section~\ref{sec:cross-njl}. The inclusion of the pion pole contribution renders more appealing the modeling. Moreover, its study is   justified, from  the experimental point of view, by the enhancement of the cross section for $\pi^+\pi^-$ production in $\gamma^{\ast}\gamma$ scattering due to the presence of the pion pole amplitude.
\\

\begin{figure}
[ptb]
\begin{center}
\includegraphics[height=5cm]{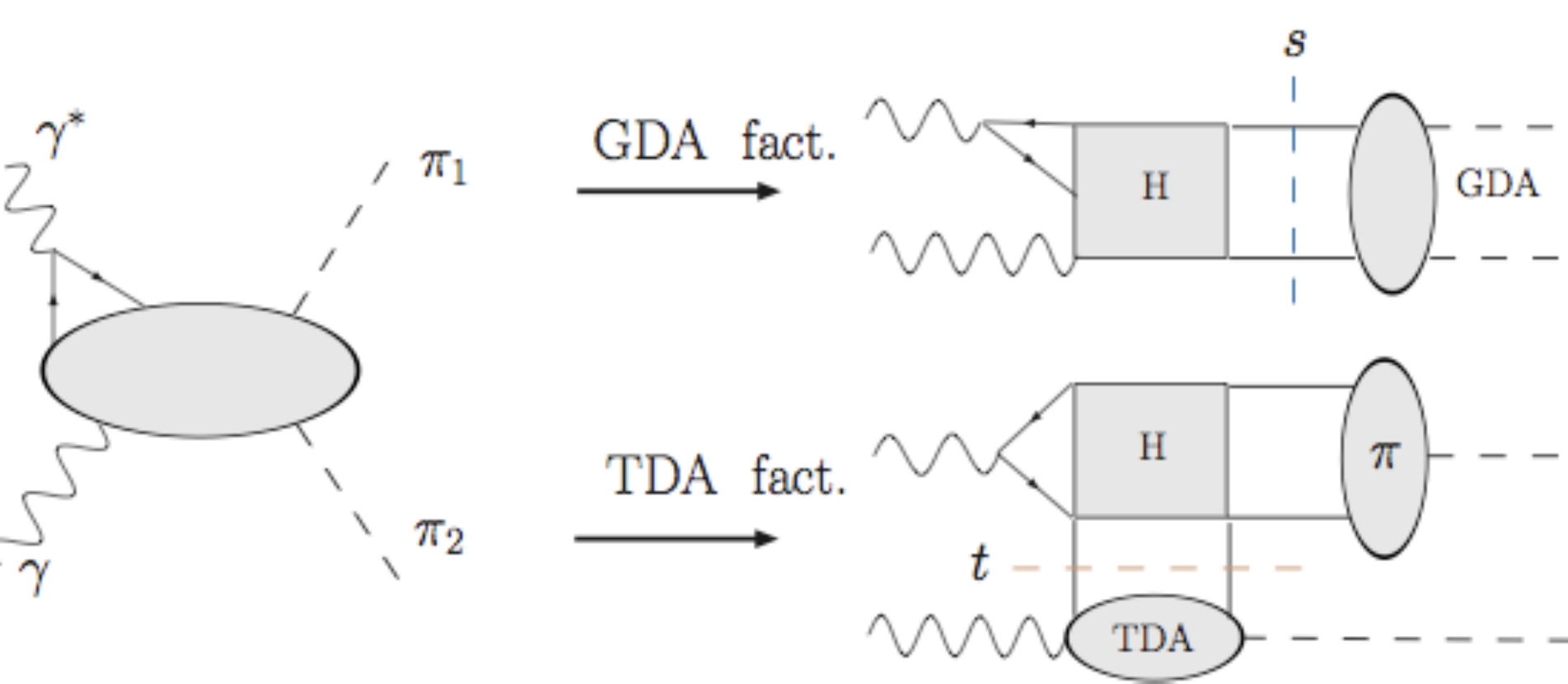}
\caption{GDA vs. TDA factorization. From Ref.~ \cite{Anikin:2008bq}.}
\end{center}
\label{facto-s-t}
\end{figure}

Among the processes involving the TDAs as the large-distance part, the process $\gamma^{\ast}\gamma\rightarrow\pi\pi$ is particularly interesting because different kinematical regimes lead to different mechanisms. This implies a description of the process through either the pions GDA or the pion-photon TDAs \cite{Anikin:2008bq}.
As it is depicted on Fig.~\ref{facto-s-t}, there is a duality in the factorization mechanisms in describing the fusion of a real photon with highly-virtual and longitudinally polarized photon. According to the kinematical regime, i.e.,
\begin{enumerate}

\item[(a)] $s \ll Q^2$ while $t$ is of order $Q^2$;
\item[(b)] $t \ll Q^2$ while $s$ is of order $Q^2$;
\item[(c)] $s, t \ll Q^2$;
\item[(d)]  $s, t \sim Q^2$;
\end{enumerate}
the process will factorize following the GDA\footnote{The twist-3 GDA here is selected by the $\gamma_L^{\ast}\gamma_T$ initial state.} (regime $(a)$) or the TDA (regime $(b)$) mechanism.
In addition  there exists a kinematical domain, namely the regime $(c)$, in which both mechanisms act in parallel, giving rise to duality in this region. The model-dependency of the GDA and its impact on duality is nevertheless pointed out. The regime $(d)$ corresponds to the large angle scattering {\it \`a la} Brodsky-Lepage.

This interplay between kinematical regimes leads us to think that it  would be  interesting to deepen our understanding of the description of this process through the analysis of the twist-3 pion GDA, e.g. in the NJL model.
\\

In conclusion, the meson-photon transitions lead to many possible applications that 
represent a natural follow-up of the work displayed in this thesis.

\subsubsection{The  Internal  Spin Structure of the Pion}

Complementarily to what has been explained in the previous Chapter, it has been shown that, even if the target is unpolarized, the momentum distribution of its quarks ejected in SIDIS can already exhibit a left-right asymmetry of the intrinsic transverse quark momentum $k_T$ relative to their own transverse spin $s_T$~ \cite{Boer:1997nt}.
The Boer-Mulders function is defined as the modulation of the probability density due to the correlation between these two variables, via the $\Phi^{[\sigma^{+j}\gamma_5]}$ of the decomposition~(\ref{decom-bm}), 
\begin{eqnarray}
\Phi^{[\sigma^{+j}\gamma_5]\,q}(x,\vec k_T,s) & = & f_1^q(x, k_T) - {\epsilon_T^{ij}s_{Ti} k_{Tj} \over M} h_{1}^{\perp q} (x, k_T)
\label{def-bm}
\end{eqnarray}
Like the Sivers function, the function $h_{1}^{\perp q} (x, k_T)$ requires the presence of Final State Interactions to allow different orbital angular momentum combinations.
\\

Obviously, one of the outlooks of this thesis would be to calculate the Boer-Mulders function in both the $1$ and $3$-body models in which we already have calculated the Sivers function.  
This calculation is going to be completed very soon~\cite{nous-bm}.\footnote{This work has been published after the thesis has been presented.}
\\

Another future perspective is related to the first part of this thesis, and it concerns the non trivial spin structure of the pion.

\begin{wrapfigure}[16]{r}{8.4cm}
\centering
 \includegraphics [height=4.5cm]{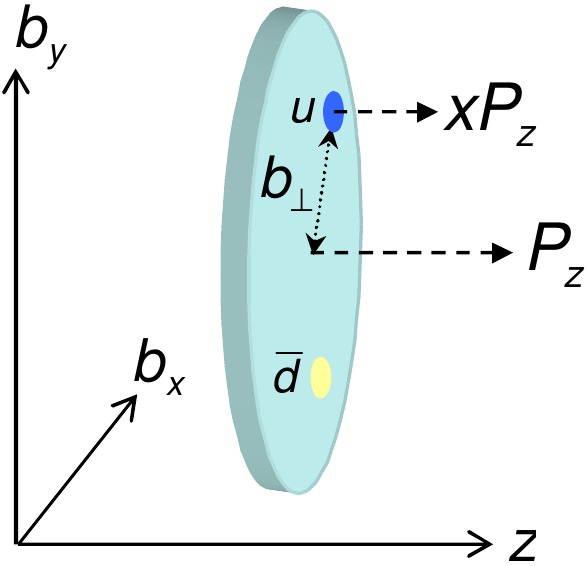}
 \caption{Illustration of the GPDs of the $\pi^+$ in the impact parameter space.}
\label{IPD}
\end{wrapfigure}

Since the pion has spin zero, its longitudinal spin structure in terms of quark and gluon degrees of freedom is trivial.
However, it would be of interest to investigate  the transverse spin structure of the pion, which, in the case of a spinless hadron, would come from a modulation of the number density due to the interplay between the transverse spin of the quarks and another intrinsic variable. The latter is namely either the momentum $k_T$ or the Impact parameter $b_{\perp}$, which gives the distance between the center of momentum of the hadron in the plane transverse to its motion as shown on Fig.~\ref{IPD}. 

Following the example of the intuitive determination of the sign of the Sivers function from the $E$ GPD~(\ref{GPD}), one expects that the sign of the Boer-Mulders function can be related to the distribution of transversely polarized quarks in the Impact Parameter space~\cite{Burkardt:2005hp}. 
\\
Little is known about both the Impact Parameter Dependent (IPD) distribution under scrutiny and the Boer-Mulders function. This holds both from the experimental and theoretical point of views. However, what motivates our project  is the existence of the first lattice results for the Form Factors in momentum space of the concerned pion GPD~\cite{Brommel:2007xd}.
\\
Our proposal  idea is basically to apply the formalism developed in  Chapter~\ref{sec:pionda} for the calculation of pion distribution in the NJL model to the calculation of the pion  IPD distribution of transversely polarized quarks.
 In this way, we hope to gain some insight about  the possible ansatz linking  the pion GPD calculated in the impact parameter space, i.e. correlations between the impact parameter and the intrinsic transverse spin of the quarks,  to the Boer-Mulders function. However, such relations involve a non trivial association of a spatial variable $b_{\perp}$ and a momentum $k_{\perp}$. It is henceforth a  tantalizing goal to obtain model independent relations. So that  one should restrain itself to global considerations, such as the relative signs. The latter can be afterwards  compared with the result in the Lattice.
\\

The IPD  distributions have been proven to be related to the GPDs at zero skewness~\cite{Burkardt:2000za}. 
The Fourier transform of the charge density defined with respect to the impact parameter  can be associated, without ambiguity, with the charge density as a function of the impact parameter  in the Infinite Momentum Frame~\cite{Burkardt:2000za}. The same can be done for the GPDs, but only for $\xi=0$.
In other words,  the GPDs at zero skewness allow a simultaneous measurement of the light-cone momentum and transverse position, i.e. impact parameter, distributions of partons in a hadron
\beq
F(x;\vec{b}_{\perp})&=& \int \frac{d^2 \Delta_{\perp}}{(2\pi)^2}\,e^{-i\vec{\Delta}_{\perp}\cdot \vec{b}_{\perp}}\, F(x, \xi=0 -;\vec{\Delta}_{\perp}^{\,2})\quad.
\label{burk-xi0}
\eeq

We now consider the transverse quark polarization, which is the only one allowed since the helicity-flip of the state is forbidden for the pion.
Quarks with transverse polarization $s_T$ are projected out by the operator  $\frac{1}{2}\bar q\, [ \gamma^+ -s^j\, i\sigma^{+j}\gamma_5]q$. Their density is therefore
\beq
\frac{1}{2}\left[
F^{\pi}(x; \vec{b}_{\perp})+s^i\, F_T^{i\,\pi}(x; \vec{b}_{\perp})
\right]
\quad,
\nonumber
\eeq
i.e. a combination of the unpolarized~(\ref{GPDNJL}) and polarized one. The latter IPD parton distribution is  given by  the  chiral-odd twist-2 pion GPD, which  is allowed by flipping the quarks' helicity. It is defined as
\begin{eqnarray}
F^{j \pi }_T(x, \xi, t)&=& \frac{-i}{2}\int \,\frac{dz^-}{2\pi}\,e^{ixz^-p^+}\left\langle \pi(P')\left	\vert\bar q\left(-\frac{z}{2}\right)\,\sigma^{+j}\gamma_5\,\,q\left(\frac{z}{2}\right)
\right\vert \pi(P)\right\rangle |_{z^{\perp}=z^+=0}\quad,\nonumber\\
\nonumber\\
&=& \frac{1}{2\,p^+}\,\frac{\epsilon^{+j\alpha\beta}\Delta_{\alpha}p_{\beta}}{\Lambda}\, \tilde H^{\pi}_T (x, \xi, t) \quad,
\label{odd-gpd}
\end{eqnarray}
where a natural choice for  some hadronic scale, say $\Lambda$, would be $m_N$ in  the nucleon case.~\footnote{ Note that, by analogy to the nucleon case, the  standard nomenclature is $E_T^{\pi}$  or
  $\bar{E}_T^{\pi}$ instead of  $\tilde{H}_T^{\pi}$. }This scale is rather chosen to be $\Lambda=4\pi\, f_{\pi}$ in Ref.~\cite{Burkardt:2007xm} in order to avoid  discussion  within the chiral limit. However, the chiral limit has been studied in Ref.~\cite{Diehl:2006js} showing that the choice  $\Lambda=m_{\pi}$ is totally justified.
\\

The Fourier transform~(\ref{burk-xi0}) of the chiral-odd distribution $F^{j \pi }_T(x, \xi, t)$ leads to a derivative of the IPD distribution with respect to $b^2$;
\beq
\frac{1}{p^+} \left(H^{\pi} (x; \vec{b}_{\perp}) +\epsilon^{ij}\, s^i\,b^j\, \frac{2}{m_{\pi}}\,\frac{\partial}{\partial b^2}\,\tilde H_T^{\pi^+} (x; \vec{b}_{\perp})\right)
\quad.
\nonumber
\eeq
The dipole term, as expected, leads to a dependence on the direction of $b_{\perp}$ for fixed $s_T$. The sign of this contribution seems to be opposite the the one of the Boer-Mulders function~({def-bm}). In other words, the IPD $\tilde H_T^{\pi^+}$ describes a sideways shift in the distribution of transversely polarized quarks in an unpolarized proton which has the  opposite sign than the Boer-Mulders function.
\\
\begin{figure}[ptb]
\centering
 \includegraphics [height=9.5cm]{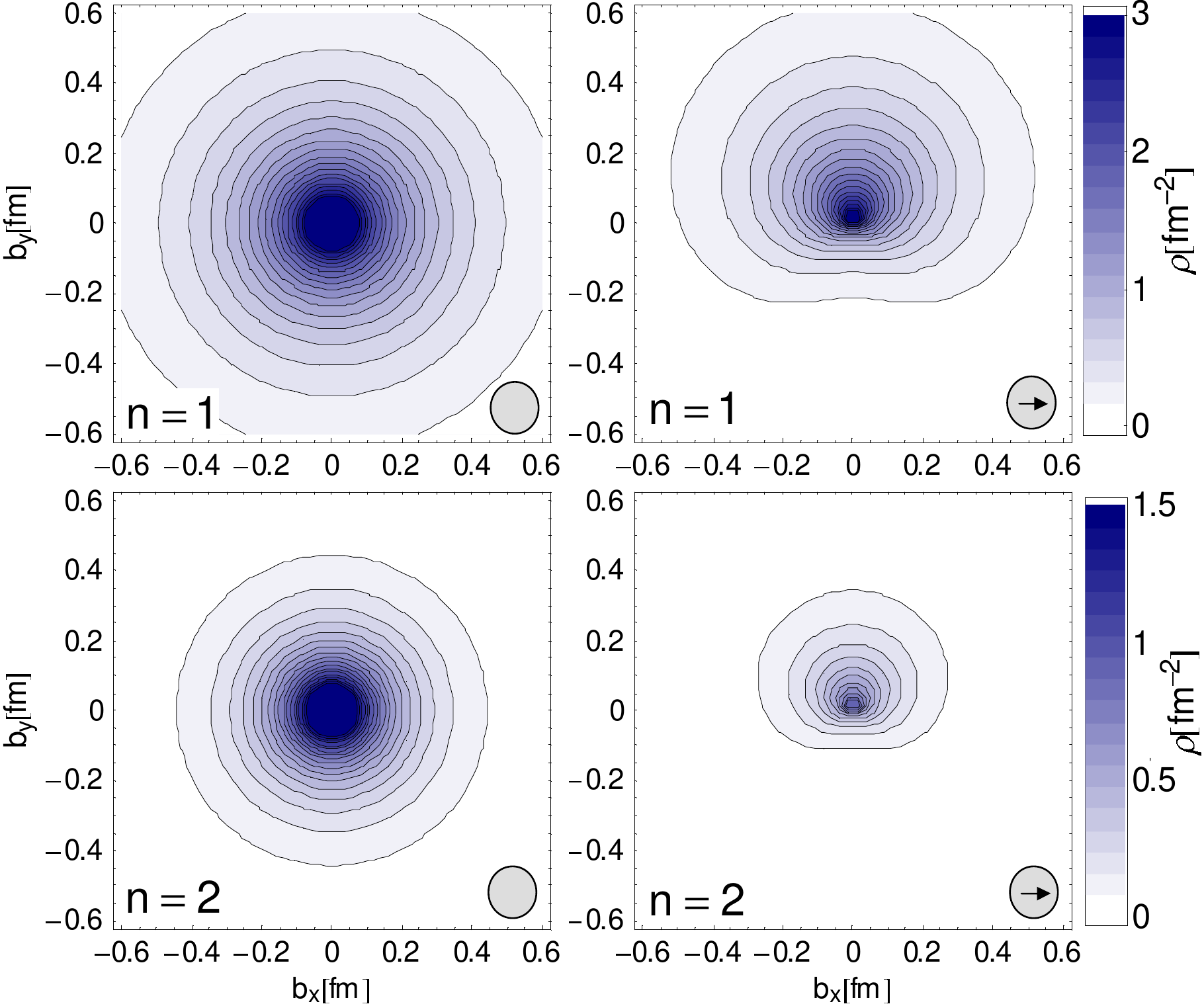}
 \caption{ The lowest moment of the densities of unpolarized (left) and transversely polarized (right) up quarks in 
a $\pi^+$ in the lattice~\cite{Brommel:2007xd}. The quark spin is oriented in the transverse plane, as indicated by the arrow.}
\label{dens-latt}
\end{figure}

In order to give such an estimation in the NJL model, 
the chiral-odd pion GPD defined in Eq.~(\ref{odd-gpd}) has to be calculated using the formalism described in Section~\ref{sec:da-cal}. In particular, the result reads
\beq
\tilde H_T^{\pi^+} (x,\xi,t)&=& i\, 8 m^3 N_c\, g^2_{\pi qq}\, \tilde I_{3}^{GPD}(x, \xi,t)\quad.
\eeq
After that, the procedure is as described above: we first Fourier transform to the Impact Parameter space, and  then match the sign conventions with the one used by, e.g.,  the Lattice Collaborations to finally yield our prediction for the sign of the Boer-Mulders function for SIDIS off  pions.
\\

A study of the Boer-Mulders function in the quark-spectator-antiquark {\it NJL-like} model is  provided in Ref.~\cite{Lu:2004hu}. The sign for the Boer-Mulders function is found to be negative.  On the other hand, a na\"ive approach of  IPD pion distribution function has been given in Ref.~\cite{Burkardt:2007xm}, where a positive sign for the distribution function was found. We thus expect a negative $h_{1\, \pi}^{\perp}$, a result that our calculation should either confirm or invalidate~\cite{courtoy:spinpion}.
\\
Lattice calculations nevertheless yield the same sign. For example, the latest lattice calculations~\cite{Brommel:2007xd} give access to $x$-moments of the quark spin densities $\rho^n( \vec{b}^{\,2}_{\perp}, s_{\perp})$. They are obtained from the generalized Form Factors in momentum space by applying the Fourier transform~(\ref{burk-xi0}). In turn, the generalized Form Factors are obtained through the usual recipe of taking the Mellin moments of the GPDs. The results are shown on Fig.~\ref{dens-latt}.
\\

\subsubsection{A Final Word\ldots}

We conclude this extended outlook by saying that the possibilities of learning something more about the partonic structure of pions and protons seem to be infinite! 
%
%
Hopefully, we will be able to unveil many secrets of the hadron's spin structure by performing simple and intuitive calculations, which can be implemented by phenomenological insights.


\appendix
\pagestyle{empty}
\chapter{Notations and Conventions}
\label{sec:convention}
\pagestyle{fancy}

In this work, natural units are adopted,
\begin{eqnarray}
\hbar=c=1\quad.
\end{eqnarray}
\\
\\
The metric tensor is
\begin{eqnarray}
g^{\mu\nu}&=&g_{\mu\nu}=\mbox{diag}(+1,-1,-1,-1)\quad.
\end{eqnarray}
\\
\\

\section{Light-Cone Vectors}

Here we first define
  the vector components in terms of light-cone variables  and then, the Sudakov parametrization \cite{baronepredazzi}.
\\
A four-vector is generically written as
\begin{eqnarray}
a^{\mu}&=&(a^0,a^1,a^2,a^3)\quad.
\label{4vector}
\end{eqnarray}
\\
We define the light-cone components 
\begin{eqnarray}
a^{\pm}&=& \frac{1}{\sqrt{2}}(a^0\pm a^3)\quad,
\end{eqnarray}
hence $a^{\mu}$ has the form
\begin{eqnarray}
a^{\mu}&=&(a^+,\vec{a}^{\perp},a^-)
\quad,
\end{eqnarray}
with $\vec{a}^{\perp}=(a^1,a^2)$.
\\
\\
The scalar product of two vectors is
\begin{eqnarray}
a.b&=& a^+b^-+a^-b^+-\vec{a}^{\perp}.\vec{b}^{\perp}\quad,
\end{eqnarray}
while the norm is
\begin{eqnarray}
a^2&=&2 a^+a^--\vec{a}^{\perp 2}\quad.
\end{eqnarray}
\\
\\
We introduce the light-like vectors, i.e. vectors which norm is zero,
\begin{eqnarray}
\bar{p}^{\mu}&=&\frac{p^+}{\sqrt{2}}(1,0,0,1)\nonumber\quad,\\
&=&p^+\,(1,\vec{0}^{\perp},0)\quad,\nonumber\\
n^{\mu}&=&\frac{1}{p^+\,\sqrt{2}}(1,0,0,-1)\nonumber\quad,\\
&=&\frac{1}{p^+}(0,\vec{0}^{\perp},1) \quad,
\label{lightvector}
\end{eqnarray}
and those vectors have the following properties
\begin{eqnarray}
\bar{p}^2=n^2&=&0\quad,\nonumber\\
\bar{p}.n&=&1\quad,\nonumber\\
n^+=p^-&=&0\quad.
\label{light-cone-vector}
\end{eqnarray}
\\
\\
It is convenient to introduce the light-cone decomposition (also called Sudakov decomposition) for a vector
\begin{eqnarray}
a^{\mu}&=& \alpha \bar{p}^{\mu}+a^{\perp \mu}+\beta n^{\mu}\nonumber\quad,\\
&=& (a.n)\bar{p}^{\mu}+a^{\perp \mu}+(a.\bar{p})n^{\mu}\quad,
\label{lightconedec}
\end{eqnarray}
with $a^{\perp \mu}=(0,\vec{a}^{\perp \mu},0)$.
\\
\\
In a ``light-cone basis'' the metric tensor $g^{\mu\nu}$ has non-zero components $g^{+-}=g^{-+}=1$ and $g^{11}=g^{22}=-1$.
Moreover we can construct the perpendicular metric tensor 
\begin{eqnarray}
g^{\mu\nu\perp}&=&g^{\mu\nu}-(\bar{p}^{\mu}n^{\nu}+\bar{p}^{\nu}n^{\mu})\quad.
\end{eqnarray}

\section{Pauli Matrices}

The Pauli matrices will often be used in the form $\vec{\tau}$, where 
\begin{equation}
\underline{\tau}=(\tau^+,\tau^3, \tau^-)\quad,
\end{equation}
with
\begin{eqnarray}
\tau^+=\frac{1}{2}(\tau^1+i \tau^2)&\,\,\,\mbox{and}\,\,\, \tau^-=\frac{1}{2}(\tau^1-i \tau^2)\quad,
\end{eqnarray}
 where $\tau^i=\sigma^i$, i.e. the usual Pauli matrices.
In the text, the notation $\vec{\tau}$ refers to 
\begin{eqnarray}
\vec{\tau}&=&(\tau^{\pi^+},\tau^{\pi^0}, \tau^{\pi^-})\quad,
\end{eqnarray}
with
\begin{eqnarray}
\tau^{\pi^+}=\frac{1}{\sqrt{2}}(\tau^1+i \tau^2)&\,\,\,\mbox{and}\,\,\, \tau^-=\frac{1}{\sqrt{2}}(\tau^1-i \tau^2)\quad.
\end{eqnarray}
\\

Hence, we have the following results
\begin{eqnarray}
&\tau^+= 
\begin{pmatrix}
0 & 1 \\   0 & 0  
\end{pmatrix}\quad,
& \tau^-=
\begin{pmatrix}
0 & 0 \\ 1 & 0 
\end{pmatrix},\\
&\tau^-\tau^+=
\begin{pmatrix}
0 & 0 \\ 0 & 1
\end{pmatrix}\quad,
&\tau^+\tau^-=
\begin{pmatrix}
1 & 0 \\ 0 & 0
\end{pmatrix}\quad.
\end{eqnarray}
\\

Therefore, the following relation can be written
\begin{eqnarray}
(\tau^{\pi^+},\tau^{\pi^0}, \tau^{\pi^-})&=&(\sqrt{2}\tau^+,\tau^3, \sqrt{2}\tau^-)\quad,\nonumber\\
\nonumber\\
&=& \left( \sqrt{2}
\begin{pmatrix}
0 & 1 \\   0 & 0  
\end{pmatrix},
\begin{pmatrix}
1 & 0 \\ 0 & -1
\end{pmatrix},
\sqrt{2}
\begin{pmatrix}
0 & 0 \\ 1 & 0 
\end{pmatrix}
\right)\quad.
\end{eqnarray}

\chapter{Discrete Symmetries}
\label{sec:sym}
\pagestyle{fancy}

In this Chapter, we treat the discrete symmetries that link the distributions for a same kind of particle among themselves.
General relations are given and then applied to the particular cases of pion DAs  and TDAs.
\\

For a generic particle ${\cal P}$ with momentum $\vec{p}$, spin projection $s_z$ and charge $Q_i$, we write the particle state $|{\cal P}(\vec{p}, s_z, Q_i) \rangle$. Parity, time reversal invariance and charge conjugation act on this generic state in the following way
\beq
P|{\cal P}(\vec{p}, s_z, Q_i) \rangle &=& \eta_P |{\cal P}(-\vec{p}, s_z, Q_i) \rangle\quad,\nonumber\\
\nonumber\\
T|{\cal P}(\vec{p}, s_z, Q_i) \rangle &=& \eta_T \,(-1)^{s-s_z} |{\cal P}(-\vec{p}, -s_z, Q_i) \rangle\quad,\nonumber\\
\nonumber\\
C|{\cal P}(\vec{p}, s_z, Q_i) \rangle &=& \eta_C  |{\cal P}(\vec{p}, s_z, -Q_i) \rangle\quad,
\label{CPT}
\eeq
with $\eta_P, \eta_T, \eta_C$ the respective phases. For the $\pi^+$, which is  $J^{PC}=0^{-+}$, are
$\eta_P=\eta_T=-1,\eta_C=1$.
\\

In the light-front, the $T$ symmetry as defined by the second expression of Eq.~(\ref{CPT}) changes $x^+$ to $-x^-$ and vice versa. It is therefore useful to define the $V$ symmetry has the combination $P_z T$, what gives
\beq
V|{\cal P}(\vec{p}, s_z, Q_i) \rangle &=& \eta_V  \,(-1)^{s-s_z}|{\cal P}(\vec{p}, -s_z, Q_i) \rangle\quad,
\label{v-sym}
\eeq
with
$\eta_V=\eta_P\eta_T$.
\\

The distribution functions for a generic transition between the state $ {\cal P}(\vec{p}, s_z, Q_i)$ to the state ${\cal P}'(\vec{p}', s'_z, Q_i)$  are associated to
\beq
&&\frac{1}{2P^+}\,{\cal G}^{{\cal P}\to{\cal P}'}_{q\to q'} (x, \vec{p}, s_z, \vec{p}', s'_z)\nonumber\\
&&=
\frac{1}{2}\,\int \,\frac{dz^-}{2\pi}\,e^{ixP^+z^-} \,\langle {\cal P}'(\vec{p}', s'_z, Q_i)| \bar \psi_{q'}\left(-\frac{z^-}{2} \right)\,\Gamma \psi_q \left(\frac{z^-}{2} \right)|{\cal P}(\vec{p}, s_z, Q_i) \rangle ,\nonumber\\
\label{generic-trans}
\eeq
where ${\cal P}$ and ${\cal P}'$ can be different particles as well as the vacuum.
\\

{\bf Let us now apply the definitions of the discrete symmetries to the generic distribution (\ref{generic-trans}) :}
\begin{itemize}

\item 
\underline{Time reversal}, as given in its light-cone version by  Eq.~(\ref{v-sym}), transforms our generic distribution  in the following way
\beq
{\cal G}^{{\cal P}\to{\cal P}'}_{q\to q'} (x, \vec{p}, s_z, \vec{p}', s'_z)
&=& \left ( \zeta_V \, \eta_V\eta'_V (-1)^{s+s'-s_z-s'_z}  \right)^{\ast}{\cal G}^{{\cal P}'\to{\cal P}}_{q'\to q} (x, \vec{p}', -s'_z, \vec{p}, -s_z)\quad,\nonumber
\eeq
with $\zeta_V$ related to the nature of the current, i.e. $\zeta_V=1$ for $\Gamma=1,i\gamma_5, \gamma_{\mu}$ while $\zeta_V=-1$ for $\Gamma=\gamma_{\mu} \gamma_5, \sigma_{\mu\nu}$.

\item
Similarly, applying \underline{charge conjugation}, Eq.~(\ref{CPT}), we find
\beq
{\cal G}^{{\cal P}\to{\cal P}'}_{q\to q'} (x, \vec{p}, s_z, \vec{p}', s'_z)
&=&  \zeta_C \, \eta_C\eta'_C  \,{\cal G}^{\bar{\cal P}\to\bar{\cal P}'}_{\bar q\to \bar q'} (-x, \vec{p}, s_z, \vec{p}', s'_z)\quad,\nonumber
\eeq
with  $\zeta_C=1$ for $\Gamma=1,i\gamma_5, \gamma_{\mu}\gamma_5$ while $\zeta_C=-1$ for $\Gamma=\gamma_{\mu} , \sigma_{\mu\nu}$.

\item
By applying both the above results, we obtain, for the \underline{$CPT$ symmetry}
\beq
{\cal G}^{{\cal P}\to{\cal P}'}_{q\to q'} (x, \vec{p}, s_z, \vec{p}', s'_z)
&=&  \zeta_{CPT}   \,{\cal G}^{\bar{\cal P}'\to\bar{\cal P}}_{\bar q'\to \bar q} (-x, \vec{p}', -s'_z, \vec{p}, -s_z)\quad,\nonumber
\eeq
where $\zeta_{CPT}=\zeta_C\zeta_V$.

\item
Let us finally give the hermitian of the generic transition (\ref{generic-trans})
\beq
\left({\cal G}^{{\cal P}\to{\cal P}'}_{q\to q'} (x, \vec{p}, s_z, \vec{p}', s'_z) \right)^{\ast}
&=&{\cal G}^{{\cal P}'\to{\cal P}}_{q'\to q} (x, \vec{p}', s'_z, \vec{p}, s_z)\quad.\nonumber
\eeq
\end{itemize}

{\bf We can now go to particular distribution functions:}

\begin{itemize}
\item 
\subsubsection{Pion Distribution Amplitude}

The pion DA would be 
\beq
{\cal G}_{u\to d}^{\pi^+ \to 0}((x-1/2), \vec{p},0)
&=&\int\frac{dz^{-}}{2\pi}e^{i(x-\frac{1}{2})p^{+}z^{-}}\left.
\left\langle 0\right\vert
\bar{q}\left(  -\frac{z}{2}\right) \gamma^+\gamma_{5}%
\tau^{-}q\left(  \frac{z}{2}\right)  \left\vert \pi^+\left( p\right)
\right\rangle \right\vert
_{z^{+}=z^{\bot}=0}\nonumber\\
&=&
i\,\sqrt{2} f_{\pi}\, \phi^{\pi^+}(x)\quad.
\eeq
\\
The hermitian conjugate of ${\cal G}^{\pi^+ \to 0}((x-\frac{1}{2}), \vec{p},0)$ is given by
\beq
{\cal G}_{d\to u}^{0\to\pi^+ }((x-1/2), 0,\vec{p})&=& -i\,\sqrt{2} f_{\pi}\, \left( \phi^{\pi^+}(x)\right)^{\ast}\quad.\nonumber
\eeq
\\
 Using the relations Eq.~(\ref{CPT}), the time reversal invariance leads to
 \beq
\left ( \phi^{\pi^+}(x) \right)^{\ast} &=& \phi^{\pi^+}(x)\quad;\nonumber
 \eeq
 charge conjugation to
  \beq
 \phi^{\pi^-}(x) &=& \phi^{\pi^+}(1-x)\quad;
 \label{CC}
 \eeq
and $CPT$ to
 \beq
\left ( \phi^{\pi^-}(1-x) \right)^{\ast} &=& \phi^{\pi^+}(x)\quad.\nonumber
 \eeq

\item 
\subsubsection{Vector Pion Transition Distribution Amplitudes}

The $\pi^+$-$\gamma$ and $\gamma$-$\pi^-$ vector TDAs are defined
\beq
{\cal G}^{\pi^+\to \gamma}_{u\to d} (x, \vec{p},  \vec{p}', \varepsilon)
&=& \int\frac{dz^{-}}{2\pi}e^{ixP^{+}z^{-}}  \left\langle \gamma( p_{\gamma}\varepsilon)
  \left\vert \bar{q}\left( -\frac{z}{2}\right) \gamma
  ^{+}\hspace*{-0.05cm}\tau^{-}q\left( \frac{z}{2}\right)
  \right\vert\pi^{+}(p_{\pi})
  \right\rangle \vert _{z^{+}=z^{\bot}=0} \nonumber\\ 
  \nonumber\\
  &=& \frac{1}{P^+} i\,e\,\varepsilon_{\nu}\,\epsilon^{+\nu\rho\sigma}\,P_{\rho}\,(p_{\gamma}-p_{\pi})_{\sigma
}\,\frac{V^{\pi^+\to\gamma}(  x, \xi,t)  }{\sqrt{2}f_{\pi}}\nonumber\quad,\nonumber\\
\nonumber\\
{\cal G}^{ \gamma\to \pi^-}_{\bar d\to \bar  u} (x, \vec{p}',  \vec{p}, \varepsilon)
&=&\int\frac{dz^{-}}{2\pi}e^{ixP^{+}z^{-}} \left\langle \pi^{-}(p_{\pi})
  \left\vert \bar{q}\left( -\frac{z}{2}\right) \gamma
  ^{+}\hspace*{-0.05cm}\tau^{-}q\left( \frac{z}{2}\right)
  \right\vert\gamma( p_{\gamma}\varepsilon)
  \right\rangle \vert _{z^{+}=z^{\bot}=0} \nonumber\\\nonumber\\
&=&\frac{1}{P^+} i\,e\,\varepsilon_{\nu}\,\epsilon^{+\nu\rho\sigma}\,P_{\rho}\,(p_{\pi}-p_{\gamma})_{\sigma
}\,\frac{V^{\gamma\to\pi^{-}}(  x,- \xi,t)  }{\sqrt{2}f_{\pi}}\nonumber\quad.
\nonumber
\eeq 
\\
The skewness variable is defined as
\beq
\xi&=&\left(
p_{\pi}-p_{\gamma}\right)  ^{+}/2P^{+}\quad.\nonumber
\eeq
\\
\\
The hermitian conjugate of the $\pi^+\to\gamma$ vector TDA is given by
\beq
\left({\cal G}^{\pi^+\to \gamma}_{u\to d} (x, \vec{p},  \vec{p}', \varepsilon) \right)^{\ast}
&=& {\cal G}_{d\to u}^{\gamma\to\pi^+ }(x, \vec{p}',\vec{p},\varepsilon)\quad;\nonumber\\
\nonumber
\eeq
\begin{equation}
\Rightarrow \boxed{\left(V^{\pi^+\to\gamma}(  x, \xi,t)  \right)^{\ast}=V^{\gamma\to\pi^+}(  x, -\xi,t) }\quad.\nonumber
\end{equation}
\\
Time reversal invariance leads to
\beq
{\cal G}^{\pi^+\to \gamma}_{u\to d} (x, \vec{p},  \vec{p}', \varepsilon) 
&=& \eta_V\,{\cal G}_{d\to u}^{\gamma\to\pi^+ }(x, \vec{p}',\vec{p},\varepsilon)\quad,\nonumber\\
\nonumber\\
i\,e\,\varepsilon_{\nu}\,\epsilon^{\mu\nu\rho\sigma}\,n_{\mu} &&\hspace{-6mm}P_{\rho}\,\frac{(p_{\gamma}-p_{\pi})_{\sigma
}}{\sqrt{2}f_{\pi}}\,V^{\pi^+\to\gamma}(  x, \xi,t)  \nonumber\\
&&=- i\,e\,\varepsilon_{\nu}\,\epsilon^{\mu\nu\rho\sigma}\,n_{\mu} P_{\rho}\,\frac{(p_{\pi}-p_{\gamma})_{\sigma
}}{\sqrt{2}f_{\pi}}\,V^{\gamma\to\pi^+}(  x, -\xi,t)  \quad,
\nonumber
\eeq
\begin{equation}
\boxed{V^{\pi^+\to\gamma}(  x, \xi,t)= V^{\gamma\to\pi^+}(  x, -\xi,t)}\quad;\nonumber
\end{equation}
charge conjugation to
\beq
{\cal G}^{\pi^+\to \gamma}_{u\to d} (x, \vec{p},  \vec{p}', \varepsilon) 
&=& \zeta_C\,\eta_C\,{\cal G}_{\bar u\to \bar d}^{\pi^-\to \gamma }(-x, \vec{p},\vec{p}',\varepsilon)\quad,\nonumber\\
\nonumber\\
i\,e\,\varepsilon_{\nu}\,\epsilon^{\mu\nu\rho\sigma}\,n_{\mu} &&\hspace{-6mm}P_{\rho}\,\frac{(p_{\gamma}-p_{\pi})_{\sigma
}}{\sqrt{2}f_{\pi}}\,V^{\pi^+\to\gamma}(  x, \xi,t)  \nonumber\\
&&=+ i\,e\,\varepsilon_{\nu}\,\epsilon^{\mu\nu\rho\sigma}\,n_{\mu} P_{\rho}\,\frac{(p_{\gamma}-p_{\pi})_{\sigma
}}{\sqrt{2}f_{\pi}}\,V^{\pi^-\to\gamma}( -x, \xi,t)  \quad,
\nonumber
\eeq
\begin{equation}
\boxed{V^{\pi^+\to\gamma}(  x, \xi,t)= V^{\pi^-\to\gamma}(  -x, \xi,t)}\quad;\nonumber
\end{equation}
and $CPT$ to
\beq
{\cal G}^{\pi^+\to \gamma}_{u\to d} (x, \vec{p},  \vec{p}', \varepsilon) 
&=& \zeta_{CPT}\,\eta_C\eta_V\,{\cal G}_{\bar d\to \bar u}^{\gamma \to \pi^-}(-x, \vec{p}',\vec{p},\varepsilon)\quad,\nonumber\\
\nonumber\\
i\,e\,\varepsilon_{\nu}\,\epsilon^{\mu\nu\rho\sigma}\,n_{\mu} &&\hspace{-6mm}P_{\rho}\,\frac{(p_{\gamma}-p_{\pi})_{\sigma
}}{\sqrt{2}f_{\pi}}\,V^{\pi^+\to\gamma}(  x, \xi,t)  \nonumber\\
&&=- i\,e\,\varepsilon_{\nu}\,\epsilon^{\mu\nu\rho\sigma}\,n_{\mu} P_{\rho}\,\frac{(p_{\pi}-p_{\gamma})_{\sigma
}}{\sqrt{2}f_{\pi}}\,V^{\gamma\to\pi^-}( -x, -\xi,t)  \quad,
\nonumber
\eeq
\begin{equation}
\boxed{V^{\pi^+\to\gamma}(  x, \xi,t)= V^{\gamma\to\pi^-}(  -x, -\xi,t)}\quad.\nonumber
\end{equation}

\item 
\subsubsection{Axial Pion Transition Distribution Amplitudes}

The $\pi^+$-$\gamma$ and $\gamma$-$\pi^-$ axial TDAs are defined
So that
\beq
&&e\,\left(\vec{\varepsilon}^{\perp}\cdot(\vec{p}_{\gamma}^{\perp}-\vec{p}_{\pi}^{\perp})\right) \frac{A^{\pi^+\to\gamma}(  x, \xi,t)  }{\sqrt{2}f_{\pi}}\nonumber\\
&&={\cal G}^{\pi^+\to \gamma}_{u\to d} (x, \vec{p},  \vec{p}', \varepsilon)
+i{\cal G}^{\pi^+\to 0}_{u\to d} \left(\frac{x}{2\xi}, \vec{p},  0\right)\,2e\, \epsilon(\xi)\,\frac{\varepsilon\cdot(p_{\gamma}-p_{\pi})}{m_{\pi}^2-t}\quad,\nonumber\\
\nonumber\\
&&e\,\left(\vec{\varepsilon}^{\perp}\cdot(\vec{p}_{\pi}^{\perp}-\vec{p}_{\gamma}^{\perp})\right) \frac{A^{\gamma\to\pi^-}(  x, -\xi,t)  }{\sqrt{2}f_{\pi}}\nonumber\\
&&={\cal G}^{\gamma\to\pi^-}_{\bar d\to \bar u} (x, \vec{p}',  \vec{p}, \varepsilon)
-i{\cal G}^{0\to\pi^-}_{\bar d\to \bar u} \left(\frac{x}{2\xi}, \vec{p},  0\right)\,2e\,\epsilon(-\xi)\, \frac{\varepsilon\cdot(p_{\pi}-p_{\gamma})}{m_{\pi}^2-t}\quad,\nonumber
\eeq
with $\epsilon(\xi)=1$ for $\xi>0$ and  $\epsilon(\xi)=-1$ for $\xi<0$.
\\
The skewness variable is defined as
\beq
\xi&=&\left(
p_{\pi}-p_{\gamma}\right)  ^{+}/2P^{+}\quad.\nonumber
\eeq
\\
\\
Combining the results for ${\cal G}^{\pi^+\to \gamma}_{u\to d} (x, \vec{p},  \vec{p}', \varepsilon)$ and for ${\cal G}^{\pi^+\to 0}_{u\to d} \left(\frac{x}{2\xi}, \vec{p},  0\right)$ we find the expressions for the axial TDAs.
\\
The hermitian conjugate of the $\pi^+\to\gamma$ axial TDA is given by
\beq
&&\left(e\,\left(\vec{\varepsilon}^{\perp}\cdot(\vec{p}_{\gamma}^{\perp}-\vec{p}_{\pi}^{\perp})\right) \frac{A^{\pi^+\to\gamma}(  x, \xi,t)  }{\sqrt{2}f_{\pi}} +  \,e\,\left(\varepsilon\cdot(p_{\gamma}-p_{\pi})\right) \,\frac{2\sqrt{2}f_{\pi}}{m_{\pi}^2-t}\,\epsilon(\xi) \, \phi\left(\frac{x+\xi}{2\xi}\right)\right)^{\ast}\nonumber\\
&&=e\,\left(\vec{\varepsilon}^{\perp}\cdot(\vec{p}_{\pi}^{\perp}-\vec{p}_{\gamma}^{\perp})\right) \frac{A^{\gamma\to\pi^+}(  x, -\xi,t)  }{\sqrt{2}f_{\pi}} +  \,e\,\left(\varepsilon\cdot(p_{\pi}-p_{\gamma})\right) \,\frac{2\sqrt{2}f_{\pi}}{m_{\pi}^2-t}\,\epsilon(-\xi) \, \phi\left(\frac{x+\xi}{2\xi}\right)\nonumber\quad,
\eeq
\begin{equation}
\Rightarrow \boxed{\left(A^{\pi^+\to\gamma}(  x, \xi,t)  \right)^{\ast}=A^{\gamma\to\pi^+}(  x, -\xi,t) }\quad.\nonumber
\end{equation}
\\
Time reversal invariance leads to
\begin{equation}
\boxed{A^{\pi^+\to\gamma}(  x, \xi,t)= A^{\gamma\to\pi^+}(  x, -\xi,t)}\quad;\nonumber
\end{equation}
charge conjugation to
\begin{equation}
\boxed{A^{\pi^+\to\gamma}(  x, \xi,t)= A^{\pi^-\to\gamma}(  -x, \xi,t)}\quad;\nonumber
\end{equation}
and $CPT$ to
\begin{equation}
\boxed{A^{\pi^+\to\gamma}(  x, \xi,t)= A^{\gamma\to\pi^-}(  -x, -\xi,t)}\quad.\nonumber
\end{equation}

\end{itemize}

Those relations remain unchanged under evolution.

\chapter{The Nambu - Jona-Lasinio Model}
\label{sec:njl}
\pagestyle{fancy}

A  pre-QCD model for the pion was introduced in 1961 by Y. Nambu and G. Jona-Lasinio with
nucleon degrees of freedom~\cite{Nambu:1961tp, Nambu:1961tp-2}. This study was motivated by the theory of superconductivity of Bardeen, Cooper and Schrieffer (BCS) and so built 
in analogy with it.
The basic idea of the theory is that the mass of the ``Dirac particles'', in the case of the original paper of Nambu and Jona-Lasinio a nucleon, could be due to some interaction between massless fermions that have the same 
quantum numbers  excepted chirality. In the same fashion as in the theory of superconductivity,  the gap is caused by nucleon-nucleon  interaction. Therefore, starting from a massless nucleon, i.e. an eigenstate of chirality, we arrive to a massive nucleon.
This is the ``solid state physics'' formulation of the chiral symmetry breaking.
\\
\\

The lagrangian was contructed to conserve chirality and ``nucleon number''. In quark language, more precisely in a two-flavor quark model, it has to be understood 
as $SU_V(2)\times SU_A(2)\times U_V(1) $ in the  second version of the model which exclude $U_A(1)$ in accordance with experiment. 
 Transformations under the isospin 
$SU_V(2)$  symmetry and
the chiral  $SU_A(2)$ symmetry  are respectively given by
\begin{eqnarray} 
\psi\rightarrow e^{-i\tau.\frac{\omega}{2}}\psi\quad,\quad
\psi\rightarrow e^{-i\tau.\frac{\theta}{2}\gamma_5}\psi\quad,
\end{eqnarray}
with $\omega$ and $\theta$ the respective generators.
\\
The NJL lagrangian density  in its 2-flavor version is
\begin{eqnarray}
{\cal L}_{NJL}&=&i \bar \psi \thru \partial \psi+ G\Big[ (\bar \psi \psi)^2 +(\bar \psi i \gamma_5 \vec{\tau}\psi)^2\Big] \quad,
\label{NJL}
\end{eqnarray}
with $\psi=
\begin{pmatrix}
u \\  d  
\end{pmatrix}$.

\section{The Chiral Symmetry Breaking}
\label{sec:ChiSB}

Nowadays it makes sense to speak about quark degrees of freedom instead of nucleons.
The quark mass topic is quite puzzling: quark models give us high values while chiral symmetry seems to impose small quark mass.
This can be understood by considering the average interaction of a quark with the other quarks in the mean-field approximation.
The concept of constituent quark mass can then be introduced in opposition to current quark mass.
The quark self-energy in a mean-field approximation is given by the Hartree-Fock (HF) approximation, \cf{hf}.
 This consists in considering perturbations around the free ``massive''
lagrangian $ {\cal L}'_0$,  i.e.
\begin{eqnarray}
{\cal L}'_0&=& {\cal L}_0+{\cal L}_{\Sigma}\quad,
\end{eqnarray}
where the self-energy lagrangian ${\cal L}_{\Sigma}=-\Sigma\, \bar \psi\psi$ is chosen in such a way that the interaction lagrangian does
 not contain any self-energy effect.

\begin{figure}[H]
\centering
 \includegraphics [height=2.5cm]{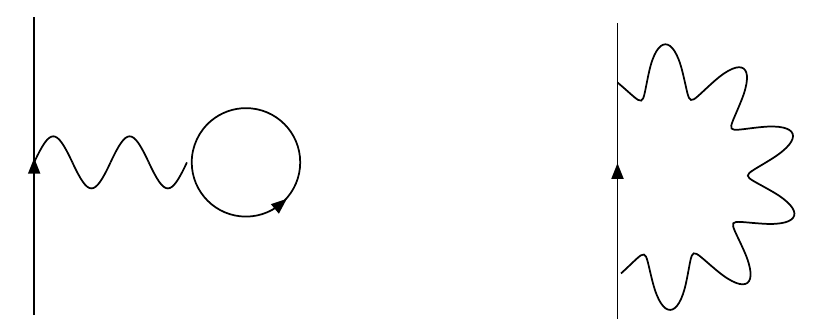}
 \caption[Self-energy]{\small{The Hartree and Fock, respectively direct and exchange, contributions to the quark self-energy.}}
\label{hf}
\end{figure}

The total self-energy $\Sigma$ is the sum of the scalar and pseudoscalar ones and is given by, in the HF approximation,
\begin{eqnarray}
\Sigma&=& \Sigma^s+\Sigma^{ps}\quad,\nonumber\\
&=& 2G\,\Big\{
\alltrace \,iS(x,x)-iS(x,x)+(i\gamma_5 \vec{\tau})\alltrace\, iS(x,x)(i\gamma_5 \vec{\tau})-(i\gamma_5 \vec{\tau}) iS(x,x)(i\gamma_5 \vec{\tau})
\Big\}\quad,
\label{self}
\end{eqnarray}
where $S(x,x)$ is the  quark propagator and satisfies
\begin{eqnarray}
\left(i\thru\partial_x-m_0-\Sigma\right)S(x,x')&=& \delta^4(x-x')\quad.
\label{propagator}
\end{eqnarray}
The current quark mass $m_0$ comes from $ {\cal L}_0$. 
\\
Given the ``observed'' constituent quark mass $m$, the propagator  $S(x,x)$ should also satisfy the following equation of motion
\begin{eqnarray}
\left(i\thru\partial_x-m\right)S(x,x')&=& \delta^4(x-x')\quad,
\end{eqnarray}
so that $m-m_0=\Sigma$. The solution of \ce{propagator} in momentum space is then,
\begin{eqnarray}
S(p)&=& \frac{\thru p+m}{p^2-m^{ 2}}\quad.
\end{eqnarray}
If we insert this result into the expression for the self-energy \ce{self}, performing the traces and noting that the integral odd in $p$ does not contribute,
 we find
\begin{eqnarray}
\Sigma=m-m_0&=&2iG\,(N_c N_f+\frac{1}{2})\,\int\,\frac{d^4 p}{(2\pi)^4}\frac{m}{p^2-m^{ 2}}\quad.
\label{gap1}
\end{eqnarray}
At this point, it is a reasonable assumption to set the small current quark mass to $m_0=0$.  
Even if the exact chiral symmetry forbids mass terms in the Lagrangian, there exists a solution of the type $m\neq0$ when the current quark mass is zero.
Actually
the equation \ce{gap1} admits two solutions. The trivial one is $m=0$ and is in agreement with symmetry requirements. The second and nontrivial one is %
\begin{eqnarray}
1&=&2iG\,N_cN_f\,\int\,\frac{d^4 p}{(2\pi)^4}\frac{1}{p^2-m^{ 2}}\quad.
\label{gap}
\end{eqnarray}
This equation  is called the self-consistency or gap equation in analogy with the nontrivial solution giving rise to the gap in the BCS theory.
The non-trivial solution, as solution of the gap equation \ce{gap}, exists if the value of the coupling strength $G$ exceeds a critical value \cite{Nambu:1961tp, Nambu:1961tp-2}.

Regarding the coefficient in front of the integral in~(\ref{gap1}), we observe that the first term, proportional to $G\, N_c$, comes from the direct diagram of Fig.~\ref{hf} (Hartree contribution), and that the second one, proportional to $G$, is the contribution of the exchange diagram (Fock contribution). We can understand the NJL model as an effective theory connected to QCD in the large $N_c$ limit. In that case, $G\, N_c\sim {\cal O}(1)$ and $G\sim N_c^{-1}$. Therefore, since we can consider that we work in a $N_c^{-1}$-expansion,  it is consistent to ignore the Fock contribution.
\\

It would make sense to work out which of the two solutions is the ground-state solution. By analysing the vacua of both the trivial and non-trivial solutions,
we can related  them in  a BCS variational vacuum fashion. The vacuum for the non-trivial solution $|0\rangle_{m}$
 is expressed as a superposition of pairs of zero total momentum and zero total helicity over the trivial vacuum $|0\rangle_0$ \cite{Nambu:1961tp, Nambu:1961tp-2}.
 We consider the Dirac equation in both cases, and with similar initial conditions,
\begin{eqnarray}
i\thru\partial \psi^0(x)&=&0\quad,\nonumber\\
\left(i\thru\partial+m\right)\psi^{m}(x)&=&0\quad,
\end{eqnarray}
and decompose the respective fields into their Fourier's components
\begin{eqnarray}
\psi^{\alpha}(x)&=& \sum_s\int\,\frac{d^4 p}{(2\pi)^4}\,\Big\{
b^{\alpha}(\vec{p},s)u^{\alpha}(\vec{p},s)e^{ipx}+d^{\dagger \alpha}(\vec{p},s)v^{\alpha}(\vec{p},s)e^{-ipx}
\Big\}\quad,
\end{eqnarray}
for $\alpha=0$ or $ m $.
The spinors for particles and antiparticles are respectively $u^{\alpha}(\vec{p},s)$ and  $v^{\alpha}(\vec{p},s)$. The annihilation operators for particles
 and antiparticles are respectively $b^{\alpha}(\vec{p},s)$ and  $d^{\alpha}(\vec{p},s)$.
\\
The annihilation operators for the fields $\psi^0$ and $\psi^{m}$ are related to each other according to they have the same initial conditions.
Therefore the vacua defined with respect to both fields, i.e.
\begin{eqnarray}
b^0(\vec{p},s)|0\rangle_0=d^0(\vec{p},s)|0\rangle_0&=&0\nonumber\quad,\\
b^{m}(\vec{p},s)|0\rangle_{m}=d^{m}(\vec{p},s)|0\rangle_{m}&=&0\quad,
\end{eqnarray}
can also be related one to each other. It can be found \cite{Klevansky:1992qe}
\begin{eqnarray}
|0\rangle_{m}&=& \prod_{p,s} \Big\{
\left(
\cos \left[\frac{\phi (p)}{2}\right]+ s\, \sin \left[\frac{\phi (p)}{2}\right]
\right)\,b^{\dagger 0}(\vec{p},s) d^{\dagger 0}(-\vec{p},s)
\Big\}|0\rangle_0\quad.
\end{eqnarray}
This is a BCS-like variational vacuum: a superposition of pairs of zero total momentum and zero total helicity. This is also called the Bogoliubov-Valatin approach 
because the operators that annihilate such a vacuum are found through a Bogoliubov-Valatin transformation \cite{Klevansky:1992qe}.
The angle $\phi (p)$ is determined by minimizing the vacuum energy $W[\phi (p)]_m=_m\langle 0|H|0\rangle_m$, obtaining $\tan\left[\phi (p)\right]= m/p$.
\\

The energy difference of the two vacua  $W_{m}-W_0$ is found to be negative \cite{Nambu:1961tp, Nambu:1961tp-2}. So $|0\rangle_{m}$ is the true vacuum and the nontrivial solution
 is the solution that minimizes the energy. In particular, minimizing the vacuum expectation value in the nontrivial solution case leads
to the gap equation \ce{gap}.
\\

The true vacuum $|0\rangle_{m}$ is not invariant under $SU_A(2)$. The consequence is a (dynamical) breaking of the chiral symmetry  $SU_A(2)$. In other words
the interaction given by \ce{NJL} between two massless quark with the same quantum numbers excepted chirality gives rise to the gap equation \ce{gap} and generates the 
constituent quark mass $m$.

\section{Regularization}
\label{sec:regunjl}

The gap equation \ce{gap}, 
\begin{eqnarray}
1&=&2G\,N_cN_f\,I_1\quad,
\label{gap2}
\end{eqnarray}
 has not been resolved yet.
In fact the integral $I_1$, given by \ce{i1}, is divergent  and we need to define a regularization scheme.
The NJL interaction is a contact interaction that renders the theory nonrenormalizable.
Therefore the parameters introduced by the regularization scheme are integrally part of the model in itself. The NJL model plus its regularization scheme is regarded as
 an effective theory of QCD.
\\

Several regularization schemes have been studied. We mention  the noncovariant 3-momentum cutoff scheme, the covariant 4-momentum cutoff scheme or the proper-time 
regularization \cite{Klevansky:1992qe}.
\\

In the study of PDFs, GPDs and TDAs, it is important to preserve Lorentz covariance and gauge invariance. The Pauli-Villars scheme fulfills these conditions.
The idea of this scheme is to introduce the minimum number of regulating masses $m_j$ and constants $c_j$ in such a way that the result becomes finite.
The following replacement is made
\begin{eqnarray}
\int \,\frac{d^4p}{(2\pi)^4}\,f(p,m^{ 2})&\rightarrow& \int \,\frac{d^4p}{(2\pi)^4}\,\sum_{j=0}^2\,c_j\,f(p,m^2_j)\quad,
\label{PV}
\end{eqnarray}
with $m^2_j=m^{ 2}+j\Lambda^2$. We use the standard values given in \cite{Itzykson:1980rh} for the $c_j$ coefficients, i.e. $c_0=c_2=1$ and $ c_1=-2$.
 $\Lambda$ and $m$ are the regularization parameters. They are determined by calculating the pion decay constant $f_{\pi}$ and the quark condensate
\cite{Klevansky:1992qe}.
For $f_{\pi}=93$MeV and $\langle \bar u u \rangle=\langle \bar d d \rangle=-(250\,\mbox{MeV})^3$, we obtain
\begin{eqnarray}
m&=& 241 \quad\mbox{MeV}\quad,\nonumber\\
\Lambda&=& 859\quad\mbox{MeV}\quad.
\end{eqnarray}

\section{The Pion and the Sigma}
\label{sec:pion}

In its two-flavor version, the NJL model  \ce{NJL} gives rise to 4 mesonic modes.
\\

 The scalar interaction is responsible for a  isoscalar ($I=0$) scalar
 $J^{PC}=0^{++}$ mode.
The scalar mode is associated with the $\sigma$-meson, also known as $f_0(600)$ \cite{Amsler:2008zz}. The $0^{++}$ sector is both theoretically and experimentally
 complex. The scalar modes are difficult to solve because of their large decay widths.

The pseudoscalar channel is responsible for the isovector ($I=1$) pseudoscalar  $J^{PC}=0^{-+}$ modes identified as the pions which are then the  three Goldstone
 bosons coming from the chiral symmetry breaking.
\\

There are two ways of expressing the pion. In the original paper \cite{Nambu:1961tp, Nambu:1961tp-2}, Nambu and Jona-Lasinio consider the pion as a bound-state of two nucleons, what
is nowadays reinterpreted as a bound-state of two quarks, and study it through the use of the Bethe-Salpeter (BS) equation in the ladder approximation.
Another approach is to consider the effective interaction resulting from the exchange of a pion in the Random Phase Approximation (RPA).
The two methods are identical.
\\

Even though both approaches will be used in the next chapters, according to we are interested in pions as bound-states or in pion exchange,
 we choose to introduce here the pion as a bound-state through the Bethe-Salpeter equation.
\\
\\

The Bethe-Salpeter (BS) amplitude for the bound-state  $\pi$ with four-momentum $P$ is given by
\begin{eqnarray}
\vec{\chi}_P(x_1,x_2)&=& \langle 0|T q(x_1)\bar q(x_2)|\vec{\pi}(P)\rangle  \quad.
\label{bsampl}
\end{eqnarray}
The NJL model contains  a four-fermion interaction. In the ladder approximation, only direct diagrams are considered.
 In our case, this approximation is equivalent to considering the iteration of the simplest closed loop \cf{ladder} with the kernel
$V(p,p';P)=2i G (i\, \gamma_5 \vec{\tau})^2$ . The equation of Bethe-Salpeter is easy to handle due to the point-like interaction of the NJL model.
\begin{figure}[H]
\centering
 \includegraphics [height=1.5cm]{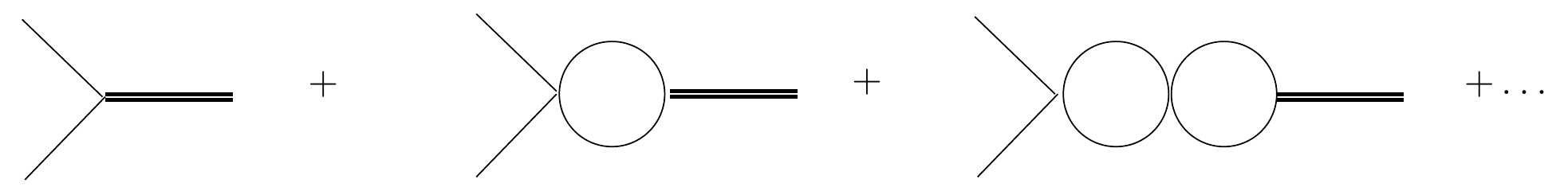}
\caption[Ladder Approximation]{\small{Graphs corresponding to the BS equation in the ladder approximation. The thick line represents the bound-state.}}
\label{ladder}
\end{figure}

The integral equation generated by the chain of diagrams \cf{ladder} is the Bethe-Salpeter equation \cite{Itzykson:1980rh}
\begin{eqnarray}
 i S^{-1}(p)\,\,\vec{\chi}_P(p)\,\, iS^{-1}(p-P)+\int\,\frac{d^4p''}{(2\pi)^4}\,\alltrace \,\left(V(p,p'';P)\vec{\chi}_P(p'')\right)&=&0\quad,
\label{bs}
\end{eqnarray}
here in momentum space.  $S^{-1}(k)=\thru k-m$ is the inverse of the Feynman quark propagator.
\\

 Solving the Bethe-Salpeter equation in the ladder approximation
\begin{eqnarray}
 i S^{-1}(p)\,\vec{\chi}_P(p)\, iS^{-1}(p-P)&=& G\gamma_5 \vec{\tau} \,\int\,\frac{d^4p''}{(2\pi)^4}\,2\alltrace\,\left(i\,\gamma_5 \vec{\tau}.\vec{\chi}_P(p'')\right)
\quad,
\label{bs-njl}
\end{eqnarray}
we find the expression for the Bethe-Salpeter amplitude
\begin{eqnarray}
\vec{\chi}_P(p)&=&-g_{\pi qq} \,iS(p)\,\gamma_5 \vec{\tau} \,iS(p-P)\quad.
\label{chi}
\end{eqnarray}
\\

Inserting the Bethe-Salpeter amplitude \ce{chi} back into the BS equation \ce{bs} gives a self-consistency condition
\begin{eqnarray}
1&=& 2G\,(-i)\int \frac{d^4p}{(2\pi)^4} \alltrace\left(\gamma_5\tau^{\pi^-}\,S(p)\,\gamma_5\tau^{\pi^+}\,S(p-P)\right)\quad.
\label{selfconsist}
\end{eqnarray}
The integral represents the pseudoscalar proper polarization $\Pi_{ps}$ shown in \cf{properpol}. The mass of the pseudoscalar mode is obtained for $P^2=m_{\pi}^2$.
That is, the mass of the pion is found by solving the following equation%
\begin{eqnarray}
1-2G\,\Pi_{ps}(m_{\pi}^2)=0\quad.
\label{polo}
\end{eqnarray}
\\

We have completly defined the pion. By resolving \ce{polo} at $k^2=m_{\pi}^2$,  the mass $m_{\pi}$ is found. %
Now we would like to show that the pion described in such a way is the researched Goldstone boson, i.e. that it has zero mass in absence of the current mass quarks.
In order to solve such a problem, we allow a small current quark mass in the Lagrangian \ce{NJL} of the type $-m_0\bar\psi\psi$ and  solve \ce{polo}.
\begin{figure}[H]
\centering
 \includegraphics [height=2cm]{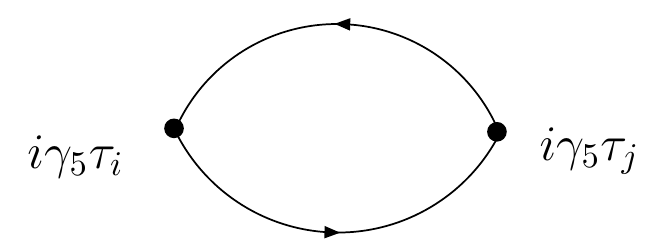}
\caption[$\Pi_{ps}$]{\small{The  pseudoscalar proper polarization  $\Pi_{ps}(k^2)$.}}
\label{properpol}
\end{figure}
The proper polarization  \cf{properpol}  is explicitly given by 
\begin{eqnarray}
\Pi_{ps}(k^2)&=&-i\int\,\frac{d^4p}{(2\pi)^4}\,\alltrace\Big\{
i\gamma^5\tau^{\pi^-}\,i\, \frac{{p\hspace{-0.5em}\raisebox{-0.2ex}{/}}+\frac{1}{2}{k\hspace{-0.5em}\raisebox{-0.2ex}{/}}+m}{(p+\frac{k}{2})^2-m^{ 2}}\,
\,i\gamma^5\tau^{\pi^+}\,i\, \frac{{p\hspace{-0.5em}\raisebox{-0.2ex}{/}}-\frac{1}{2}{k\hspace{-0.5em}\raisebox{-0.2ex}{/}}+m}{(p-\frac{k}{2})^2-m^{ 2}}
\Big\}\quad,\nonumber\\\nonumber\\
&=& -i\,4N_c N_f \,\int\,\frac{d^4p}{(2\pi)^4}\,\frac{m^{ 2}-p^2+\frac{k^2}{4}}{[(p+\frac{k}{2})^2-m^{ 2}][(p-\frac{k}{2})^2-m^{ 2}]}\quad,
\label{polar}
\end{eqnarray}
where the change of variables $p=p+k/2$, with  $P=k$,  has been performed.
We bring \ce{polar} into elementary integrals so that we can easily extract the pion mass $m_{\pi}$.
The numerator can be reexpressed like $(m^{ 2}-p^2-\frac{k^2}{4})+\frac{k^2}{2}$ and the denominator
\begin{equation}
  \frac{1}{[(p+\frac{k}{2})^2-m^{ 2}][(p-\frac{k}{2})^2-m^{ 2}]}=\frac{1}{2[p^2+\frac{k^2}{4}-m^{ 2}]}\,\left(\frac{1}{[(p+\frac{k}{2})^2-m^{ 2}]}+
\frac{1}{[(p-\frac{k}{2})^2-m^{ 2}]} \right)\quad.
\end{equation}
So the pseudoscalar polarization \ce{polar} takes the form\footnote{A shift has been performed in order to obtain the $2I_1$ term. This  still holds if one uses 
the Pauli-Villars regularization scheme. What is not obvious in other schemes.}
\begin{eqnarray}
\Pi_{ps}(k^2)&=& 2N_c N_f[2I_1-k^2I_2(k^2)]\quad,
\label{pol21}
\end{eqnarray}
with $I_1$ and $I_2(k^2)$ respectively one and two-propagator integrals, see Appendix \ref{sec:integral}.  Moreover, the self-consistency 
condition \ce{selfconsist} gives us the gap equation \ce{gap2} when the mass of the bound-state is zero, i.e. $k^2=0$.
\\

The first term on the right-hand side of \ce{pol21} can be replaced by its expression in the gap equation \ce{gap}.
Adding a small current quark mass term to the gap equation, we find for $I_1$
\begin{eqnarray}
8\,G\,N_c N_f\,I_1&=&1-\frac{m_0}{m^{}}\quad.
\end{eqnarray}
In doing so we obtain, using \ce{polo},
\begin{eqnarray}
1-2G\Pi_{ps}(k^2)&=&\frac{m_0}{m^{}}+4\,G\,N_c N_f\, k^2\,I_2(k^2)\quad,\\
&=&0\quad.
\end{eqnarray}
For $k^2=m^2_{\pi}$,  we may deduce the mass of the pion
\begin{eqnarray}
 m_{\pi}^2&=& - \frac{m_0}{m}\,\frac{1}{4\,G\,N_c N_f\,I_2( m_{\pi}^2)}\quad.
\label{mpi}
\end{eqnarray}
\\
In the chiral limit, i.e. in the absence of current quark mass $m_0$, the mass of the pseudoscalar mode vanishes, as expected.
\\

The quark-pion coupling constant $g_{\pi qq}$ has been determined by the standard normalization of
 the Bethe-Salpeter equation \cite{Itzykson:1980rh}
\begin{eqnarray}
g_{\pi qq}^2&=& \frac{-1}{12 \left(I_2(m_{\pi}^2)+m_{\pi}^2\left( \frac{\partial I_2(P)}{\partial P^2}\right)|_{P^2=m_{\pi}^2}\right)}\quad,
\label{gpiqq}
\end{eqnarray}
with $I_2(p)$ given in \ce{i2}.
\\

Alternatively, we can determine the interaction properties of the bound-states in the NJL model through the study of $q\bar q$-scattering amplitudes \cite{Nambu:1961tp, Nambu:1961tp-2}.
In the ladder approximation the bound-states are expected to be real stable particles. So should be the poles corresponding to the virtual exchange of, in the
pseudoscalar channel example, pions. The bound-state is therefore the iteration of the bubble diagrams $\Pi_{ps}(k^2)$ in the pseudoscalar channel \ce{properpol}.
\\

 The scattering matrix \cf{rpafig} generated by the exchange of a pion is then
\begin{eqnarray}
iU_{ij}(k^2)&=&(i\gamma_5\tau^i)\,\Big[2iG+2iG(-i\Pi_{ps}(k^2))2iG+\ldots\Big](i\gamma_5\tau^j)\quad,\nonumber\\
&=&(i\gamma_5\tau^i)\,\frac{2iG}{1-2G\Pi_{ps}(k^2)}\,(i\gamma_5\tau^j)\quad.
\label{ladderint}
\end{eqnarray}
The $\tau^i$ selects the appropriate channel. For exchanging a $\pi^+$, $\tau^i=\tau^{\pi^-}$ and $\tau^j=\tau^{\pi^+}$, vice versa for a $\pi^-$ and
 $\tau^i=\tau^j=\tau^{\pi^0}$ for a $\pi^0$.
This scattering matrix can be compared to the one coming from the minimal local interaction Lagrangian that describes the coupling of a pion field to quark fields.
Namely, the effective pion exchange is  
 \begin{eqnarray}
iU_{ij}(k^2)&=&(i\gamma_5\tau^i)\,\frac{-ig_{\pi qq}^2}{k^2-m_{\pi}^2}\,(i\gamma_5\tau^j)\quad.
\label{effect}
\end{eqnarray}
\begin{figure}
\centering
 \includegraphics [height=1.6cm]{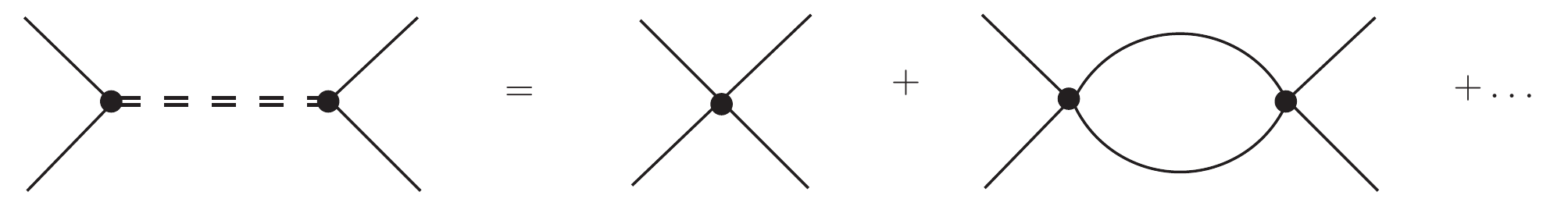}
\caption[Scattering matrix]{\small{Scattering matrix generated by the bubble diagram \cf{properpol}.}}
\label{rpafig}
\end{figure}
Hence, in the light of \ce{effect}, we can immediately determines the mass of the pion by finding  the pole of \ce{ladderint} at $k^2=m_{\pi}^2$
.  In other words, the self-consistency
equation \ce{selfconsist} of the Bethe-Salpeter equation comes to be the pole of \ce{ladderint}
\begin{eqnarray}
1-2G\,\Pi_{ps}(k^2=m_{\pi}^2)=0\quad.
\label{polo2}
\end{eqnarray}
In the same fashion, we can relate the quark-pion coupling constant  to the residue at the pole \ce{polo2}
\begin{eqnarray}
g_{\pi qq}^2&=& \left(\frac{\partial \Pi_{ps}(k^2)}{\partial k^2}\right)^{-1}\big|_{k^2=m_{\pi}^2}\quad.
\end{eqnarray}
With $\Pi_{ps}(k^2)$ given by \ce{polar}, the expression \ce{gpiqq} is recovered.
\begin{wrapfigure}[7]{l}{6cm}
\centering
 \includegraphics [height=1.8cm]{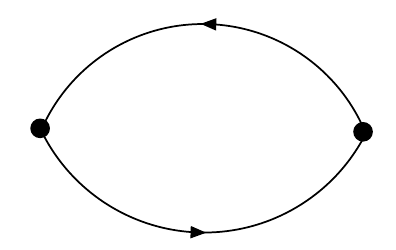}
\caption{}
\label{properpolscal}
\end{wrapfigure}

\vskip 15pt

The mass of the scalar mode $\sigma$ associated with the term $(\bar \psi \psi)^2$ can  be obtained in the same fashion. The zero of the function
 $1-2G\Pi_{s}(k^2)$,
with $\Pi_{s}(k^2)$ the scalar bubble diagram \cf{properpolscal}, 
defines its mass $m_{\sigma}$. Also its coupling strength can be related to the residue of the pole at $k^2=m_{\sigma}^2$.

\vskip 20pt

The scalar loop is given as
\begin{eqnarray}
-i\Pi_{s}(k^2)&=&-\int\,\frac{d^4p}{(2\pi)^4}\,\alltrace\Big\{
\,i\, \frac{{p\hspace{-0.5em}\raisebox{-0.2ex}{/}}+\frac{1}{2}{k\hspace{-0.5em}\raisebox{-0.2ex}{/}}+m}{(p+\frac{k}{2})^2-m^{ 2}}\,
\,\,i\, \frac{{p\hspace{-0.5em}\raisebox{-0.2ex}{/}}-\frac{1}{2}{k\hspace{-0.5em}\raisebox{-0.2ex}{/}}+m}{(p-\frac{k}{2})^2-m^{ 2}}
\Big\}\quad,\nonumber\\
&=& -4N_c N_f \int\,\frac{d^4p}{(2\pi)^4}\,\frac{m^{ 2}+p^2-\frac{k^2}{4}}{[(p+\frac{k}{2})^2-m^{ 2}][(p-\frac{k}{2})^2-m^{ 2}]}\quad.
\label{polarscal}
\end{eqnarray}
\\
Using the same trick as in the pseudoscalar case, i.e. writing \ce{polarscal} in terms of elementary integrals, we find
\begin{eqnarray}
\Pi_{s}(k^2)&=& 2 N_c N_f\,[2I_1-(k^2-4m^{ 2}) \,I_2(k^2)]\quad.
\end{eqnarray}
As before the one-propagator integral is extracted from the gap equation and we find
\begin{eqnarray}
1-2G\Pi_{s}(k^2)&=& \frac{m_0}{m}+4\,G\,N_c N_f\,(k^2-4m^{ 2})\,I_2( k^2)\quad,
\end{eqnarray}
which is exactly zero.
On the pole $k^2=m_{\sigma}^2$, we find
\begin{eqnarray}
m_{\sigma}^2&=&  - \frac{m_0}{m}\,\frac{1}{4\,G\,N_c N_f\,I_2( m_{\sigma}^2)}+4m^{ 2}\quad,\nonumber\\
&\simeq& 4m^{ 2}+m^2_{\pi}\quad.
\end{eqnarray}
This result linking the pion and the sigma masses is independent of the regularization scheme employed.
In the chiral limit, the mass of the $\sigma$-meson is non-zero and its value is
\begin{eqnarray}
m_{\sigma}^2&=& 4m^{ 2}\quad.
\end{eqnarray}

\vspace*{.6cm}

We have described a model for the pion. In the NJL model the pion arises through the dynamical breaking of the chiral symmetry. As a consequence its mass $m_{\pi}$
should be zero in the chiral limit.
\\

The NJL Lagrangian contains a point-like interaction that renders the theory nonrenormalizable. The model should always be completed by a regularization scheme
fulfilling the conditions imposed by the theory: the existence of a ``gap'' and, therefore, zero-modes or Goldstone bosons.
Another shortcoming is that the model is not confining.
\\
The Nambu - Jona-Lasinio model together with its regularization scheme is treated like an effective theory and should therefore be used in a characteristic scale 
imposed by the same regularization scheme.

\section{Partial Conservation of the  Axial Current}
\label{sec:pcac}

We showed  in Section~\ref{sec:ChiSB} that the true vacuum is not invariant under $SU_A(2)$. One might wonder what happens for the axial transformations. Familiar conservation laws  are derived from the conservation of the currents associated with  manifest symmetries of the, e.g.,  strong interactions. So  the vector $SU_V(2)\times U_V(1)$ transformations, which are manifest symmetries,  imply baryon number and isospin conservation.

The  $ SU_A(2)$ symmetry  being broken, we restore the chiral symmetry by  assuming that the axial currents
\beq
A_{\mu}^a(x)&=& \bar q(x)\,\gamma_{\mu}\gamma_5\,\frac{\tau^a}{2}\,q(x)\quad,
\eeq
are partially / approximately conserved. The symmetry is implemented in the Goldstone mode with the pion as massless particles. Besides the fact that it is spontaneously broken, the chiral symmetry is explicitly broken by the mass term in the  Lagrangian. That is, when the current quark mass are non zero, the pion which is effectively massless, acquires a mass, just as explained in Section~\ref{sec:pion}.

Consider that the axial current matrix element between a pion state and the vacuum can be parameterized by
\beq
\langle 0| A_{\mu}^j (x)|	\pi^k(p)\rangle &=& ip_{\mu} \delta^{jk} \, f_{\pi} \, e^{-i p\cdot x}\quad,
\label{axialme}
\eeq
with $f_{\pi}$ the pion decay constant. It is experimentally estimated to $f_{\pi}=93$ MeV~\cite{Amsler:2008zz} through the decay $\pi^-\to \mu^- \nu$ 
\beq
\Gamma_{\pi^-\to \mu^- \nu}&=& \frac{G^2\, m_{\mu}^2 f_{\pi}^2\, (m_{\pi}^2-m_{\mu}^2)^2}{4\pi \, m_{\pi}^3}\, \mbox{cos}^2\theta_C\quad,\nonumber
\eeq
with $\theta_C$ the Cabibbo angle and $G$ the Fermi constant.
\\

The conservation of the axial current implies that
\beq
\langle 0| \partial^{\mu}A_{\mu}^j (x)|	\pi^k(p)\rangle &=& m_{\pi}^2\, \delta^{jk} \, f_{\pi} \, e^{-i p\cdot x}=0\quad.
\nonumber
\eeq
It means that  $m_{\pi}^2\,  f_{\pi}=0$, so that,  $m_{\pi}^2$ is chosen to be zero since the Goldstone realization of the symmetry is in agreement with nearly massless pions.
\begin{figure}[H]
\centering
 \includegraphics [height=2.6cm]{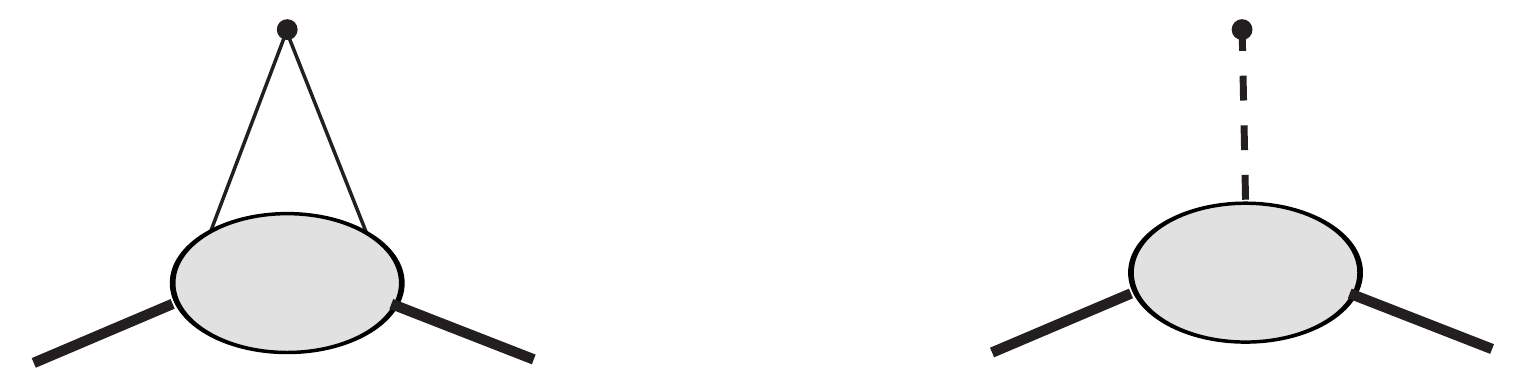}
\caption{Axial current in the nucleon. On the right panel, the pion pole contribution.}
\label{axialnn}
\end{figure}
However, when restoring the quark (current) mass term, the axial current is no longer exactly conserved. The pions acquire a mass as  given by Eq.~(\ref{mpi}).
Hence, it is worth finding an expression for the current conservation when  extrapolating from $p^2=0$ to $p^2=m_{\pi}^2$, with $m_{\pi}^2\neq 0$. This can be done by identifying the divergence of the current with a smooth interpolating pion field $\pi^j$,
\beq
\partial^{\mu}A_{\mu}^j (x)&=& m_{\pi}^2 \, f_{\pi}\,\pi^j(x)\quad.
\label{pcac}
\eeq

This behavior of this interpolating field can be guessed from the analysis of the matrix element of the axial current between nucleon states\footnote{See, for instance, classical notebooks Refs.~\cite{Peskin:1995ev,Itzykson:1980rh}.}, Fig.~\ref{axialnn}.
Current conservation in the exact chiral limit effectively implies the exchange of a massless particle, identified here, to be the pion. It therefore means that the matrix elements of the divergence of the axial current are  dominated by a pion pole (on the right panel of Fig.~\ref{axialnn}) for small momentum transfers $|q^2|<m_{\pi}^2$
\beq
\langle A (p+q)| \partial^{\mu}A_{\mu}^j |B(q)\rangle &=& \frac{\mbox{Res}}{q^2-m_{\pi}^2}+\ldots
\eeq

\begin{wrapfigure}[6]{r}{8cm}
\centering
 \includegraphics [height=3cm]{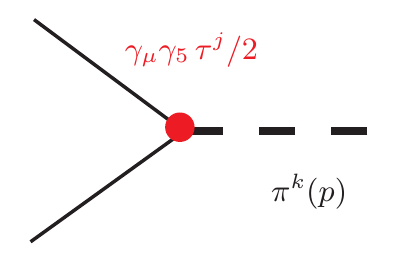}
\caption{}%
\label{fpi-fig}
\end{wrapfigure}

\vskip 15pt

By consistency with the model calculation, it is important to evaluate the pion decay  with the technology developed in the previous Sections of this Appendix.
In the NJL model,  the expression Eq.~(\ref{axialme}) becomes, see Fig.~\ref{fpi-fig}
\\
\\
\beq
 ip_{\mu} \delta^{jk} \, f_{\pi} \, e^{-i p\cdot x} &=&\langle 0| \sum_c    \bar \psi_c(x) \gamma_{\mu}\gamma_5\frac{\tau^j}{2}\psi_c(x)  |	\pi^k(p)\rangle\quad,\nonumber\\
 &=& -e^{-i p\cdot x} \int\,\frac{d^4k}{(2\pi)^4}\,\alltrace \left( iS_F(k)\, ig_{\pi qq} i\gamma_5\tau^j\, iS_F(-p+k)\gamma_{\mu}\gamma_5\frac{\tau^j}{2}\right)\quad,
 \nonumber\\
\eeq
which gives us the expression for $f_{\pi}$ in terms of 2-propagator integrals
\beq
f_{\pi}&=& -2N_cN_f\, g_{\pi qq}\, m\, I_2(m, p)\quad.
\label{piondecaycst}
\eeq

\chapter{Transition Form Factors}
\label{sec:trans-ff}

In the NJL model, the vector form factor is
\begin{equation}
F_{V}^{\pi^{+}}(t)=\frac{8\,N_{c}}{3}\,m\,m_{\pi}\,\frac{g_{\pi qq}}{\sqrt{2}%
}\,I_{3}(p,p^{\prime})\qquad. \label{fv}%
\end{equation}
In order to calculate the form factors we need the expression for the
three-propagator integral. In the particular case where $p^{2}=m_{\pi}^{2}$
and $p^{\prime}{}^{2}=0$, the expression for $I_{3}(p,p^{\prime})$ given by Eq.~(\ref{i3}).
\\

The axial form factor involves the three-propagator integral as well, but also
the two-propagator integral given by Eq.~\ref{i2}
\begin{equation}
F_{A}^{\pi^{+}}(t)=4N_{c}\,m\,m_{\pi}\,g_{\pi qq}\sqrt{2}\,\left(
I_{3}(p,p^{\prime})+\frac{2}{m_{\pi}^{2}-t}\left[  I_{2}(m_{\pi}^{2}%
)-I_{2}(t)\right]  \right)  \qquad, \label{faalltwists}%
\end{equation}
\\

The $\pi^{0}\rightarrow\gamma^{\ast}\gamma$ form factor can be obtained from
the vector form factor through a isospin rotation
\beq
F_{\pi\gamma^{\ast}\gamma}\left(  t\right)  &=&\frac{\sqrt{2}}{m_{\pi}}\, F_{V}^{\pi^{+}}\left(  t\right)  \quad.
\eeq
\\

 In the
chiral limit and for $t=0$ all the considered form factors have the simple
expression:%
\begin{align}
\left.  \frac{F_{\pi\gamma^{\ast}\gamma}\left(  0\right)  }{\sqrt{2}}\right\vert _{m_{\pi}\rightarrow0}  
&  =\left.  \frac{F_{V}^{\pi^{+}}\left(0\right)  }{m_{\pi}}\right\vert _{m_{\pi}\rightarrow0}
=\left.  \frac{F_{A}^{\pi^{+}}\left(  0\right)  }{m_{\pi}}\right\vert _{m_{\pi}\rightarrow0}
=\frac{1}{\sqrt{2}\,4\pi^{2}f_{\pi}}\left[  1-\frac{2m^{2}}{m^{2}+\Lambda^{2}%
}+\frac{m^{2}}{m^{2}+2\Lambda^{2}}\right] \nonumber\\
&  =0.192\times\left(  1-0.108\right)
\operatorname{GeV}%
^{-1}=0.171%
\operatorname{GeV}%
^{-1}%
\label{ff-t0}
\end{align}
The first coefficient of the right hand side is what is expected from the
axial anomaly contribution to $\pi\rightarrow\gamma^{\ast} \gamma$ decay. The term
between brackets has a small correction to the expected value of 1 due to the
finiteness of the regularization masses. In the NJL model not only the quarks,
but also the counter-terms run in the triangle diagram of the axial anomaly. In
a proper renormalizable theory this correction disappears in the limit
$\Lambda\rightarrow\infty.$
\chapter{Elementary Integrals}
\label{sec:integral}


Regularization of the following integrals is insured by the Pauli-Villars regularization scheme, i.e. see Section~\ref{sec:regunjl}.

\section{One, two and three-propagator Integrals}

The one-propagator integral $I_1$ is given by
\begin{eqnarray}
I_1&=& i \int \,\,\frac{d^4k}{(2\pi)^4}\,\,\frac{1}{k^2-m^2+i\epsilon}\qquad,\nonumber\\
&=& \frac{1}{(4\pi)^2}\,\,\sum_{j=0}^2 c_j\,m_j^2\,\,\mbox{ln}\frac{m^2_j}{m^2}\qquad.
\label{i1}
\end{eqnarray}
\\
\\
The two-propagator integral $I_2(p)$ is given by, in a general case,
\begin{eqnarray}
I_2(p^2)&=& i  \int \,\,\frac{d^4k}{(2\pi)^4}\,\,\frac{1}{(k^2-m^2+i\epsilon)((p+k)^2-m^2+i\epsilon)}\qquad,\nonumber\\
&=& \frac{1}{(4\pi)^2}\,\,\sum_{j=0}^2 c_j\,\Big\{\,\mbox{ln}\frac{m^2_j}{m^2}+ \sqrt{\frac{p^2-4m_j^2}{p^2}}\,\mbox{ln}
\frac{1-\sqrt{\frac{p^2}{p^2-4m_j^2}}}{1-\sqrt{\frac{p^2}{p^2-4m_j^2}}}\Big\}
\qquad.
\label{i2}
\end{eqnarray}
\\
In particular, for $ 0<p^2<4m^2$, \ce{i2} becomes
\begin{eqnarray}
I_2(p^2)&=&\frac{1}{(4\pi)^2}\,\,\sum_{j=0}^2 c_j\,\Big\{\,\mbox{ln}\frac{m^2_j}{m^2}+ 
2\sqrt{\frac{4m_j^2-p^2}{p^2}}\,\mbox{arctg}\frac{1}{\sqrt{\frac{4m_j^2-p^2}{p^2}}}\Big\}\qquad.
\end{eqnarray}
\\
\\
The three-propagator integral $I_3(p_1,p_2)$ is given by
\begin{eqnarray}
I_3(p_1,p_2)&=& i  \int \,\,\frac{d^4k}{(2\pi)^4}\,\,\frac{1}{(k^2-m^2+i\epsilon)((p_1+k)^2-m^2+i\epsilon)((p_2+k)^2-m^2+i\epsilon)}\qquad,\nonumber\\
&=& \frac{1}{(4\pi)^2}\,\,\sum_{j=0}^2 c_j\,\int_0^1\,dz\,\frac{z}{\sqrt{D}}\,\mbox{ln}\,\frac{E-\sqrt{D}}{E+\sqrt{D}}\qquad,
\label{i3}
\end{eqnarray}
with, in the particular case where $p_1^2=p_2^2=m^2_{\pi}$,
\begin{eqnarray}
D&=& t^2z^4-4z^2t(m_j^2+m^2_{\pi}z(z-1))\qquad;\nonumber\\
E&=& z^2t-2m_j^2-2m^2_{\pi}z(z-1)\qquad;
\end{eqnarray}
and with, in the particular case where $p_1^2=m^2_{\pi}$ and $p_2^2=0$, we obtain
\begin{eqnarray}
D&=& t^2z^4+z^2(1-z)^2m^4_{\pi}+2\,t\,z^2(m^2_{\pi}z(1-z)-2m_j^2)\qquad,\\
E&=& z^2t+z(1-z)m^2_{\pi}-2m_j^2\qquad.
\end{eqnarray}
\section{Light-front Integrals}

\subsection{Two-propagator  light-front integrals}

We can now calculate the three 2-propagator integrals, $\tilde{I}_{2,\Delta}(x, \xi,t)$, $\tilde{I}_{2,P'}$ and $\tilde{I}_{2,P}(x, \xi)$ defined as
\begin{eqnarray}
\tilde{I}_{2,\Delta}(x, \xi,t)&=&i\,p^+\,\,\int \,\,\frac{d^4k}{(2\pi)^4}\,\,
\frac{\delta\left((x-1)p^++k^+\right)}{((P-k)^2-m^2+i\epsilon)((P'-k)^2-m^2+i\epsilon)}\qquad,\nonumber\\
\nonumber\\
\tilde{I}_{2,P'}&=&i\,p^+\,\,\int \,\,\frac{d^4k}{(2\pi)^4}\,\,
\frac{\delta\left((x-1)p^++k^+\right)}{(k^2-m^2+i\epsilon)((P'-k)^2-m^2+i\epsilon)}\qquad,\nonumber\\
\nonumber\\
\tilde{I}_{2,P}(x, \xi)&=&i\,p^+\,\,\int \,\,\frac{d^4k}{(2\pi)^4}\,\,
\frac{\delta\left((x-1)p^++k^+\right)}{(k^2-m^2+i\epsilon)((P-k)^2-m^2+i\epsilon)}\qquad.
\label{i2p}
\end{eqnarray}
%
%
%
%
%
The first step of this integration 
consists in decomposing the momenta in their light-cone components. The delta function selects the $+$-momenta carried by the quarks so that
 the integral over $dk^+$ is straightforward. The integral over $dk^-$ is performed with the help of the Residue Theorem so that we are left with an integration over
$dk^{\perp}$. The latter is straightforward.


\subsubsection{GPDs' kinematics}

The result is
\begin{eqnarray}
\tilde{I}_{2,\Delta}(x, \xi,t)
&=& \frac{1}{(4\pi)^2}\frac{1}{2\xi}\,\sum_{j=0}^2\,c_j\,\mbox{ln}\frac{m_j^2+\frac{t}{4}\frac{x^2-\xi^2}{\xi^2}}{m^2}\,\,\,\, \theta(x+\xi)\theta(\xi-x)\qquad.
\label{i2delta}
\end{eqnarray}
\\
The same steps can be done with the $\tilde{I}_{2,P'}$ and $\tilde{I}_{2,P}(x, \xi)$ integrals:
\begin{eqnarray}
\tilde{I}_{2,P'}&=&\frac{1}{(4\pi)^2}\frac{1}{1-\xi}\,\sum_{j=0}^2\,c_j\,\mbox{ln}\frac{m_j^2+m_{\pi}^2\frac{(x-\xi)(x-1)}{(1-\xi)^2}}{m^2}
\,\,\,\,  \theta(x-\xi)\theta(1-x)\qquad.
\label{i2ppr}
\end{eqnarray}
\\
The $\tilde{I}_{2,P}(x, \xi)$ integral is defined on $-\xi<x<1$.
\begin{eqnarray}
\tilde{I}_{2,P}(x, \xi)
&=& \frac{1}{(4\pi)^2}\frac{1}{1+\xi}\,\sum_{j=0}^2\,c_j\,\mbox{ln}\frac{m_j^2+m_{\pi}^2\frac{(x+\xi)(x-1)}{(1+\xi)^2}}{m^2}\,\,\,\, \theta(x+\xi)\theta(1-x)\qquad.
\label{i2p}
\end{eqnarray}

\subsubsection{TDAs' kinematics}

In the TDA's kinematics, the  Eqs.~(\ref{i2delta}, \ref{i2p}) do not change but $\tilde{I}_{2,P'}$  becomes
\begin{eqnarray}
\tilde{I}_{2,P'}
&=& \frac{1}{(4\pi)^2}\frac{1}{1-\xi}\,\sum_{j=0}^2\,c_j\,\mbox{ln}\frac{m_j^2}{m^2}\,\,\,\,  \theta(x-\xi)\theta(1-x)\qquad.
\end{eqnarray}
\\

\subsection{Three-propagator  light-front integrals}
\label{sec:3-prop}

The three-propagator light-front integral is given by
\subsubsection{GPDs' kinematics}

\begin{eqnarray}
&&\hspace*{-1cm}\tilde I_{3}^{GPD} (x, \xi,t)\nonumber\\
&=& i\,p^+\,\int \,\,\frac{d^4k}{(2\pi)^4}\,\,\frac{\delta\left((x-1)p^++k^+\right)}{(k^2-m^2+i\epsilon)
((P-k)^2-m^2+i\epsilon)
((P'-k)^2-m^2+i\epsilon)}\qquad,\nonumber\\
\nonumber\\
&=&\frac{1}{2\,(4\pi)^2}\sum_{j=0}^2 c_j\,\nonumber\\
&&\Big\{
\theta(1-x)\theta(x-\xi)\frac{1}{\sqrt{D}}\,\mbox{ln}\,\frac{-m_{\pi}^2x(1-x)+m_j^2(1-\xi^2)-\frac{t}{2}(1-x)^2+(1-x)\sqrt{D}}
{-m_{\pi}^2x(1-x)+m_j^2(1-\xi^2)-\frac{t}{2}(1-x)^2-(1-x)\sqrt{D}}\nonumber\\
\nonumber\\
&&+\theta(\xi-x)\theta(x+\xi)\frac{1}{\sqrt{D}}\mbox{ln}\,\frac{-m_{\pi}^2\xi(\xi+x)+2m_j^2\xi(1+\xi)-\frac{t}{2}(1-x)(x+\xi)+(x+\xi)\sqrt{D}}
{-m_{\pi}^2\xi(\xi+x)+2m_j^2\xi(1+\xi)-\frac{t}{2}(1-x)(x+\xi)-(x+\xi)\sqrt{D}}
\Big\}\,,\nonumber\\
&&
\label{i3pgpd-a}
\end{eqnarray}
with 
\begin{eqnarray}
D&=& m^4_{\pi}\xi^2+\frac{t^2}{4}(1-x)^2-4m_j^2\xi^2 m_{\pi}^2+tm^2_{\pi}x(1-x)-m_j^2t(1-\xi^2)\qquad.
\end{eqnarray}
\\
\\
\subsubsection{TDAs' kinematics}

In the TDA's kinematics, the skewness variable is bounded by $t/(2m_{\pi}^2-t)<\xi<1$. We therefore have to consider both the cases with positive and negative values of $\xi$.
\\
\\
For positive values of $\xi$, the three-propagator light-front integral becomes
\begin{eqnarray}
&&\hspace*{-1cm}\tilde I_{3}^{TDA} (x, \xi,t)\nonumber\\
&=& i\,p^+\,\int \,\,\frac{d^4k}{(2\pi)^4}\,\,\frac{\delta\left((x-1)p^++k^+\right)}{(k^2-m^2+i\epsilon)
((P-k)^2-m^2+i\epsilon)
((P'-k)^2-m^2+i\epsilon)}\qquad,\nonumber\\\nonumber\\
\nonumber\\
&=&\frac{1}{(4\pi)^2}\sum_{j=0}^2 \,c_j\nonumber\\
&&\Big\{
\theta(x+|\xi|)\theta(|\xi|-x)\frac{1}{\sqrt{D}}\,\,\mbox{ln}\,\frac{m_{\pi}^2(x-\xi)+t(x-1)+4\xi\frac{1+\xi}{x+\xi}m_j^2+\sqrt{D}}
{m_{\pi}^2(x-\xi)+t(x-1)+4\xi\frac{1+\xi}{x+\xi}m_j^2-\sqrt{D}}\nonumber\\\nonumber\\
&&+\theta(x-|\xi|)\theta(1-x)\frac{1}{\sqrt{D}}\,\,\mbox{ln}\,\frac{m_{\pi}^2(x-\xi)-t(x-1)+2\frac{1-\xi^2}{x-1}m_j^2-\sqrt{D}}
{m_{\pi}^2(x-\xi)-t(x-1)+2\frac{1-\xi^2}{x-1}m_j^2+\sqrt{D}}
\Big\}\qquad,
\label{i3tda}
\end{eqnarray}
where we have introduced the Pauli-Villars regularization parameters and with
\begin{eqnarray}
D&=& (m_{\pi}^2\,(x-\xi)+t(1-x))^2\,+\,4\,m_j^2\,(2m_{\pi}^2\xi(1-\xi)-t(1-\xi^2))\qquad.
\end{eqnarray}
\\
\\
On the other hand, for negative values of the skewness variable, the three-propagator light-front integral becomes
\begin{eqnarray}
&&\hspace*{-1cm}\tilde I_{3}^{TDA} (x, \xi,t)\nonumber\\
&=& i\,p^+\,\int \,\,\frac{d^4k}{(2\pi)^4}\,\,\frac{\delta\left((x-1)p^++k^+\right)}{(k^2-m^2+i\epsilon)
((P-k)^2-m^2+i\epsilon)
((P'-k)^2-m^2+i\epsilon)}\qquad,\nonumber\\\nonumber\\
\nonumber\\
&=&\frac{1}{(4\pi)^2}\sum_{j=0}^2 \,c_j\nonumber\\
&&\Big\{
\theta(x+|\xi|)\theta(|\xi|-x)\frac{1}{\sqrt{D}}\,\,\mbox{ln}\,\frac{-m_{\pi}^2(x+\xi)+t(x-1)-4\xi\frac{1-\xi}{x-\xi}m_j^2+\sqrt{D}}
{-m_{\pi}^2(x+\xi)+t(x-1)-4\xi\frac{1-\xi}{x-\xi}m_j^2-\sqrt{D}}\nonumber\\\nonumber\\
&&+\theta(x-|\xi|)\theta(1-x)\frac{1}{\sqrt{D}}\,\,\mbox{ln}\,\frac{m_{\pi}^2(x-\xi)-t(x-1)+2\frac{1-\xi^2}{x-1}m_j^2-\sqrt{D}}
{m_{\pi}^2(x-\xi)-t(x-1)+2\frac{1-\xi^2}{x-1}m_j^2+\sqrt{D}}
\Big\}\qquad.
\label{i3tda-neg}
\end{eqnarray}

\chapter{Double Distributions and the $\alpha$-representation}
\label{sec:dd}
\label{sec:dd-tda-kin}
\pagestyle{fancy}

This Appendix is based on the work by Broniowski, Ruiz Arriola and Golec-Bernat~\cite{Broniowski:2007si}.
\\

In this Appendix we aim to give an explicit calculation of the Double Distribution of the form given by reduction formula~(\ref{reduc-form}). This is conveniently  done through the Feynman $\alpha$-representation of the scalar propagators.
\\
So the scalar propagator reads
\beq
S_k&=& \frac{1}{k^2-m^2+i\epsilon}=\int_0^{\infty} d\alpha e^{-\,\alpha (k^2-m^2+i\epsilon)}\quad.
\nonumber
\eeq

\subsubsection{The 3-propagator Integral}

We write the 3-propagator integral in the asymmetric notation for it is more appropriate for our purposes, omitting to write the $+i\epsilon$.
It reads~\cite{Broniowski:2007fs}
\beq
\tilde I_3(X, \zeta,t)&=&i\, \int\frac{d^4k}{(2\pi)^4}\, \frac{\delta(X-k\cdot n)}{\left(k^2-m^2\right)\left((k+\Delta)^2-m^2\right)\left((k-p_{\pi})^2-m^2\right)}\quad,
\nonumber
\eeq
with $\Delta=p_{\gamma}-p_{\pi}$. In terms of exponential representation of the propagator, the vector TDA becomes
\beq
\tilde I_3(X, \zeta,t)&=&i\,\int\frac{d^4k}{(2\pi)^4}\,\int\,\frac{d\lambda}{2\pi}\,e^{i\lambda (k\cdot n-X)}\nonumber\\
&&\int_0^{\infty} d\alpha\int_0^{\infty} d\beta \int_0^{\infty} d\gamma\,
e^{-\alpha\left(k^2-m^2\right)-\beta\left ((k+\Delta)^2-m^2\right)-\gamma\left((k-p_{\pi})^2-m^2\right)}\quad.
\nonumber
\eeq

We perform the following change of variables 
\beq
s=\alpha+\beta+\gamma\quad, \quad y=\frac{\beta}{s}\quad, \quad z=\frac{\gamma}{s}\quad,
\nonumber
\eeq
with $a, b, c>0$ so that we can constrain $y, z$ to have support between $[0,1]$ and also $y+z\leq 1$. The constrains on the limits for $y, z$ are automatically imposed by the $\alpha$-representation of the scalar propagator.

Concentrating on the argument of the exponential, we find
\beq
&&-(\alpha+\beta+\gamma)(k^2-m^2)-\beta \,\Delta^2-\gamma\,p_{\pi}^2-2 \beta\,\Delta \cdot k+2\gamma\, p_{\pi}\cdot k+i\lambda k\cdot n-i\lambda X\nonumber\\
&&= -s\left( (k^2-m^2)+y\, \Delta^2 +z p_{\pi}^2 +2 y \,\Delta\cdot k - 2z\, p_{\pi} \cdot k -i\lambda \frac{k\cdot n}{s}+i\lambda \frac{X}{s}\right)\quad.\nonumber\\
\label{expo}
\eeq
\\
The following change of variable is performed,
\beq
&&k=k'+z\,p_{\pi} -y\,\Delta+i\frac{\lambda n}{2s}\quad,\label{change-var-dd}\\
\nonumber\\
\Rightarrow&&
k^{' 2}= k^2+z^2\, p_{\pi}^2+y^2\, \Delta^2-2z \,p_{\pi} \cdot k +2y \,\Delta\cdot k -i\lambda \frac{k\cdot n}{s}- 2\,zy\, p_{\pi} \cdot \Delta\nonumber\\
&&\hspace{1.1cm}+i\lambda z\frac{p_{\pi}\cdot n}{s}-i\lambda y \frac{\Delta\cdot n}{s}\quad.
\nonumber
\eeq
\\
The argument of the exponential~(\ref{expo}) becomes
\beq
&&\hspace{-.9cm}-s\left( (k^{'2}-m^2)+ \Delta^2\, y(1-y)+p_{\pi}^2 \, z(1-z)+2zy\, p_{\pi}\cdot \Delta\right .\nonumber\\
&&\left. -i\lambda z\frac{p_{\pi}\cdot n}{s}+i\lambda y \frac{\Delta\cdot n}{s}+i\lambda \frac{X}{s}\right)\quad.
\eeq
\\

The integral over $dk$ is a Gaussian-integral when going to the Euclidean space.
Rotating the contour of the $k^0$ integration by performing a Wick rotation, i.e. $k^0\equiv i k_E^0$ while ${\bf k}={\bf k}_E $, we are left with
\beq
\tilde I_3(X, \zeta,t)&=& \frac{1}{(2\pi)^4}\,\int\frac{d\lambda}{2\pi}\, \int_0^{\infty}ds\, s^2\,\frac{\pi^2}{s^2}\,\int_0^1 dy \int_0^1 dz\,\theta \left(  1-y-z\right) \nonumber\\
&&e^{-s\left( -m^2+ \Delta^2\, y(1-y)+p_{\pi}^2 \, z(1-z)+2zy\, p_{\pi}\cdot \Delta
 -i\lambda z\frac{p_{\pi}\cdot n}{s}+i\lambda y \frac{\Delta\cdot n}{s}+i\lambda \frac{X}{s}\right)}\nonumber
\eeq
\\
\\
We can now perform the integral over $d\lambda$. The dependence upon $\lambda$ comes only from the exponential
\beq
e^{-i\lambda\left(  X+y\,\Delta\cdot n-z\,p_{\pi}\cdot n \right)}\quad.
\nonumber
\eeq
\\

The expression for the 3-propagator integral hence becomes
\beq
\tilde I_3(X, \zeta,t)&=&\frac{1}{(4\,\pi)^2} \,\int_0^{\infty} ds\int_0^1 dy \int_0^1 dz\theta \left(  1-y-z\right) \nonumber\\
&&\,\delta\left(X-z\, p_{\pi}\cdot n-y \Delta\cdot n\right)
e^{-s m^2} e^{- p_{\pi}^2 z(1-z) s} e^{-\Delta^2y(1-y) s } e^{-2 yz s \,p_{\pi}\cdot \Delta}\quad.
\nonumber
\eeq
\\
In turn, the integral over $ds$ gives %
\beq
\tilde I_3(X, \zeta,t)&=&\frac{1}{(4\,\pi)^2}\,\int_0^1 dy \int_0^1 dz\theta \left(  1-y-z\right)\, \nonumber\\
&&\frac{\delta\left(X-z\, p_{\pi}\cdot n-y \Delta\cdot n\right)}{m^2- p_{\pi}^2 \,z(1-z) - \Delta^2\,y(1-y)-2\, yz\,p_{\pi}\cdot \Delta  }\quad.
\label{i3-gen}
\eeq
The latter expression can be reduced to a Double Distribution following Eq.~(\ref{reduc-form})
\beq
\tilde I_3(X, \zeta,t)&=&\frac{1}{(4\,\pi)^2}\,\int_0^1 dy \int_0^1 dz\,\delta\left(X-z\, p_{\pi}\cdot n-y \Delta\cdot n\right)\, {\cal I}_{3}(z,y,  t)\quad;
\label{reduc}
\eeq
where  the Double Distribution ${\cal F}_{\pi}^q( z,y, t)$ has be determined to be
\beq
{\cal I}_{3}(z, y,  t)&=& \,\frac{\theta \left(  1-y-z\right)}{m^2- p_{\pi}^2 \,z(1-z) - \Delta^2\,y(1-y)-2\, yz\,p_{\pi}\cdot \Delta   }\quad.
\label{dd-3pf}
\eeq
\\

By definition, the DD is not constrained by the kinematics. When linking the 2-dimensional Fourier transform that defines the DD to the 1-dimensional Fourier transform defining the skewed parton distributions, one has to use the kinematics imposed by the process and to relate the two variables $p\cdot n$ and $\Delta \cdot n$. This is done through the $\delta$-function in Eq.~(\ref{reduc}). In other words, the function $\tilde I_3$ depends also on the skewness variable.
\\

 So far, we have hardly used the typical kinematics of the transition and the result is somehow general.
We can now
 apply the previous developments to the TDA's kinematics.
We will use the asymmetric notation introduced by Radyushkin. The kinematics are slightly changed with respect to the one described in Section~\ref{sec:kin-tda}. In Ref.~\cite{Broniowski:2007fs}, the following conventions are used
\beq
n^2=0\, ,\qquad p_{\pi}\cdot n=1\,,\qquad \Delta \cdot n=-\zeta\,,\nonumber\\
\nonumber\\
p_{\pi}^2=m_{\pi}^2\,, \qquad \Delta^2=t\,,\qquad 2\,p_{\pi}\cdot \Delta = m_{\pi}^2+t\quad,
\label{kin-asym-tda}
\eeq
with the {\it asymmetric} skewness variable a priori defined between $\zeta\in[0,1]$.
The couple of symmetric variables $(x, \xi)$ is related the the asymmetric ones $(X,\zeta)$ by
\beq
X=\frac{x+\xi}{1+\xi}\, ,\qquad \zeta=\frac{2\xi}{1+\xi}\quad.
\label{asym-not}
\eeq
\\
Also, in this asymmetric notation, one  should change the couple of variables 
\beq
(\beta, \alpha)&\longrightarrow&(z, (2y+z-1))\quad,
\nonumber
\eeq
 what is equivalent to changing $(x, \xi)$ for $(X,\zeta)$.
\\

The  expression~(\ref{i3-gen}) can be written in the form of a reduction formula for the TDA's kinematics
\beq
\tilde I_3(X, \zeta,t)&=&\frac{1}{(4\,\pi)^2}\,\int_0^1 dy \int_0^1 dz\, \delta\left(X-z-y \zeta\right)\, {\cal I}_{3}(z,y,t)\quad,
\eeq
with the double distribution in the same  kinematics 
\beq
{\cal I}_{3}(z,y,t)&=&\,\frac{\theta \left(  1-y-z\right)}{m^2- m_{\pi}^2 \,z(1-z-y) - t\,y(1-y-z) }\quad.
\label{dd-tda}
\eeq
\\

Performing the $z$-integration, its value is set at $z=X-y\zeta$. In the TDA's kinematics, the value of the skewness variable is not constrained to be positive, see~(\ref{xi-lim-tda}).  This leads to two support decompositions according to the sign of $\zeta$.  For positive $\zeta$ values
\beq
\tilde I_3(X, \zeta,t)&=&\frac{1}{(4\,\pi)^2}\, \left( \theta(X)\theta(\zeta-X)\, \int_0^{X/\zeta} + \theta(1-X)\theta(X-\zeta)\, \int_0^{(1-X)/(1-\zeta)}\right) \nonumber\\
&&\frac{dy}{m^2- m_{\pi}^2 \,(X-y\zeta)(1-(X-y\zeta)+y) - t\,y(1-y+(X-y\zeta))}\quad (\zeta>0)\quad;
\nonumber\\
\label{supp-xi-pos}
\eeq
leading to a support $x\in[0,\zeta]$ and $x\in[\zeta, 1]$.
On the other hand,  for negative values of $\zeta$, the support is decomposed in the following way
\beq
\tilde I_3(X, \zeta,t)&=&\frac{1}{(4\,\pi)^2}\, \left( \theta(X)\theta(\zeta-X)\, \int^{(1-X)/(1-\zeta)}_{X/\zeta} + \theta(X)\theta(1-X)\, \int_0^{(1-X)/(1-\zeta)}\right) \nonumber\\
&&\frac{dy}{m^2- m_{\pi}^2 \,(X-y\zeta)(1-(X-y\zeta)+y) - t\,y(1-y+(X-y\zeta))}\quad (\zeta<0)\quad,
\nonumber\\
\label{supp-xi-neg}
\eeq
which corresponds to a support in $x$ between $[\zeta,0]$ and $[0, 1]$.
\\

\subsubsection{The $k^{\nu}$-dependent 3-propagator Integral}

In the case of an integral proportional to the loop variable, i.e.
\beq
\tilde I^{\nu}_3(X, \zeta,t)&=&i\, \int\frac{d^4k}{(2\pi)^4}\, \frac{\delta(X-k\cdot n)\, k^{\nu}}{\left(k^2-m^2\right)\left((k+\Delta)^2-m^2\right)\left((k-p_{\pi})^2-m^2\right)}\quad,
\nonumber
\eeq
the shifting operation changes to $y, z$ dependence of the DD. 
\\
\\
In the particular case of the axial TDA,  the Dirac trace taken over the axial matrix element brings a factor of $-2\,k^{\nu}$ into the integral.
The change of variable (\ref{change-var-dd}) produces a shift of 
\beq
-2\, k^{\nu}&\longrightarrow& -\left(2k+2z\,p_{\pi} -2y\,\Delta\right)^{\nu}\quad.
\nonumber
\eeq
Using the fact that the photon polarization implies $\varepsilon\cdot p_{\gamma}=0$, we obtain $\varepsilon\cdot p_{\pi}=-\varepsilon\cdot \Delta$;
\beq
\varepsilon\cdot\tilde I_3 (X, \zeta,t)&=& i\, \int\frac{d^4k}{(2\pi)^4}\,\int_0^{\infty}ds\, s^2\,
\int_0^1 dy \int_0^1 dz\,\theta \left( 1-y-z\right)\, \nonumber\\
\nonumber\\
&&\varepsilon_{\nu} \left(2k^{\nu}-(2y+2z)\,\Delta^{\nu} \right)\nonumber\\
\nonumber\\
&&e^{-s\left( (k^2-m^2)+ \Delta^2\, y(1-y)+p_{\pi}^2 \, z(1-z)+2zy\, p_{\pi}\cdot \Delta
 -i\lambda z\frac{p_{\pi}\cdot n}{s}+i\lambda y \frac{\Delta\cdot n}{s}+i\lambda \frac{X}{s}\right)}\quad.\nonumber\\
\eeq
The Gaussian integration over $dk$ carries out a $\left(\sqrt{\pi/s}\right)^4$. The term proportional to $k$ cancels out. 
 \\
The following steps are as described in above for $\tilde I_3$, we obtain
\beq
\varepsilon\cdot\tilde I_3 (X, \zeta,t)&=& \frac{ \varepsilon\cdot\Delta}{2\pi^2}\int_0^1 dy \int_0^1 dz\, \delta\left(X-z-y \zeta\right)\, \left(2y+2z \right)\,D(z,y,t)\quad,
\eeq
with the double distribution $D(z,y,t)$ defined by Eq.~(\ref{dd-tda}).
\\
\\

\subsubsection{Polynomiality Property}

Turning our attention to the polynomiality condition~(\ref{oddeven}) resulting from Lorentz invariance, we find that it is an inherent property of the Double Distribution parameterization,
\beq
\int _{0}^1\, dX\, X^{n-1}\, \tilde I_3(X, \zeta, t)
&=&\frac{1}{2\,\pi^2}\,\int_0^1 dy \int_0^1 dz\, \theta\left(z+y \zeta\right)\, \theta\left(1-z-y \zeta\right)\left(\zeta\, y+z\right)^{n-1}\, {\cal I}_3(z,y,t)\quad,
\nonumber\\
&=&\sum_{i=0}^{n-1}\,c_{n,i}(t)\,\zeta^{i}\quad.
\nonumber
\eeq
The latter expression implies  the expression for the general form factors as follows
\beq
c_{n,i}(t)&\propto&\frac{1}{2\pi^2}\, \int_0^1dy\,\int_0^1 dz\, y^i\, z^{n-1-i}\,  \,{\cal I}_3(z,y,t)\quad.
\label{dd-gen-ff}
\eeq
Since  the kinematical range for the skewness variable is $-1\leq \zeta \leq 1$, the condition on the integration limits imposed by the $\theta$-function are verified.
\\

The expression~(\ref{dd-gen-ff}) corresponds to the Double Distribution form of the General Moments~(\ref{gen-mmt-ff}), as expected.

%
\clearpage
\addcontentsline{toc}{chapter}{Bibliography}

\bibliographystyle{plain}

\end{document}